\documentclass{jfm}
\usepackage{graphicx}
\usepackage{epstopdf, epsfig}
\usepackage{psfrag}
\usepackage{mathrsfs}
\usepackage{amsmath}
\usepackage{bm}
\usepackage{natbib}
\usepackage{color}
\usepackage{ulem}
\usepackage{pifont}
\usepackage{verbatim}
\usepackage{booktabs}
\usepackage{stfloats}
\usepackage{multirow}
\usepackage{rotfloat}
\usepackage{hyperref}
\usepackage{tikz}
\usetikzlibrary{decorations.pathreplacing}
\hypersetup{colorlinks=true,
            linkcolor=blue,
            anchorcolor=blue,
            citecolor=blue}

\ifCUPmtlplainloaded \else
  \checkfont{eurm10}
  \iffontfound
    \IfFileExists{upmath.sty}
      {\typeout{^^JFound AMS Euler Roman fonts on the system,
                   using the 'upmath' package.^^J}%
       \usepackage{upmath}}
      {\typeout{^^JFound AMS Euler Roman fonts on the system, but you
                   dont seem to have the}%
       \typeout{'upmath' package installed. JFM.cls can take advantage
                 of these fonts,^^Jif you use 'upmath' package.^^J}%
      }
  \else
  \fi
\fi

\ifCUPmtlplainloaded \else
  \checkfont{msam10}
  \iffontfound
    \IfFileExists{amssymb.sty}
      {\typeout{^^JFound AMS Symbol fonts on the system, using the
                'amssymb' package.^^J}%
       \usepackage{amssymb}%
         \let\leq=\leqslant
         \let\geq=\geqslant
      }{}
  \fi
\fi

\newcommand{\ri}{\mathop{\rm i}\nolimits}
\newcommand{\re}{\mathop{\rm e}\nolimits}

\title[Excitation of hypersonic non-modal perturbations]{Excitation of non-modal perturbations in hypersonic boundary layers by freestream forcing:  shock-fitting harmonic linearised Navier-Stokes approach}

\author[L. Zhao and M. Dong]%
{ Lei\ns ZHAO$^{1}$
and
Ming\ns DONG$^{2}$%
  \thanks{Email address for correspondence: dongming@imech.ac.cn},}

\affiliation{$^1$Department of Mechanics, Tianjin University, Tianjin, 300072, China\\
$^2$State Key Laboratory of Nonlinear Mechanics, Institute of Mechanics, Chinese Academy of Sciences, Beijing 100190, China
}

\begin{document}

\maketitle

\begin{abstract}
In this paper, we study the receptivity of non-modal perturbations in hypersonic boundary layers over a blunt wedge subject to freestream vortical, entropy and acoustic perturbations. Due to the absence of the Mack-mode instability and the rather weak growth of the entropy-layer instability within the domain under consideration, the non-modal perturbation is considered as the dominant  factor triggering laminar-turbulent transition. This is a highly intricate problem, given the complexities arising from the presence of the bow shock, the entropy layer, and their interactions with oncoming disturbances.
To tackle this challenge, we develop a  highly efficient numerical tool, the shock-fitting harmonic linearized Navier-Stokes (SF-HLNS) approach, which offers a comprehensive investigation on the dependence of the receptivity efficiency on the nose bluntness and properties of the freestream forcing. The numerical findings suggest that the non-modal perturbations are more susceptible to freestream acoustic and entropy perturbations compared to the vortical perturbations, with the optimal spanwise length scale being  comparable with the downstream boundary-layer thickness.  Notably, as the nose bluntness increases, the receptivity  to the acoustic and entropy perturbations intensifies, reflecting the transition reversal phenomenon observed experimentally in configurations with relatively large bluntness. In contrast, the receptivity to freestream vortical perturbations weakens with increasing bluntness.
Additionally, through the SF-HLNS calculations, we examine the credibility of the optimal growth theory (OGT) on describing the evolution of non-modal perturbations. While the OGT is able to predict the overall streaky structure in the downstream region, its accuracy in predicting the early-stage evolution and the energy amplification proves  to be unreliable. Given its high-efficiency and high-accuracy nature, the SF-HLNS approach shows great potential as a valuable tool for conducting future research on hypersonic blunt-body boundary-layer transition.

\end{abstract}

\begin{keywords}
receptivity, non-modal perturbation, hypersonic boundary layer
\end{keywords}

\section{Introduction}
Laminar-turbulent transition in hypersonic boundary layers stands as a critical concern in the design of high-speed flying vehicles, given its association with substantial rises in surface friction and heat flux. However, accurately predicting the onset of transition remains a substantial challenge due to its intricate  connection to various influencing factors and complex physical mechanisms. Among these factors, the environmental perturbations may emerge as the most influential one.

When environmental perturbations are relatively weak,  transition is typically triggered through a natural route \citep{Fedorov2011transition,Zhong2012direct}, for which the exponential amplification of normal instability modes, such as the Mack mode  in supersonic and hypersonic boundary layers, dominates the majority of the laminar phase. The initial amplitude of the normal mode is determined by a receptivity process \citep{Fedorov2001}.
As the normal modes evolve to a finite-amplitude state, nonlinearity takes over, triggering the rapid growth of additional perturbations and the distortion of the mean flow \citep{Mayer2011direct,sivasubramanian2015direct,song2024principle}. This  ultimately ensures the breakdown of the laminar phase.
In contrast, when the environmental perturbations are sufficiently strong, the bypass transition route would emerge. Due to the non-orthogonality of the Orr-Sommerfeld operator, a set of stable normal modes can display transient, algebraic growth, known as the non-modal perturbations \citep{schmid2007nonmodal}. These perturbations often manifest as longitudinal streaky structures in boundary-layer flows \citep{fransson2005transition} and, when influenced by strong environmental perturbations, can progress to the nonlinear phase before the transient growth saturates. Subsequently, the streaky base flow facilitates the rapid amplification of secondary instability modes \citep{andersson2001breakdown,zhang2018}, generating sufficient Reynolds stress to distort the mean flow towards the turbulent phase.

Transition in hypersonic boundary layers is recognized for its heightened complexity compared to that in low-speed boundary layers. First, the hypersonic Mack instability exhibits multiple branches, with the Mack second mode emerging as the most unstable one \citep{Mack1987}. 
Second, the role of the forebody shock wave is pivotal  to the hypersonic receptivity process. In physics, any type of oncoming perturbations, after interacting with the forebody shock, would stimulate all the three types of perturbation components, including the acoustic, entropy and vorticity waves, in the potential region sandwiched between the shock and the boundary layer.  These waves serve as the seeds for the receptivity process with ample mechanisms \citep{Fedorov2001,Qin2016,wan2018response,Liu2020,Dong2020,schuabb2024hypersonic}. 
Third, in scenarios where the flying vehicle's leading-edge is blunt, the forebody shock wave becomes detached. As the hypersonic flow encounters a nearly normal shock wave around the nose region, a noticeable increase in entropy occurs, leading to the formation of an entropy layer in the downstream potential region characterized by a substantial entropy gradient. This entropy layer gradually merges with the expanding boundary layer, with the merging point delaying as the nose radius increases. While the presence of the entropy layer  stabilises the growth of the normal Mack mode \citep{kara2007receptivity}, it may also give rise to a new entropy-layer instability within the inviscid entropy layer above the viscous boundary layer. Nonetheless, this entropy-layer instability alone is less likely to trigger transition in hypersonic blunt boundary layers \citep{Fedorov2004evolution}. The impact of the entropy layer on hypersonic blunt-body boundary-layer transition remains an unresolved question to date.

In the review of hypersonic transition over sharp and blunt cones \citep{schneider2004hypersonic}, it was highlighted that the sharp-cone transition is predominantly  driven by the amplification of the Mack second mode, and increase of the nose radius  results in a stabilising effect. However, a notable upstream shift of the transition onset  is observed as the bluntness increases beyond a certain threshold, as demonstrated in the wind-tunnel experiments of \citet{stetson1967shock,stetson1983nosetip}. Such a transition reversal phenomenon was further supported  in blunt-plate experiments by \citet{lysenko1990influence}, \citet{kosinov1990experiments} and \citet{Borovoy2021laminar}. Recently, detailed measurements conducted by \citet{grossir2014hypersonic} and \citet{kennedy2022} revealed inclined disturbances extending beyond the boundary-layer edge, which are distinguished from the usual rope-like structures associated with the  Mack second mode. 
To elucidate  the transition reversal phenomenon, extensive numerical studies have emerged. Through direct numerical simulations (DNSs) on Mach 6 hypersonic boundary-layer transition over cones with different nose radii, \citet{Kara2011effects} particularly studied the receptivity of Mack second mode to freestream acoustic perturbations. The simulations confirmed the stabilising effect of the nose bluntness and the role of the entropy layer in the delay of boundary-layer transition. Subsequent DNS studies \citep{cerminara2017acoustic,wan2018response,Wan2020} showed detailed perturbation field in the second-mode receptivity process. However, these Mack-mode receptivity studies are inadequate to show the transition reversal phenomenon. \citet{paredes2018blunt,paredes2019nonmodal} proposed that the transient growth of  non-modal perturbations might be linked to this transition reversal. Particularly, \cite{paredes2020mechanism} conducted an analysis on the linear optimal non-modal perturbation in hypersonic boundary layers over cones with moderate to large bluntness. It was   revealed that these linear optimal perturbations peak in the entropy layer, showing a rather weak signature in the boundary layers. Interestingly, a pair of finite-amplitude oblique non-modal perturbations could generate stationary streaks that penetrate and intensify within the boundary layer, ultimately leading to  downstream transition. However, the origin of such optimal perturbations is out of the scope of the optimal growth theory, and to establish the link between the entropy-layer perturbation and the freestream forcing is of particular interest.

Given the challenges associated with acquiring freestream perturbation data in wind tunnels, \citet{hader2018towards} employed a random forcing technique featuring a broadband spectrum of frequencies and wavenumbers as inflow perturbations for DNS, and simulated the fundamental breakdown  of a hypersonic boundary layer over a flare cone at Mach 6. In a subsequent study, \citet{goparaju2021effects} adopted a similar methodology to explore transition reversal on blunt plates with varying nose radii. Their findings revealed that at lower bluntness,  the dominant boundary-layer perturbations exhibited characteristics aligned with the Mack second mode, while beyond a critical nose radius, the dominant perturbations, peaking within the entropy layer, exhibited frequencies distinct from the Mack frequency band. Recognizing that the noise levels and spectral distribution in random forcing may not reflect the real situation, \citet{balakumar2018transition} extracted perturbation spectra from wind tunnel experiments to model the external forcing. Their DNS investigation on the receptivity of blunt-cone boundary layers with different nose radii unveiled weaker receptivity efficiency for configurations with larger bluntness compared to sharp-nose-tip designs. \citet{duan2019characterization} particularly discussed the construction of a tunnel-like perturbation field based on experimental and numerical data for further receptivity studies in specific wind tunnel conditions. In subsequent works by \citet{liu2022interaction} and \citet{schuabb2024hypersonic}, a tunnel-like acoustic field replicating both frequency-wavenumber spectra and temporal evolution of broadband tunnel noise emitted from tunnel walls was introduced as external forcing for DNS of a hypersonic blunt-cone boundary layer. The calculated perturbations in the entropy layer resembles the predictions of the non-modal perturbations in \citet{paredes2020mechanism}. More recently, \cite{guo2024transition} performed DNS study to show the entire transition process over hypersonic blunt plates with various nose radii. The numerical observations clearly demonstrated the transition reversal phenomenon and confirmed the presence of the non-modal perturbations for configurations with large bluntness.

However, the story does not end here.  DNS investigations concentrating solely on particular oncoming  conditions may not provide adequate predictive capabilities for transition in diverse scenarios, particularly within flight conditions. In the latter cases, the significance of freestream vorticity or entropy perturbations may be more relevant. Given the low efficiency of the DNS methodology, simplified approaches are emerging. Recognizing the low-level nature of freestream perturbations, the linear assumption is often employed in receptivity studies. Due to the rapid distortion of the mean flow in the nose region, traditional linear stability theory (LST) or parabolized stability equation (PSE) are insufficient to describe such receptivity problems. Therefore, a methodology that preserves the ellipticity of the original Navier-Stokes (N-S) system is required. Neglecting the nonlinear terms in the N-S equations, and performing Fourier transform with respect to time such that the time derivative is replaced by $-\ri\omega$, we arrive at the harmonic linearised Navier-Stokes (HLNS) equations, where $\ri=\sqrt{-1}$ and $\omega$ denotes the perturbation frequency. The first implementation of this methodology was by \citet{guo1997solution}, who analyzed the excitation and evolution of TS and cross-flow modes in smooth-surface boundary-layer flows. More recently, the HLNS approach has been utilized to illustrate the evolution of boundary-layer instability within a rapidly distorted mean flow induced by surface irregularities such as roughness, steps and gaps \citep{franco2018effect,Zhao2019,franco2020influence}, demonstrating its superiority over PSE in accuracy and over DNS in efficiency. The HLNS calculations were also used to verify the asymptotic theories on the receptivity and scattering of Mack modes by roughness and thermal spots \citep{Dong2021,Zhao2022,zhao2023asymptotic}. In \citet{paredes2019nonmodal}, the HLNS approach was also used to study the linear evolution of modal and non-modal perturbations over blunt-cone boundary layers, which were recently extended to include the nonlinear effect to calculate the nonlinear non-modal perturbations. Considering the weakly nonlinear interactions, \citet{song2024influence} also extended this methodology to study the impact of wall vibration to the evolution of the non-modal perturbations.

While the HLNS approach shows promise as a valuable tool for conducting systematic parametric studies on hypersonic receptivity,  it faces a notable challenge when the bow shock ahead of the nose tip emerges. This obstacle primarily arises from the fact that the numerical schemes commonly employed in the HLNS approach lack stability in handling  shock-perturbation interactions. Drawing inspiration from the shock-fitting approach \citep{zhong1998high}, this paper develops a shock-fitting HLNS (SF-HLNS) method. By positioning the upper boundary of the computational domain at the bow shock, the SF-HLNS method ensures the stability of the numerical scheme across the entire computational domain. From a physical perspective, there is a notable gap in our understanding of how non-modal perturbations are excited in hypersonic boundary layers with moderate bluntness subject   to various freestream perturbations, and so our primary emphasis will be directed towards addressing this specific issue.

The rest of this paper is structured as follows. In $\S$\ref{sec:methodology}, we present the physical model and the numerical methodology to be employed. The SF-HLNS approach will be particularly illustrated in $\S$\ref{sec:SF-HLNS}. The numerical results will be introduced in $\S$\ref{sec:results}, including the base flow in $\S$\ref{sec:base_flow}, the entropy-layer mode in $\S$\ref{sec:modal_analysis}, the excitation  of non-modal perturbations for varying controlling parameters  in $\S$\ref{sec:SF-HLNS_calculations}, the comparison of the receptivity efficiency across different freestream forcing in $\S$\ref{sec:summary}, and the comparison with the prediction by the optimal growth theory in $\S$\ref{sec:compare_optimal}. Finally, the conclusion and discussion are present in $\S$\ref{sec:conclusion}.

\section{Physical problem and numerical methodology}
\label{sec:methodology}
\begin{figure}
    \begin{center}
    \includegraphics[width = \textwidth] {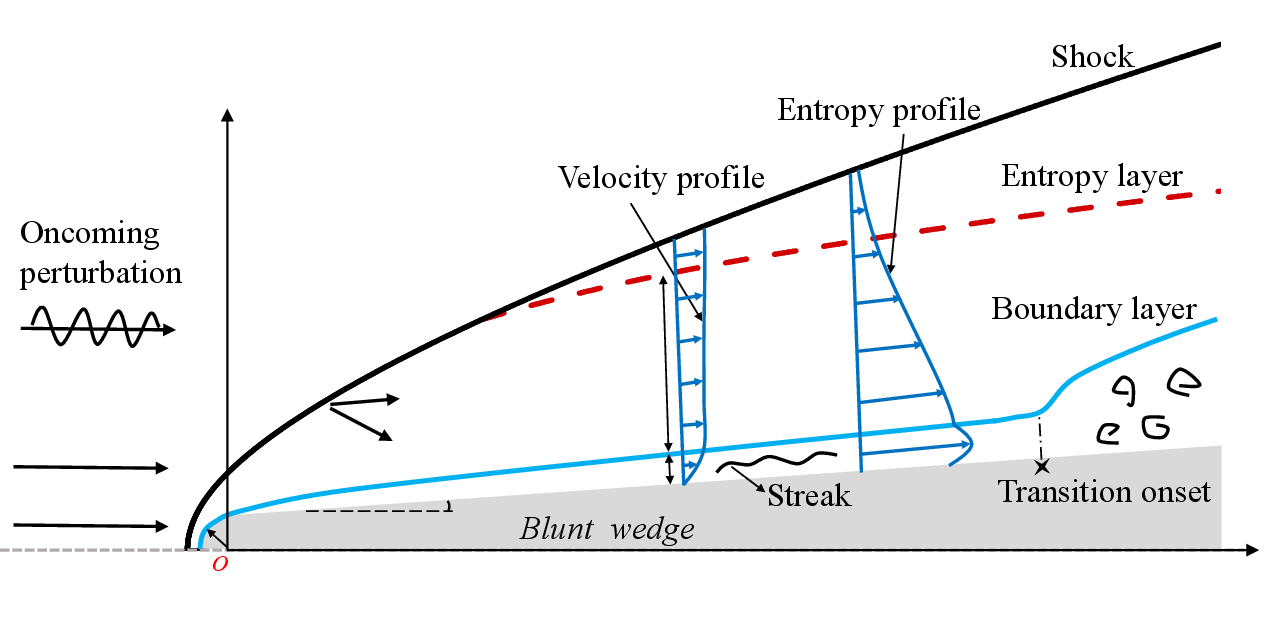}
    \put(-325,152){{\color{red}$y^*$}}
    \put(-360,50){$U^*_{\infty}$}
    \put(-10,25){{\color{red}$x^*$}}
    \put(-245,33){$\theta$}
    \put(-199,70){$\delta_{EL}$}
    \put(-199,42){$\delta_{BL}$}
    \put(-315,23){{\color{black}$r^*$}}
    \put(-260,65){$v$,~$e$}
    \put(-260,53){$a$}
    \caption{Sketch of the physical model, where ‘$a$', '$v$' and '$e$' denote acoustic, vortical and entropy disturbances, respectively.}
    \label{fig:model_sketch}
    \end{center}
\end{figure}

The physical model to be considered is a blunt wedge with a semi-tip-angle $\theta$ and a radius $r^*$ embedded in a hypersonic stream with zero angle of attack, as sketched in Figure \ref{fig:model_sketch}. A detached bow shock wave forms from the leading edge region, and a boundary layer and an entropy layer form in the region sandwiched between the shock and the wall. In the free stream, any small perturbations can be decomposed into a superposition of   acoustic, vortical and entropy disturbances. Any  of the three components introduced in the oncoming stream can interact with the bow shock and transmit to all the three components behind the shock. These perturbations can further penetrate the entropy and boundary layers, generating either the modal or non-modal perturbations.

\begin{figure}
    \begin{center}
    \includegraphics[width = 0.48\textwidth] {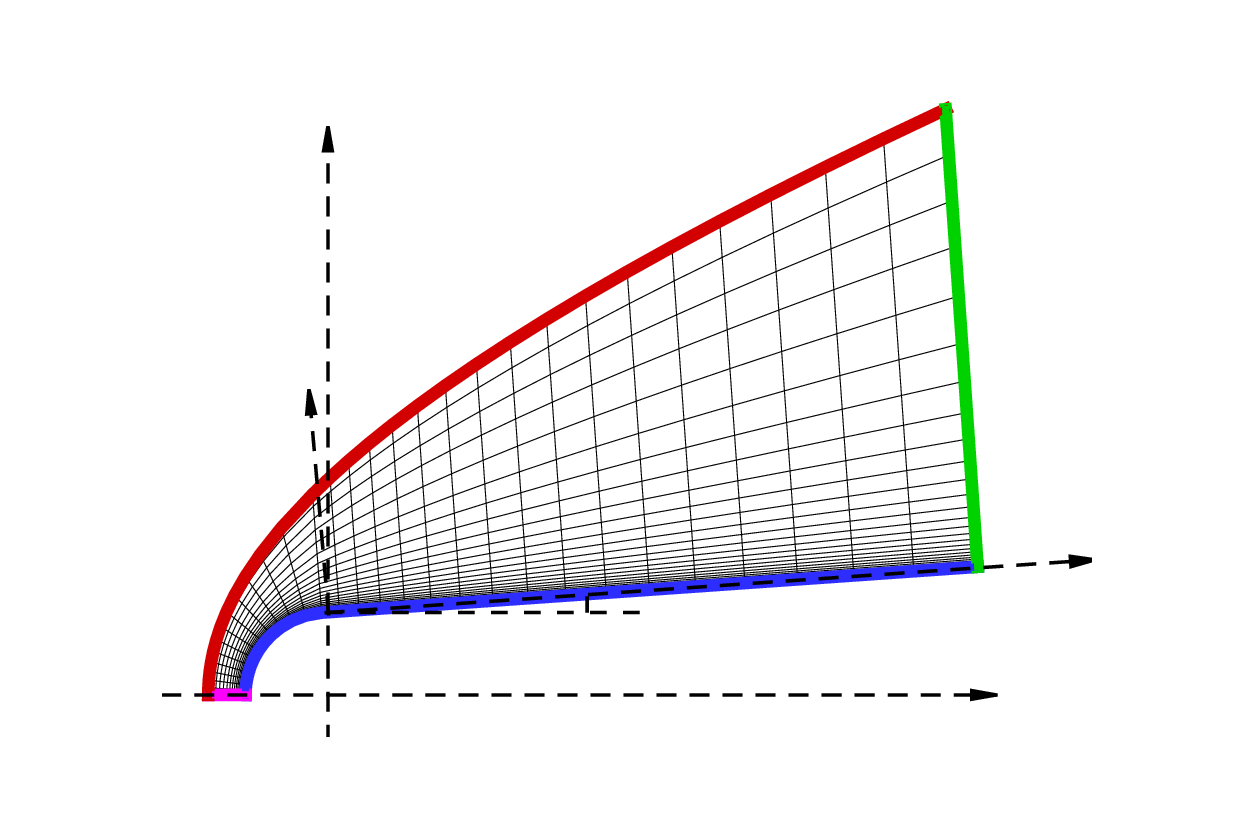}
    \put(-185,100){$(a)$}
    \put(-143,100){$y$}
    \put(-40,13){$x$}
    \put(-144,12){$O$}
    \put(-143,24){$o'$}
    \put(-87,29){$\theta$}
    \put(-151,60){$y_n$}
    \put(-30,33){$x_s$}
    \includegraphics[width = 0.48\textwidth] {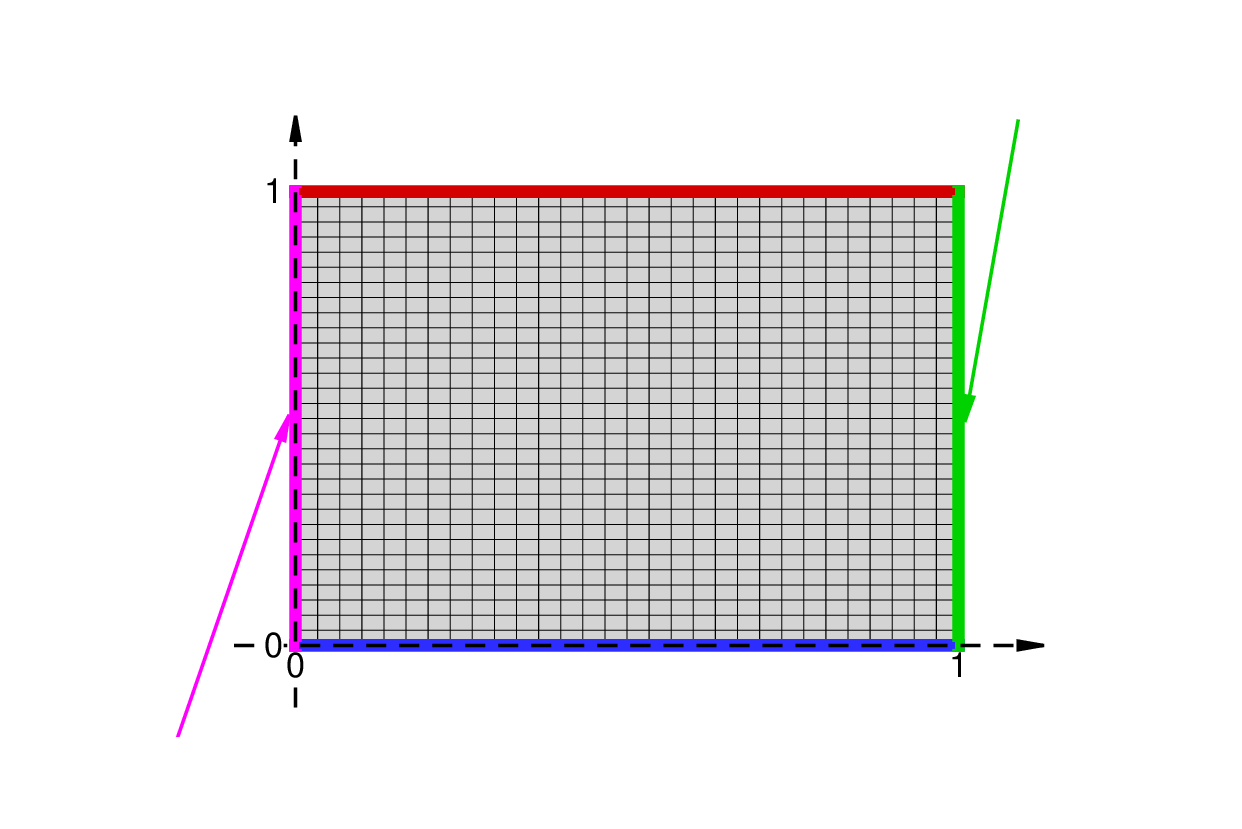}
    \put(-185,100){$(b)$}
    \put(-148,100){$\eta$}
    \put(-34,17){$\xi$}
    \put(-170,6){{\color{magenta}$\xi=0$: B.C.(2.17)}}
    \put(-110,110){{\color{green}$\xi=1$: extrapolated B.C.}}
    \put(-125,18){{\color{blue}$\eta=0$: B.C.(2.16)}}
    \put(-125,98){{\color{red}$\eta=1$: B.C.(2.21)}}
    \caption{Demonstration of the mesh systems $(x,y)$ ($a$) and $(\xi,\eta)$ ($b$) at a fixed $z$ position. In ($a$), the body-fitted system ($x_s$-$o'$-$y_n$) is also shown for convenience of results presentation.
    The boundary conditions (B.C.) for the base-flow calculations are shown in ($b$).}
    \label{fig:mesh_sketch}
    \end{center}
\end{figure}
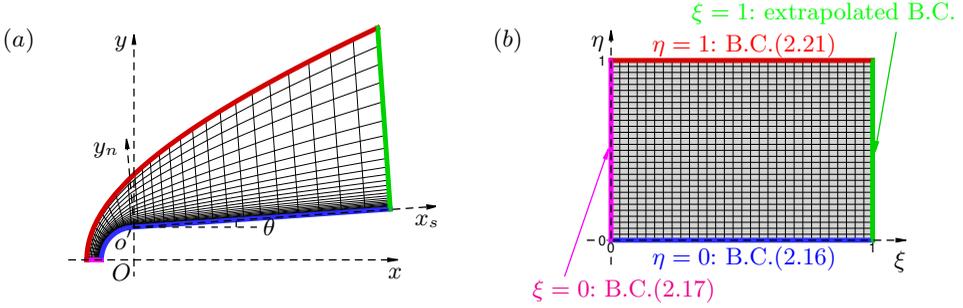
To describe the problem, two coordinate systems are employed, namely, the Cartesian coordinate system $(x^*, y^*, z^*)$ and the computational coordinate system $(\xi,\eta,\zeta)$ with $\xi$ and $\eta$ ranging from 0 to 1, as demonstrated in figure \ref{fig:mesh_sketch}. Here,
the $\zeta$ axis aligns with the $z^*$ axis, and $\eta$ is perpendicular to the wall. The lines corresponding to $\eta=0$ and 1 are located at the body surface and the bow shock, respectively. 
The origin of the Cartesian coordinate system $o$ is located at the center of the blunt nose, while that of the computational coordinate system is located at the nose tip of the blunt wedge. The relation of the two coordinate systems will be introduced in (\ref{eq:xyz_definition}) in $\S$\ref{sec:SFDNS}. Throughout this paper, the superscript $*$ denotes the dimensional quantities. Using the nose radius as the reference length, the Cartesian coordinate is normalised as $(x,y,z)= (x^*,y^*,z^*) /r^*$. The shape of the blunt wedge $(x_w,y_w)=(x(\xi,0),y(\xi,0))$ is represented by 
\begin{equation}
y_w=\left\{
  \begin{aligned}
  &\sqrt{1-x_w^2}, \quad &x_w&<-\sin\theta,\\
  &\tan\theta x_w+1/\cos\theta,\quad
  &x_w&\geq-\mathrm{sin}\theta.
  \end{aligned}
  \right.
  \label{eq:shape_model}
\end{equation}
 In figure \ref{fig:mesh_sketch}($a$), the body-fitted system $(x_s,y_n)$ with the origin at $o'$ is also presented, which will be used in $\S$\ref{sec:results} for convenience of presentation. 

The density $\rho$, velocity field $\pmb u = (u, v, w)$, temperature $T$, pressure $p$, dynamic viscosity
$\mu$ are normalised by the freestream quantities $\rho^*_{\infty}$, $U^*_{\infty}$, $T^*_{\infty}$, $\rho^*_{\infty}U^{*2}_{\infty}$ and $\mu^*_{\infty}$, respectively, where the velocity vector is defined in the Cartesian coordinate. The   time is normalised as  $t = t^*U^*_{\infty}/ r^*$. The Reynolds number and the Mach number are defined as
\refstepcounter{equation}
$$
    Re=\rho^*_{\infty}U^*_{\infty}r^*/\mu^*_{\infty},\quad
    M=U^*_{\infty}/c^*_{\infty},\eqno{(\theequation{\mathit{a,b}})}
    \label{eq:Re_Ma}
$$
where $c^*_{\infty}$ denotes the acoustic speed in the free stream.

\subsection{Governing equations}
The dimensionless conservative compressible Navier-Stokes (N-S) equations in the Cartesian coordinate system $(x, y, z)$ are expressed as
\begin{equation}\label{eq:NS_Cartesian}
    \frac{\partial \pmb{U}}{\partial t} + \frac{\partial \pmb{E}^c}{\partial x}
    +\frac{\partial \pmb{F}^c}{\partial y} + \frac{\partial \pmb{G}^c}{\partial z}
    =\frac{\partial \pmb{E}^v}{\partial x}
    +\frac{\partial \pmb{F}^v}{\partial y} + \frac{\partial \pmb{G}^v}{\partial z},
\end{equation}
where the conservative variables $\pmb{U}$, convective fluxes $\pmb{E}^c,\,\pmb{F}^c,\,\pmb{G}^c$ and viscous terms $\pmb{E}^v,\,\pmb{F}^v,\,\pmb{G}^v$
read 

\begin{equation*}
    \begin{split}
    &\pmb{U}=(\rho, \rho u, \rho v, \rho w, \rho e_t)^{\mathrm{T}},\quad
    \pmb{E}^c=(\rho u, \rho u^2+p, \rho uv, \rho uw, \rho uh_t)^{\mathrm{T}},\\
    &\pmb{F}^c=(\rho v, \rho uv, \rho v^2+p, \rho vw, \rho vh_t)^{\mathrm{T}},\quad
    \pmb{G}^c=(\rho w, \rho uw, \rho vw, \rho w^2+p, \rho wh_t)^{\mathrm{T}},\\
    &\pmb{E}^v=\frac{1}{Re}(0,\tau_{11},\tau_{12},\tau_{13}, u\tau_{11} +v\tau_{12} +w\tau_{13} -q_x)^{\mathrm{T}},\quad q_x=-\kappa\frac{\partial T}{\partial x},\\
    &\pmb{F}^v=\frac{1}{Re}(0,\tau_{21},\tau_{22},\tau_{23}, u\tau_{21} +v\tau_{22} +w\tau_{23} -q_y)^{\mathrm{T}},\quad
    q_y=-\kappa\frac{\partial T}{\partial y},\\
    &\pmb{G}^v=\frac{1}{Re}(0,\tau_{31},\tau_{32},\tau_{33}, u\tau_{31} +v\tau_{32} +w\tau_{33} -q_z)^{\mathrm{T}},\quad
    q_z=-\kappa\frac{\partial T}{\partial z},\\
    &e_t=c_v T+e_k,\quad h_t=c_p T+e_k, \quad e_k=(u^2+v^2+w^2)/2,\quad \kappa=\mu c_p/Pr,
    \end{split}
\end{equation*}
\begin{equation*}
    \begin{split}
    \tau_{11}=&\mu(2\frac{\partial u}{\partial x}-\frac{2}{3}\nabla \cdot \pmb{u}),\quad
    \tau_{22}=\mu(2\frac{\partial v}{\partial y}-\frac{2}{3}\nabla \cdot \pmb{u}),\quad
    \tau_{12}=\tau_{21}=\mu(\frac{\partial u}{\partial y}+ \frac{\partial v}{\partial x}),
    \\
    \tau_{33}=&\mu(2\frac{\partial w}{\partial z}-\frac{2}{3}\nabla \cdot \pmb{u}),\quad
    \tau_{13}=\tau_{31}=\mu(\frac{\partial u}{\partial z}+ \frac{\partial w}{\partial x}),\,
    \tau_{23}=\tau_{32}=\mu(\frac{\partial v}{\partial z}+ \frac{\partial w}{\partial y}),
    \end{split}
\end{equation*}
with $\gamma= 1.4$ being the ratio of the specific heats, and $c_v=\frac{1}{\gamma(\gamma-1)M^2}$ and $c_p=\frac{1}{(\gamma-1)M^2}$ being the dimensionless constant-volume and constant-pressure specific heats, respectively. The dimensionless equation of state reads $p=\rho T/(\gamma M^2)$. The dimensionless dynamic viscosity is calculated by the Sutherland formula $\mu=T^{3/2}(1+T_s)/(T+T_s)$ with $T_s=110.4\mathrm{K}/T^*_{\infty}$, and the dimensionless heat-conductivity coefficient $\kappa$ is correlated to $\mu$ through the Prandtl number $Pr= 0.72$.

When an incident acoustic, vortical, or entropy perturbation propagates through a bow shock wave, it undergoes rapid deformation, leading to the excitation of all three types of perturbations in the downstream region. Simultaneously, the shock wave itself becomes oscillatory. To address this physical process, two numerical approaches have emerged.
(1) The shock-capturing approach involves calculating the interaction between the shock and perturbations using a shock-capturing numerical scheme, such as the weighted essentially non-oscillatory (WENO) scheme \citep{jiang1996efficient}, where the computational domain includes the shock region. This approach has been utilized in numerous previous studies \citep{balakumar2018transition,cerminara2017acoustic,wan2018response}.
(2) The shock-fitting approach positions the upper boundary of the computational domain at the shock wave, with the {physical quantities involving the mean flow and the transmitted perturbations behind the shock wave obtained from the Rankine-Hugonio (R-H) relation {being} set as the boundary condition. This approach has been applied in \cite{zhong1998high}. While the derivation of the upper boundary condition in the shock-fitting approach is complex, it offers the advantage of avoiding the need for a shock-capturing scheme. This feature is particularly favorable for the application of HLNS. Therefore, this paper aims to develop an efficient SF-HLNS approach to address the receptivity of hypersonic boundary layers over a blunt body.

\subsection{Shock-fitting direct numerical simulation (SF-DNS) method}
\label{sec:SFDNS}

Following \cite{zhong1998high}, we present a brief overview of the shock-fitting method in this subsection, {which is also the pivotal  foundation in estiblashing the SF-HLNS approach}. For the base-flow calculation, we select a computational domain from the wall to the shock position, as depicted in figure \ref{fig:mesh_sketch}($a$).
\subsubsection{Coordinates movement and the governing equations}
In the framework of shock-fitting approach, to account for the movement of the shock wave, the    mesh system becomes unsteady, and the moving coordinates are functions of $(\xi, \eta, \zeta,\tau)$. 
In the present problem, the coordinate transformation is expressed as
\begin{equation}\label{eq:xyz_definition}
    \left\{
    \begin{aligned}
        x(\xi,\eta,\zeta,\tau)&=x_w(\xi)+g(\xi,\eta)H(\xi,\zeta,\tau)e_x(\xi),\\
        y(\xi,\eta,\zeta,\tau)&=y_w(\xi)+g(\xi,\eta)H(\xi,\zeta,\tau)e_y(\xi),\\
        z(\zeta)&{=\zeta},
    \end{aligned}
    \right.
\end{equation}
where $x_w$ and $y_w$ signify the body surface with their relations  specified in (\ref{eq:shape_model}), $\pmb e=(e_x,e_y,0)$ is the direction vector along the wall-normal grid lines, the function  $g(\xi,\eta)\in [0,1]$ presents the distribution of mesh along the $\eta$ direction, and $H(\xi,\zeta,\tau)$ represents the distance between the shock wave and the body surface along the $\eta$ axis. {In the present paper, the grid lines for a fixed $\xi$ are set to be perpendicular to the wall, but the formulations are applicable for more general configurations.} Here, $(x_w,y_w)$, $\pmb e$ and $g(\xi,\eta)$ are independent of the flow field, but $H$ is coupled with the flow field due to the shock movement. Applying the chain rule to (\ref{eq:xyz_definition}), we obtain
$$
    x_\xi=x_w'+(g_\xi H+gH_\xi)e_x+gHe_x',\quad x_\eta=g_\eta H e_x,\quad x_\zeta=gH_\zeta e_x,\quad x_\tau=gH_\tau e_x,
$$
$$
y_\xi=y_w'+(g_\xi H+g H_\xi)e_y+gHe_y',\quad y_\eta=g_\eta He_y,\quad y_\zeta=gH_\zeta e_y,\quad y_\tau=gH_\tau e_y,
$$
where a prime denotes the derivative with respect to its argument. Introducing the Jacobian determinant of the coordinate transformation $J=1/(x_{\xi}y_{\eta} -x_{\eta}y_{\xi})$, we obtain 
\begin{equation}
\begin{split}
    & 
    (\xi_x,\xi_y)=J(y_{\eta},-x_\eta),\quad
    (\eta_x,\eta_y)=J(- y_{\xi},x_\xi),\\& \eta_z=J(x_{\zeta} y_{\xi}-x_{\xi}y_{\zeta}),\quad
    \eta_t=-x_{\tau}\eta_x- y_{\tau}\eta_y.
\end{split}
\end{equation}
Apparently, these quantities relies on both {the shock position $H$ and} the shock speed $H_{\tau}\equiv\partial H/\partial \tau$.

Under the coordinate transformation (\ref{eq:xyz_definition}), the governing equation (\ref{eq:NS_Cartesian}) in the computational coordinates system reads
\begin{equation}\label{eq:NS_Computational}
    \frac{1}{J}\frac{\partial \pmb U}{\partial \tau} +\pmb U\frac{\partial (1/J)}{\partial \tau}
    +\frac{\partial \pmb{\hat E}^c}{\partial \xi}
    +\frac{\partial \pmb{\hat F}^c}{\partial \eta}
    +\frac{\partial \pmb{\hat G}^c}{\partial \zeta}
    =\frac{\partial \pmb{\hat E}^v}{\partial \xi}
    +\frac{\partial \pmb{\hat F}^v}{\partial \eta}
    +\frac{\partial \pmb{\hat G}^v}{\partial \zeta},
\end{equation}
where $\frac{\partial (1/J)}{\partial \tau}$ is also dependent on {both $H$ and} $H_{\tau}$, and
\begin{equation}
    \begin{aligned}
        &\pmb{\hat E}^c=\frac{\xi_x\pmb E^c +\xi_y\pmb F^c}{J},\quad&
        &\pmb{\hat F}^c=\frac{\eta_x\pmb E^c +\eta_y\pmb F^c +\eta_z\pmb G^c +\eta_t \pmb U}{J},\quad&
        &\pmb{\hat G}^c=\frac{\pmb G^c}{J},&
        \\
        &\pmb{\hat E}^v=\frac{\xi_x\pmb E^v +\xi_y\pmb F^v}{J},\quad&
        &\pmb{\hat F}^v=\frac{\eta_x\pmb E^v +\eta_y\pmb F^v +\eta_z\pmb G^v}{J},\quad&
        &\pmb{\hat G}^v=\frac{\pmb G_v}{J}.&
    \end{aligned}
    \label{eq:EFG}
\end{equation}

The governing equation (\ref{eq:NS_Computational}) alone is insufficient to close the differential equation system, and an additional equation that describes the motion of the shock is necessary. In $\S$\ref{eq:supplementary_equation}, we will derive this equation by examining the acceleration of the shock wave, and derive the supplementary equation with the form
$ \frac{\partial^2 H}{\partial \tau^2}=a_s(\xi,\eta,\zeta,\tau)$.

\subsubsection{Derivation of the supplementary equation}\label{eq:supplementary_equation}
Considering the conservative nature of the fluids around the moving shock wave, $\eta=1$, we obtain
\begin{equation}\label{eq:RH_con}
   \Big( \pmb{\hat F}^c_s-\pmb{\hat F}^c_0\Big)_{\eta=1}=0,
\end{equation}
{where the subscripts '0' and 's' represent the quantities immediately before and behind the shock, respectively.}
Substituting the second relation of (\ref{eq:EFG}) into (\ref{eq:RH_con}), we obtain
\begin{equation}\label{eq:RH_con2}
   \Big( (\pmb{\vec F}^c_s-\pmb{\vec F}^c_0)\cdot \pmb l + (\pmb U_s-\pmb U_0)b\Big)_{\eta=1}=0,
\end{equation}
where
$\pmb{\vec F}^c=(\pmb E^c,\pmb F^c,\pmb G^c)$, $\pmb l=({\eta_x},{\eta_y},{\eta_z})/J$ and $b={\eta_t}/{J}$. 
Differentiating (\ref{eq:RH_con2}) with respective to $\tau$ leads to
\begin{equation}\label{eq:RH_tau}
    \Big(\pmb{\hat B}_s\cdot\frac {\partial \pmb U_s}{\partial \tau}-\pmb{\hat B}_0\cdot\frac {\partial \pmb U_0}{\partial \tau} +(\pmb{\vec F}^c_s-\pmb{\vec F}^c_0)\cdot \frac{\partial \pmb l}{\partial \tau} + (\pmb U_s-\pmb U_0)\frac{\partial b}{\partial \tau}\Big)_{\eta=1}=0,
\end{equation}
where 
\refstepcounter{equation}
$$
\pmb{\hat B}=\frac {\partial \pmb{\vec F}^c}{\partial \pmb U}\cdot \pmb l +b=
    \frac {\partial \pmb{\hat F}^c}{\partial \pmb U},\quad \frac{\partial b}{\partial \tau}=-\frac{(\eta_x e_x + \eta_y e_y)}JH_{\tau\tau} - (\pmb e \cdot \frac{\partial \pmb l}{\partial \tau})H_{\tau}.
\eqno{(\theequation{\mathit{a,b}})}
\label{eq:b_tau}$$

The flux Jacobian matrix $\pmb{\hat B}_s$ has the eigenvalues
\begin{equation}
    \label{eq:eigen_Bs}\lambda_{1,2,3}=(\pmb l \cdot \pmb u +b)_s,\quad \lambda_4=(\pmb l \cdot \pmb u +b-c)_s,\quad \lambda_5=(\pmb l \cdot \pmb u +b+c)_s,
\end{equation}
where $c=\sqrt{T}/M$ is the local acoustic speed. Denote their corresponding left eigenvectors by
$\pmb I_1,\pmb I_2,\pmb I_3,\pmb I_4,\pmb I_5$, such that $\pmb I_i \cdot \pmb{\hat B}_s=\lambda_i \pmb I_i$ with $i=1,\cdots,5$. Note that only $\lambda_5$ corresponds to a characteristic line outgoing from the inner flow region to the bow shock.
Its corresponding eigenvector is expressed as
\begin{equation}
    \pmb I_5=\left[
    \begin{aligned}
    -c(u\eta_x+v\eta_y+w\eta_z)+(\gamma-1)e_k|\nabla\eta|\\
    c\eta_x-(\gamma-1)u|\nabla\eta|\\
    c\eta_y-(\gamma-1)v|\nabla\eta|\\
    c\eta_z-(\gamma-1)w|\nabla\eta|\\
    (\gamma-1)|\nabla\eta|
    \end{aligned}
    \right]_s,
\end{equation}
with $|\nabla\eta|=\sqrt{\eta_x^2+\eta_y^2+\eta_z^2}$.
Multiplying  (\ref{eq:RH_tau}) by $\pmb I_5$ leads to {the compatibility relation}
\begin{equation}\label{eq:CR}
    \lambda_5 \pmb I_5 \cdot \frac {\partial \pmb U_s}{\partial \tau}+\pmb I_5\cdot[-\pmb{\hat B}_0\cdot\frac {\partial \pmb U_0}{\partial \tau} +(\pmb{\vec F}^c_s-\pmb{\vec F}^c_0)\cdot \frac{\partial \pmb l}{\partial \tau} + (\pmb U_s-\pmb U_0)\frac{\partial b}{\partial \tau}]=0.
\end{equation}
Substituting (\ref{eq:b_tau}b) into  (\ref{eq:CR}), we can express the acceleration of the shock explicitly,
\begin{equation}\label{eq:a_s}
    \begin{aligned}
    a_s\equiv H_{\tau\tau}=\frac{ \pmb I_5 \cdot \left [
     \lambda_{5}\frac{\partial \pmb U_s}{\partial \tau}
    -\pmb{\hat B}_0\cdot\frac{\partial \pmb U_0}{\partial \tau}
    +(\pmb{\vec F}^c_s-\pmb{\vec F}^c_0)\cdot \frac{\partial \pmb l}{\partial \tau}\right]}
    {\pmb I_5 \cdot (\pmb U_s-\pmb U_0)(\eta_x e_x+\eta_y e_y)/J}
    -\frac{H_{\tau}(\pmb e \cdot \frac{\partial \pmb l}{\partial \tau})}
    {(\eta_x e_x+\eta_y e_y)/J}.
    \end{aligned}
\end{equation}
This is the supplementary equation for the shock-fitting system (\ref{eq:NS_Computational}).

\subsubsection{Boundary conditions}

A summary of the boundary conditions can be found in figure \ref{fig:mesh_sketch}($b$).
The no-slip, non-penetration and isothermal boundary conditions are employed at the wall,
\refstepcounter{equation}
$$
    u=v=w=0,\quad T=T_w\quad \mbox{at } \eta=0,
    \eqno{(\theequation{\mathit{a,b}})}
    \label{eq:BC_wall_meanflow}
$$
where $T_w$ represents the dimensionless wall temperature. For the calculation of the base flow, the symmetry boundary conditions are employed at the axis before the stagnation point, 
\begin{equation}
    \frac{\partial \rho}{\partial y}=\frac{\partial u}{\partial y}=\frac{\partial w}{\partial y} =\frac{\partial T}{\partial y}=v=0\quad \mbox{at } \xi=0.
    \label{eq:BC_symmetry_meanflow}
\end{equation}
In the numerical process, these symmetry boundary conditions are realised by utilizing a few ghost points for $\xi<0$. These boundary conditions can also be employed for computing the perturbations symmetric about the centreline. However, for asymmetric perturbations, it is necessary to extend the computational domain to the lower-half plane, which ensures that the variables along the centerline are treated as part of the internal flow field.
The extrapolated interpolation boundary condition is employed at the  outlet of the computational domain $\xi=1$.

At the upper boundary where $\eta=1$, the boundary condition is derived by relating the freestream quantities and the shock wave motion via  the R-H  relation. Denote $\pmb n_s$ as the unit normal vector of the shock wave, which is expressed as
\begin{equation}\label{eq:vec_ns}
    \pmb n_s=\frac{(\eta_x,\eta_y,\eta_z)}{|\nabla\eta|}\Big|_{\eta=1}.
\end{equation}
Since the value of $\eta$ is invariant at the shock, we have
\begin{equation}
\mathrm{d}\eta\Big|_{\eta=1}=\Big(\eta_x\mathrm{d}x+\eta_y\mathrm{d}y+\eta_z\mathrm{d}z+\eta_t\mathrm{d}t\Big)_{\eta=1}=0.
\end{equation}
Then, the projection of shock velocity along its external normal direction, $v_n$, can be  derived as
\begin{equation}\label{eq:v_n}
    {v_{n}=}(\frac{\mathrm{d}x}{\mathrm{d}t},\frac{\mathrm{d}y}{\mathrm{d}t},\frac{\mathrm{d}z}{\mathrm{d}t})\Big|_{\eta=1}
\cdot \pmb n_s=-\frac{\eta_t}{|\nabla\eta|}\Big|_{\eta=1}.
\end{equation}

The R-H relation at the shock wave is expressed as \citep{zhong1998high}
\refstepcounter{equation}\label{eq:RH_DNS}
$$
    p_s= p_0[1+\frac{2\gamma}{\gamma+1}(M_{n0}^2-1)],\quad
    \rho_s= \rho_0 \frac{(\gamma+1)M_{n0}^2}{(\gamma-1)M_{n0}^2+2},\eqno{(\theequation{\mathit{a,b}})}
    \label{eq:RH_relation}
$$
$$
    u_{ns}= v_n+\frac{\rho_0}{\rho_s}(u_{n0}-v_n),\quad
 \pmb u_s=\pmb u_{ts}+u_{ns}\pmb n_s=\pmb u_0+(u_{ns}-u_{n0})\pmb n_s,\eqno{(\theequation{\mathit{c,d}})}
$$
where $M_{n0}=M(u_{n0}-v_n)/\sqrt{T_0}$ is the normal component of the freestream Mach number  relative to the  shock wave, and $\pmb u_t$ and $u_n{=\pmb u \cdot \pmb n_s}$ are the tangential and normal components of the vector  $\pmb u$ along the shock wave plane, respectively. Note that $\pmb u_{ts}=\pmb u_{t0}=\pmb u_0-u_{n0}\pmb n_s$ has been utilized in (\ref{eq:RH_relation}d). Since the oncoming quantities are given, the equations (\ref{eq:RH_relation}) indicate five relations for the upper boundary conditions. 

\subsubsection{Discretization}
The governing equation (\ref{eq:NS_Computational}) and (\ref{eq:a_s}), accompanied by the boundary conditions (\ref{eq:BC_wall_meanflow}), (\ref{eq:BC_symmetry_meanflow}) and (\ref{eq:RH_DNS}), form a closed-form temporal-evolving differential equation system, which can be solved using the third-order Runge-Kutta method.

Following \cite{zhong1998high}, the nonlinear flux terms in (\ref{eq:NS_Computational}), $\frac{\partial \pmb{\hat E}^c}{\partial \xi}$ and $\frac{\partial \pmb{\hat F}^c}{\partial \eta}$, are split into positive and negative components by using a Lax-Friedrichs scheme, and the  fifth-order partial differential scheme  is employed. The viscous terms $\frac{\partial \pmb{\hat E}^v}{\partial \xi}$ and $\frac{\partial \pmb{\hat F}^v}{\partial \eta}$ are   discretized using the sixth-order central differential scheme.

For the calculations of the  evolution of three-dimensional perturbations, we can employ the Pseudo-spectral method to discretize the derivatives of physical quantities with respect to $\zeta$. This method guarantees high accuracy while requiring a minimal number of grid points in the spanwise direction.

\subsection{Freestream disturbances}
\label{sec:freestream_perturbation}
The freestream disturbance in the uniform stream $\pmb u_\infty=(1,0,0)$ can be expressed as
\begin{equation}\label{eq:gust}
    \pmb\varphi'_{\infty}(x,y,z,t)=\frac{\varepsilon_\infty}{2}\hat{\pmb\varphi}_{\infty}e^{\ri(k_1x+k_2y+k_3z-\omega t)}+c.c.,
\end{equation}
where $\pmb k=(k_1,k_2,k_3)$ and $\omega$ are the wave-number vector and the frequency of the disturbance, respectively, $\varepsilon_\infty$ measures the disturbance amplitude, and $c.c.$ denotes the complex conjugate. Denote the declination angle by \begin{equation}
    \vartheta=\tan^{-1}(k_2/k_3).
\end{equation} The acoustic, entropy, and vortical perturbations exhibit distinct dispersion relations, which are outlined as follows.

(i) An acoustic wave could propagate either faster or slower than the base flow, which are referred to as the fast and slow acoustic waves, respectively. The dispersion relations of the two waves can be expressed 
\begin{equation}
    {\omega} =  k_1 \pm \frac{1}{M}{|\pmb k|},
    \label{eq:acoustic}
\end{equation}
where the plus and minus signs distinguish the fast and slow acoustic waves, respectively.  The eigenfunction, normalized by the amplitude of the pressure fluctuation, reads
\begin{equation}
    (\hat \rho,\hat u,\hat v,\hat w,\hat T,\hat p)_{\infty}=
    (M^2,\pm \frac{k_1}{|\pmb k|}  M,\pm \frac{k_2}{|\pmb k|}  M,\pm \frac{k_3}{|\pmb k|} M,(\gamma-1)M^2,1).
    \label{eq:acoustic_fun}
\end{equation}

(ii)
For an entropy wave, we have the dispersion relation 
\begin{equation}
\omega=k_1,\label{eq:dispersion_entropy}
\end{equation}
and the eigenfunction is 
\refstepcounter{equation}
$$
       (\hat \rho,\hat u,\hat v,\hat w,\hat T,\hat p)_{\infty}=(1,0,0,0,-1,0)^T .
    \eqno{(\theequation)}
    \label{eq:gust_entropy}
$$

(iii)
For a vortical wave, the dispersion relation is the same as (\ref{eq:dispersion_entropy}), but the engenfunction is changed to
\refstepcounter{equation}
$$
       (\hat \rho,\hat u,\hat v,\hat w,\hat T,\hat p)_{\infty}=(0,\hat u_\infty,\hat v_\infty,\hat w_\infty,0,0)^T,
    \eqno{(\theequation)}
    \label{eq:gust_vortices}
$$
where $k_1\hat u_{\infty}+k_2\hat v_{\infty}+k_3\hat w_{\infty}=0$. For normalization, we let $\sqrt{\hat u_{\infty}^2+\hat v_{\infty}^2+\hat w_{\infty}^2}=1$, but we still need another condition to uniquely determine the values of $\hat u_\infty$, $\hat v_\infty$ and $\hat w_\infty$. Introduce the vertical vorticity $\hat\Omega_2\equiv \ri k_3\hat u_\infty-\ri\ k_1\hat w_\infty$, then the oncoming perturbation field can be expressed as
\refstepcounter{equation}
$$
    (\hat u_{\infty},\hat v_{\infty},\hat w_{\infty})=\frac{(-k_1k_2+k_3\tilde A,k_1^2+k_3^2,-k_2k_3-k_1\tilde A)}{\tilde B} \mbox{sgn}(k_2).
    \eqno{(\theequation)}
    \label{eq:gust_vortex_uvw}
$$
where $\tilde A=-\ri\hat \Omega_2|\pmb k|\mbox{sgn}(k_2)/\sqrt{k_1^2+k_3^2+\hat\Omega_2^2}$, $\tilde B= (k_1^2+k_3^2)|\pmb k|/\sqrt{k_1^2+k_3^2+\hat\Omega_2^2}$, and $\mbox{sgn}(k_2)=1$ for $k_2\geq 0$ and $=-1$ for $k_2<0$.

\subsection{Shock-Fitting harmonic linearised Navier-Stokes equation (SF-HLNS) approach}
\label{sec:SF-HLNS}
To characterize the development of an infinitesimal disturbance with a given frequency $\omega$, the harmonic approach can be utilized, wherein the time derivative $\partial_t$ is substituted with $-\ri\omega$ in the Fourier space. This technique, known as the harmonic linearised Navier-Stokes approach \citep{Zhao2019}, allows for the analysis of the evolution of linear  perturbations. In the case of the shock-fitting system, a similar methodology can be applied, but the movement of the shock must be considered as an  additional component of the solution. Therefore, the new approach is referred to as the shock-fitting harmonic {linearised} Navier-Stokes (SF-HLNS) approach.

\subsubsection{Linearised governing equations and compatibility relation}
To characterize the perturbation evolution, the instantaneous flow field $\pmb\varphi=(\rho,u,v,w,T)^{\mathrm{T}}$ and the shock position $H$ are
 decomposed into a two-dimensional (2-D) mean flow field $(\bar{\pmb\varphi},\bar H)$ and a three-dimensional (3-D) perturbation field $(\pmb\varphi', H')$,
\begin{equation}\label{eq:varphi}
    \pmb\varphi(\xi,\eta,\zeta,\tau)= \bar{\pmb\varphi}(\xi,\eta)+\varepsilon\pmb\varphi'(\xi,\eta,\zeta,\tau),
\end{equation}
\begin{equation}\label{eq:H}
    H(\xi,\zeta,\tau)=\bar H(\xi)+\varepsilon H'(\xi,\zeta,\tau),
\end{equation}
where $\varepsilon$ measures the perturbation amplitude.
Substituting  (\ref{eq:varphi}) and (\ref{eq:H}) into  (\ref{eq:NS_Computational}) and collecting the $O(\varepsilon)$ terms, we arrive at a set of linear differential equations,
\begin{equation}\label{eq:LNS_compact}
    \begin{aligned}
   (
    \frac{\pmb{\Gamma}}{J} \frac{\partial}{\partial \tau} + \pmb A \frac{\partial}{\partial \xi} +
    \pmb B \frac{\partial}{\partial \eta} +\pmb C \frac{\partial}{\partial \zeta} + \pmb D +
    \pmb V_{\xi\xi} \frac{\partial^2}{\partial \xi^2} + \pmb V_{\eta\eta} \frac{\partial^2}{\partial \eta^2} +
    \pmb V_{\zeta\zeta} \frac{\partial^2}{\partial \zeta^2} +
    \pmb V_{\xi\eta} \frac{\partial^2}{\partial \xi \partial \eta}
    \\
    +\pmb V_{\xi\zeta} \frac{\partial^2}{\partial \xi \partial \zeta} + \pmb V_{\eta\zeta} \frac{\partial^2}{\partial \eta \partial \zeta}
    ) \pmb \varphi' +
    (\pmb \Gamma^H \frac{\partial}{\partial \tau} + \pmb A^H \frac{\partial}{\partial \xi} +
    \pmb C^H \frac{\partial}{\partial \zeta} + \pmb D^H)H' =0,
    \end{aligned}
\end{equation}
where $\pmb \Gamma$, $\pmb A$, $\pmb B$, $\pmb C$, $\pmb D$, $\pmb V_{\xi\xi}$, $\pmb V_{\eta\eta}$, $\pmb V_{\zeta\zeta}$, $\pmb V_{\xi\eta}$, $\pmb V_{\xi\zeta}$ and $\pmb V_{\eta\zeta}$ are  $5\times 5$ dimensional matrices, and $\pmb \Gamma^H$, $\pmb A^H$, $\pmb C^H$ and $\pmb D^H$ are 5 dimensional {column} vectors. Their nonzero elements are introduced in Appendix \ref{Appendix:matrix}.

Now we express the perturbation field at a given frequency $\omega$ and a given spanwise wavenumber $k_3$   as
\refstepcounter{equation}
$$
\pmb\varphi'(\xi,\eta,\zeta,\tau)=\frac{1}{2} \hat{\pmb\varphi}(\xi,\eta)e^{\ri(k_3\zeta-\omega \tau)}+c.c.,
\eqno{(\theequation{\mathit{a}})}
    \label{eq:travelling}
$$
$$H'(\xi,\zeta,\tau)=\frac{1}{2} \hat{H}(\xi)e^{\ri(k_3\zeta-\omega \tau)}+c.c..\eqno{(\theequation{\mathit{b}})}$$
Substitute  (\ref{eq:travelling}) into the equation system (\ref{eq:LNS_compact}), then, we arrive at 
\begin{equation}\label{eq:HLNS}
    \mathcal L_1 \hat{\pmb \varphi}(\xi,\eta) +
    \mathcal L_2 \hat{H}(\xi)=0,
\end{equation}
where
\begin{equation}
    \begin{aligned}
    \mathcal L_1=&
    \hat{\pmb A} \frac{\partial}{\partial \xi} +
    \hat{\pmb B} \frac{\partial}{\partial \eta} + \hat{\pmb D} +
    \pmb V_{\xi\xi} \frac{\partial^2}{\partial \xi^2} + \pmb V_{\eta\eta} \frac{\partial^2}{\partial \eta^2} +
    \pmb V_{\xi\eta} \frac{\partial^2}{\partial \xi \partial \eta},
    \\
    \mathcal L_2=&
    \pmb A^H \frac{\partial}{\partial \xi}
    + (\pmb D^H - \ri\omega \pmb{\Gamma}^H +\ri k_3\pmb C^H),
    \end{aligned}
\end{equation}
with
\refstepcounter{equation}\label{eq:HLNS_matrix}
$$
    \hat{\pmb A}= \pmb A +\ri k_3\pmb V_{\xi\zeta},\quad \hat{\pmb B}= \pmb B +\ri k_3\pmb V_{\eta\zeta},\quad
    \hat{\pmb D}= \pmb D - \ri \omega \pmb{\Gamma}/J +\ri k_3\pmb C -k_3^2\pmb V_{\zeta\zeta}.
    \eqno{(\theequation{\mathit{a,b,c}})}
$$

There are two main distinctions between the classical linearised N-S system and the present  shock-fitting linearised N-S system. Firstly, in the present system, the  upper boundary of the computational domain is located at the bow shock, necessitating the consideration of the linearised R-H relation as a boundary condition. Secondly, in  (\ref{eq:HLNS}), the perturbation field $\hat{\pmb\varphi}$ is coupled with the shock-movement perturbation $\hat H$, {therefore, a supplementary equation is needed.

To address the second issue, we consider (\ref{eq:HLNS}) at the shock wave position, $\eta=1$. The compatibility relation determines that the projection of (\ref{eq:HLNS}) in the ${\pmb I_5}$ space should be zero, namely,
\begin{equation}\label{eq:HLNS_Linear_CR}
    \begin{aligned}
    \bar{\pmb I}_5\cdot [\mathcal L_1 \hat{\pmb \varphi}(\xi,\eta) +
    \mathcal L_2 \hat{H}(\xi)]_{\eta=1} =0.
    \end{aligned}
\end{equation}
This is  the supplementary equation to the shock-fitting HLNS system (\ref{eq:HLNS}).

\subsubsection{Boundary conditions}
To derive the upper boundary condition, we express the perturbation of the unit normal vector $\delta \pmb n_s$ and the unsteady shock velocity $v_n$ in terms of $H'$,
\begin{equation}\label{eq:n_perturbation_compact}
    \begin{aligned}
    (\delta \pmb n_{s,x},\delta \pmb n_{s,y},\delta \pmb n_{s,z})^{\mbox T}=
    (\pmb A^{n} \frac{\partial}{\partial \xi} +
    \pmb C^{n}\frac{\partial}{\partial \zeta} + \pmb D^{n})H',\quad  v_n=\Gamma^{v_n} \frac{\partial H'}{\partial \tau},
    \end{aligned}
\end{equation}
where $\pmb A^{n}$, $\pmb C^{n}$ and $\pmb D^{n}$ are three-dimensional {column} vectors, and $\Gamma^{v_n}$ is a scalar. The expressions for these quantities are provided in the Appendix \ref{Appendix:RH}.

From the conservation law, the perturbations at the shock-wave {position} are governed by the following linearised system,
\begin{equation}\label{eq:RH_Linear}
    \begin{split}
    [\bar u_n \rho'+\bar\rho(\bar {\pmb n}_{s}\cdot \pmb u' +\bar {\pmb u}_0\cdot \delta \pmb n_s-v_n)]=&0,
    \\
    [\bar p \delta \pmb n_{s}+ p' \bar{\pmb n}_{s}+\bar u_n(\bar \rho \pmb u'+\bar{\pmb u}\rho')
    +\bar \rho \bar{\pmb u}(\bar {\pmb n}_{s}\cdot \pmb u' +\bar {\pmb u}_0\cdot \delta \pmb n_s-v_n) ]=&0,
    \\
    [\bar p v_n +\bar u_n(c_p\bar \rho T'+\bar h_t\rho'+\bar \rho\bar{\pmb u}\cdot \pmb u')
    +\bar \rho \bar h_t(\bar {\pmb n}_{s}\cdot \pmb u' +\bar {\pmb u}_0\cdot \delta \pmb n_s-v_n)]=&0,
    \end{split}
\end{equation}
where $[\cdot]\equiv (\cdot)_s-(\cdot)_0$ represents the difference between the values {immediately} before and behind the shock wave, and $\bar u_n=\bar {\pmb u}\cdot \bar {\pmb n}_s$. Note that the relation
$\bar{\pmb u}_s\cdot \delta \pmb n_s=\bar{\pmb u}_0\cdot \delta \pmb n_s$ has been applied because $\bar{\pmb u}_s=\bar{\pmb u}_0+(\bar u_{ns}-\bar u_{n0})\bar {\pmb n}_s$ (see (\ref{eq:RH_DNS})) and $\delta \pmb n_s\cdot \bar{\pmb n}_s=0$.

Combining with (\ref{eq:n_perturbation_compact}), we can rewrite (\ref{eq:RH_Linear}) as a compact form,
\begin{equation}\label{eq:HLNS_RH_Linear}
    \pmb D_s^{RH} {\pmb\varphi'}_s + (\pmb \Gamma^{RH} \frac{\partial}{\partial \tau} + \pmb A^{RH} \frac{\partial}{\partial \xi} + \pmb C^{RH} \frac{\partial}{\partial \zeta} +
    {\pmb D}^{RH}) H'=\pmb D_0^{RH} {\pmb \varphi'}_{\infty},
\end{equation}
where the vector ${\pmb \varphi'}_{\infty}(\xi,1,\zeta,\tau)$ represents the known freestream disturbance, and the matrices $\pmb D_s^{RH}$ and $\pmb D_0^{RH}$ are provided in Appendix \ref{Appendix:RH}. The five-dimensional column vectors $\pmb \Gamma^{RH}$, $\pmb A^{RH}$, $\pmb C^{RH}$ and $\pmb D^{RH}$   are defined as
\begin{equation}
    \begin{split}
    &\pmb \Gamma^{RH}=-\Gamma^{v_n}\left([\bar\rho],[\bar\rho\bar{\pmb u}],[\bar\rho\bar h_t-\bar p]\right)^{\mbox T},\\
    &\pmb A^{RH}=\bar{\pmb u}_0\cdot\pmb A^n \left([\bar\rho],[\bar\rho\bar{\pmb u}],[\bar\rho\bar h_t]\right)^{\mbox T}+[\bar p](0,\pmb A^n,0)^{\mbox T},\\
    &\pmb C^{RH}=\bar{\pmb u}_0\cdot\pmb C^n \left([\bar\rho],[\bar\rho\bar{\pmb u}],[\bar\rho\bar h_t]\right)^{\mbox T}+[\bar p](0,\pmb C^n,0)^{\mbox T},\\
    &\pmb D^{RH}=\bar{\pmb u}_0\cdot\pmb D^n \left([\bar\rho],[\bar\rho\bar{\pmb u}],[\bar\rho\bar h_t]\right)^{\mbox T}+[\bar p](0,\pmb D^n,0)^{\mbox T}.
    \end{split}
\end{equation}

Substituting (\ref{eq:gust}) and (\ref{eq:travelling}) into (\ref{eq:HLNS_RH_Linear}) leads to the upper boundary condition
\begin{equation}\label{eq:HLNS_RH_Linear2}
    \pmb D_s^{RH}\hat{\pmb\varphi}_s(\xi,\eta) + (\pmb A^{RH} \frac{\partial}{\partial \xi} +
    \hat{\pmb D}^{RH})\hat H(\xi)=\pmb D_0^{RH}\hat{\pmb \varphi}_{\infty}e^{\ri(k_1x+k_2y)},
\end{equation}
where $\hat{\pmb D}^{RH}= \pmb D^{RH} - \ri\omega \pmb{\Gamma}^{RH} +\ri k_3\pmb C^{RH}$.

To construct the boundary condition at the centreline  before the stagnation point, $\xi=0$, we decompose any perturbation field $\hat{\pmb\varphi}(\xi,\eta)$  into a linear superposition of a symmetric component $\hat{\pmb\varphi}^s(\xi,\eta)$ and an anti-symmetric component $\hat{\pmb\varphi}^a(\xi,\eta)$. Thus, for a perturbation with a particular vertical wavenumber $k_2$ or $-k_2$, we have 
\begin{equation}
\hat{\pmb\varphi}(\xi,\eta;\pm k_2)= \hat{\pmb\varphi}^s(\xi,\eta;k_2) \pm \hat{\pmb\varphi}^a(\xi,\eta;k_2).
\label{eq:convert_sa}
\end{equation}
For the two perturbation components, the oncoming perturbations are expressed as
\refstepcounter{equation}\label{eq:gust_cs}
$$
    \begin{aligned}
\varepsilon\hat{\pmb\varphi}^{s,a}_{\infty}e^{\ri(k_1x+k_3z-\omega t)}+c.c.,
    \end{aligned}
    \eqno{(\theequation)}
$$
where 
$$
    \begin{aligned}
    &\hat{\pmb\varphi}^{s}_{\infty}=[\hat \rho_{\infty}\cos(k_2y),\hat u_{\infty}\cos(k_2y),\hat v_{\infty}\mathrm{i}\sin(k_2y),\hat w_{\infty}\cos(k_2y),\hat T_{\infty}\cos(k_2y)]^{\mathrm{T}},\\
    &\hat{\pmb\varphi}^{a}_{\infty}=[\hat \rho_{\infty}\mathrm{i}\sin(k_2y),\hat u_{\infty}\mathrm{i}\sin(k_2y),\hat v_{\infty}\cos(k_2y),\hat w_{\infty}\mathrm{i}\sin(k_2y),\hat T_{\infty}\mathrm{i}\sin(k_2y)]^{\mathrm{T}}.
    \end{aligned}
$$
For the symmetric perturbation $\hat{\pmb\varphi}^s(\xi,\eta)$, the  boundary conditions at the centreline read
\refstepcounter{equation}
$$
    \frac{\partial \hat \rho^s}{\partial \xi}=\frac{\partial \hat u^s}{\partial \xi}=\hat v^s =\frac{\partial \hat w^s}{\partial \xi}=\frac{\partial \hat T^s}{\partial \xi}=0,\quad \mbox{at }\xi=0,
    \eqno{(\theequation)}\label{eq:symmetric_BC}
$$
while for the anti-symmetric perturbation  $\hat{\pmb\varphi}^a(\xi,\eta)$, the  boundary conditions are
\refstepcounter{equation}
$$
    \hat \rho^a=\hat u^a=\frac{\partial \hat v^a}{\partial \xi}=\hat w^a=\hat T^a=0,\quad \mbox{at }\xi=0.
    \eqno{(\theequation)}
    \label{eq:anti_symmetreic_BC}
$$

At the wall, the no-slip, non-penetration, and isothermal conditions are imposed,
\refstepcounter{equation}
$$
    \hat u=\hat v=\hat w=\hat T=0,\quad \mathrm{at}\,\eta=0.
    \eqno{(\theequation)}\label{eq:HLNS_wall_BC}
$$
The boundary condition at the outlet of the computational domain $\xi=\xi_{I}$ is obtained by applying the governing equation system (\ref{eq:HLNS}) with $\partial_\xi$ being discretised by a forward finite difference scheme.

\subsubsection{Summary of SF-HLNS}
In the numerical process, the partial derivatives of $\hat{\pmb\varphi}$, i.e., $\partial_\xi$, $\partial_\eta$, $\partial_{\xi\xi}$, $\partial_{\eta\eta}$ and $\partial_{\xi\eta}$, {and the partial derivative $\partial_\xi \hat H$} are discretised  using the 4$^{th}$-order central finite-difference scheme. Then, the linearised governing equation system (\ref{eq:HLNS}), along with the compatibility relation (\ref{eq:HLNS_Linear_CR}) and the boundary conditions (\ref{eq:HLNS_RH_Linear2}), (\ref{eq:symmetric_BC}) (or (\ref{eq:anti_symmetreic_BC})) and (\ref{eq:HLNS_wall_BC}), form a closed-form system of algebraic equations, which can be solved using the numerical approach illustrated in Section II.B.4 in \cite{Zhao2019}. 

\section{Numerical results}
\label{sec:results}
\subsection{Flow parameters and  mean-flow calculations}
\label{sec:base_flow}
The oncoming condition is selected  to agree with that of the experimental study by \cite{qiang2020experimental} and  the numerical study by \cite{peicheng2022study}. Additionally, three nose radii are chosen, as outlined in Table \ref{tab:flow_Parameters}. The mean flow is calculated using an in-house SF-DNS code, validated for accuracy through comparison with SF-DNS results presented in \cite{zhong1998high}, as shown in figure \ref{fig:pressure_zhong} in Appendix \ref{Appendix:code_validation}.
\begin{table}
  \begin{center}
    \def~{\hphantom{0}}
    \begin{tabular}{ccccccc}
    \vspace{.2cm}
  Case& $\begin{array}{c}
      \mbox{Mach}  \\
        \mbox{number}\\M
  \end{array}$   & $\begin{array}{c}
      \mbox{Oncoming}  \\
        \mbox{temperature}\\T^*_\infty
  \end{array}$  & $\begin{array}{c}
      \mbox{Wall}  \\
        \mbox{temperature}\\T^*_w
  \end{array}$  & $\begin{array}{c}
      \mbox{Reynolds}  \\
        \mbox{number}\\Re
  \end{array}$  & $\begin{array}{c}
      \mbox{Nose}  \\
        \mbox{radius}\\r^*
  \end{array}$ & $\begin{array}{c}\\
      \mbox{semi-tip-angle}  
         \\ \theta
  \end{array}$\\ \vspace{.2cm}
  A &         &     &    & $3.34\times 10^4$&   1 mm & \\ \vspace{.2cm}
  B & 5.96         &87K     & $\begin{array}{c}
      290\mbox K  \\ (3.33T^*_\infty)
  \end{array}$
  & $1.67\times 10^5$&   5 mm&$4^\circ$\\ \vspace{.2cm}
 C &        &     &   &$3.34\times 10^5$&10 mm&
    \end{tabular}
    \caption{Controlling parameters for case studies.}
    \label{tab:flow_Parameters}
  \end{center}
\end{table}

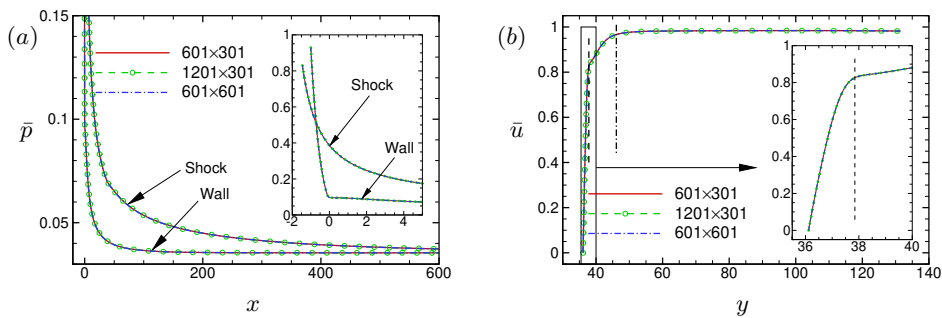
\begin{figure}
    \begin{center}
    \includegraphics[width = 0.48\textwidth] {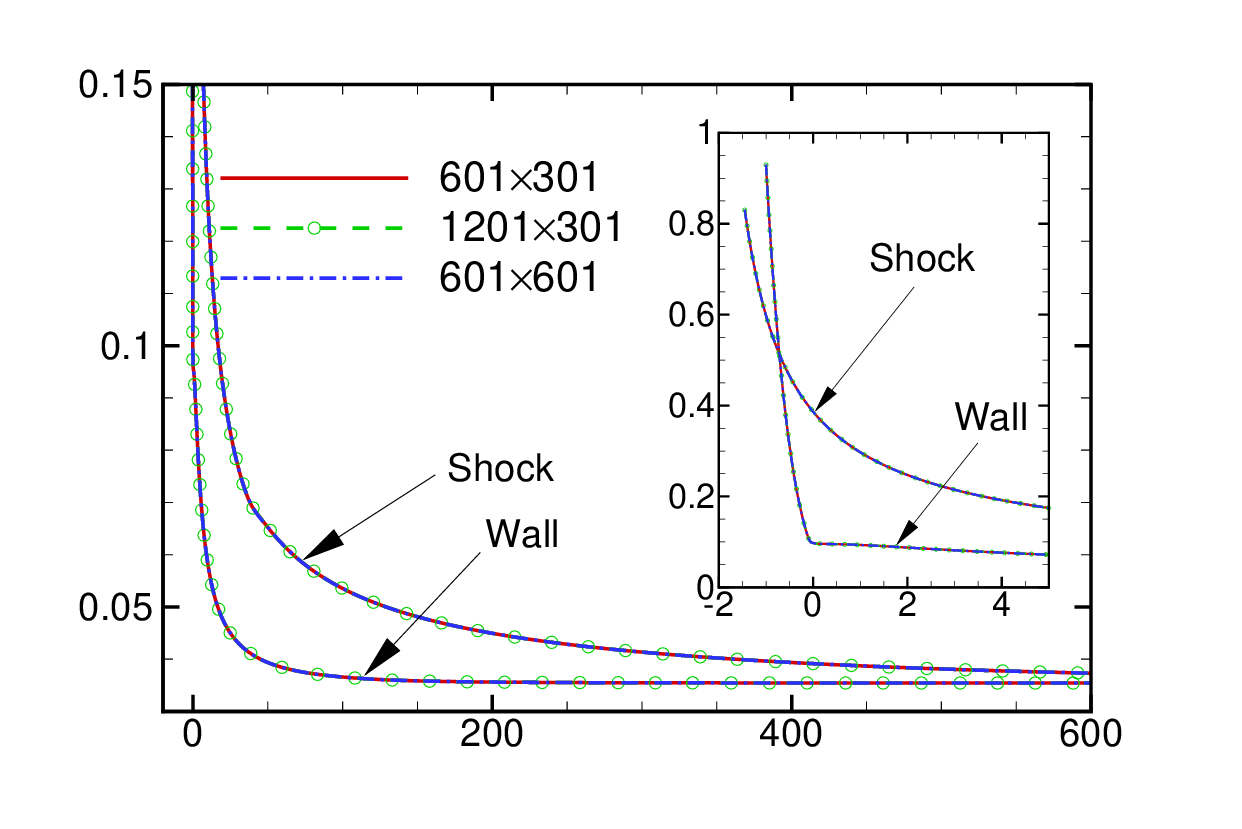}
    \put(-185,100){$(a)$}
    \put(-180,65){$\bar p$}
    \put(-95,0){$x$}
    \includegraphics[width = 0.48\textwidth] {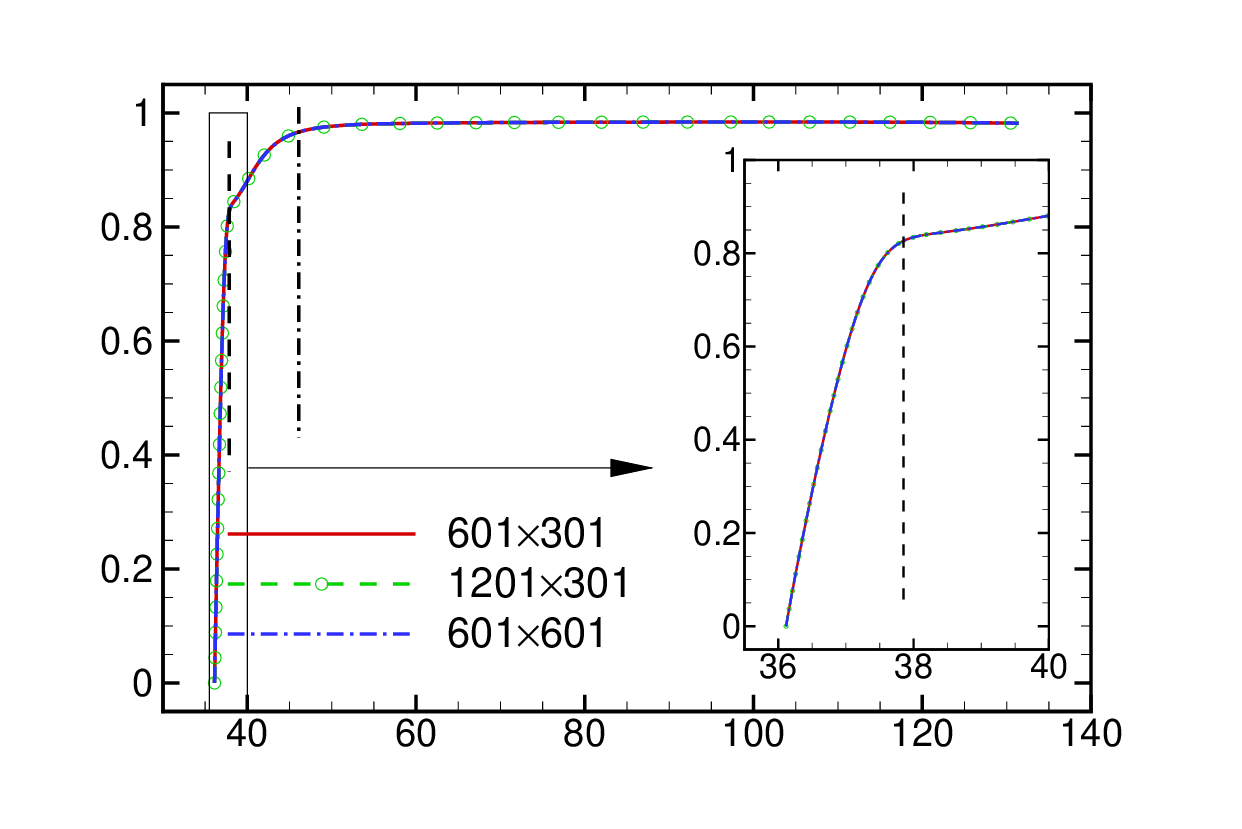}
    \put(-185,100){$(b)$}
    \put(-180,65){$\bar u$}
    \put(-95,0){$y$}
    \caption{Resolution study for three mesh scales for case A. ($a$): Comparison of the streamwise distributions of pressure at the shock and wall; ($b$) {wall-normal} profiles of $\bar u$ at $x_w=502$. In ($b$), the edges of the boundary layer and entropy layer are marked by vertical dashed and dash-dotted lines, respectively.}
    \label{fig:mesh_p_u}
    \end{center}
\end{figure}
We choose case A for resolution study, with the computational domain specified as $x_w\in[-1,635]$. Figure \ref{fig:mesh_p_u} presents a comparison of the base-flow calculations across three  mesh scales: $601\times 301$, $601\times {601}$ and $1201\times 301$. Panel ($a$) illustrates the pressure at both the shock position and the wall, showing a sharp decline  with $x$ in the nose region, followed by a more gentle decrease in the downstream region. This behaviour is attributed to the rapid changes of the oblique angle of the bow shock in  the nose region. Panel ($b$) displays the velocity profile at $x_w=502$, with vertical lines denoting the boundary-layer thickness $\delta_{BL}$ and entropy-layer thickness $\delta_{EL}$. These two layers are determined based on the profiles of the total enthalpy $h_t=\bar T/[(\gamma-1)M^2]+|\textbf{u}|^2/2$ and the normalised entropy increment relative to the free stream $\Delta s=\frac{\gamma}{\gamma-1}\ln \bar T-\ln(\gamma M^2 \bar p)$, respectively. By observing the numerical data obtained for various case, we set the threshold of $h_t$ to be 0.995 times its freestream value, and the threshold of $\Delta s$  to be 0.698, approximately 0.2 times the entropy increment at the normal shock position.  
Notably, a significant velocity gradient is observed within the boundary layer, accompanied by  a mild gradient  in the entropy layer above the boundary layer. The perfect agreement among the three sets of curves in both panels indicates the adequacy of the $601\times 301$ mesh scale, which will be utilized in the subsequent mean-flow calculations.

\begin{figure}
    \begin{center}
    \includegraphics[width = 0.48\textwidth] {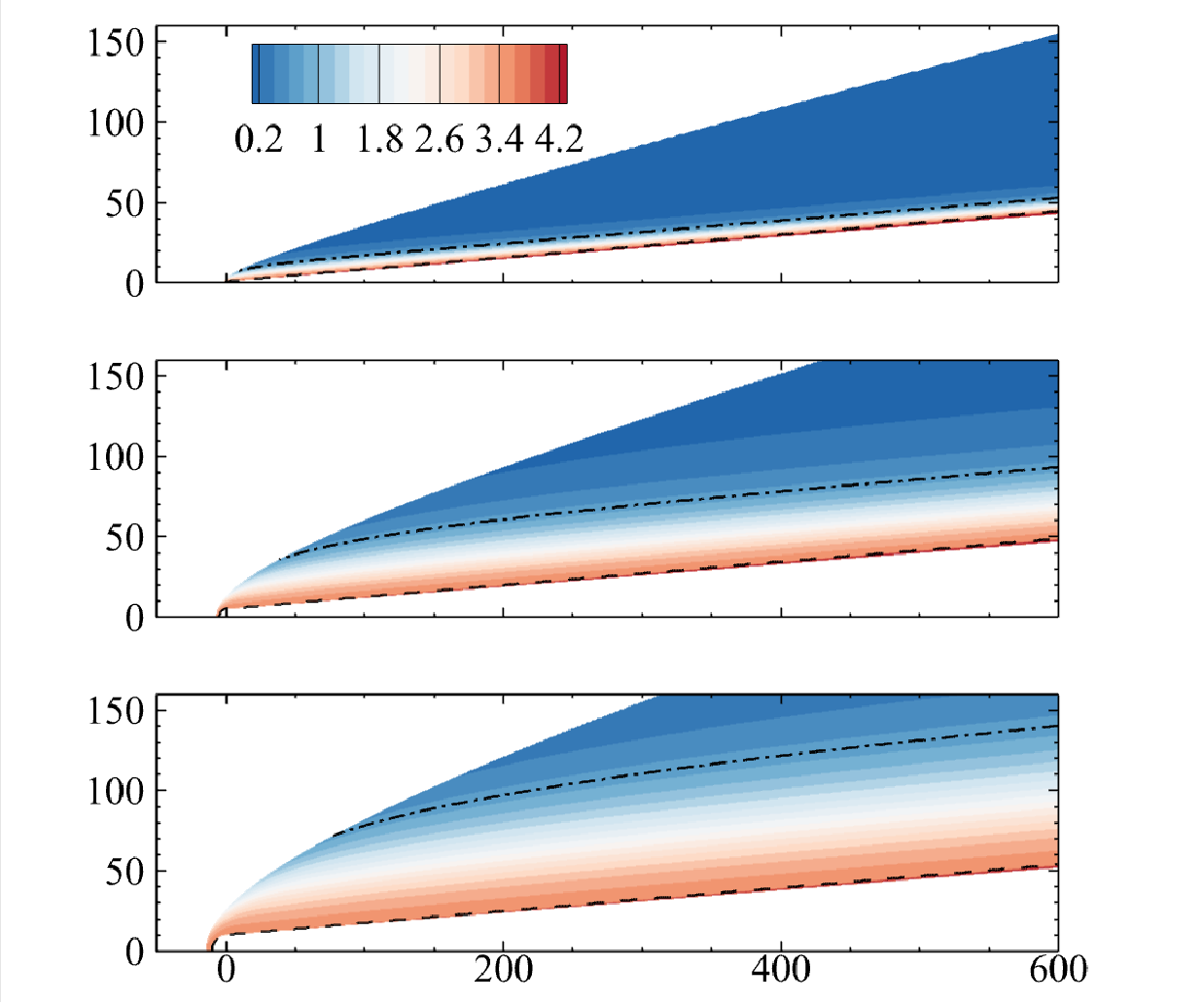}
    \put(-188,150){$(a)$}
    \put(-160,140){$\Delta s$}
    \put(-158,120){Case A}
    \put(-158,85){Case B}
    \put(-158,35){Case C}
    \put(-175,152){(mm)}
    \put(-32,-5){(mm)}
    \put(-180,113){\rotatebox{90}{$y^*=yr^*$}}
    \put(-180,65){\rotatebox{90}{$y^*=yr^*$}}
    \put(-180,16){\rotatebox{90}{$y^*=yr^*$}}
    \put(-105,-4){$x^*=xr^*$}
    \includegraphics[width = 0.48\textwidth] {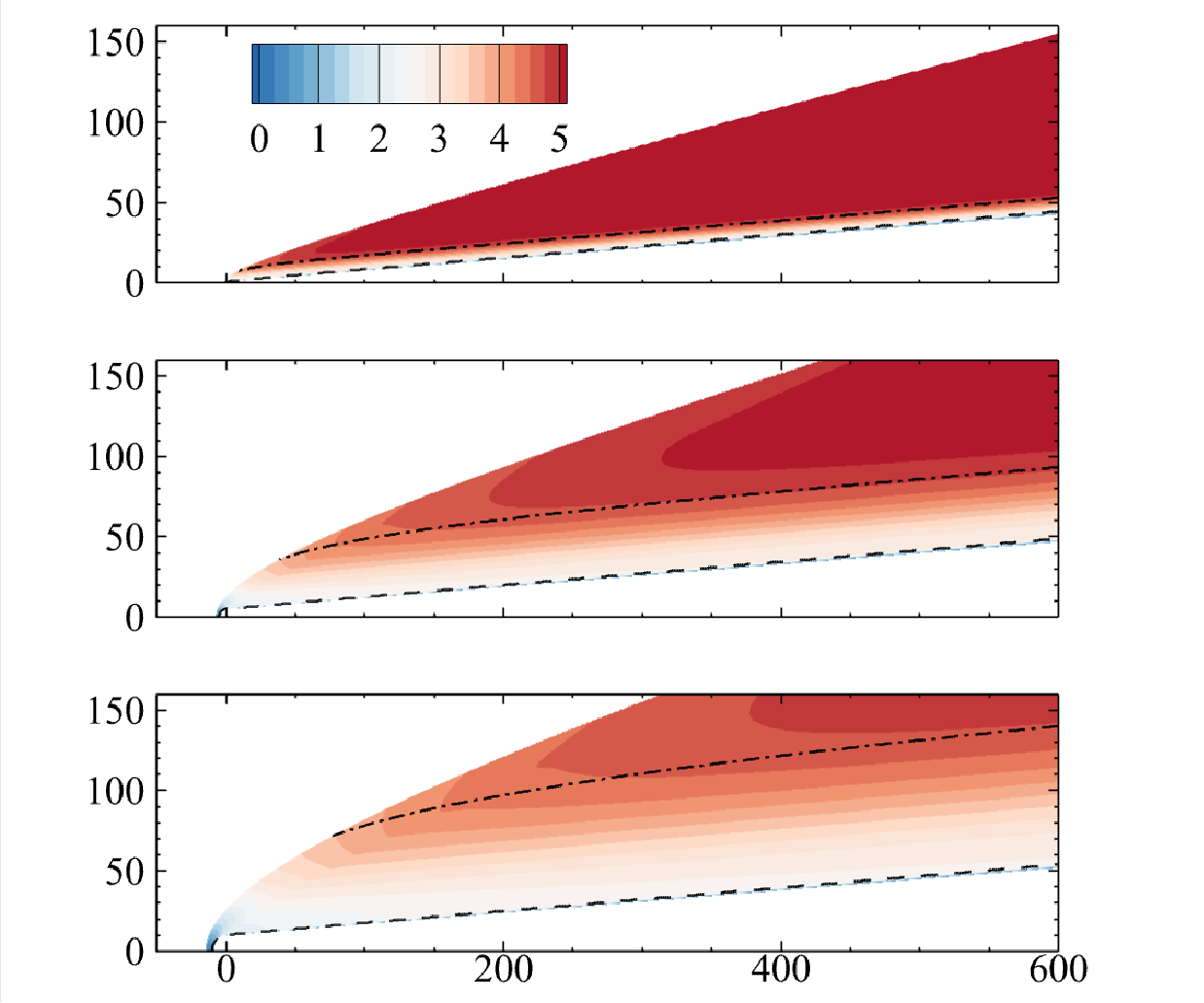}
    \put(-188,150){$(b)$}
    \put(-160,138){$\bar M$}
    \put(-158,120){Case A}
    \put(-158,85){Case B}
    \put(-158,35){Case C}
    \put(-175,152){(mm)}
    \put(-32,-5){(mm)}
    \put(-105,-4){$x^*=xr^*$}
    \caption{Contours of the entropy increment ($a$) and local Mach number ($b$) in the $x^*$-$y^*$ plane for the three cases. The edges of the boundary layer and entropy layer are marked by dashed and dash-dotted lines, respectively. The dimensional coordinates $(x^*,y^*)$ are converted  for clarity and consistency across various cases. }
    \label{fig:entropy_c_R}
    \end{center}
\end{figure}
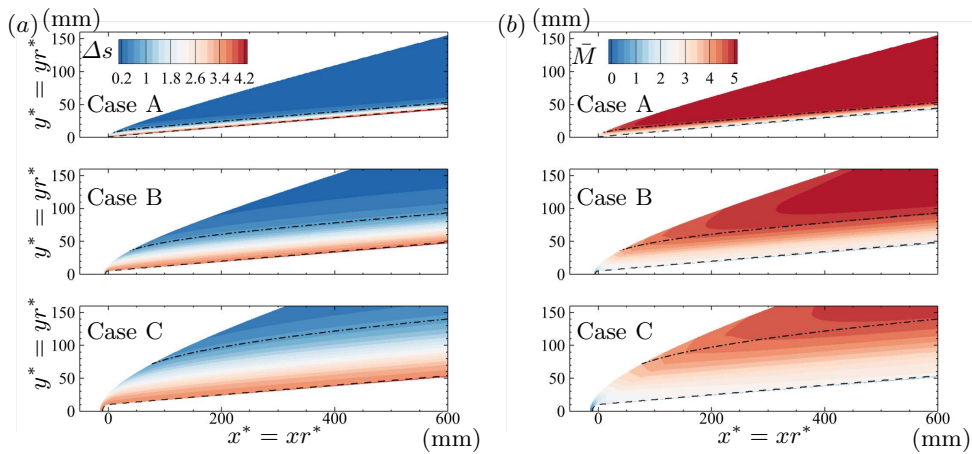
Figure \ref{fig:entropy_c_R} displays the contours of the normalised entropy {increment} $\Delta s$ and the local Mach number $\bar M=M\sqrt{(\bar u^2+\bar v^2)/\bar T}$ for different cases. As mentioned before, the contour line of $\Delta s=0.698$, denoted by the dash-dotted line in each subfigure, represents the entropy-layer {edge}. For each case, the entropy layer maintains almost a consistent thickness  as $x$ progresses downstream. Notably, an increase in the nose radius results in a notable expansion of the entropy layer. Specifically, at $x^*=600$mm, the dimensional entropy-layer thicknesses $\delta_{EL}^*$ are measured at 10.1mm, 46.2mm, and 87.7mm for cases A, B, and C, respectively.  Within the computational domain, the boundary layer  is much thinner than the entropy layer  for each case, with  boundary-layer thicknesses  $\delta_{BL}^*$ at $x^*=600$mm measuring only 1.81mm, 1.93mm and 1.98mm for cases A, B and C, respectively. It is important to note that the nose radius has a minimal impact on the downstream boundary-layer thickness. As the boundary layer successively grows downstream, it is anticipated that the entropy layer may be swallowed by the expanding boundary layer in far downstream positions.

\subsection{Modal instability analysis}
\label{sec:modal_analysis}
Based on the LST analysis, we have confirmed the absence of unstable Mack mode in the computational domain for all three cases, consistent with the findings in \cite{qiang2020experimental}, where the onset of Mack-mode instability for case A was identified at  $x^*_s\simeq 1000\mbox{mm}$.
On the other hand, the mean-flow profile in the entropy layer for each case exhibits a generalized inflectional point, potentially supporting an inviscid instability in the entropy layer.
  By conducting LST analysis, we have identified the unstable entropy-layer instabilities in all three cases, with  the 2-D modes displaying a more unstable feature compared to 3-D modes. In figure \ref{fig:neutral_sig-N-x_2d-entropy-mode}$(a,c,e)$, we present the contours of the growth rate $-\alpha_i$ of the 2-D entropy-layer mode across the three cases. Notably, the computational domain is intentionally shortened for a larger nose radius, facilitating a more straightforward comparison of  perturbation fields at equivalent dimensional positions. For the case with the smallest bluntness, case A, the entropy-layer instability shows a maximum growth at around $x_s=100$, which decays as $x_s$ further increases. For cases with larger nose radii, cases B and C, we observe higher growth rates in the region $x_s\leq 100$ than in case A.
By integrating the growth rate along the $x_s$-axis starting from the neutral position, we obtain the $N$-factor, as shown in figure \ref{fig:neutral_sig-N-x_2d-entropy-mode}$(b,d,f)$ for the three cases. Up to the end of the computational domain for case A, the highest $N$-factor is around 0.5,  with even smaller values for the other two cases. These low $N$ factors are unlikely to trigger transition to turbulence in the traditional sense of  natural transition. Therefore, we explore an alternative transition route  by monitoring  the excitation of the non-modal perturbations in the subsequent subsection.

\begin{figure}
    \begin{center}
    \includegraphics[width = 0.48\textwidth] {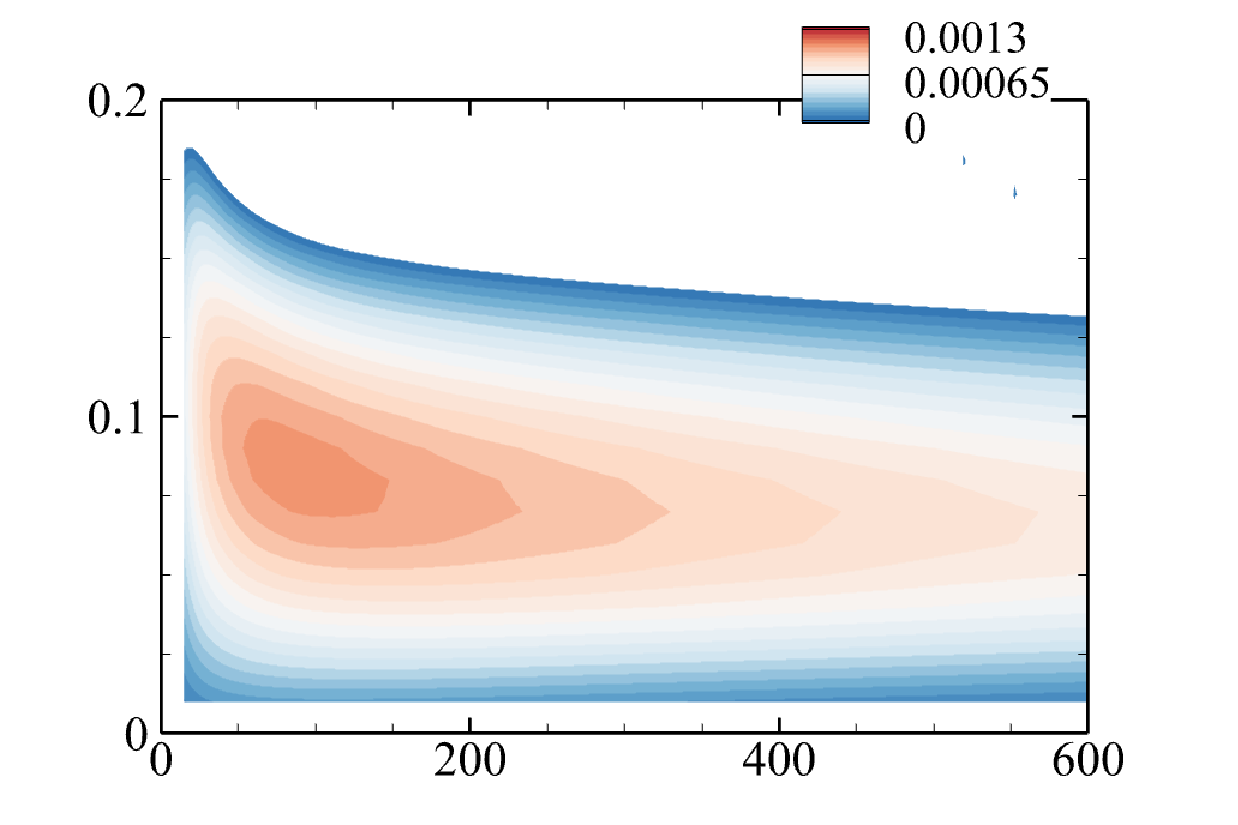}
    \includegraphics[width = 0.48\textwidth] {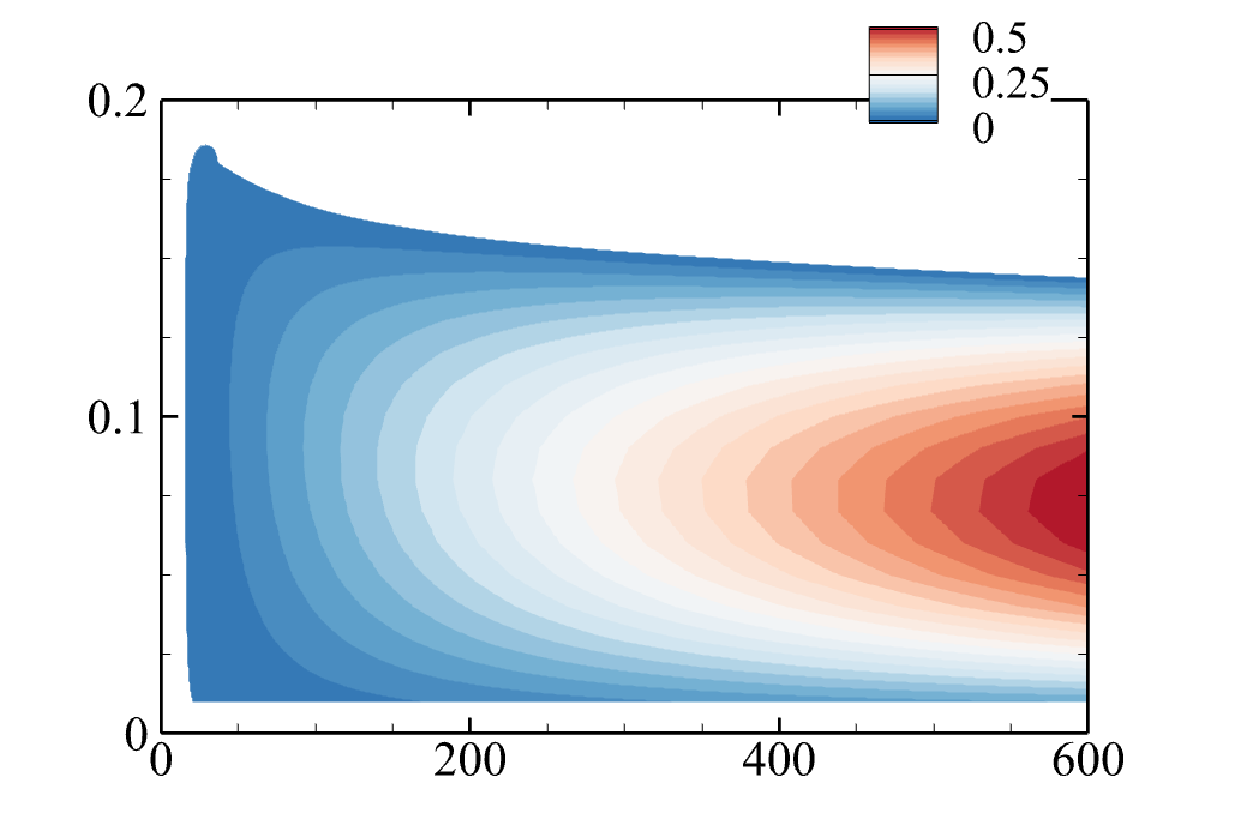}    
    \put(-375,110){$(a)$}
    \put(-370,60){$\omega$}
    \put(-275,112){$-\alpha_i$}
    \put(-185,110){$(b)$}
    \put(-70,112){$N$}\\
    \includegraphics[width = 0.48\textwidth] {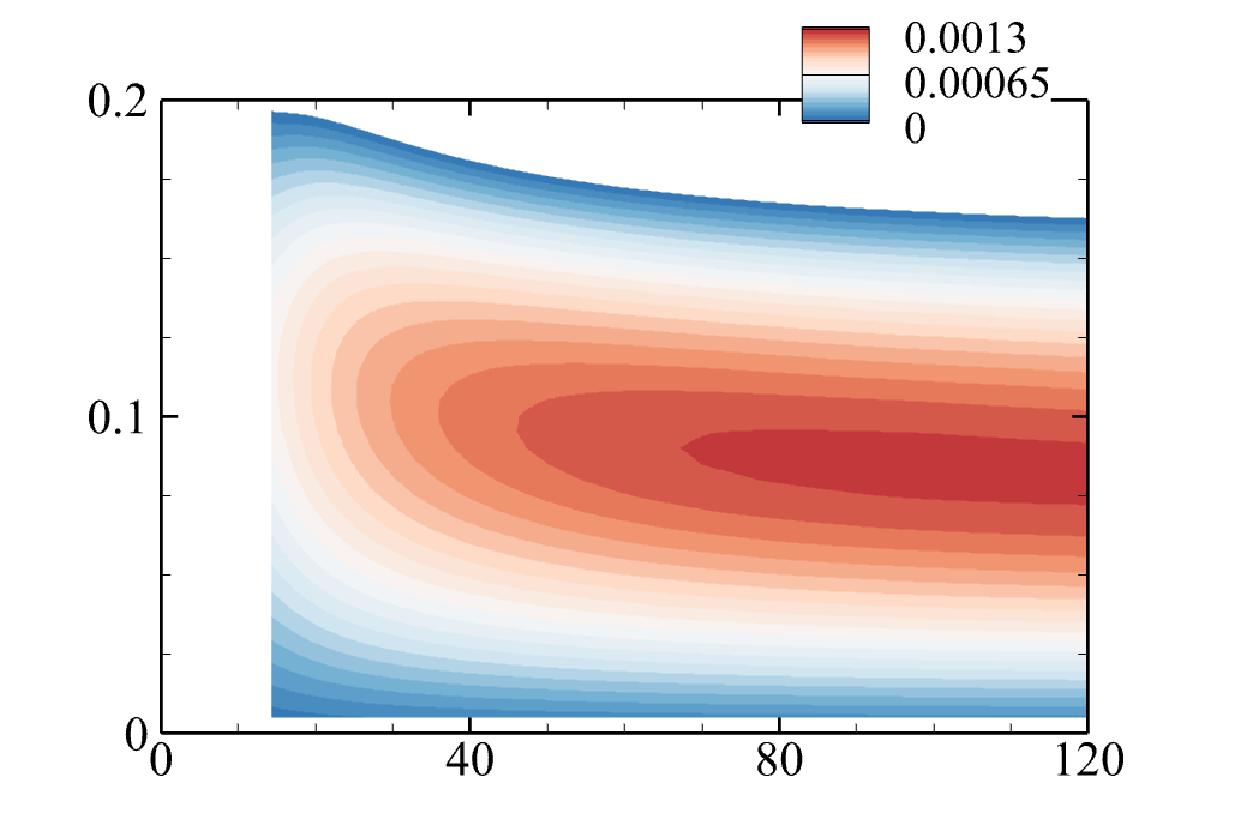}
    \includegraphics[width = 0.48\textwidth] {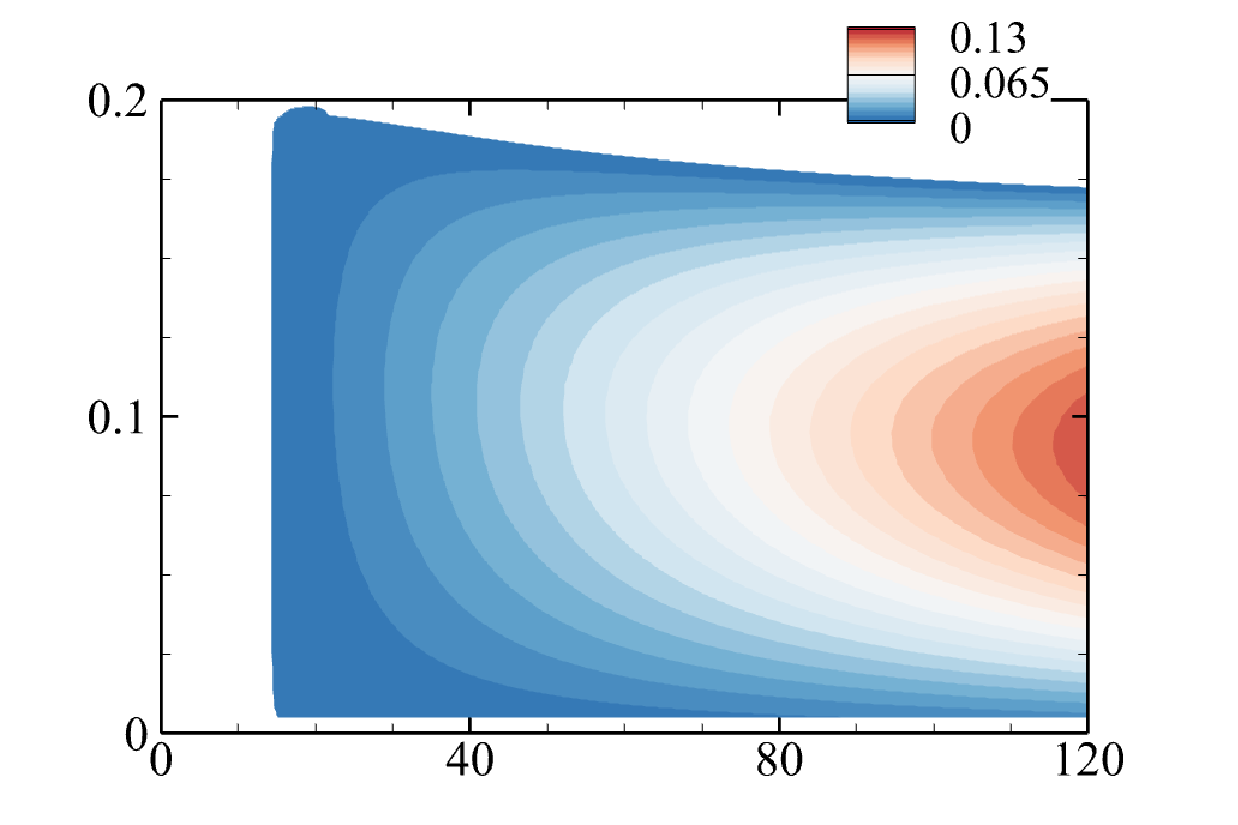}
    \put(-375,110){$(c)$}
    \put(-370,60){$\omega$}
    \put(-275,112){$-\alpha_i$}
    \put(-185,110){$(d)$}
    \put(-70,112){$N$}\\
    \includegraphics[width = 0.48\textwidth] {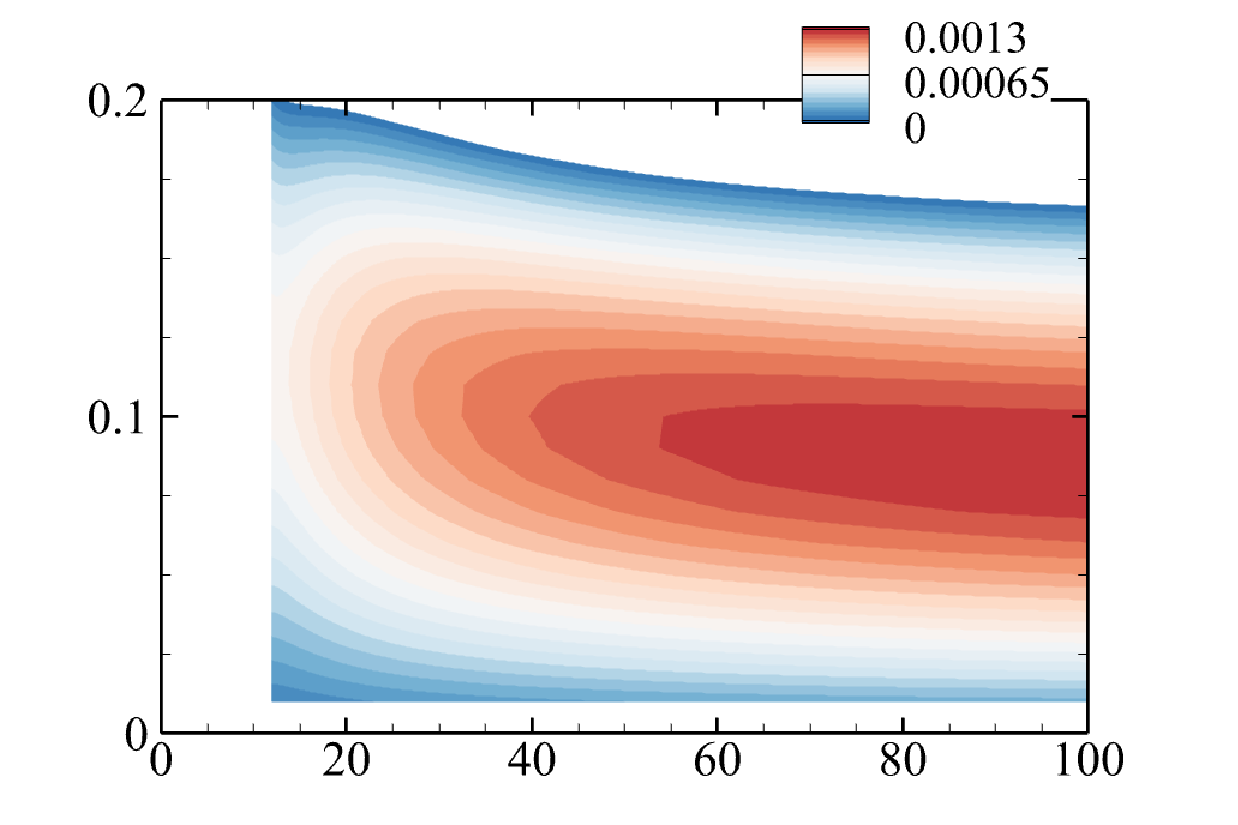}
    \includegraphics[width = 0.48\textwidth] {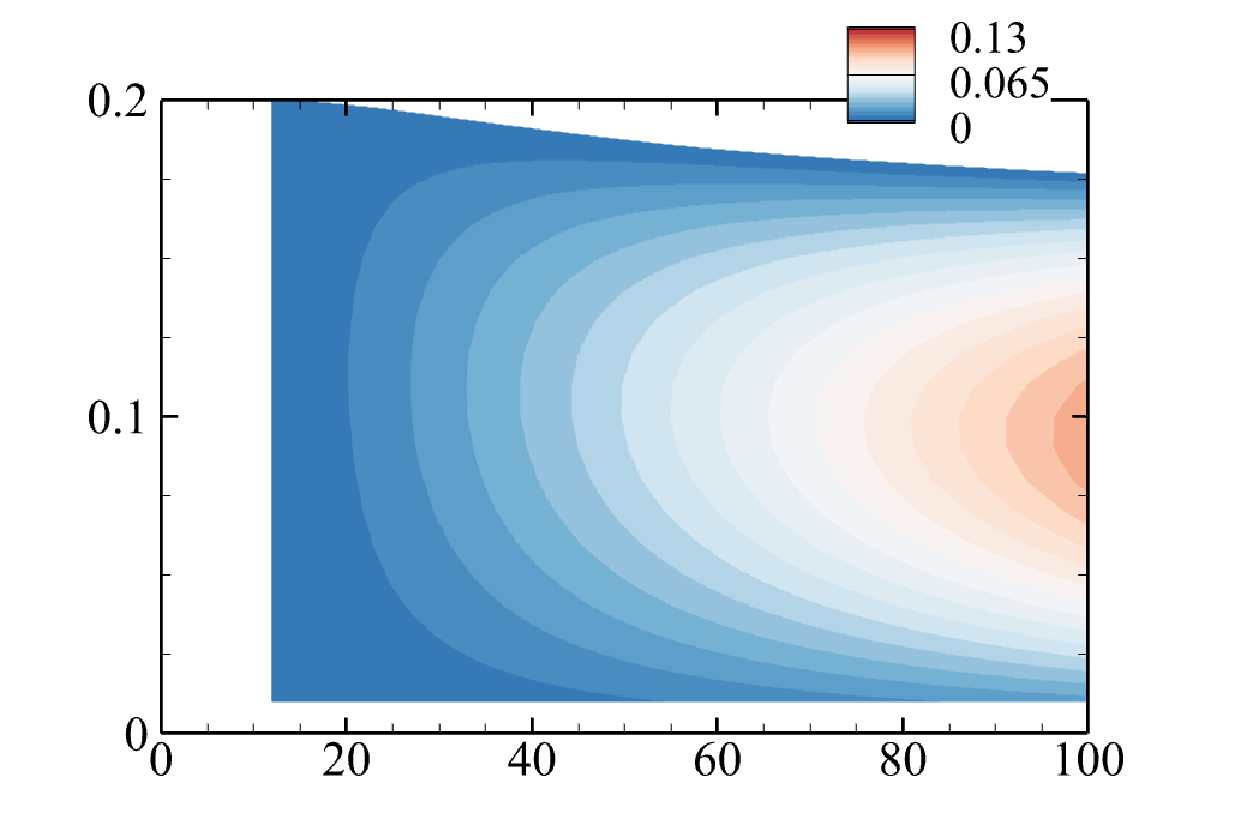}
    \put(-375,110){$(e)$}
    \put(-370,60){$\omega$}
    \put(-275,112){$-\alpha_i$}
    \put(-280,0){$x_s$}
    \put(-185,110){$(f)$}
    \put(-70,112){$N$}
    \put(-95,0){$x_s$}    
    \caption{Contours of growth rate $-\alpha_{\ri}$ $(a,c,e)$ and $N$-factor $(b,d,f)$ of the 2-D entropy-layer mode, for case A $(a,b)$, case B $(c,d)$ and case C $(e,f)$.}
    \label{fig:neutral_sig-N-x_2d-entropy-mode}
    \end{center}
\end{figure}

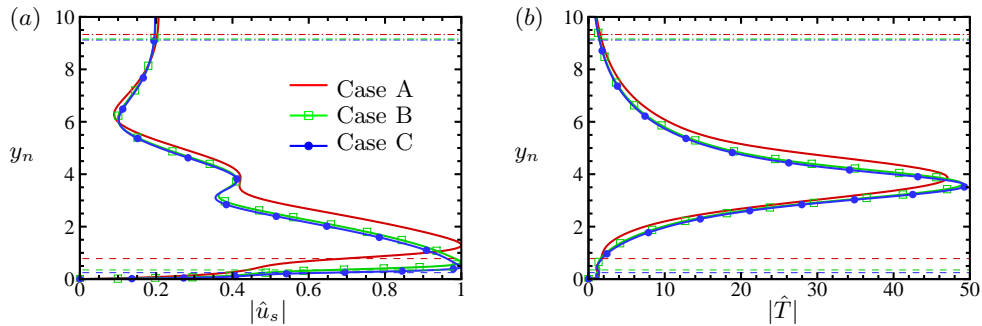
\begin{figure}
    \begin{center}
    \includegraphics[width = \textwidth] {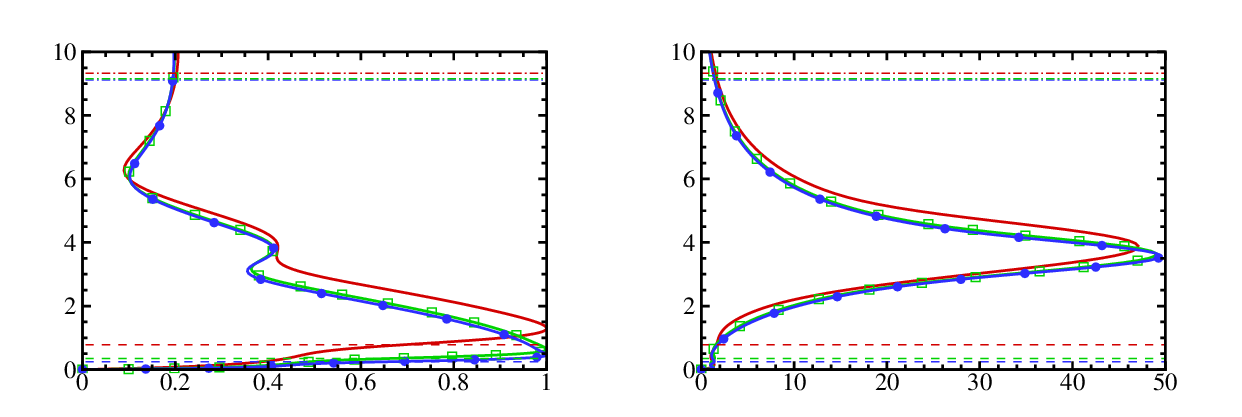}
    \put(-385,110){$(a)$}
    \put(-385,60){${y_n}$}
    \put(-295,0){$|\hat u_s|$}
    \put(-195,110){$(b)$}
    \put(-195,60){${y_n}$}
    \put(-100,0){$|\hat T|$}
    \put(-280,63){\begin{tikzpicture}
    \draw[red,thick] (0,0) -- (0.5,0);
    \draw[green,thick] (0,-0.4) -- (0.5,-0.4);
    \node[draw,green,rectangle,inner sep=1.5pt] at (0.25,-0.4) {};
    \draw[blue,thick] (0,-0.8) -- (0.5,-0.8);
    \node[draw,blue,circle,fill,inner sep=1pt] at (0.25,-0.8) {};
    \end{tikzpicture}}
    \put(-262,83){Case A}
    \put(-262,73){Case B}
    \put(-262,63){Case C}  
    \caption{Wall-normal profiles of  $\hat u_s$ ($a$) and $\hat T$ ($b$) for  2-D entropy-layer mode with $ \omega=0.08$
at $x_s=100$. The profiles are normalised by the peak of $|\hat u_s|$. }
    \label{fig:eigenf_entropymode_xs=600}
    \end{center}
\end{figure}
The wall-normal profiles of $\hat u_s$ and $\hat T$ for 2-D entropy-layer mode across the three cases are compared in figure \ref{fig:eigenf_entropymode_xs=600}, where $\hat u_s$ is the projection of  $\hat{\pmb u}$ along the $x_s$ axis.
At the same dimensionless position, $x_s=100$, the entropy-layer thicknesses across the three cases are almost identical, whereas the boundary-layer thickness diminishes with increasing the nose radius. 
Remarkably, the $\hat T$ profiles across the three cases show similar behaviour in the entropy layer, with the peaks located at around $y_n=4$. 
In contrast, the $\hat u_s$ profile displays a peak just above the boundary-layer edge. The magnitude of $\hat T$  significantly surpasses that of $\hat u_s$. Profiles for a 3-D  mode also exhibit peaks in the entropy layer, which are omitted here for brevity.

\subsection{Excitation of non-modal perturbations by various freestream disturbances}
\label{sec:SF-HLNS_calculations}
The SF-HLNS approach serves as an effective tool for conducting comprehensive investigations on the excitation of both modal and non-modal perturbations in supersonic or hypersonic boundary layers under various freestream forcing. To ensure the reliability of our code, we have verified its performance through comparisons with the existing DNS data   on the Mack-mode receptivity \citep{zhong1998high} and with our SF-DNS calculations on the excitation of non-modal perturbations, as detailed in Appendix \ref{Appendix:code_validation}.

In the analysis of each case listed in Table \ref{tab:flow_Parameters}, we investigate the perturbation evolution under all three types of freestream perturbations outlined in $\S$\ref{sec:freestream_perturbation}. For clarity, in the following, we adopt a two-digit identifier to denote different case studies.  The first digit distinguishes the nose radius (Reynolds number)  selected from Table \ref{tab:flow_Parameters}, while the second digit indicates the freestream perturbation, with 'f', 's', 'e' and 'v' denoting the fast acoustic, slow acoustic, entropy and vorticity waves, respectively. For instance, case Bf signifies a case study involving a nose radius $r^*=5$mm subject to a fast acoustic forcing.

For each case, we compute both symmetric and anti-symmetric configurations with centerline boundary conditions (\ref{eq:symmetric_BC}) and (\ref{eq:anti_symmetreic_BC}), respectively. By utilizing the results from these calculations, we can determine the perturbation evolution for a specific $k_2$ through the relationship (\ref{eq:convert_sa}). 

In addition to the nose radius, the downstream boundary-layer thickness  serves as another representative length scale, particularly relevant to the receptivity of non-modal perturbations. At $x^*=600$mm, the dimensional boundary-layer thicknesses $\delta_{BL}^*$ are almost identical across the three cases considered. This indicates that the optimal spanwise wavelength of the freestream perturbation $2\pi/k_3^*$ could be comparable, as will be elaborated in the following. Therefore, for convenience of comparison across different cases, we may use coordinate systems $(k_3x,k_3y)$ and $(k_3x_s,k_3y_n)$ to demonstrate the SF-HLNS results in the subsequent discussion when necessary.

\subsubsection{Excitation of non-modal perturbations by the freestream vortical disturbance}
As pointed out by \citet{trefethen1993hydrodynamic} and \citet{schmid2007nonmodal}, the most amplified non-modal perturbations in boundary-layer flows often exhibit a longitudinal streaky structure attributed to the lift-up mechanism, commonly known as streaks.  In figure \ref{fig:cont-sn-u3ddis-zx-y=1.2}, we present the  contours of the velocity and temperature perturbation in the $z-x_s$ plane at $y_n=1.2$ for case Av at $\omega=0$, illustrating prominent streaky structures.

\begin{figure}
    \begin{center}
    \includegraphics[width = 0.48\textwidth] {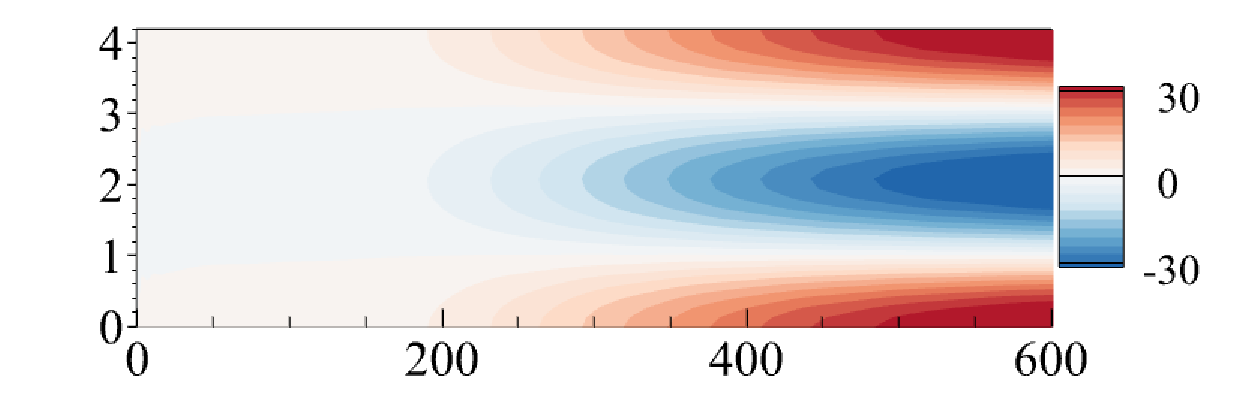}
    \includegraphics[width = 0.48\textwidth] {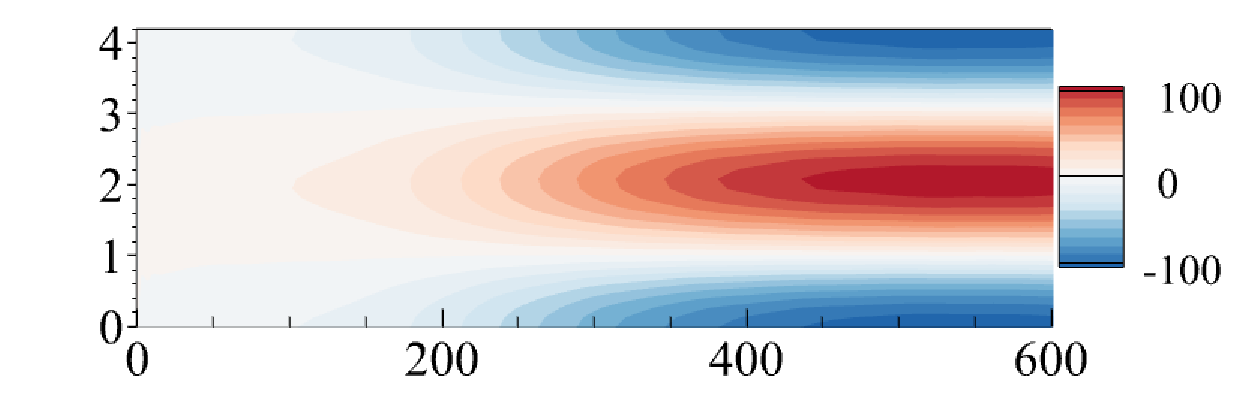}
    \put(-370,55){$(a)$}
    \put(-367,32){$z$}
    \put(-285,0){$x_s$}
    \put(-212,52){$u_s'$}
    \put(-185,55){$(b)$}
    \put(-97,0){$x_s$}
    \put(-25,52){$T'$}
    \caption{Contours of the streaky structure for case Av with  $\omega=0$, $k_3=1.5$ and $\vartheta=15^{\circ}$ in $z-x_s$ plane at a fixed $y_n=1.2$. ($a$): $u_s'$; ($b$): $T'$.}
    \label{fig:cont-sn-u3ddis-zx-y=1.2}
    \end{center}
\end{figure}

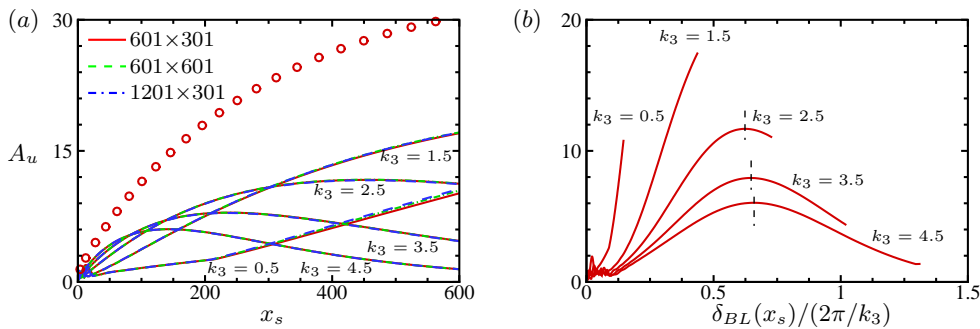
\begin{figure}
    \begin{center}
    \includegraphics[width = \textwidth] {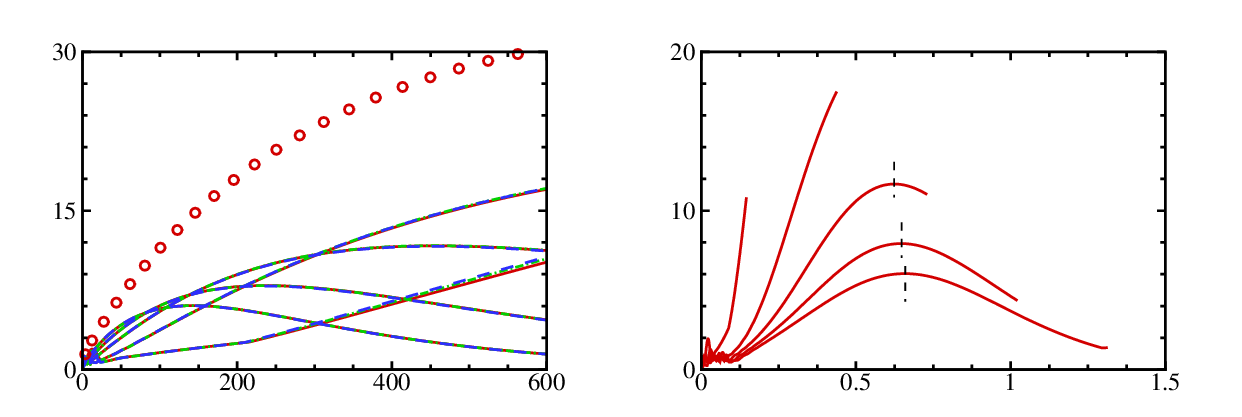}
    \put(-385,110){$(a)$}
    \put(-385,60){$A_u$}
    \put(-290,0){$x_s$}
    \put(-339,103){\fontsize{8pt}{8pt}\selectfont 601$\times $301}
    \put(-339,93) {\fontsize{8pt}{8pt}\selectfont 601$\times $601}
    \put(-339,83) {\fontsize{8pt}{8pt}\selectfont 1201$\times $301}
    \put(-355,85){\begin{tikzpicture}
    \draw[red,thick] (0,0) -- (0.5,0);
    \draw[green,thick,dashed] (0,-0.35) -- (0.5,-0.35);
    \draw[blue,thick,dash dot] (0,-0.7) -- (0.5,-0.7);
    \end{tikzpicture}}
    \put(-310,18){\fontsize{6pt}{6pt}\selectfont $k_3=0.5$}
    \put(-245,60){\fontsize{6pt}{6pt}\selectfont $k_3=1.5$}
    \put(-270,47){\fontsize{6pt}{6pt}\selectfont $k_3=2.5$}
    \put(-250,26){\fontsize{6pt}{6pt}\selectfont $k_3=3.5$}
    \put(-275,18){\fontsize{6pt}{6pt}\selectfont $k_3=4.5$}
    \put(-195,110){$(b)$}
    \put(-120,0){$\delta_{BL}(x_s)/(2\pi/k_3)$}
    \put(-165,75){\fontsize{6pt}{6pt}\selectfont $k_3=0.5$}
    \put(-140,105){\fontsize{6pt}{6pt}\selectfont $k_3=1.5$}
    \put(-105,75){\fontsize{6pt}{6pt}\selectfont $k_3=2.5$}
    \put(-90,51){\fontsize{6pt}{6pt}\selectfont $k_3=3.5$}
    \put(-60,30){\fontsize{6pt}{6pt}\selectfont $k_3=4.5$}
    \caption{The evolution of amplitude $A_u$ for case Av with $\omega=0$, $\vartheta=0$  for different $k_3$, where the resolution test is also provided in ($a$). The curves in (a) represent results for $\hat\Omega_2=0$, and the red circles denote the result for $\hat \Omega_2=1$. The vertical dashed lines in ($b$) mark the peaks.}
    \label{fig:Au_mesh}
    \end{center}
\end{figure}
The curves in figure \ref{fig:Au_mesh}-(a) display the streamwise evolution of the perturbation velocity amplitude $A_u(x_s)=\displaystyle \max_{y_n}|\hat u_s(x_s,y_n)|$ for case Av with $\omega=0$, $\vartheta=0$, $\hat\Omega_2=0$, and different $k_3$ values. For $k_3\geq 2.5$, each curve exhibits a successive increase until reaching a saturation point, followed by a gradual decay. With increasing $k_3$, the saturation amplitude decreases, and the saturation position shifts upstream. Consequently, for lower values like $k_3=1.5$ and 0.5, the saturation position extends beyond  the selected domain. While $A_u$ for $k_3=1.5$ may not be the highest in the early region ($x_s<300$), it becomes the most amplified perturbation at the end of the domain under consideration, $x_s=600$. 
Each curve in figure \ref{fig:Au_mesh}-(a) is calculated across three mesh scales, and they align well. This  suggests that the mesh scale $601\times 301$ is adequately accurate for the current computational domain, which is utilized in the following SF-HLNS calculations.
Additionally, to probe the impact of $\hat \Omega_2$, we  calculate the case for $\hat\Omega_2=1$ and $k_3=1.5$, represented by the circles in panel (a). While a quantitative discrepancy is observed between the results for $\hat \Omega_2=0$ and 1, they remain the same order of magnitude. Thus, we will choose $\hat \Omega_2=0$ for representative demonstration in the following.
In figure \ref{fig:Au_mesh}-(b), we replace the horizontal axis by the local boundary-layer thickness $\delta_{BL}$ normalised by the spanwise wavelength of the external forcing $2\pi/k_3$. This adjustment reveals that  the saturation position appears at  $\delta_{BL}/(2\pi/k_3)\approx 0.6$, indicating the preferential regime of the length scale of the freestream forcing determined by the local boundary-layer thickness. Given the comparable dimensional boundary-layer thicknesses in the downstream region across the three nose radii,  the optimal spanwise wavenumbers within the same computational domain for cases Bv and Cv  also  appear to be $k_3^*\approx $1.5 mm$^{-1}$, as will be shown in figure \ref{fig:LNS_A-k3_vortex}-(a). Thus, this specific spanwise wavenumber is selected as a representative wavenumber for the subsequent analysis. Note that for alternative configurations of the controlling parameters, the optimal spanwise wavenumber may vary in response to changes in the boundary-layer thickness.

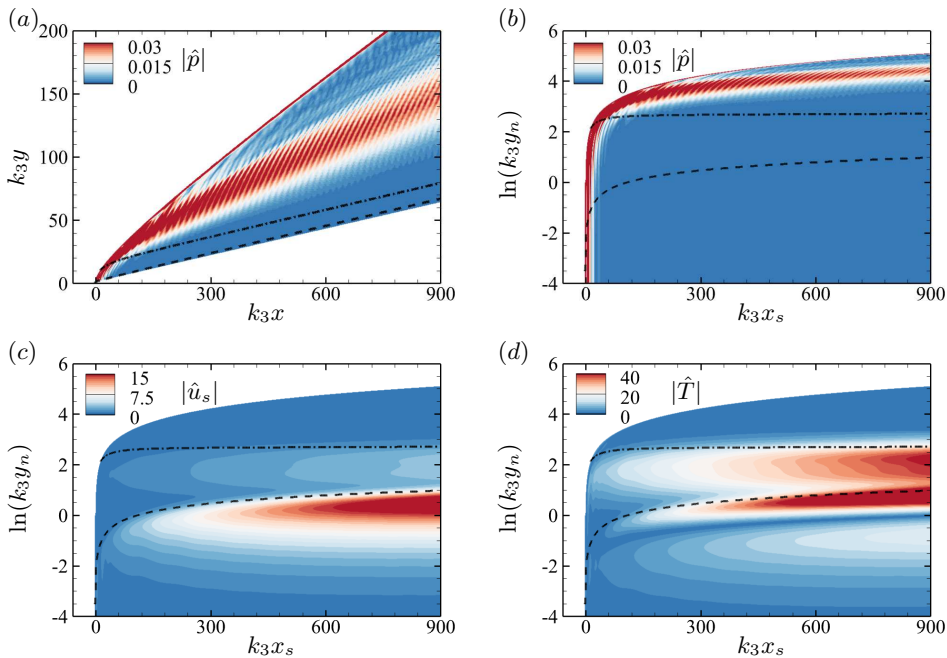
\begin{figure}
    \begin{center}
    \includegraphics[width = 0.48\textwidth] {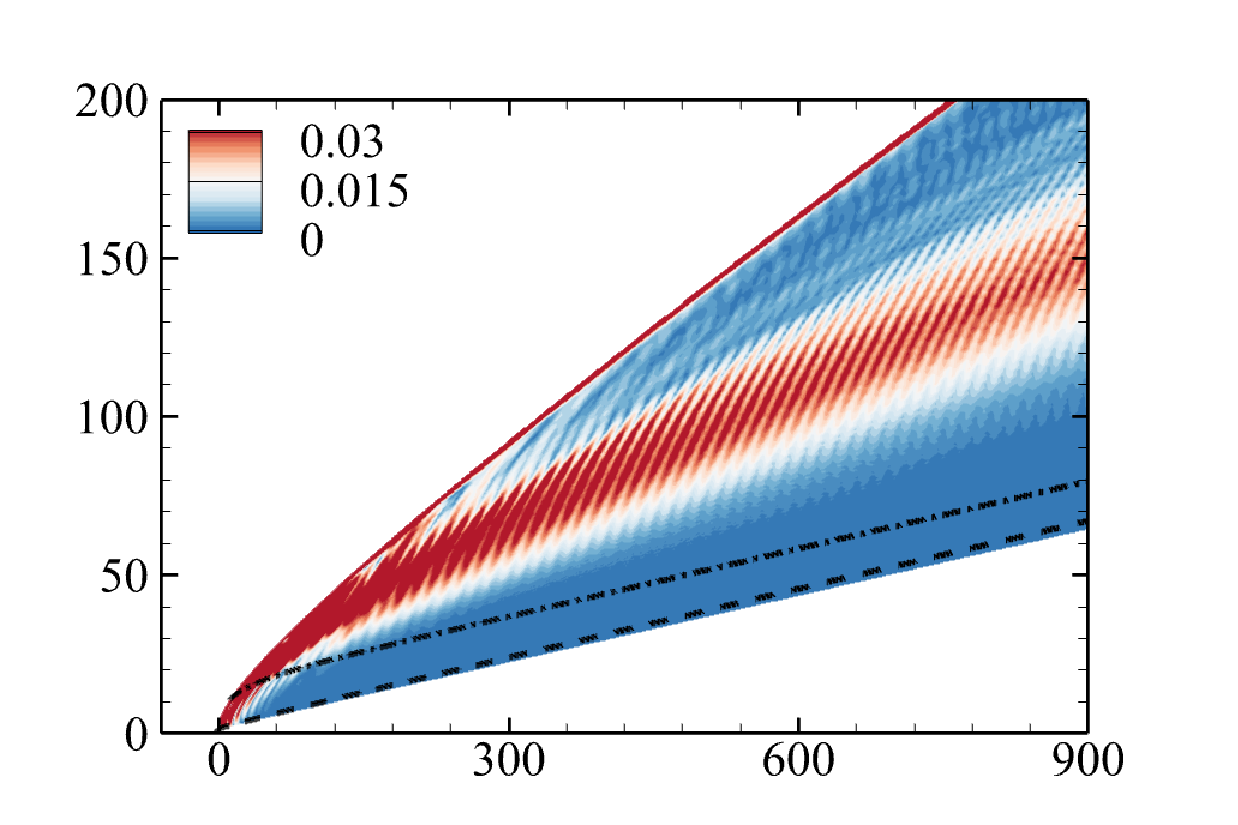}
    \put(-185,50){\rotatebox{90}{$k_3 y$}}
    \put(-95,0){$k_3 x$}
    \put(-120,95){$|\hat p|$}
    \put(-185,110){$(a)$}
    \includegraphics[width = 0.48\textwidth] {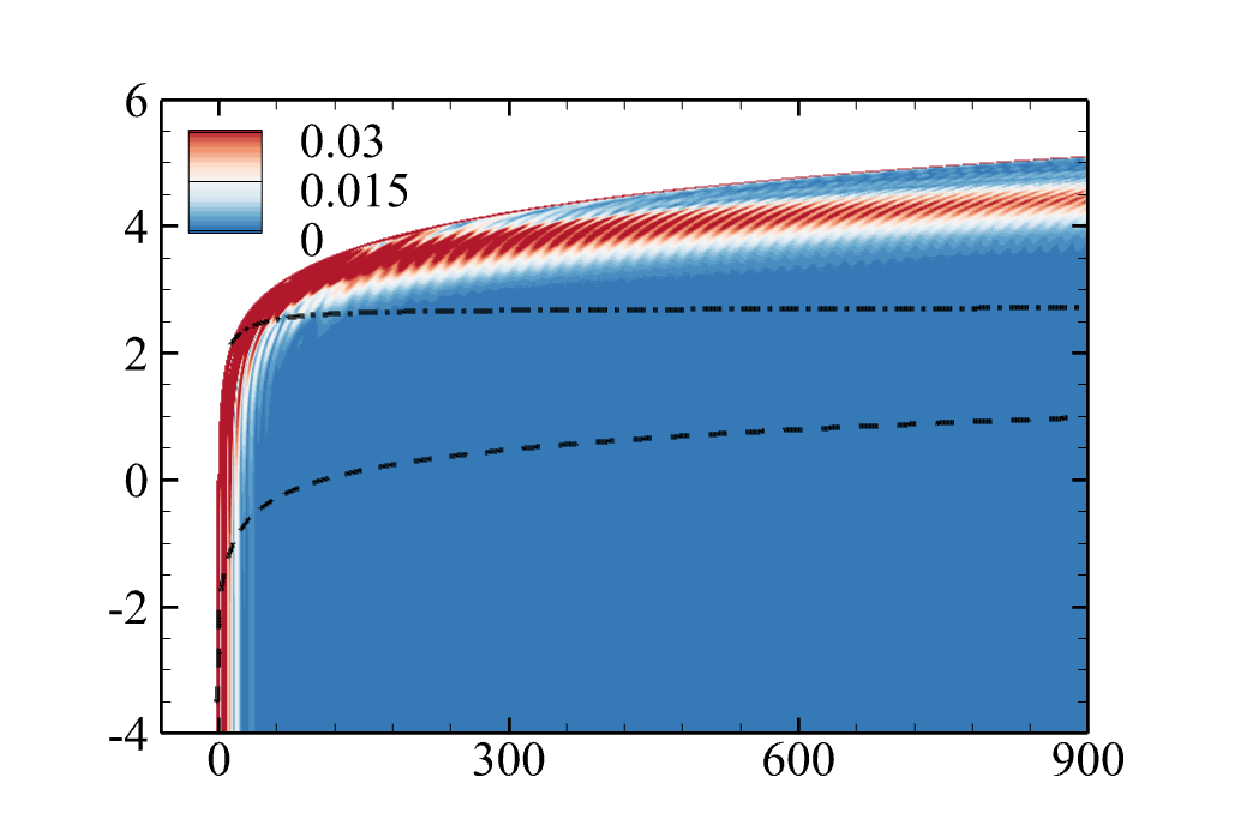}
    \put(-185,45){\rotatebox{90}{$\ln(k_3 y_n)$}}
    \put(-95,0){$k_3 x_s$}
    \put(-120,95){$|\hat p|$}
    \put(-185,110){$(b)$}\\
    \includegraphics[width = 0.48\textwidth] {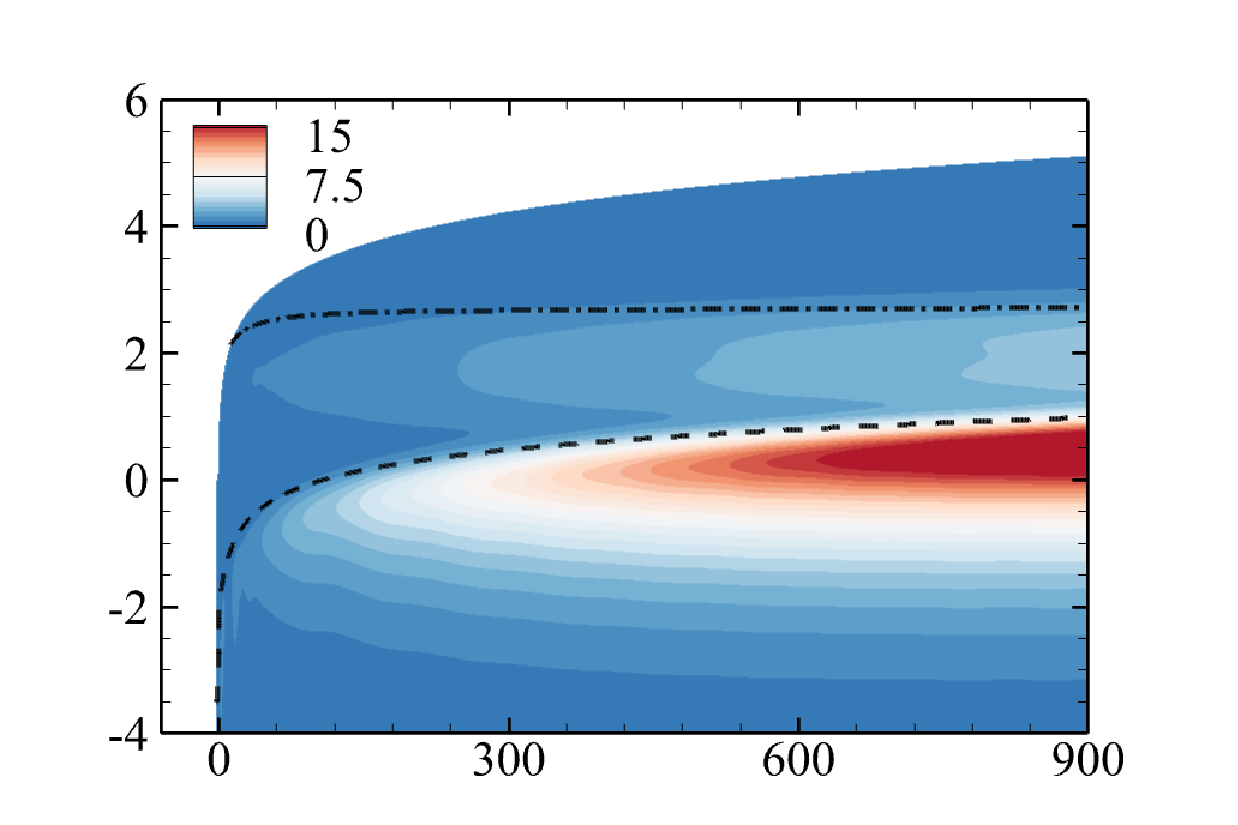}
    \put(-185,45){\rotatebox{90}{$\ln(k_3 y_n)$}}
    \put(-95,0){$k_3 x_s$}
    \put(-120,95){$|\hat u_s|$}
    \put(-185,110){$(c)$}
    \includegraphics[width = 0.48\textwidth] {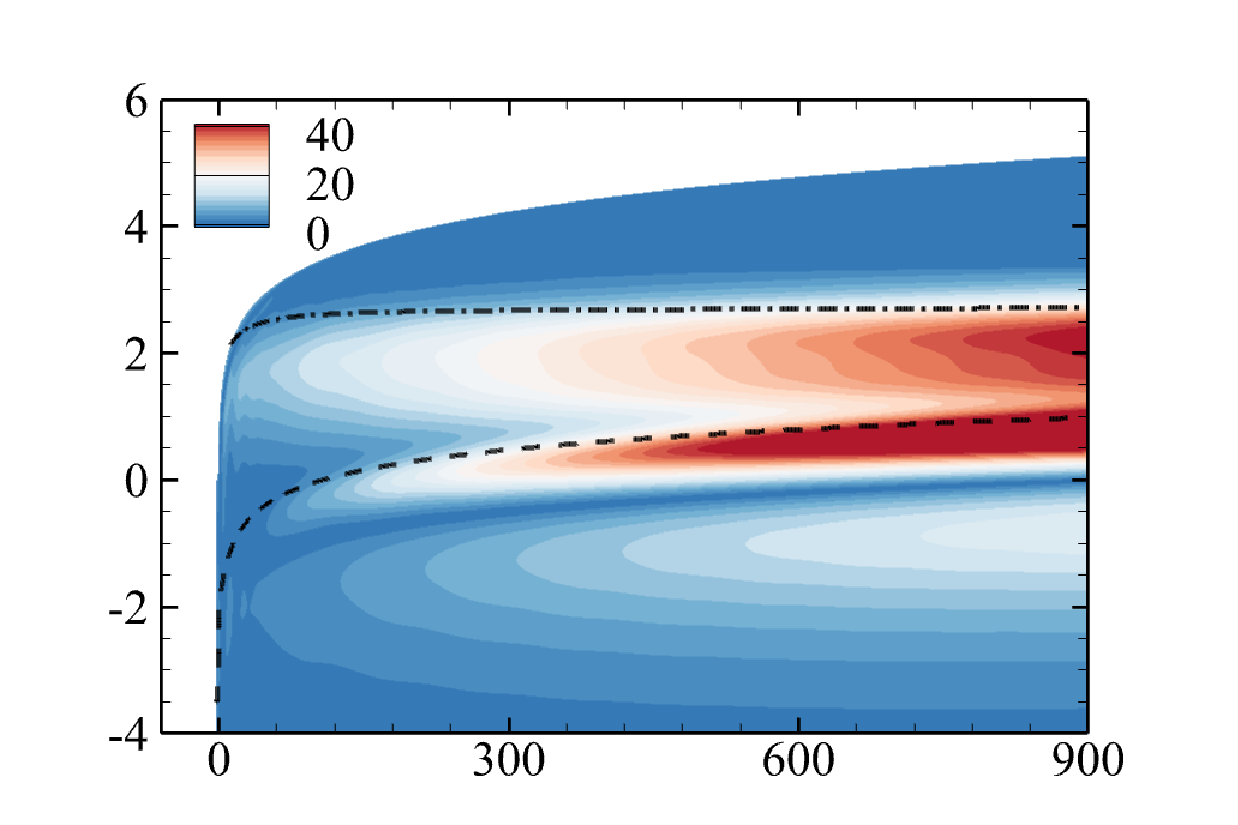}
    \put(-185,45){\rotatebox{90}{$\ln(k_3 y_n)$}}
    \put(-95,0){$k_3 x_s$}
    \put(-120,95){$|\hat T|$}
    \put(-185,110){$(d)$}
    \caption{Contours of the perturbation field for case Av with  $\omega=0$, $k_3=1.5$  and $\vartheta=15^{\circ}$.
    ($a$) is for $\hat p$ in the $k_3 x$-$k_3 y$ plane; ($b,c,d$) are for $\hat p$, $\hat u_s$ and $\hat T$ in the $k_3 x_s$-$\ln(k_3 y_n)$ plane. The dash-dotted and dashed lines mark the entropy layer and boundary layer edges, respectively. }
    \label{fig:Contour_vortex_r=1_w=0}
    \end{center}
\end{figure}
Figure \ref{fig:Contour_vortex_r=1_w=0} displays the contours of  perturbation field for case Av with $\omega=0$, $k_3=1.5$ and $\vartheta=15^\circ$. To facilitate comparison with the results for cases Bv and Cv, the coordinate system is represented by $(k_3x,k_3 y)$ or $(k_3x_s,k_3 y_n)$. The  pressure contours in panel (a) reveal a narrow band of high-pressure perturbation in the inviscid region above the entropy layer, suggesting an acoustic beam propagating in the potential flow. 
To observe clearly the near-wall behaviour of the perturbation, panel (b) displays the contours of $\hat p$ by plotting the vertical axis $k_3y_n$ in logarithmic form.
 A rather weak signature of $\hat p$ is observed  in the downstream entropy and boundary layers, indicating that non-modal perturbations do not show acoustic feature as the Mack modes do. Additionally, we have also perform HLNS calculations using a smaller computational domain, with the upper boundary positioned below the acoustic beam. 
 It is found that even when the upper boundary perturbations are set to zero, the downstream amplitude of the non-modal perturbation remains consistent with that in the present SF-HLNS calculations, indicating that the downstream evolution of the non-modal perturbation is not primarily influenced by the acoustic beam propagating in the potential region.
In contrast, the velocity perturbation $|\hat u_s|$, shown in panel ($c$), displays a dominant peak in the boundary layer and a secondary peak in the entropy layer, indicating the high-vorticity nature of the excited boundary-layer perturbation. Observing the temperature perturbation $|\hat T|$  in panel ($d$), it is seen that two peaks emerge in the boundary layer and one peak emerges in the entropy layer, with the dominant peak located at the edge of the boundary layer.
As $x_s$ approaches downstream, both $|\hat u_s|$ and $|\hat T|$ amplifies progressively, with the magnitude of   $|\hat T|$ surpassing that of $|\hat u_s|$, consistent with the observation for the entropy-layer mode depicted in figure \ref{fig:eigenf_entropymode_xs=600}. However, the observation that the dominant peaks of both $\hat u_s$ and $\hat T$ of the vorticity-induced perturbation are located inside the viscous boundary layer starkly contrasts with   the entropy-layer mode in figure \ref{fig:eigenf_entropymode_xs=600}. This contrast indicates that the boundary-layer perturbation excited  by the freestream vortical forcing  does not resemble a typical normal mode.

Figure \ref{fig:Contour_vortex_r=5-10_w=0} presents  the contours of $\hat p$ in the $k_3x$-$k_3y$ plane and $\hat T$ in the $k_3x_s$-$\ln(k_3y_n)$ plane  for cases Bv and Cv.  Comparing figures \ref{fig:Contour_vortex_r=5-10_w=0}$(a,c)$ with figure \ref{fig:Contour_vortex_r=1_w=0}$(a)$, it is evident that the origin of the acoustic beam shifts downstream as the entropy layer thickens due to the increased nose radius. Despite this shift, the acoustic signature within the boundary layer for all the cases remains relatively weak, indicating its limited influence on the evolution of boundary-layer streaks. 
Comparing the temperature perturbations among figures \ref{fig:Contour_vortex_r=1_w=0}$(b)$ and \ref{fig:Contour_vortex_r=5-10_w=0}$(b,d)$, we observe that as the nose radius expands, both the dominant peak of $\hat T$ at the boundary-layer edge and the secondary peak within the entropy layer  decrease. Conversely, the near-wall peak in the downstream region remains at the same magnitude across all cases.
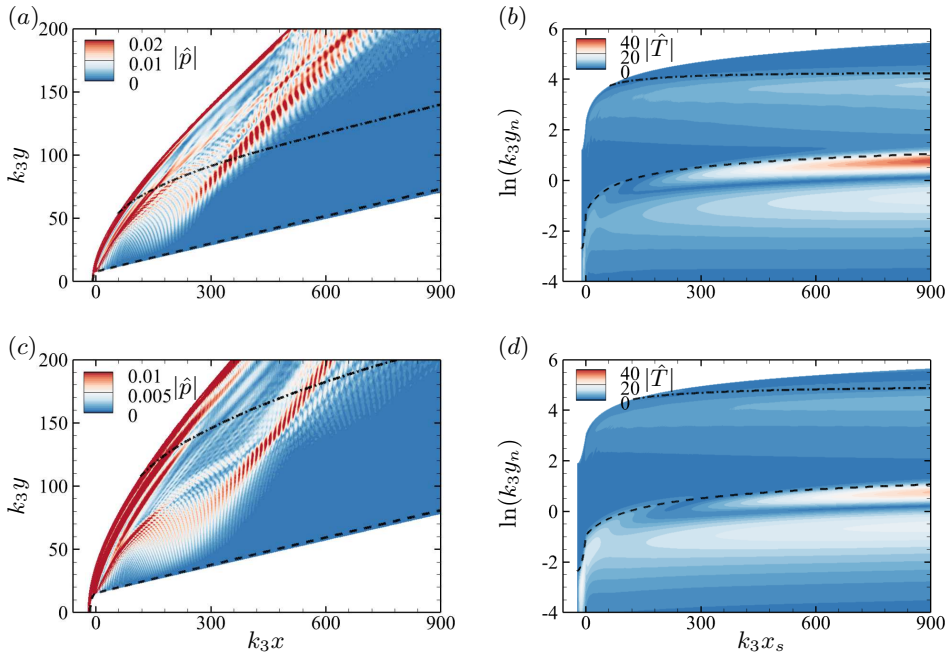
\begin{figure}
    \begin{center}
    \includegraphics[width = 0.48\textwidth] {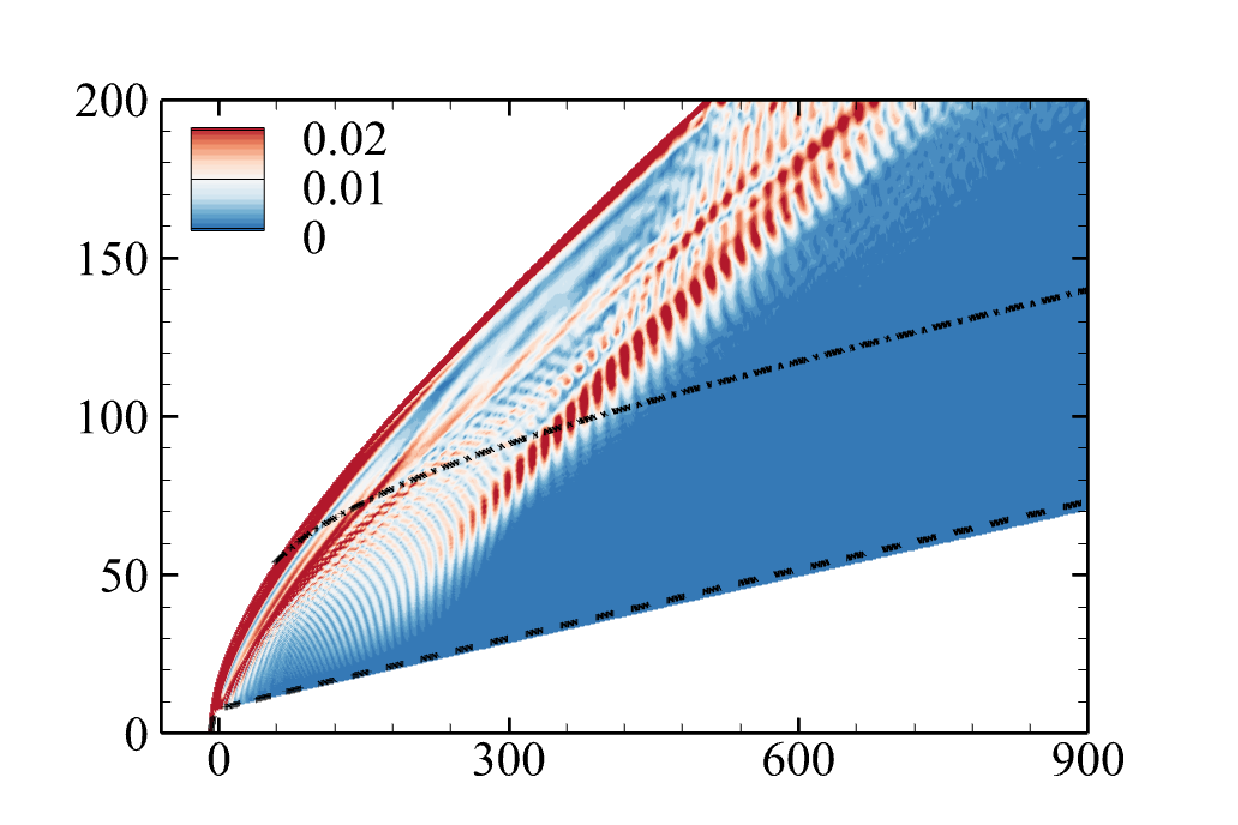}
    \put(-185,50){\rotatebox{90}{$k_3 y$}}
    \put(-123,95){$|\hat p|$}
    \put(-185,110){$(a)$}
    \includegraphics[width = 0.48\textwidth] {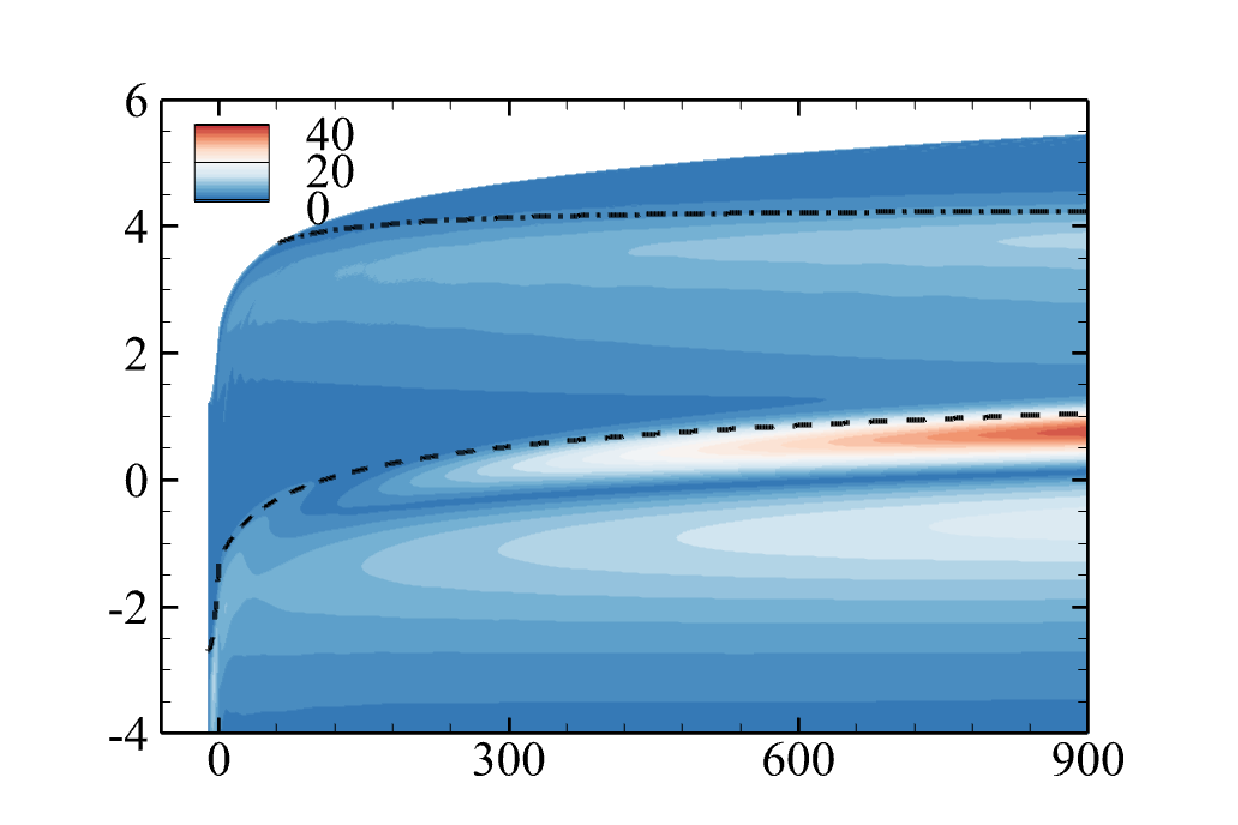}
    \put(-185,45){\rotatebox{90}{$\ln(k_3 y_n)$}}
    \put(-130,98){$|\hat T|$}
    \put(-185,110){$(b)$}\\
    \includegraphics[width = 0.48\textwidth] {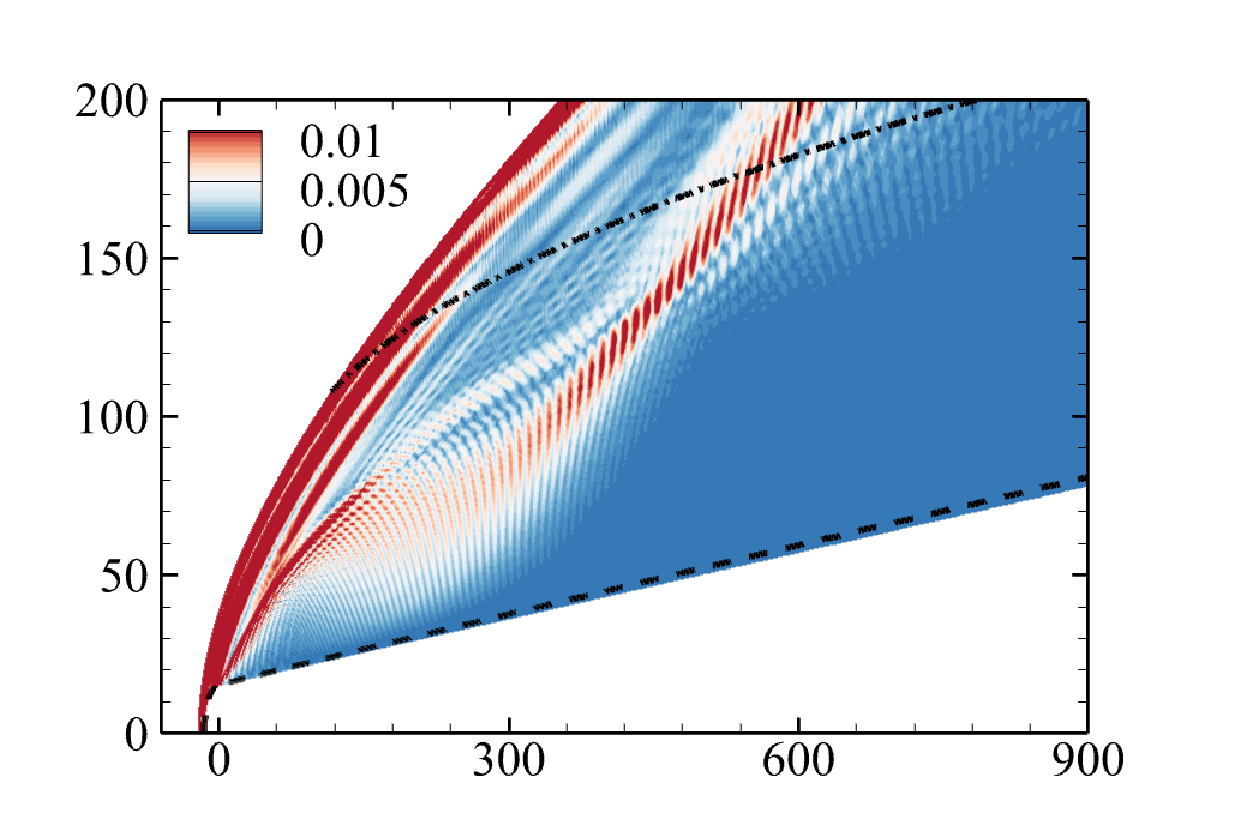}
    \put(-185,50){\rotatebox{90}{$k_3 y$}}
    \put(-95,0){$k_3 x$}
    \put(-123,95){$|\hat p|$}
    \put(-185,110){$(c)$}
    \includegraphics[width = 0.48\textwidth] {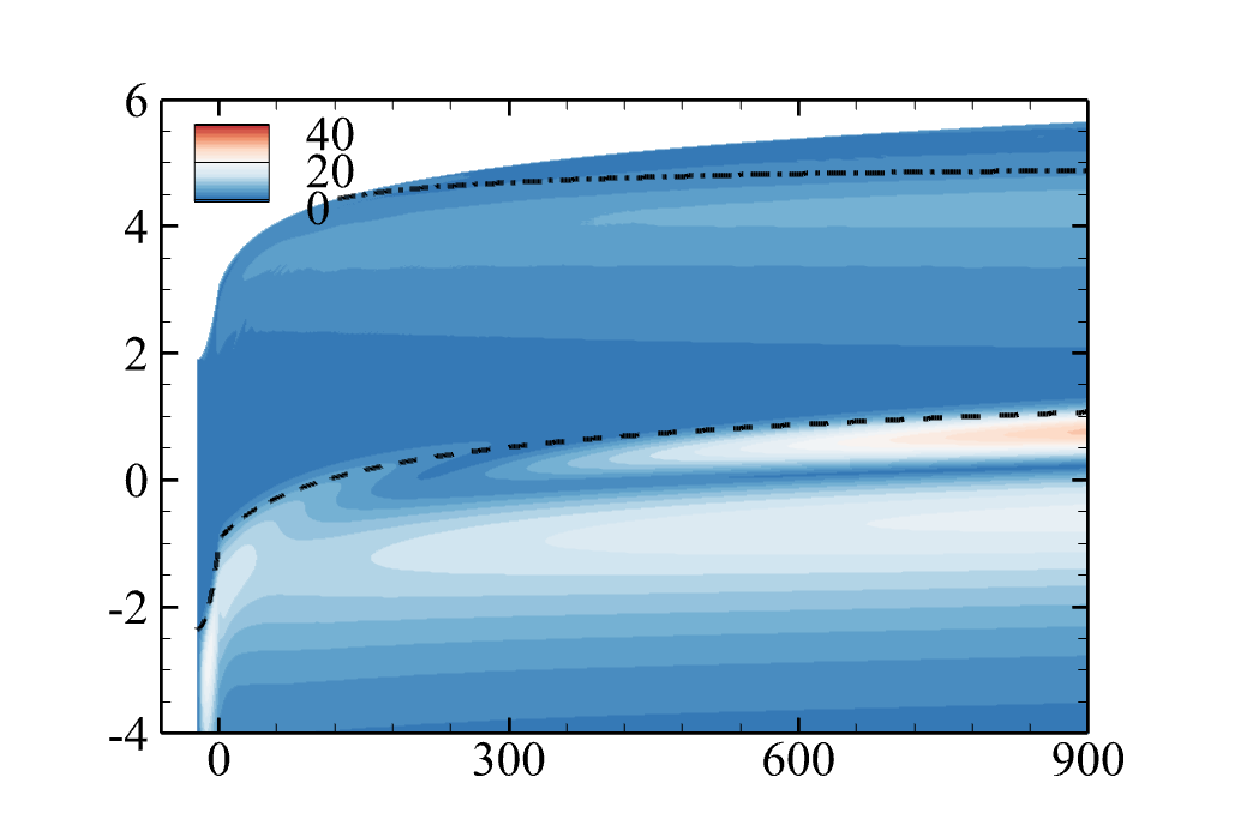}
    \put(-185,45){\rotatebox{90}{$\ln(k_3 y_n)$}}
    \put(-95,0){$k_3 x_s$}
    \put(-130,98){$|\hat T|$}
    \put(-185,110){$(d)$}
    \caption{Contours of the perturbation field for cases Bv ($a,b$) and Cv ($c,d$) with  $\omega=0$, $k_3^*=1.5$ mm$^{-1}$ and $\vartheta=15^{\circ}$. ($a,c$): $|\hat p|$ in the $k_3 x$-$k_3 y$ plane; ($b,d$): $|\hat T|$ in the $k_3 x_s$-$\ln(k_3 y_n)$ plane. The dash-dotted and dashed lines mark the entropy layer and boundary layer, respectively.}
    \label{fig:Contour_vortex_r=5-10_w=0}
    \end{center}
\end{figure}

The wall-normal profiles of $\hat u_s$ and $\hat T$ at $k_3x_s=900$ for cases Av, Bv and Cv are compared in figure \ref{fig:profile_vortex_2}. The profile of $\hat u_s$ exhibits a notable similarity in both the magnitude and the distribution for cases Bv and Cv, while the profile for case Av displays higher values at the peak locations in both the boundary layer and the entropy layer. 
In contrast, the amplitude of $\hat T$ reduces consistently as the nose radius expands, particularly in the regions of the boundary-layer edge and the entropy-layer core. Once again, these profiles show great distinction from those of the entropy-layer mode depicted in figure \ref{fig:eigenf_entropymode_xs=600}.
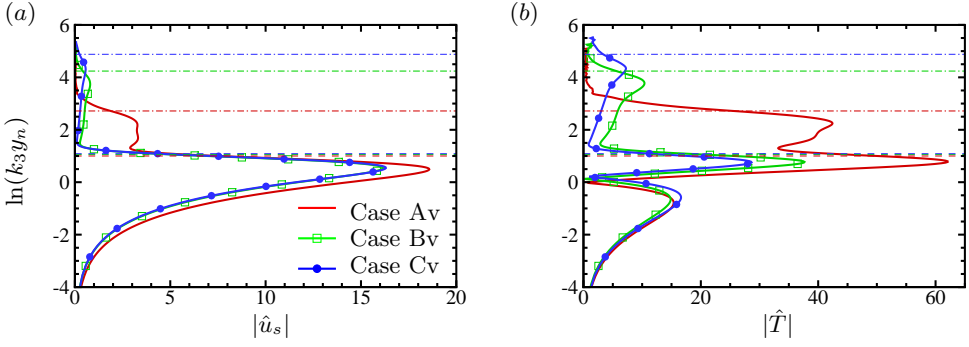
\begin{figure}
    \begin{center}
    \includegraphics[width = \textwidth] {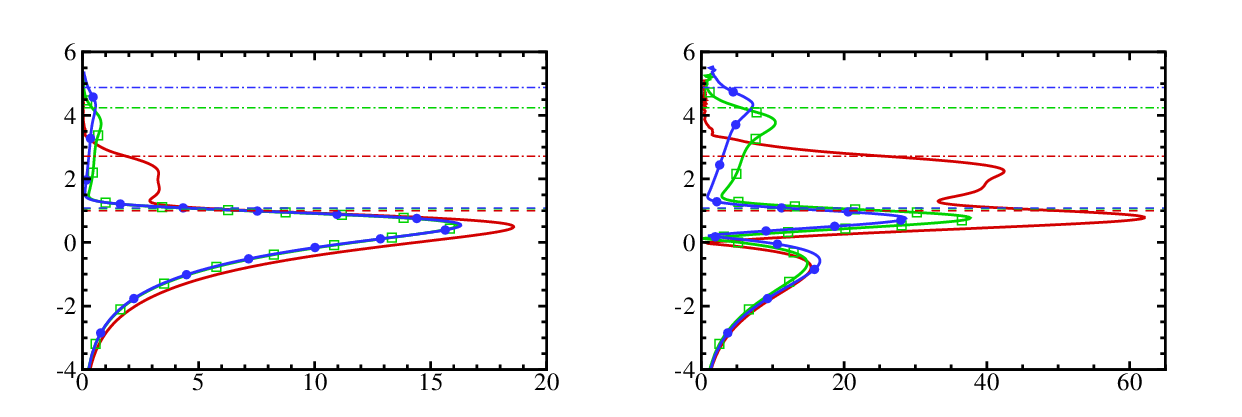}
    \put(-385,45){\rotatebox{90}{$\ln(k_3 y_n)$}}
    \put(-292,-2){$|\hat u_s|$}
    \put(-275,20){\begin{tikzpicture}
    \draw[red,thick] (0,0) -- (0.5,0);
    \draw[green,thick] (0,-0.4) -- (0.5,-0.4);
    \node[draw,green,rectangle,inner sep=1.5pt] at (0.25,-0.4) {};
    \draw[blue,thick] (0,-0.8) -- (0.5,-0.8);
    \node[draw,blue,circle,fill,inner sep=1pt] at (0.25,-0.8) {};
    \end{tikzpicture}}
    \put(-255,40){Case Av}
    \put(-255,30){Case Bv}
    \put(-255,20){Case Cv}
    \put(-385,115){$(a)$}
    \put(-100,-2){$|\hat T|$}
    \put(-195,115){$(b)$}
    \caption{Wall-normal profiles of $|\hat u_s|$ ($a$) and $|\hat T|$ ($b$)  for cases Av, Bv and Cv with $\omega=0$, $k_3^*=1.5$ mm$^{-1}$ and $\vartheta=15^{\circ}$ at $k_3 x_s=900$. The dash-dotted and dashed lines mark the entropy-layer and boundary-layer edges, respectively.}
    \label{fig:profile_vortex_2}
    \end{center}
\end{figure}

\begin{figure}
    \begin{center}
    \includegraphics[width = \textwidth] {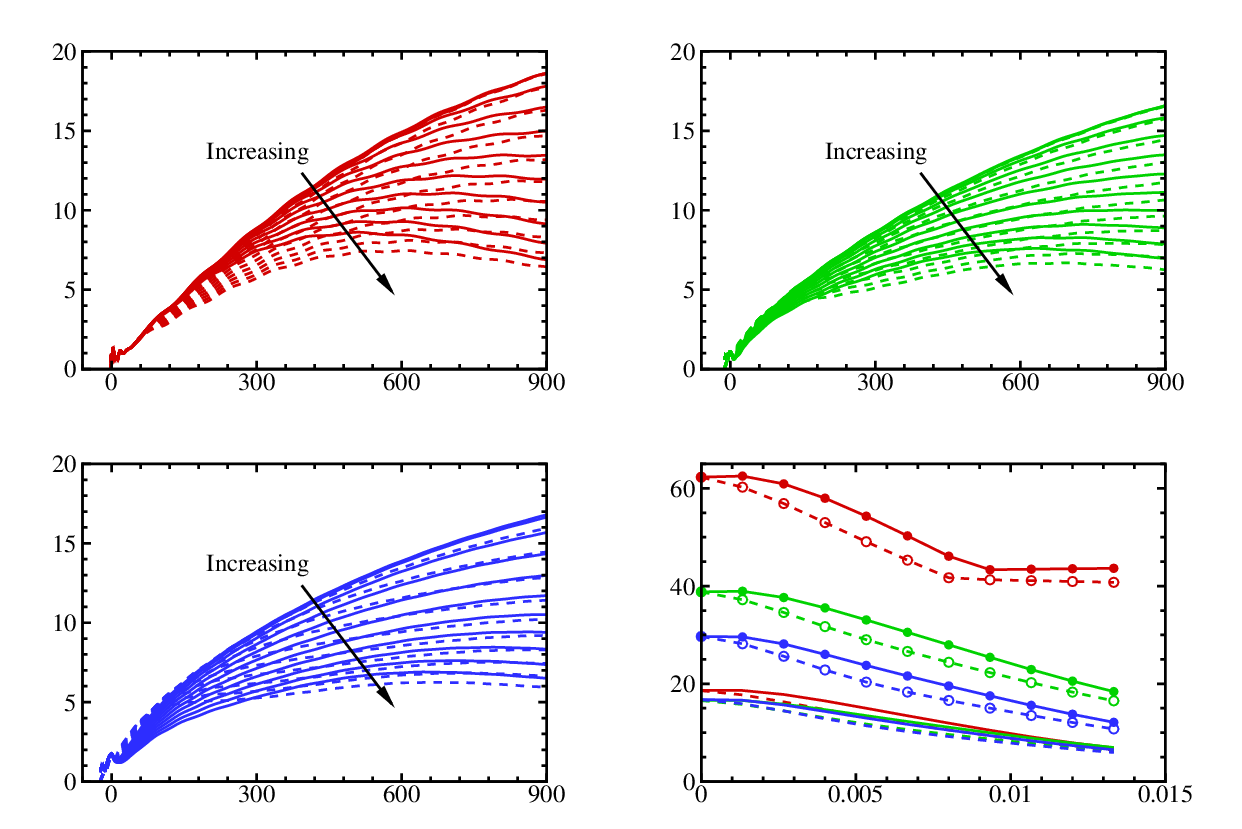}
    \put(-380,240){$(a)$}
    \put(-295,128){$k_3 x_s$}
    \put(-383,190){$A_u$}
    \put(-285,207){$\omega$}
    \put(-190,240){$(b)$}
    \put(-105,128){$k_3 x_s$}
    \put(-193,190){$A_u$}
    \put(-93,207){$\omega$}
    \put(-380,120){$(c)$}
    \put(-295,0){$k_3 x_s$}
    \put(-383,62){$A_u$}
    \put(-285,79){$\omega$}
    \put(-190,120){$(d)$}
    \put(-100,1){$\omega/k_3$}
    \put(-193,50){\rotatebox{90}{$A_u$, $A_T$}}
    \put(-240,285){Case Av}
    \put(-190,285){Case Bv}
    \put(-140,285){Case Cv}
    \put(-300,273){$A_u$}
    \put(-300,253){$A_T$}
    \put(-280,276){$\vartheta=15^\circ$}
    \put(-280,268){$\vartheta=-15^\circ$}  
    \put(-280,256){$\vartheta=15^\circ$}
    \put(-280,248){$\vartheta=-15^\circ$} 
    \put(-285,247){\begin{tikzpicture}
    \draw[decorate, decoration={brace, amplitude=3pt, mirror}, thick](0,0) -- (0,-0.5);
    \draw[decorate, decoration={brace, amplitude=3pt, mirror}, thick](0,-0.7) -- (0,-1.2);
    \end{tikzpicture}}
    \put(-230,250){\begin{tikzpicture}
    \draw[red,thick] (0,0) -- (0.5,0);
    \draw[green,thick] (1.75,0) -- (2.25,0);
    \draw[blue,thick] (3.5,0) -- (4.0,0);
    \draw[red,thick,dashed] (0,-0.3) -- (0.5,-0.3);
    \draw[green,thick,dashed] (1.75,-0.3) -- (2.25,-0.3);
    \draw[blue,thick,dashed] (3.5,-0.3) -- (4.0,-0.3);
    \draw[red,thick] (0,-0.6) -- (0.5,-0.6);
    \node[draw,red,circle,fill,inner sep=1.2pt] at (0.25,-0.6) {};
    \draw[green,thick] (1.75,-0.6) -- (2.25,-0.6);
    \node[draw,green,circle,fill,inner sep=1.2pt] at (2.0,-0.6) {};
    \draw[blue,thick] (3.5,-0.6) -- (4.0,-0.6);
    \node[draw,blue,circle,fill,inner sep=1.2pt] at (3.75,-0.6) {};
    \draw[red,thick,dashed] (0,-0.9) -- (0.5,-0.9);
    \node[draw,red,circle,inner sep=1.2pt] at (0.25,-0.9) {};
    \draw[green,thick,dashed] (1.75,-0.9) -- (2.25,-0.9);
    \node[draw,green,circle,inner sep=1.2pt] at (2.0,-0.9) {};
    \draw[blue,thick,dashed] (3.5,-0.9) -- (4.0,-0.9);
    \node[draw,blue,circle,inner sep=1.2pt] at (3.75,-0.9) {};
    \end{tikzpicture}}
    \caption{Impact of the frequency on the perturbation evolution. (a-c):
    Streamwise evolution of $A_u$ for cases Av, Bv and Cv, respectively, where $\omega/k_3=0,0.0013,0.0027,0.004,0.0053,0.0067,0.008,0.0093,0.0107,0.012$ and $0.0133$.   (d): Dependence  on $\omega/k_3$ of $A_u$ (lines) and $A_T$ (symbolized lines) at $k_3 x_s=900$, respectively.
    In these plots, we fix $k_3^*=1.5$ mm$^{-1}$. }
    \label{fig:LNS_A-k1_vortex}
    \end{center}
\end{figure}
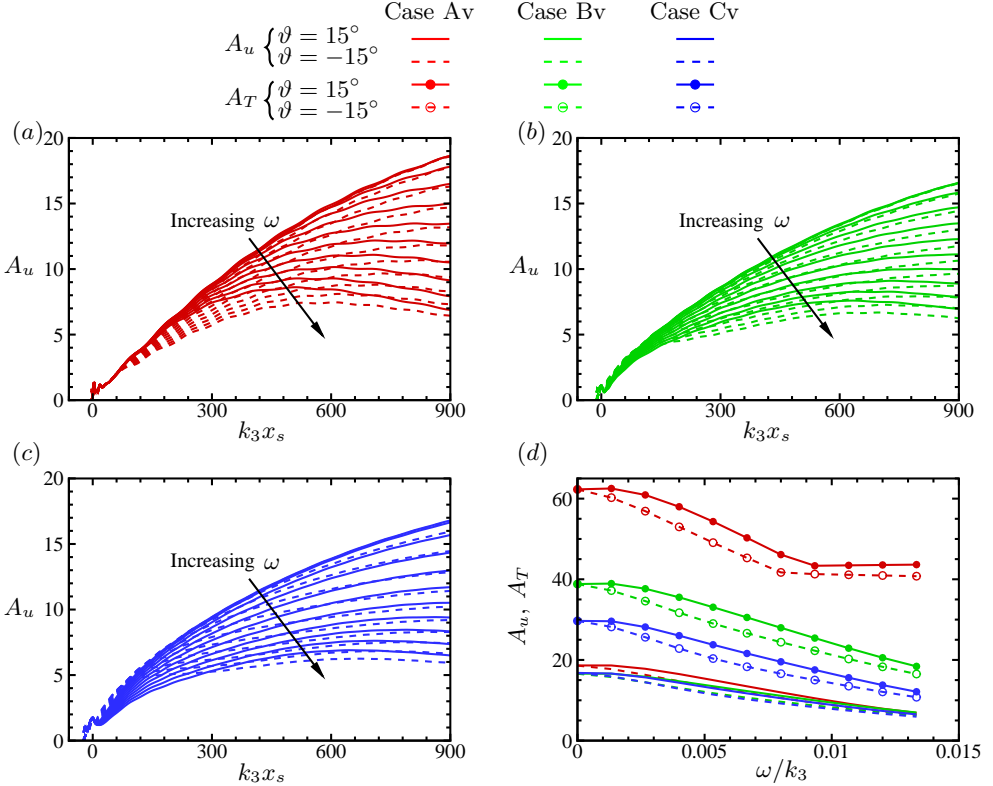
Figures \ref{fig:LNS_A-k1_vortex}$(a$-$c)$ show the streamwise evolution of the perturbation amplitude $A_u$ for different nose radii, considering different frequencies  at both $\vartheta=15^\circ$ and $-15^\circ$ configurations.  Overall, across each nose radius, a higher frequency leads to an earlier saturation point, with the perturbation amplitude decreasing as $\omega$ increases. As a consequence, the curves for $\omega=0$ acquire the greatest downstream amplitude  among curves for various frequencies.
Notably, the amplitude for a positive $\vartheta$  marginally surpasses that for a negative $\vartheta$. The solid lines in panel ($d$) summarises the curves from panels ($a$-$c$) at the same position $k_3x_s=900$, corresponding to the same dimensional position $x_s^*=600$mm, while  the symbolised lines in panel ($d$) further illustrate the variation of the temperature perturbation $A_T(x_s)=\displaystyle \max_{y_n}|\hat T(x_s,y_n)|$ at $x_s=900/k_3$ for the three cases. These curves unveil a consistent decrease in downstream amplitude with an increase in $\omega$. Moreover, for a vortical forcing with given frequency, receptivity proves to be the most effective for the smallest nose radius, particularly when measured by  $A_T$. Upon revisiting the perturbation field depicted in figures \ref{fig:Contour_vortex_r=1_w=0} and \ref{fig:Contour_vortex_r=5-10_w=0}, we observe that although the temperature perturbation for case Cv exhibits a higher value in the near-wall nose region, its downstream magnitude, determined by the peak at outer reach of the boundary layer, is the lowest among the three cases.

\begin{figure}
    \begin{center}
    \includegraphics[width = 0.48\textwidth] {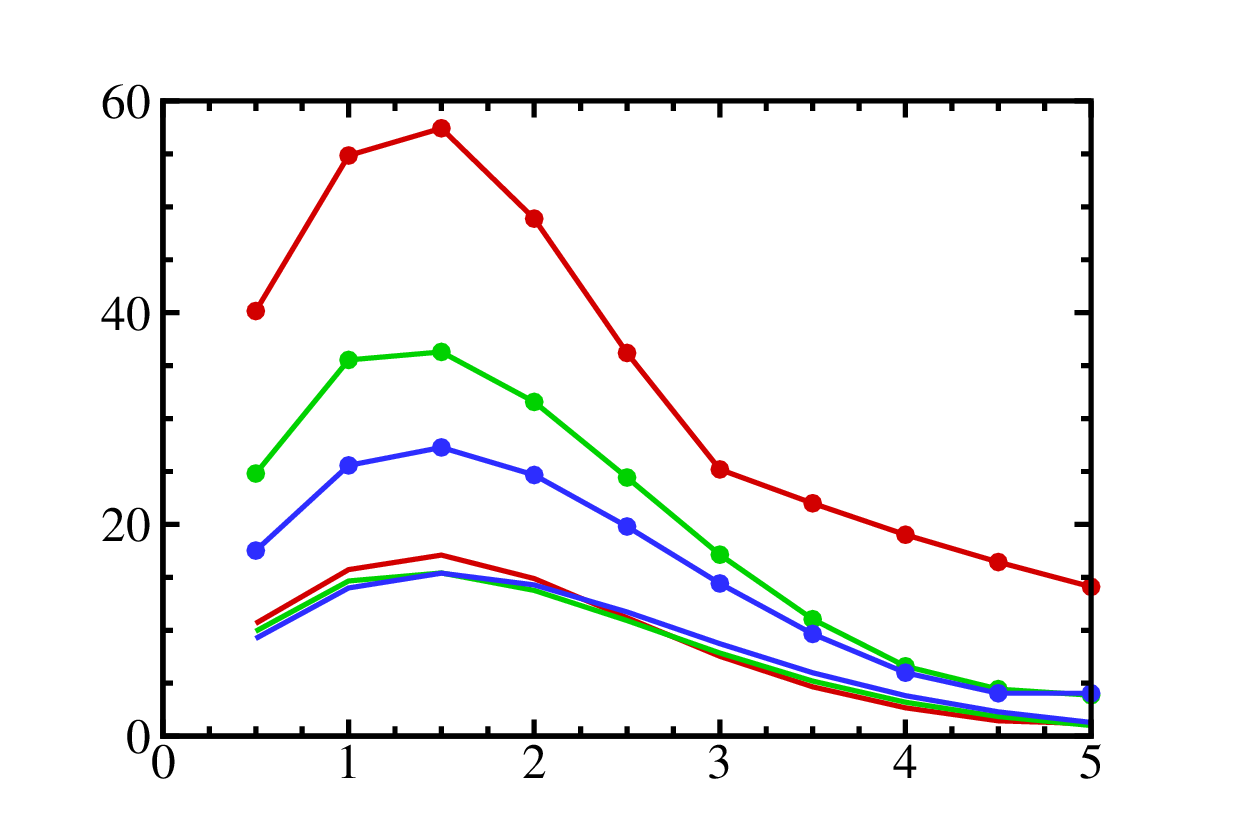}
    \includegraphics[width = 0.48\textwidth] {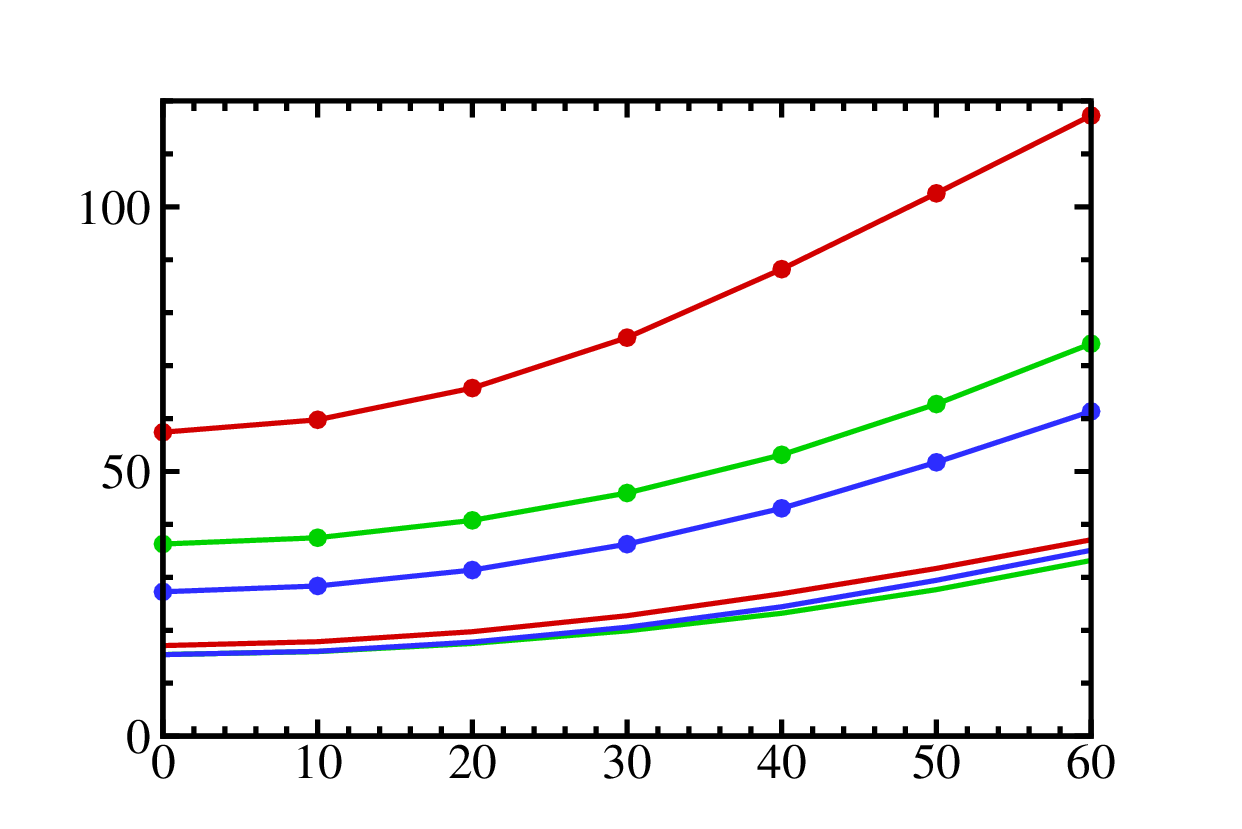}
    \put(-380,110){$(a)$}
    \put(-185,110){$(b)$}
    \put(-280,0){$k_3^*$}
    \put(-225,0){(mm$^{-1}$)}
    \put(-372,50){\rotatebox{90}{$A_u$, $A_T$}}
    \put(-95,0){$\vartheta$}
    \put(-185,50){\rotatebox{90}{$A_u$, $A_T$}}
    \put(-28,0){($^{\circ}$)}
    \put(-260,130){$A_u$}
    \put(-260,120){$A_T$}  
    \put(-240,140){Case Av}
    \put(-190,140){Case Bv}
    \put(-140,140){Case Cv}
    \put(-230,120){\begin{tikzpicture}
    \draw[red,thick] (0,0) -- (0.5,0);
    \draw[green,thick] (1.75,0) -- (2.25,0);
    \draw[blue,thick] (3.5,0) -- (4.0,0);
    \draw[red,thick] (0,-0.4) -- (0.5,-0.4);
    \node[draw,red,circle,fill,inner sep=1pt] at (0.25,-0.4) {};
    \draw[green,thick] (1.75,-0.4) -- (2.25,-0.4);
    \node[draw,green,circle,fill,inner sep=1pt] at (2.0,-0.4) {};
    \draw[blue,thick] (3.5,-0.4) -- (4.0,-0.4);
    \node[draw,blue,circle,fill,inner sep=1pt] at (3.75,-0.4) {};
    \end{tikzpicture}}
    \caption{{The amplitude of the excited perturbation $A_u$ (lines) and $A_T$ (symbolised lines)} at $x_s^*=600$ mm, for cases Av, Bv and Cv. ($a$): Dependence on $k_3^*$ for $\omega=0$ and $\vartheta=0$; ($b$): Dependence on $\vartheta$ for $\omega=0$ and $k_3^*=1.5$ mm$^{-1}$.}
    \label{fig:LNS_A-k3_vortex}
    \end{center}
\end{figure}
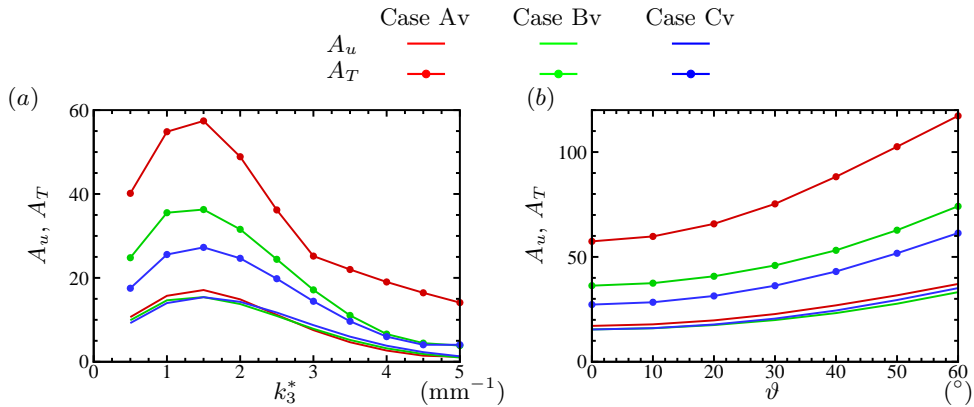

For the three cases subject to freestream vortical disturbances, figure \ref{fig:LNS_A-k3_vortex}-($a$) and ($b$) display the variations of the downstream amplitudes  with $k_3^*$ and $\vartheta$, respectively. Again, the differences of $A_u$ among the various curves in each panel are not substantial, but notable distinctions emerge when assessed using $A_T$. The curves representing case Av exhibit the most pronounced amplification. Overall, the optimal dimensional spanwise wavenumber is around 1.5mm$^{-1}$, and the receptivity efficiency increases with $\vartheta$. 

\subsubsection{Excitation of non-modal perturbations subject to freestream acoustic disturbance}
\label{sec:excitation_acoustic}
Figures \ref{fig:Contour_fast-slow_r=1_w=0_fig1}($a,c,e$) and ($b,d,f$) present  the contours of the perturbation field for cases Af and As, respectively, with $\omega=0$, $k_3=1.5$ and $\vartheta=15^{\circ}$.  In panel ($a$), the pressure perturbation for case Af illustrates the propagation of an acoustic beam in the potential region above the entropy layer, resembling that of case Av. The angles of the acoustic beams, defined by the angle between the beam direction and the 
$x$ axis, are both approximately $9^\circ$ 
for these two cases. In contrast, the acoustic field for case As, depicted in panel ($b$), exhibits a distinct feature where a series of acoustic beams emerge in the potential region, with the intensity being one order of magnitude greater than that for case Af. The  intensity of the acoustic beams  diminishes as the shock wave is approached. 
Being similar to the receptivity to freestream vortical perturbations, a rather weak pressure signature is observed within the entropy and boundary layers.
Upon comparing panels ($c,e$) with ($d,f$), we find that the  temperature and velocity perturbations for cases Af and As are  almost indistinguishable. 
This confirms that the acoustic field in the post-shock potential region has minimal impact on the evolution of non-modal perturbations.
The velocity perturbation $\hat u_s$ displays a dominant peak in the bulk boundary-layer region, whereas  the temperature perturbation $\hat T$ displays  a double peak feature in the boundary layer, with the peak at the boundary-layer edge being more pronounced. A notable distinction from figure \ref{fig:Contour_vortex_r=1_w=0} is the absence of the perturbation peaks in the entropy layer above the boundary layer. 
\begin{figure}
    \begin{center}
    \includegraphics[width = 0.48\textwidth] {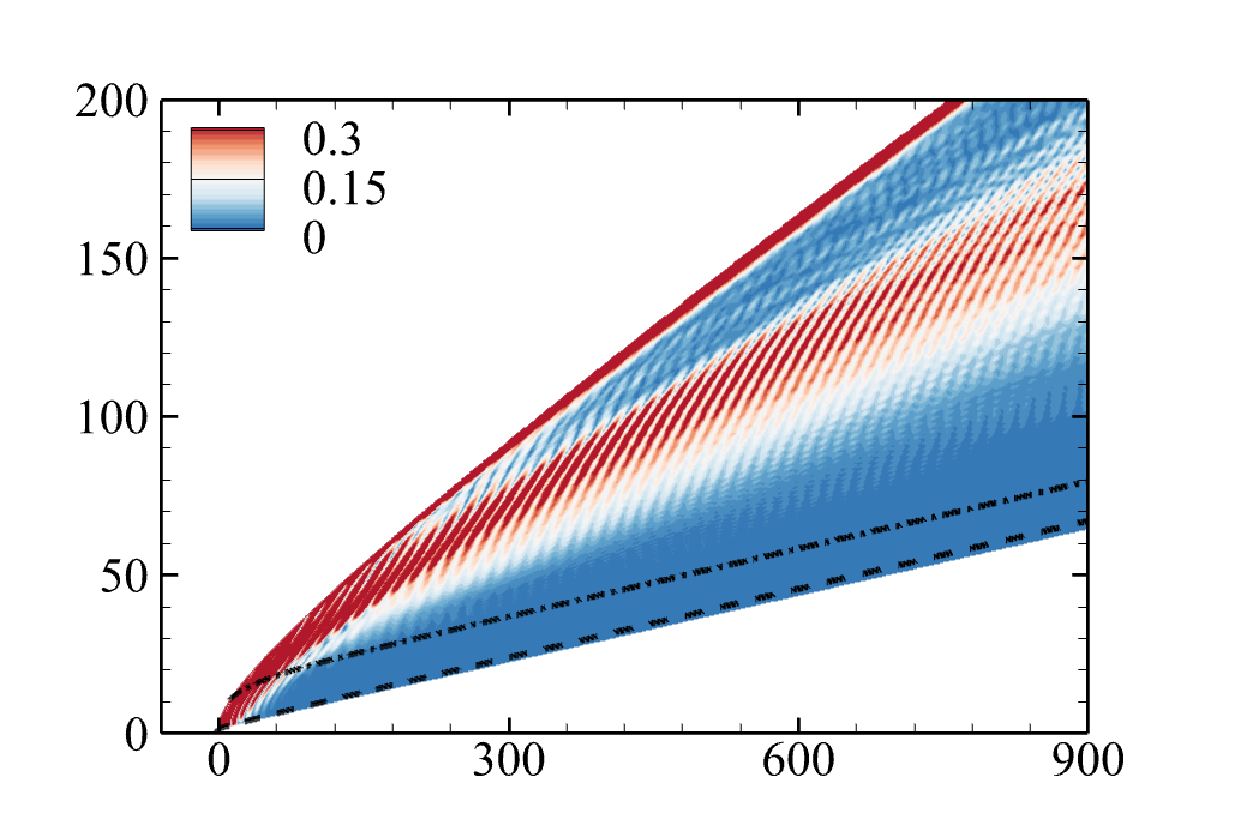}
    \put(-185,50){\rotatebox{90}{$k_3 y$}}
    \put(-95,0){$k_3 x$}
    \put(-120,95){$|\hat p|$}
    \put(-185,110){$(a)$}
    \includegraphics[width = 0.48\textwidth] {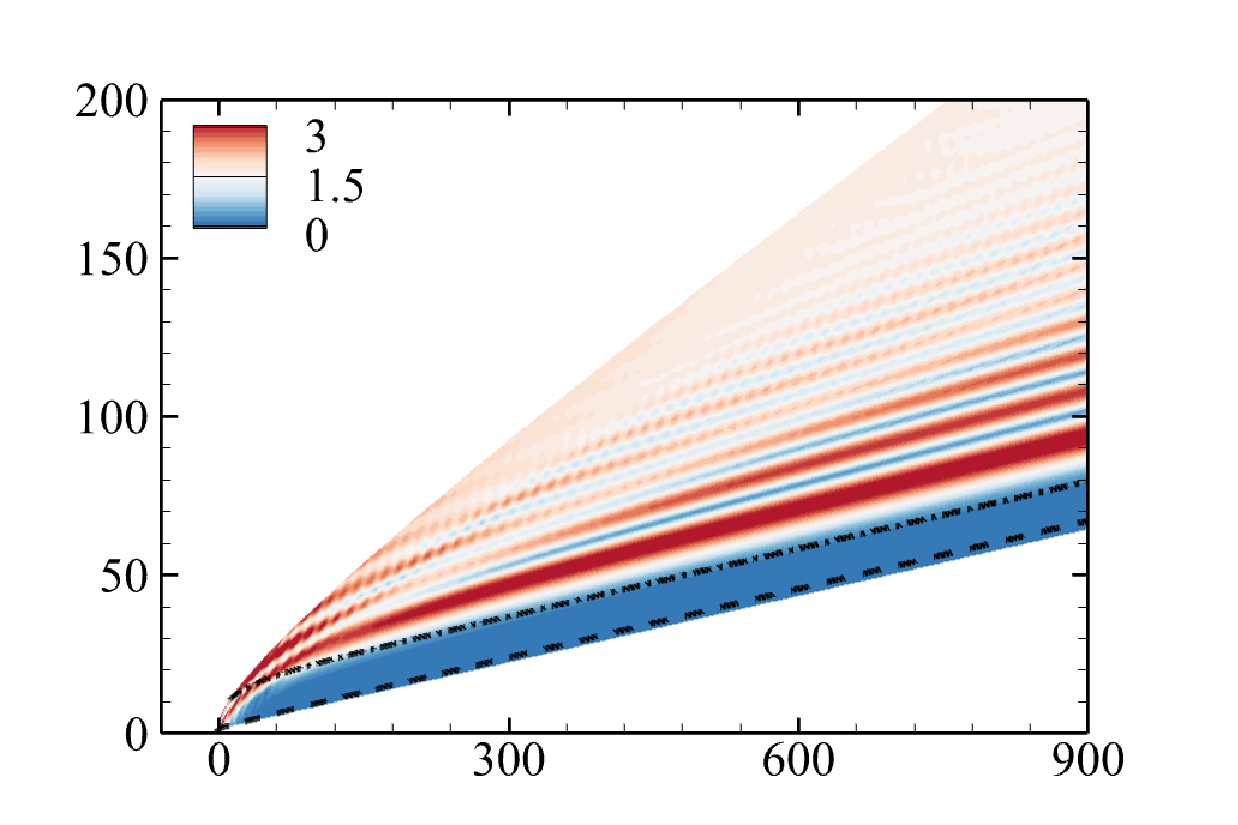}
    \put(-185,50){\rotatebox{90}{$k_3 y$}}
    \put(-95,0){$k_3 x$}
    \put(-120,95){$|\hat p|$}
    \put(-185,110){$(b)$}\\
    \includegraphics[width = 0.48\textwidth] {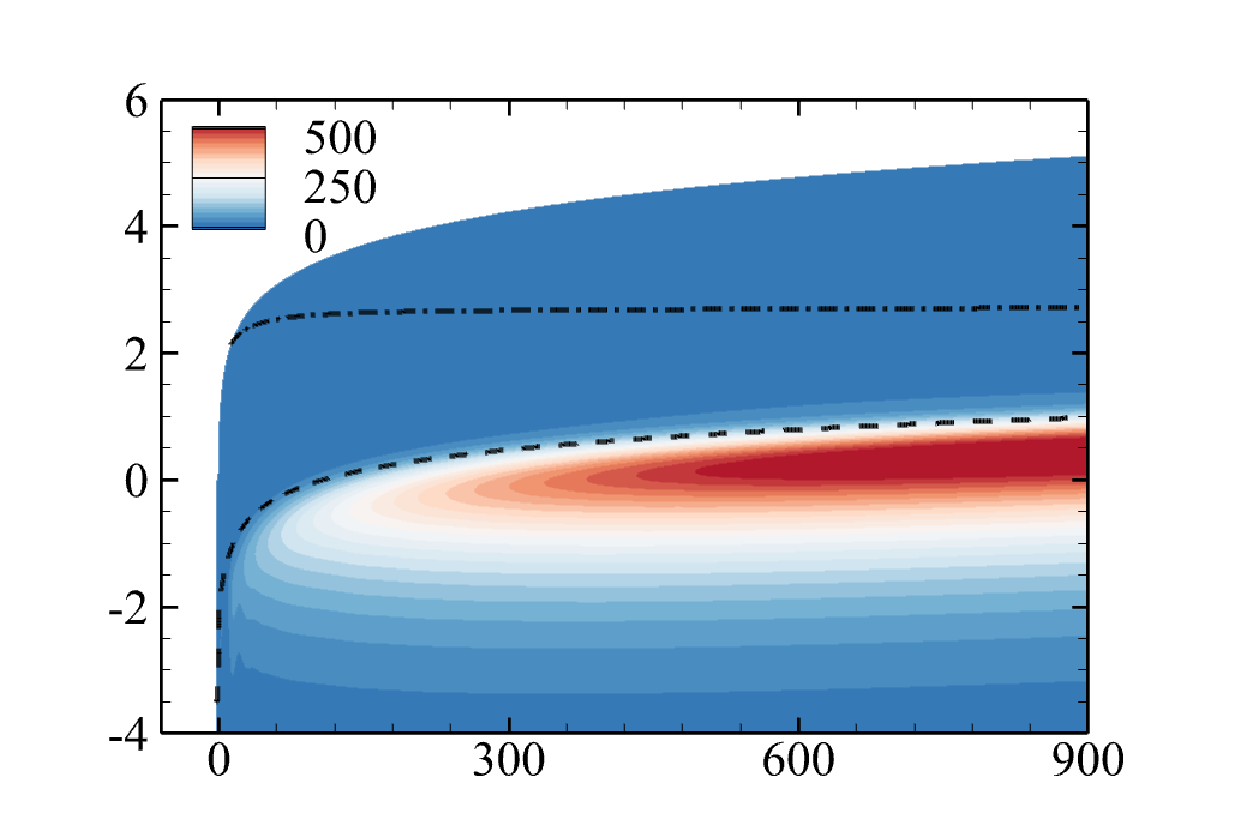}
    \put(-185,45){\rotatebox{90}{$\ln(k_3 y_n)$}}
    \put(-95,0){$k_3 x_s$}
    \put(-120,95){$|\hat u_s|$}
    \put(-185,110){$(c)$}
    \includegraphics[width = 0.48\textwidth] {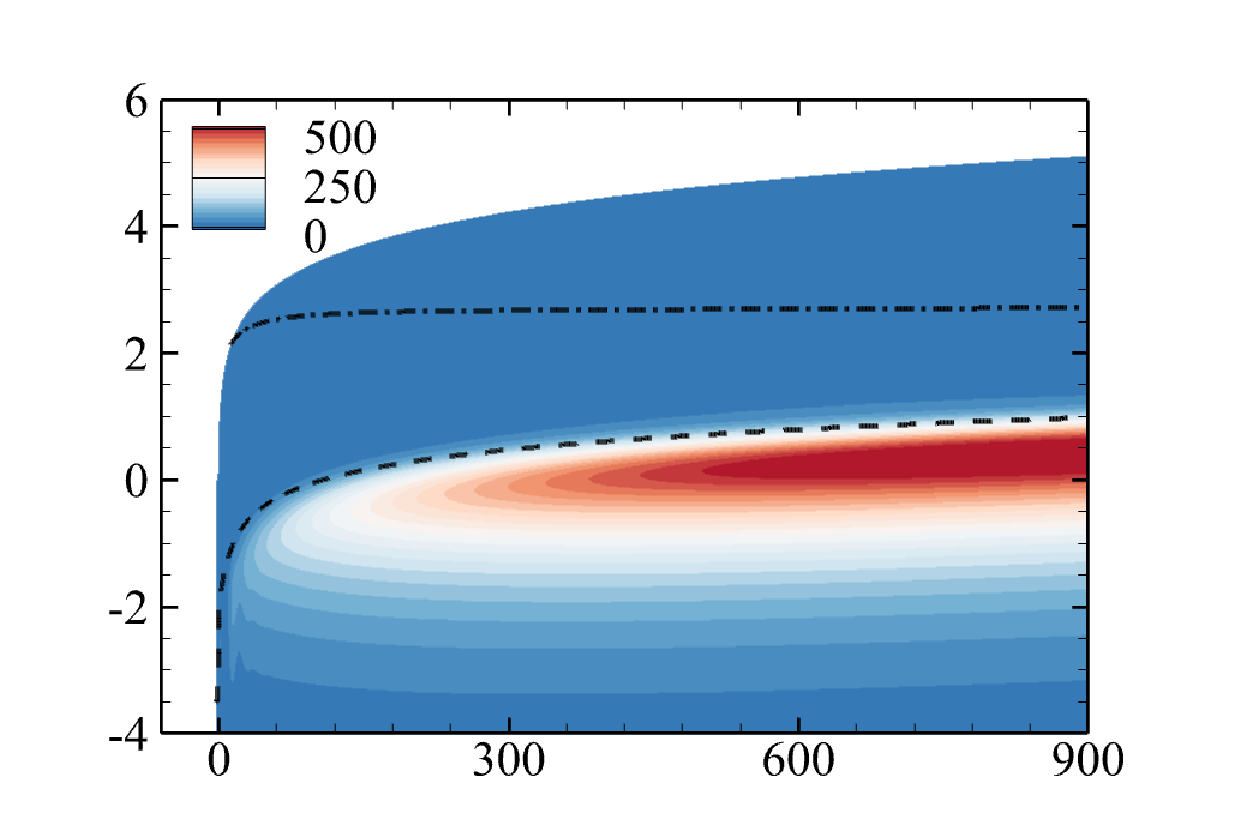}
    \put(-185,45){\rotatebox{90}{$\ln(k_3 y_n)$}}
    \put(-95,0){$k_3 x_s$}
    \put(-120,95){$|\hat u_s|$}
    \put(-185,110){$(d)$}\\
    \includegraphics[width = 0.48\textwidth] {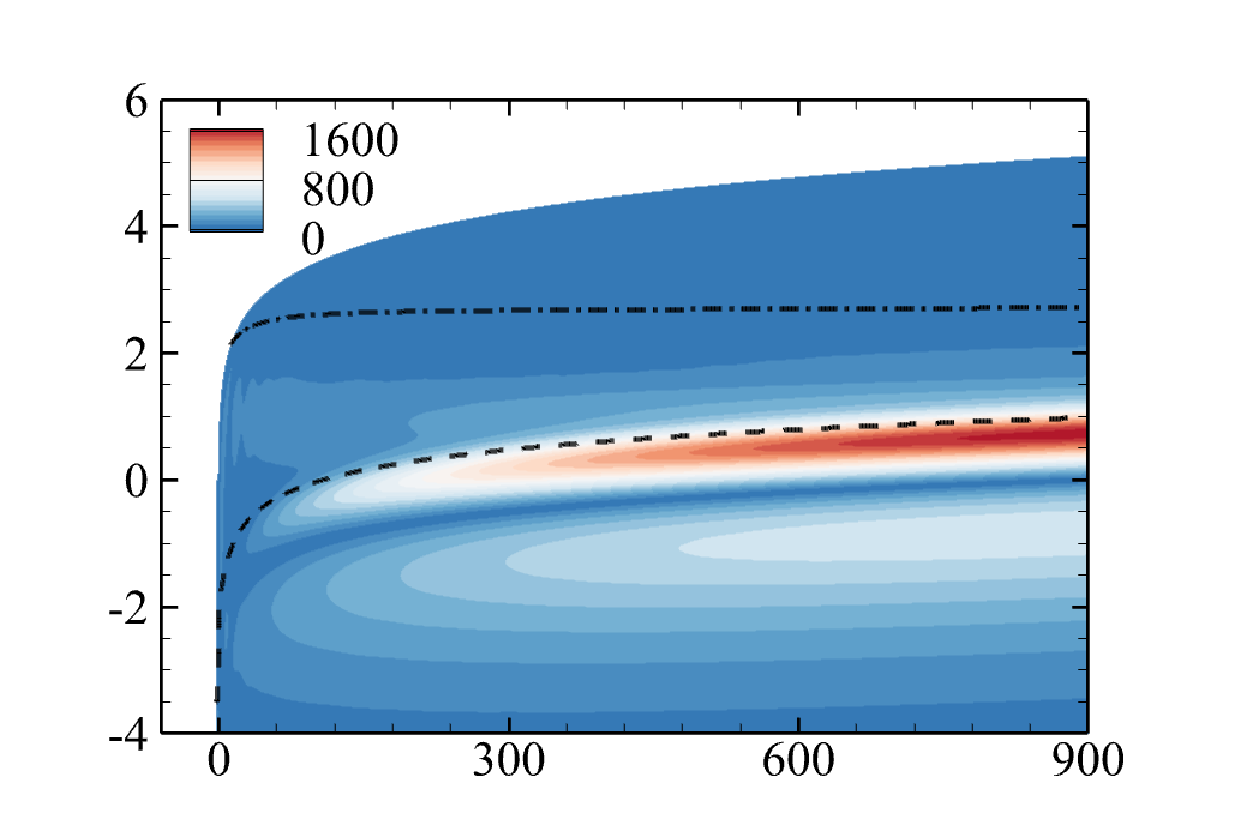}
    \put(-185,45){\rotatebox{90}{$\ln(k_3 y_n)$}}
    \put(-95,0){$k_3 x_s$}
    \put(-120,95){$|\hat T|$}
    \put(-185,110){$(e)$}
    \includegraphics[width = 0.48\textwidth] {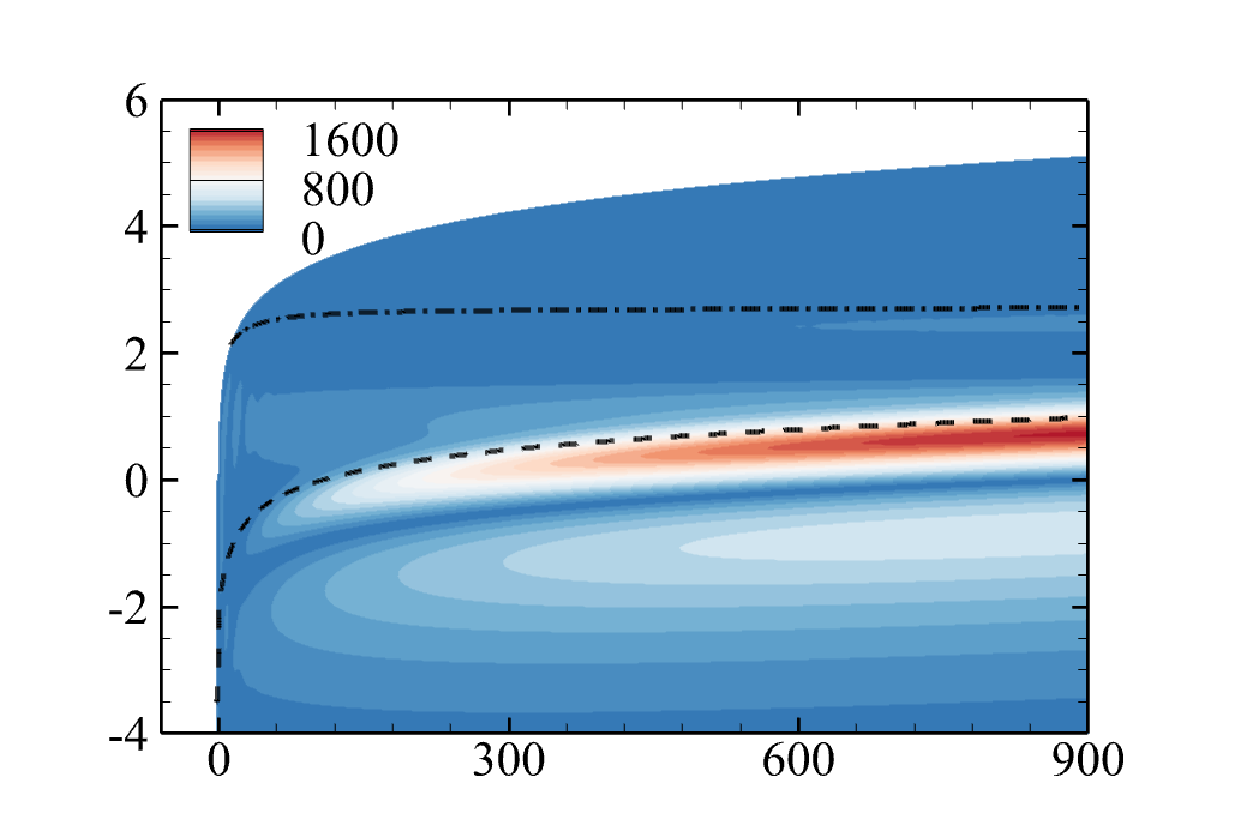}
    \put(-185,45){\rotatebox{90}{$\ln(k_3 y_n)$}}
    \put(-95,0){$k_3 x_s$}
    \put(-120,95){$|\hat T|$}
    \put(-185,110){$(f)$}
    \caption{Contours of the perturbation field for cases Af ($a,c,e$) and As ($b,d,f$) with $\omega=0$, $k_3^*=1.5$ mm$^{-1}$ and $\vartheta=15^{\circ}$.
    ($a,b$) are for $\hat p$ in the $k_3 x$-$k_3 y$ plane; ($c,d$) and ($e,f$) are $\hat u_s$ and $\hat T$ in the $k_3 x_s$-$\ln(k_3 y_n)$ plane, respectively. The dash-dotted and dashed lines mark the entropy-layer and boundary-layer edges, respectively.}
    \label{fig:Contour_fast-slow_r=1_w=0_fig1}
    \end{center}
\end{figure}

\begin{figure}
    \begin{center}
    \includegraphics[width = 0.48\textwidth] {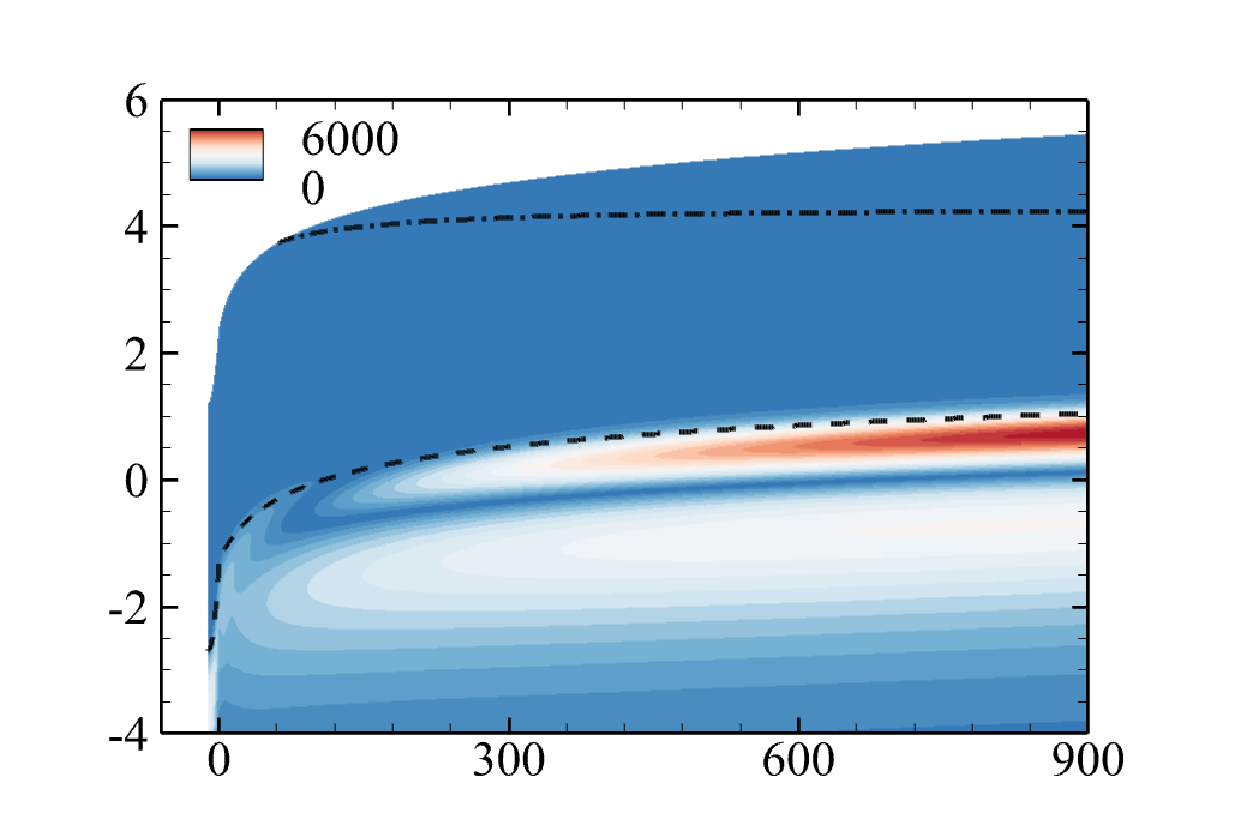}
    \put(-185,45){\rotatebox{90}{$\ln(k_3 y_n)$}}
    \put(-95,0){$k_3 x_s$}
    \put(-120,98){$|\hat T|$}
    \put(-185,110){$(a)$}
    \includegraphics[width = 0.48\textwidth] {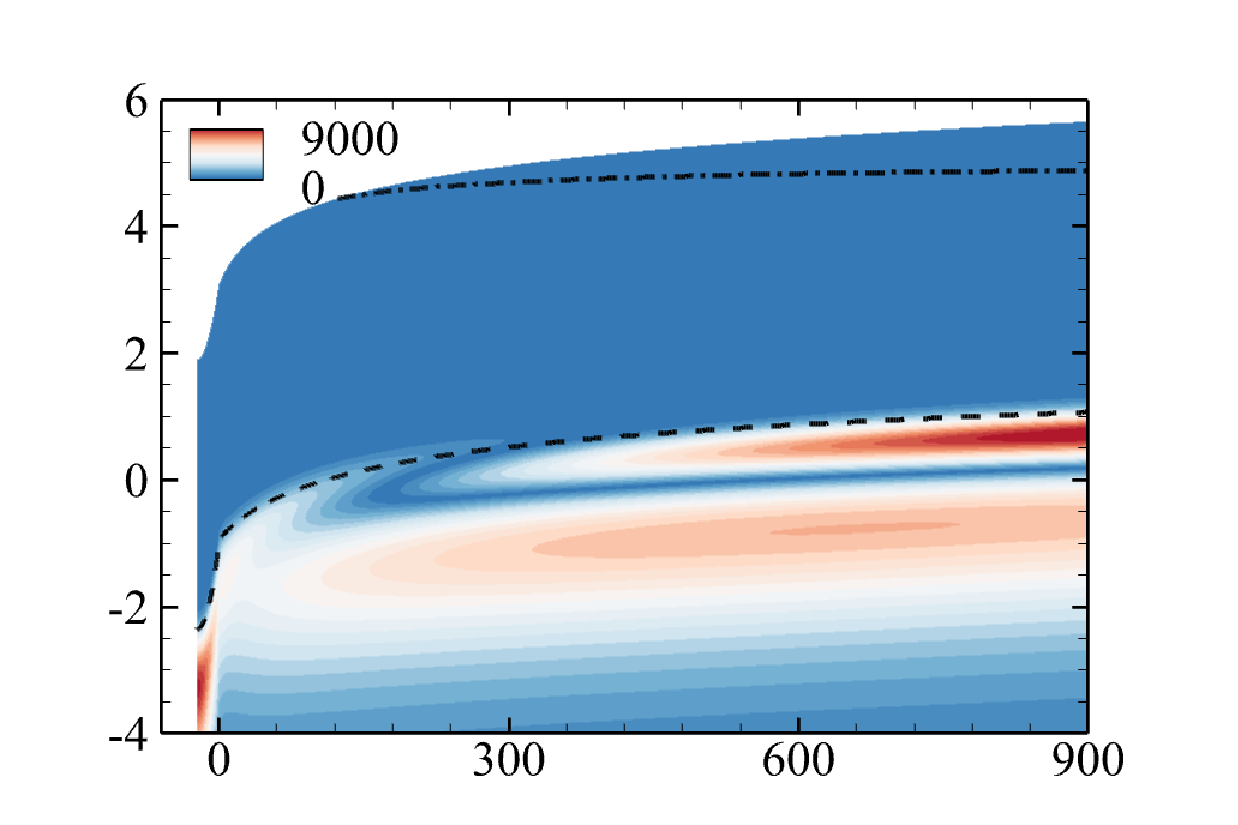}
    \put(-185,45){\rotatebox{90}{$\ln(k_3 y_n)$}}
    \put(-95,0){$k_3 x_s$}
    \put(-120,98){$|\hat T|$}
    \put(-185,110){$(b)$}
  \caption{{Contours of the perturbation field $\hat T$ for cases {Bf ($a$) and Cf ($b$)} with $\omega=0$, $k_3^*=1.5$ mm$^{-1}$ and $\vartheta=15^{\circ}$.  The dash-dotted and dashed lines mark the entropy-layer and boundary-layer edges, respectively.}}
    \label{fig:Contour_fast_r=5-10_w=0}
    \end{center}
\end{figure}
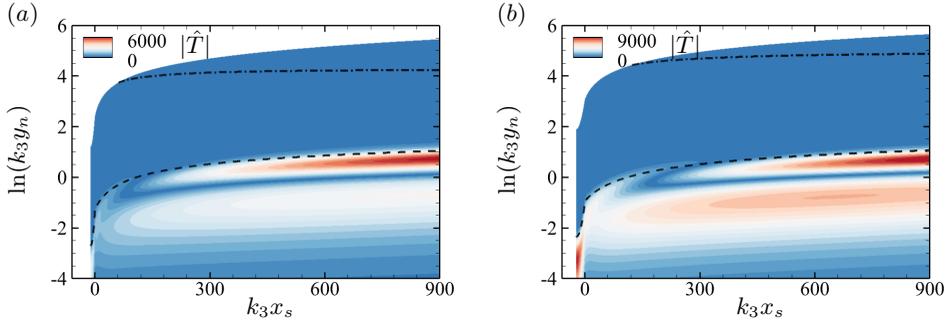
With the increase in the nose radius, a significant enhancement in the magnitude of the temperature perturbation in the downstream region is evident when compared to figures \ref{fig:Contour_fast_r=5-10_w=0}-(a) and (b). Additionally, a notable feature is observed in the nose region, where another local peak emerges, exhibiting a 'thermo spot' with a magnitude comparable to the downstream peak in case Cf.

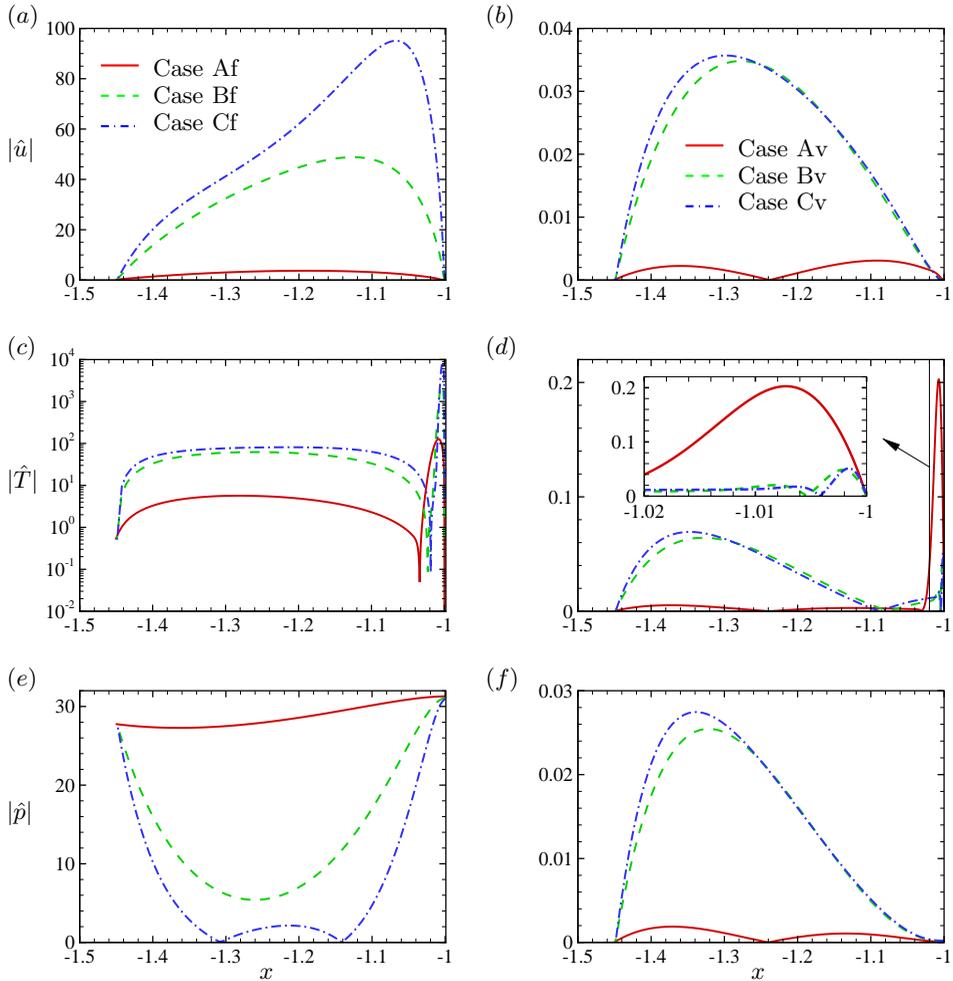
\begin{figure}
    \begin{center}
    \includegraphics[width = 0.48\textwidth] {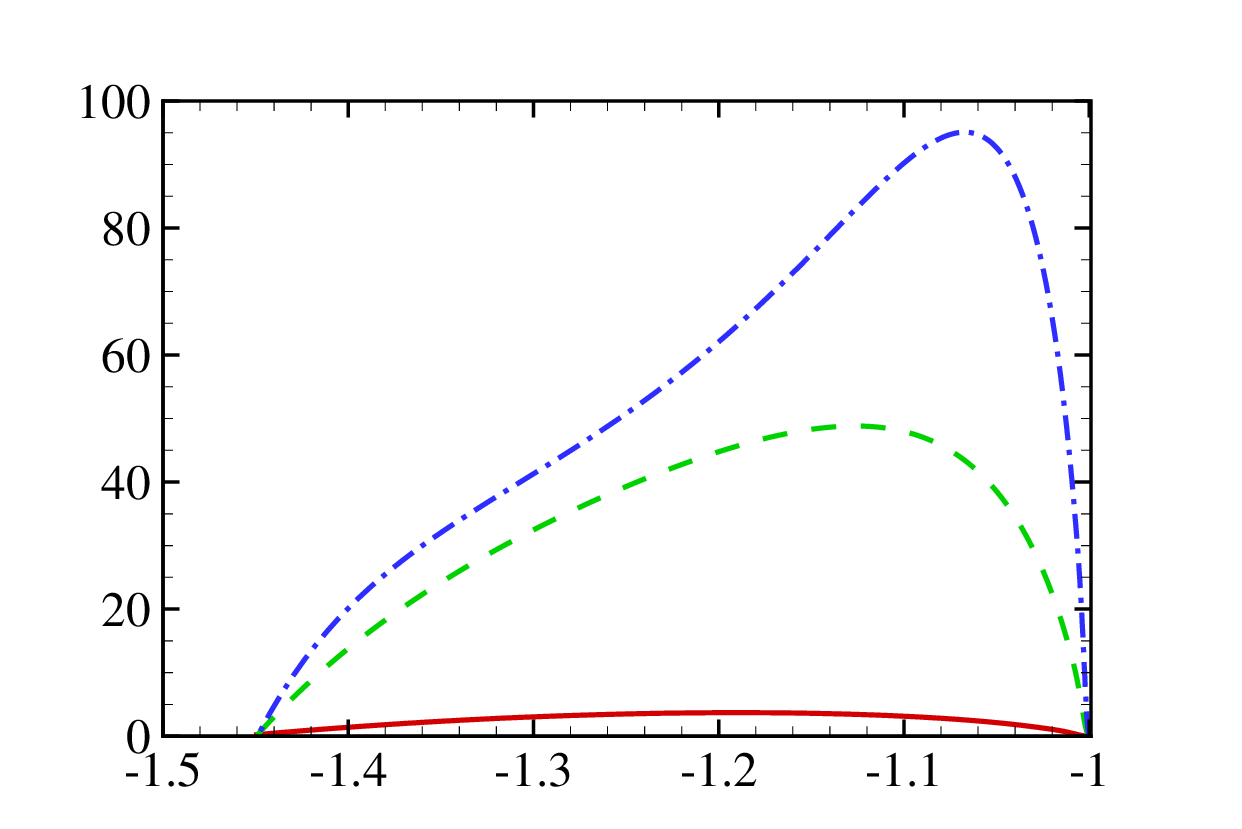}
    \includegraphics[width = 0.48\textwidth] {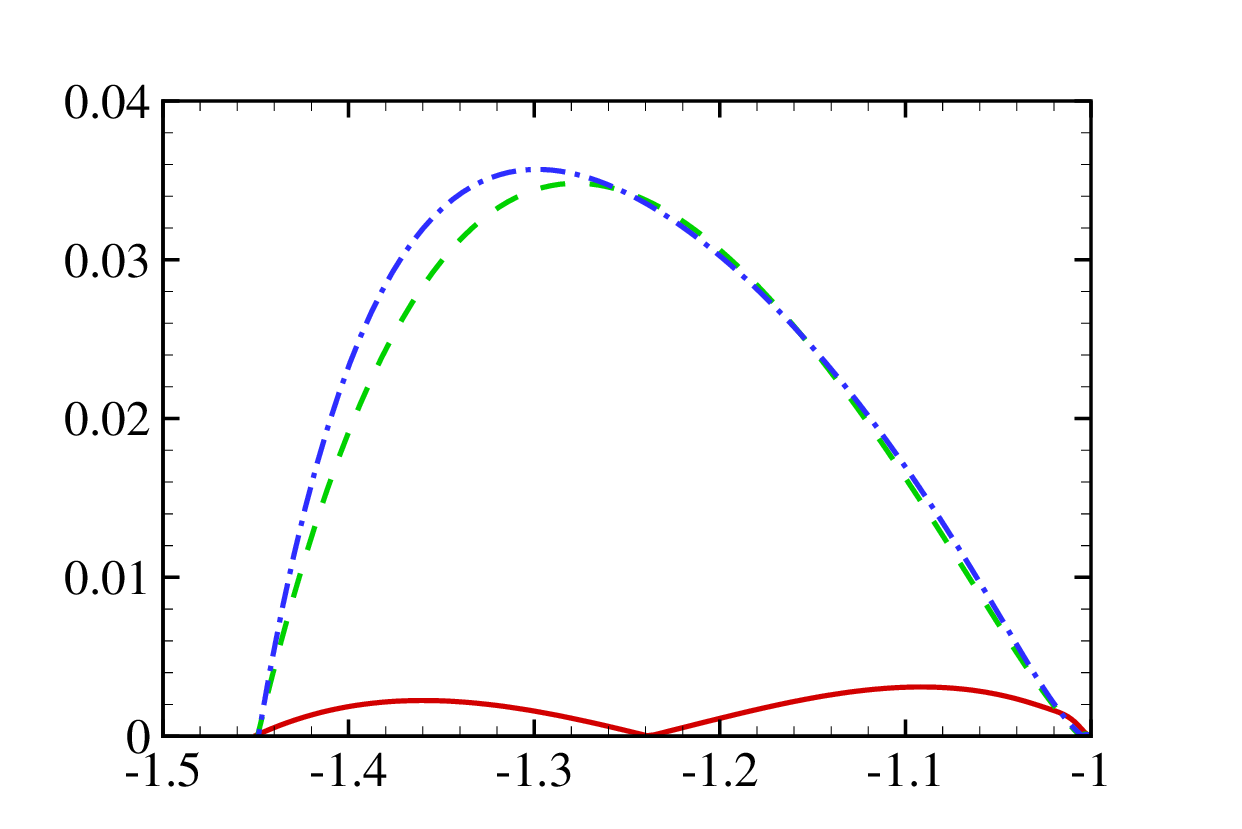}
    \put(-375,60){$|\hat u|$}
    \put(-375,110){$(a)$}
    \put(-195,110){$(b)$}
    \put(-320,90){Case Af}
    \put(-320,80){Case Bf}
    \put(-320,70){Case Cf}
    \put(-340,71){\begin{tikzpicture}
    \draw[red,thick] (0,0) -- (0.5,0);
    \draw[green,dashed,thick] (0,-0.4) -- (0.5,-0.4);
    \draw[blue,dash dot,thick] (0,-0.8) -- (0.5,-0.8);
    \end{tikzpicture}}
    \put(-100,60){Case Av}
    \put(-100,50){Case Bv}
    \put(-100,40){Case Cv}
    \put(-120,41){\begin{tikzpicture}
    \draw[red,thick] (0,0) -- (0.5,0);
    \draw[green,dashed,thick] (0,-0.4) -- (0.5,-0.4);
    \draw[blue,dash dot,thick] (0,-0.8) -- (0.5,-0.8);
    \end{tikzpicture}}\\
    \includegraphics[width = 0.48\textwidth] {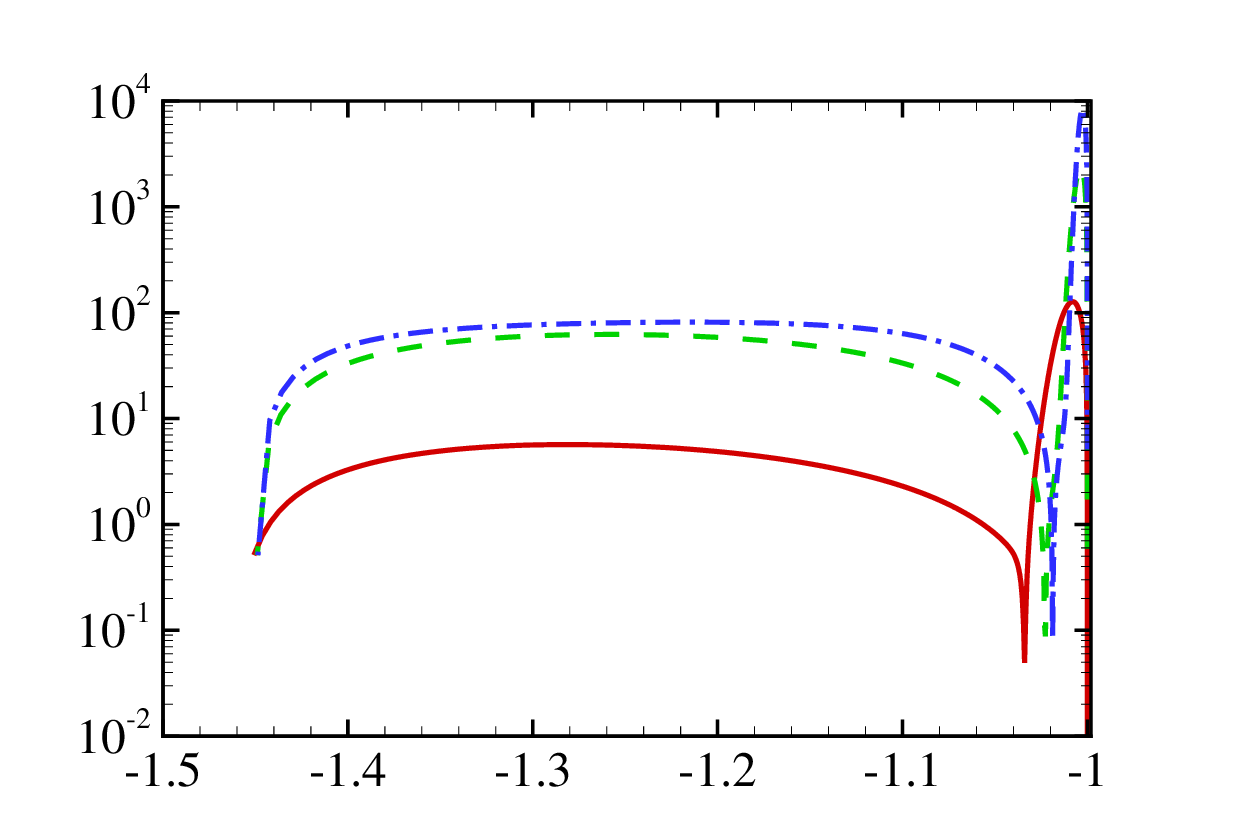} 
    \includegraphics[width = 0.48\textwidth] {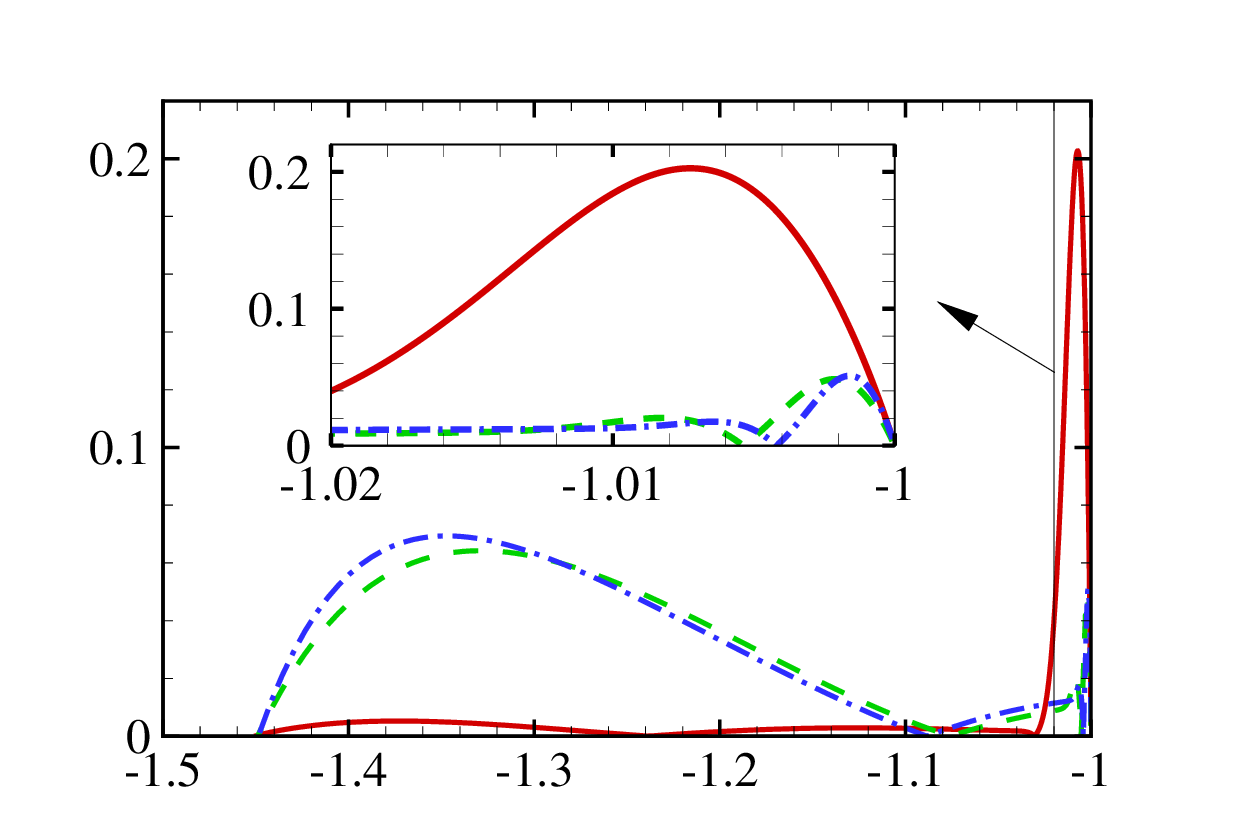}
    \put(-375,60){$|\hat T|$}
    \put(-375,110){$(c)$}
    \put(-195,110){$(d)$}\\
    \includegraphics[width = 0.48\textwidth] {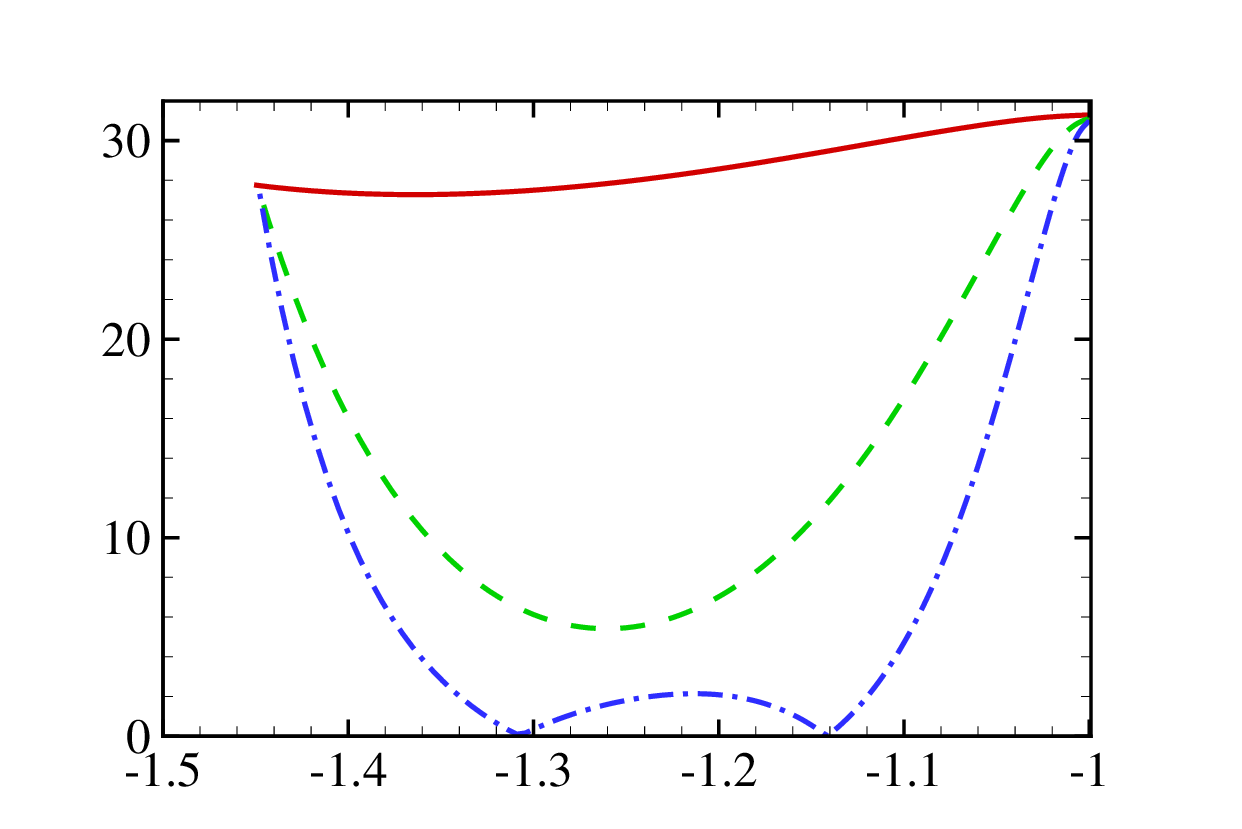} 
    \includegraphics[width = 0.48\textwidth] {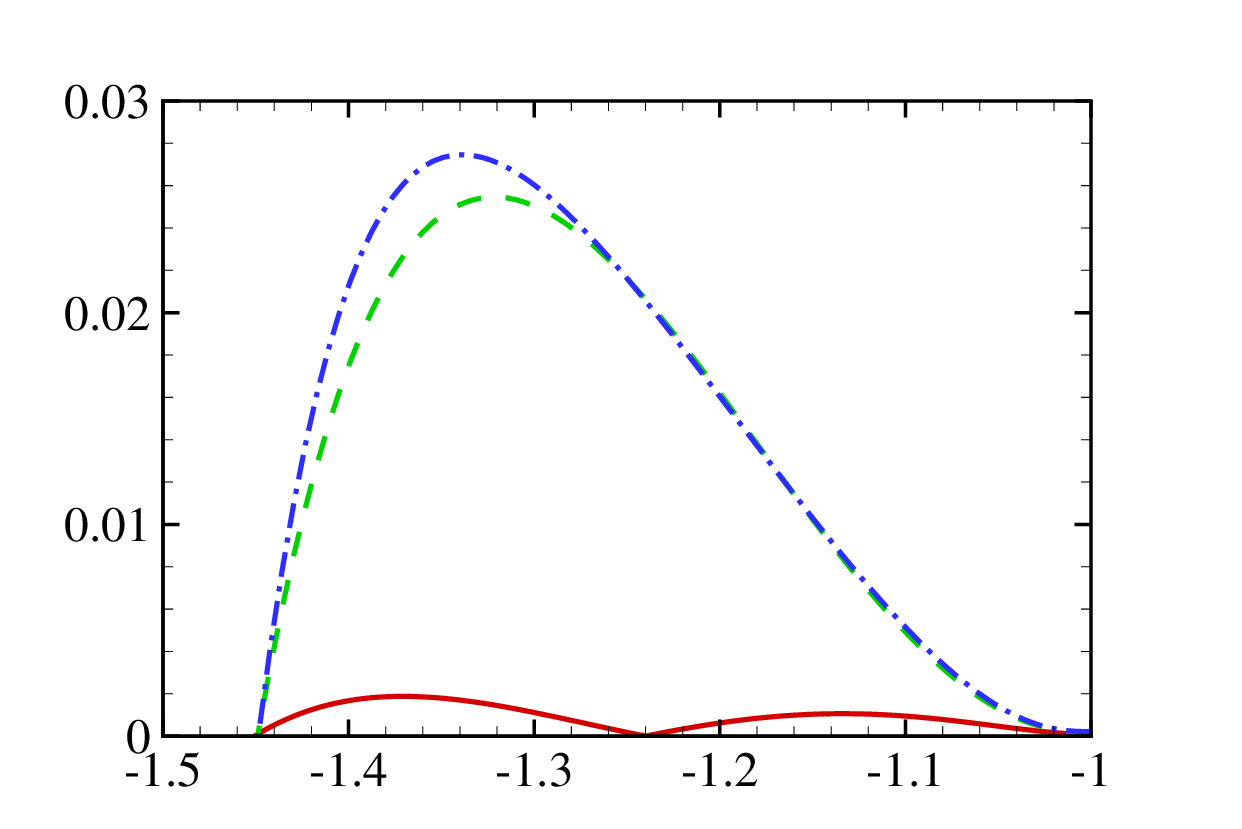}
    \put(-375,60){$|\hat p|$}
    \put(-280,0){$x$}
    \put(-375,110){$(e)$}
    \put(-95,0){$x$}
    \put(-195,110){$(f)$}\\
    \caption{{Profiles of the $\hat u$ ($a,b$), $\hat T$ ($c,d$) and $\hat p$ ($e,f$) along the centreline $y=0$ for cases (Af,Bf,Cf) ($a,c,e$) and cases (As,Bs,Cs) ($b,d,f$) with $\omega=0$, $k_3^*=1.5$ mm$^{-1}$ and $\vartheta=15^{\circ}$.}}
    \label{fig:line_uT-xs_centreline}
    \end{center}
\end{figure}
To offer a more detailed insight into the perturbation behavior within this peak region, we focus on the perturbation profiles along the centerline of the nose region $x\in [-1.45,-1]$ and $y=0$. 
For cases  with different nose radii, figures \ref{fig:line_uT-xs_centreline}($a,c,e$) compare the perturbation profiles of $\hat u$, $\hat T$ and $\hat p$ for fast acoustic forcing, respectively.
 The location of $x=-1.45$ corresponds to the normal shock position, while $x=-1$  signifies the stagnation point. Due to the substantial  magnitude variation of  $\hat T$ spanning approximately   five orders, we plot the $\hat T$ profile in the logarithmic form in panel (c). It is evident that  $\hat u$   undergoes significant  amplification in the post-shock region for each case, progressing until reaching the near-wall region, where the non-penetration condition enforces  the velocity perturbation to decay to zero. In contrast, the temperature perturbation experiences a sharp increase in the region immediately behind the shock, forming a plateau before approaching the vicinity of the stagnation point, where a thin layer with a notable peak is observed. The thickness of this layer diminishes as the nose radius expands, while  the  magnitude of $\hat T$ increases accordingly. Particularly in case Cf,  this magnitude reaches  $O(10^4)$, leading to the emergence of the local 'thermo spot' as depicted in figure \ref{fig:Contour_fast_r=5-10_w=0}-(b). Overall, in the post-shock bulk region, the perturbations of  $\hat u$ and $\hat T$ intensifies as the nose radius increases, whereas $\hat p$ demonstrates a noticeable decreasing trend. This indicates that the large bluntness suppresses the acoustic wave in the post-shock region but enhances the vorticity and entropy waves. The enhancement of the entropy wave may serve as the predominant factor contributing to the heightened receptivity efficiency for larger bluntness. 

 In figures \ref{fig:line_uT-xs_centreline} ($b,d,f$), we compare the perturbation profiles along the centreline for cases subject to freestream vortical forcing (including cases Av, Bv and Cv). The perturbation field in the nose region exhibits a distinct deviation from those subjected to acoustic forcing. A notable observation is the reduced  perturbations in the near-wall region. As the nose radius increases from 1mm to 5mm, the amplitude of the perturbations in the bulk post-shock region increases significantly. However, there is a remarkable decrease in the temperature perturbation around the stagnation point. 
 The pressure perturbation in case Av is considerably smaller compared to cases Bv and Cv, indicating a much weaker acoustic field in the post-shock nose region.

\begin{figure}
    \begin{center}
    \includegraphics[width = 0.48\textwidth] {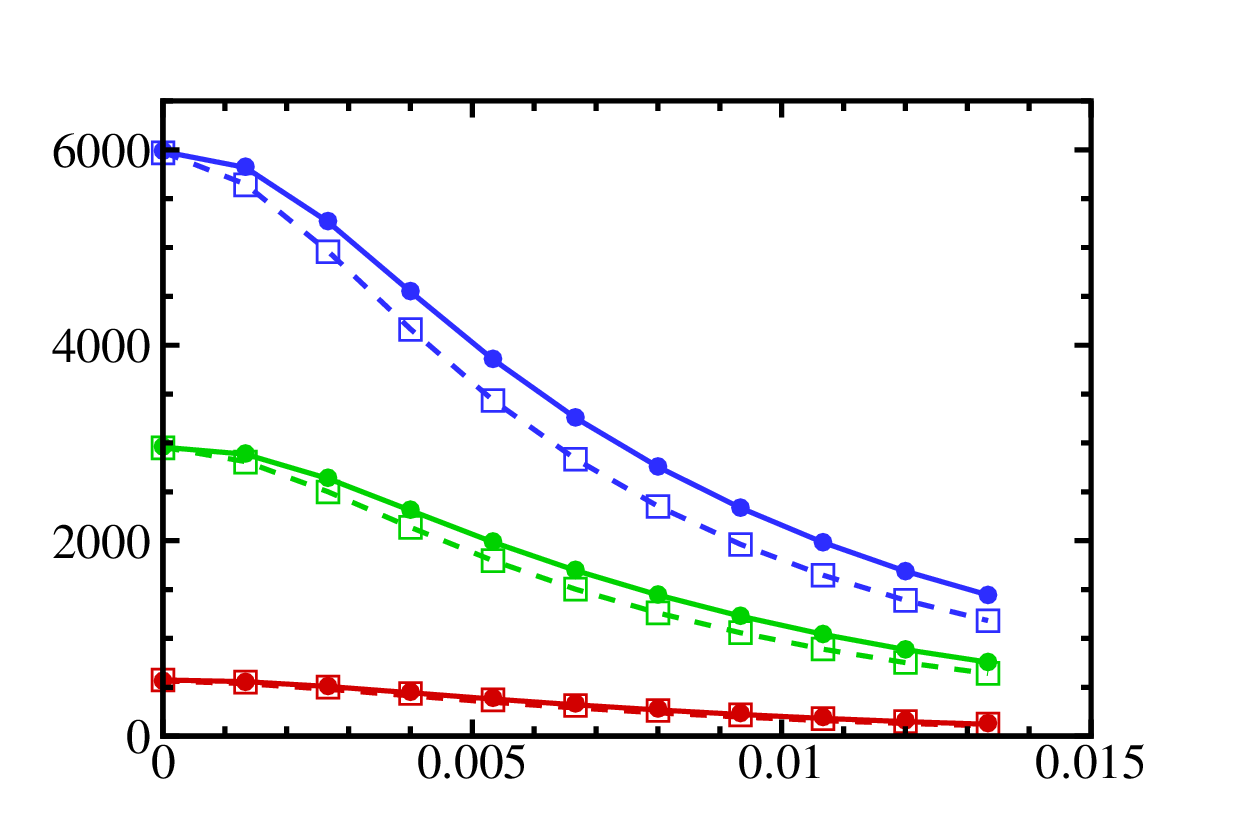}
    \includegraphics[width = 0.48\textwidth] {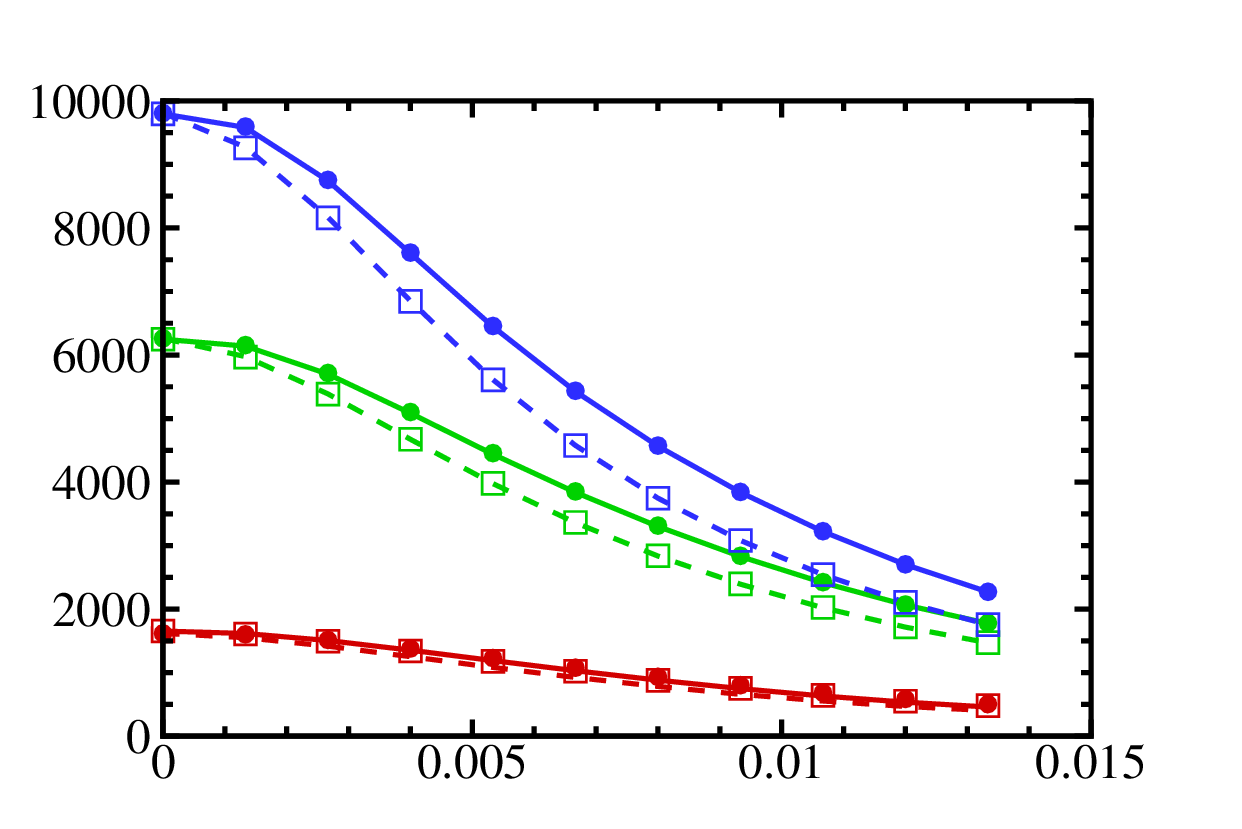}
    \put(-380,110){$(a)$}
    \put(-195,110){$(b)$}
    \put(-380,60){$A_u$}
    \put(-195,60){$A_T$}  
    \put(-285,0){$\omega/k_3$}
    \put(-100,0){$\omega/k_3$} 
    \put(-330,130){$\vartheta=15^\circ$}
    \put(-330,120){$\vartheta=-15^\circ$}  
    \put(-290,140){Case Af}
    \put(-250,140){Case Bf}
    \put(-210,140){Case Cf}
    \put(-160,140){Case As}
    \put(-120,140){Case Bs}
    \put(-80,140){Case Cs}
    \put(-280,122){\begin{tikzpicture}
    \draw[red,thick] (0,0) -- (0.5,0);
    \draw[green,thick] (1.4,0) -- (1.9,0);
    \draw[blue,thick] (2.8,0) -- (3.3,0);
    \draw[red,thick,dashed] (0,-0.4) -- (0.5,-0.4);
    \draw[green,thick,dashed] (1.4,-0.4) -- (1.9,-0.4);
    \draw[blue,thick,dashed] (2.8,-0.4) -- (3.3,-0.4);
    \end{tikzpicture}}
    \put(-150,120){\begin{tikzpicture}
    \node[draw,red,circle,fill,inner sep=1pt] at (0.25,0) {};
    \node[draw,green,circle,fill,inner sep=1pt] at (1.75,0) {};
    \node[draw,blue,circle,fill,inner sep=1pt] at (3.25,0) {};
    \node[draw,red,rectangle,inner sep=1.75pt] at (0.25,-0.4) {};
    \node[draw,green,rectangle,inner sep=1.75pt] at (1.75,-0.4) {};
    \node[draw,blue,rectangle,inner sep=1.75pt] at (3.25,-0.4) {};
    \end{tikzpicture}}\\
    \includegraphics[width = 0.48\textwidth] {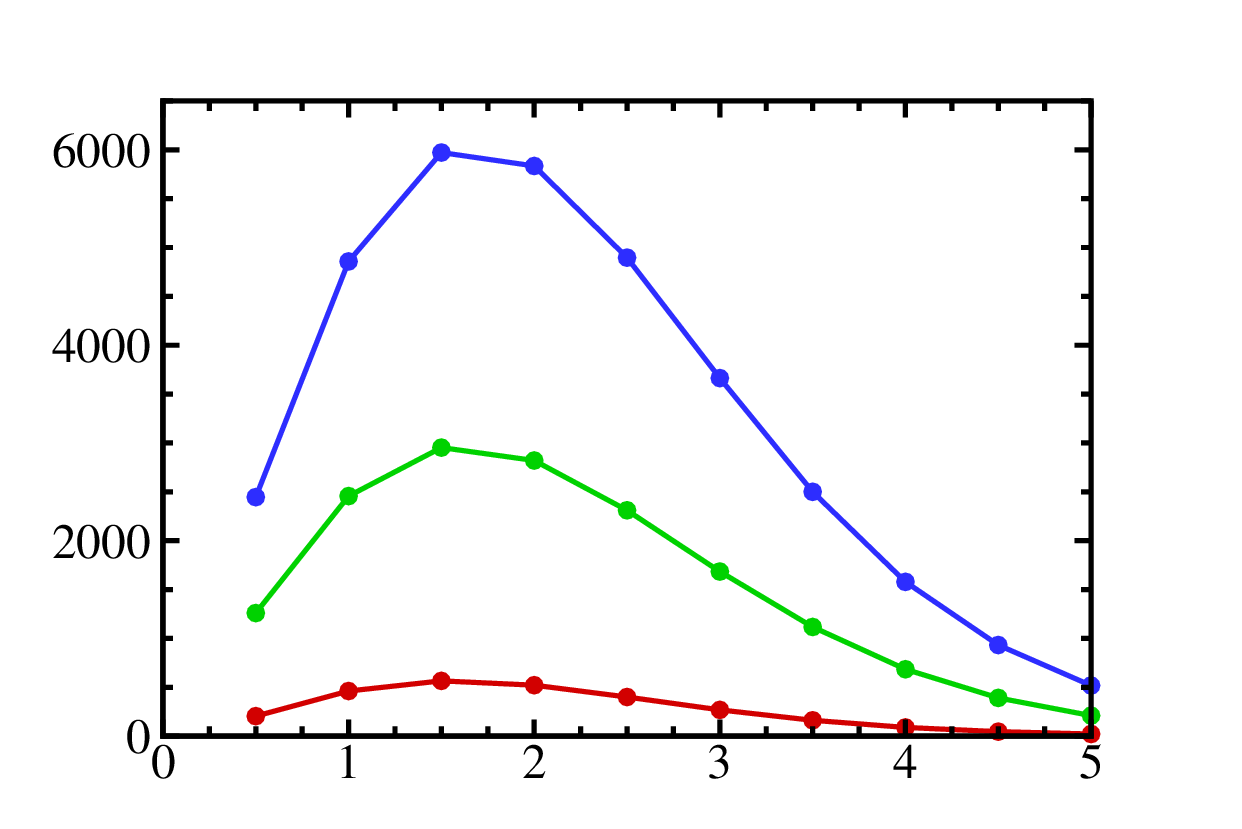}
    \includegraphics[width = 0.48\textwidth] {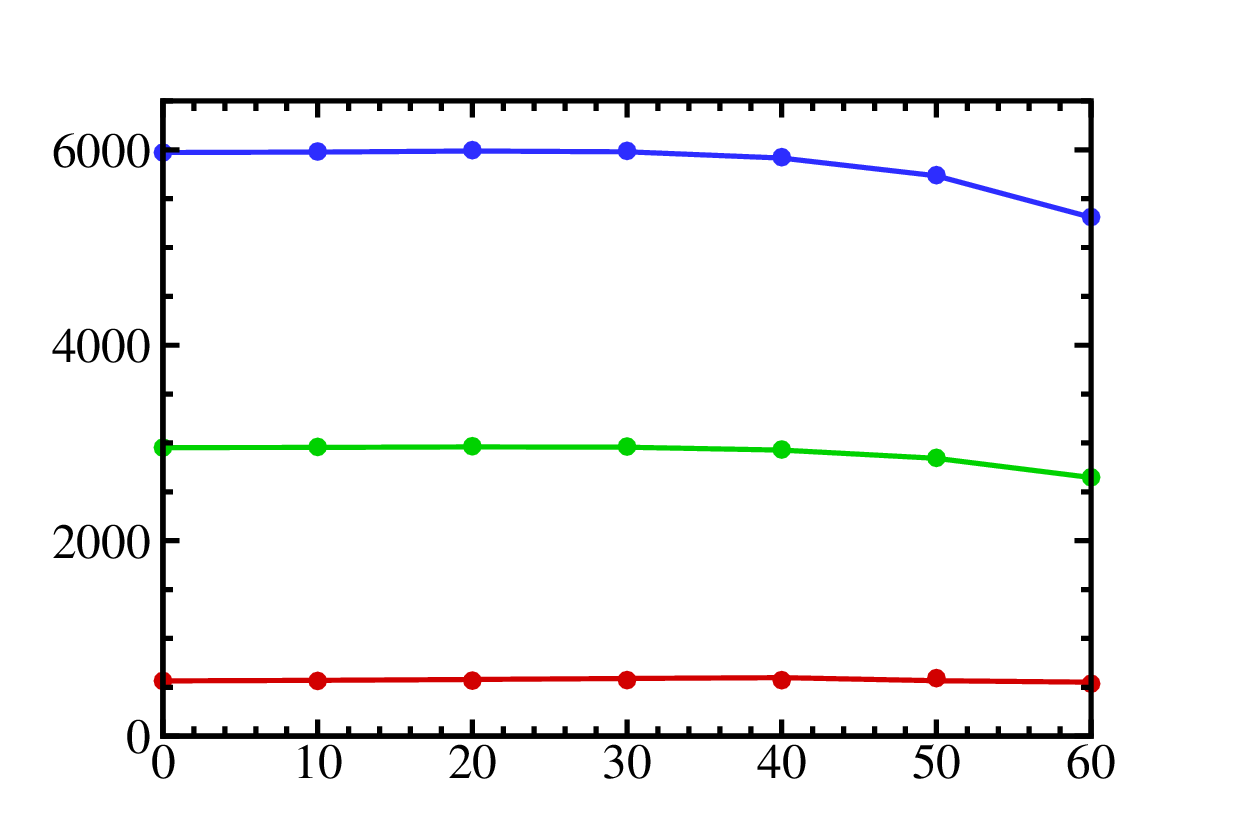}
    \put(-380,110){$(c)$}
    \put(-195,110){$(d)$}
    \put(-380,60){$A_u$}
    \put(-195,60){$A_u$}  
    \put(-280,0){$k_3^*$}
    \put(-92,0){$\vartheta$}
    \put(-225,0){(mm$^{-1}$)}
    \put(-28,0){($^{\circ}$)}
    \caption{Amplitudes $A_u$ and $A_T$ at $x_s^*=600$ mm for cases (Af, Bf, Cf) and (As, Bs, Cs). ($a,b$): dependence on $\omega/k_3$  for $k_3^*=1.5$ mm$^{-1}$ and $\vartheta=\pm 15^{\circ}$; ($c$): dependence on $k_3^*$ for $\omega=0$ and $\vartheta=0$; ($d$): dependence on $\vartheta$  for $\omega=0$ and $k_3^*=1.5$ mm$^{-1}$.}
    \label{fig:LNS_A-k1k2k3_acoustic}
    \end{center}
\end{figure}
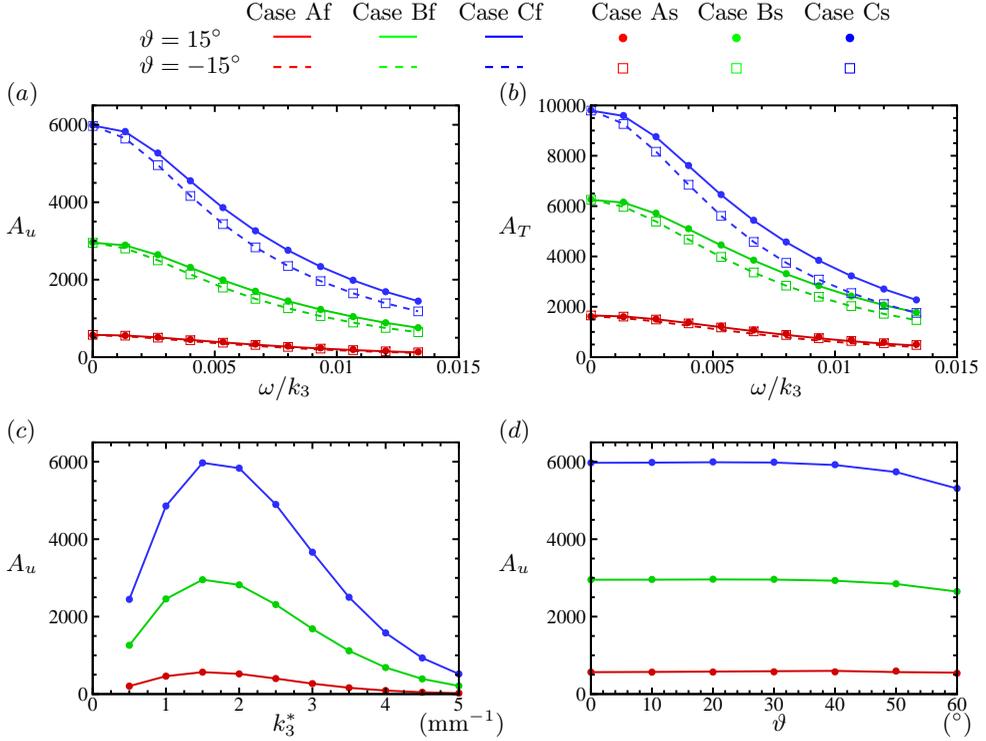
Figure \ref{fig:LNS_A-k1k2k3_acoustic}$(a)$ presents the frequency-dependent variation of the velocity amplitude $A_u$ at $x_s^*=600$ mm, comparing cases Af, Bf, Cf, As, Bs and Cs. Across all the curves, there is a consistent decline in  $A_u$ with  $\omega$ increasing, with slightly elevated values observed for a positive $\vartheta$ compared to those for a negative $\vartheta$. These trends mirror the observations from receptivity to freestream vortical disturbances (cases Av, Bv and Cv). Of particular interest is the notable distinction in the impact of the nose radius on receptivity efficiency to acoustic forcing, where greater bluntness leads to a more pronounced effect, contrasting with that to vortical forcing. In figure \ref{fig:LNS_A-k1k2k3_acoustic}($b$), the behaviours of the temperature amplitude $A_T$ at $x_s^*=600$ mm are compared, exhibiting the same trend with higher values. In panels ($c$) and ($d$), the variations of $A_u(x_s^*=600\mbox{mm})$ with respect to $k_3^*$ and $\vartheta$ are displayed. For all the nose radii, the receptivity efficiency peaks at around $k_3^*=1.5$ mm$^{-1}$ for fixed $\vartheta$, while  demonstrates a plateau for $\vartheta<50$ and  decreasing as $\vartheta$ exceeds this range.

Based on the aforementioned observations, we can deduce that the receptivity processes of non-modal perturbations to freestream fast and slow acoustic waves follow a similar pattern, albeit with a slight variance in receptivity efficiency. Furthermore, the disparities in the post-shock acoustic field for fast and slow acoustic cases do not appear to affect the evolution of non-modal perturbations in the boundary layer.

\subsubsection{Excitation of non-modal perturbations  by freestream entropy disturbances}
\label{sec:excitation_entropy}
\begin{figure}
    \begin{center}
    \includegraphics[width = 0.48\textwidth] {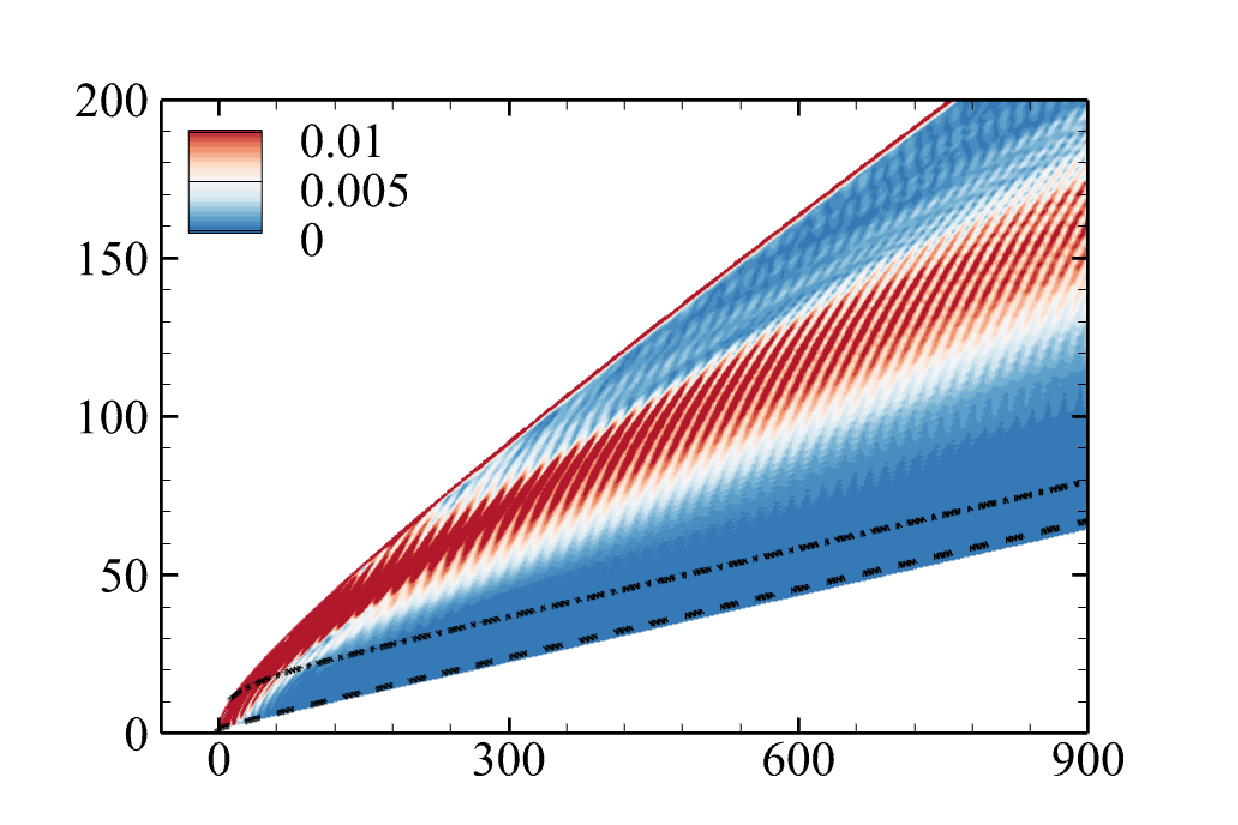}
    \put(-183,55){\rotatebox{90}{$k_3 y$}}
    \put(-90,0){$k_3 x$}
    \put(-120,95){$|\hat p|$}
    \put(-185,110){$(a)$}
    \includegraphics[width = 0.48\textwidth] {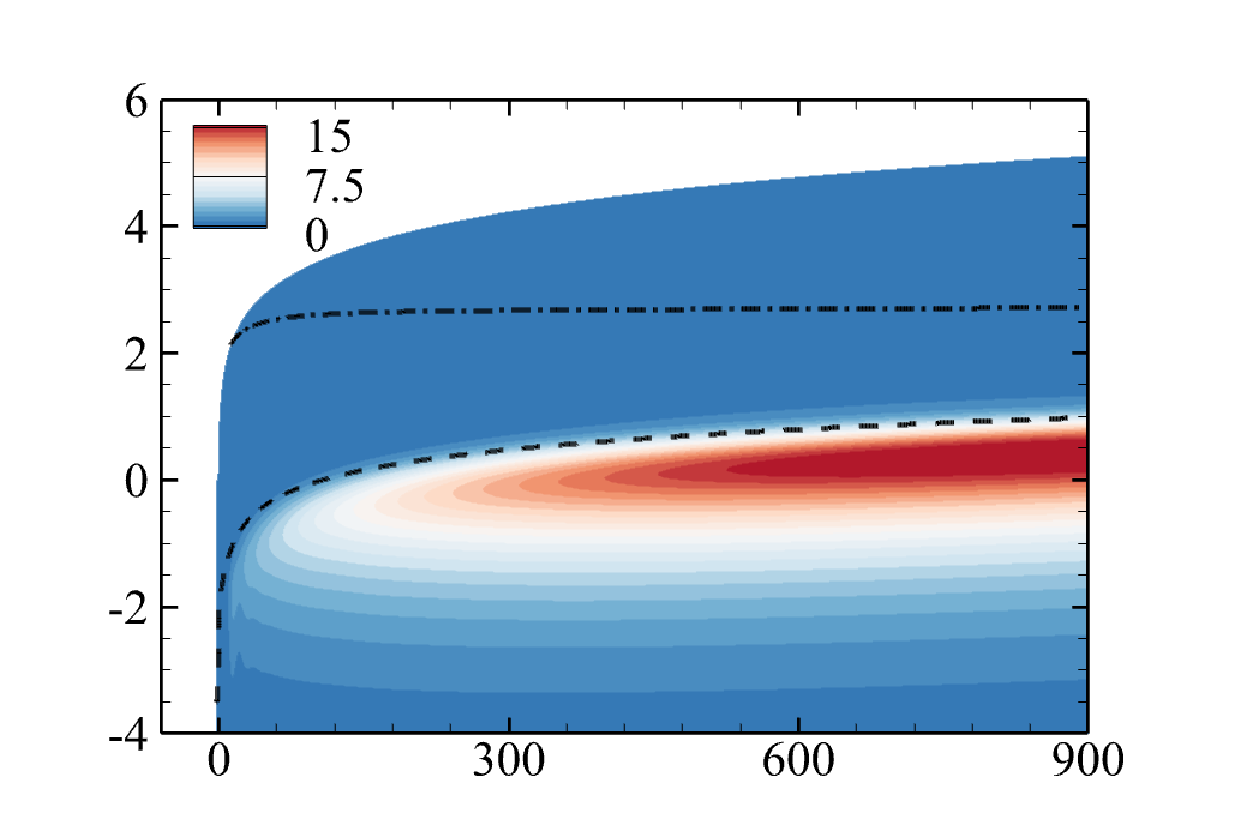}
    \put(-180,50){\rotatebox{90}{$\ln(k_3 y_n)$}}
    \put(-90,0){$k_3 x_s$}
    \put(-120,95){$|\hat u_s|$}
    \put(-185,110){$(b)$}\\
    \includegraphics[width = 0.48\textwidth] {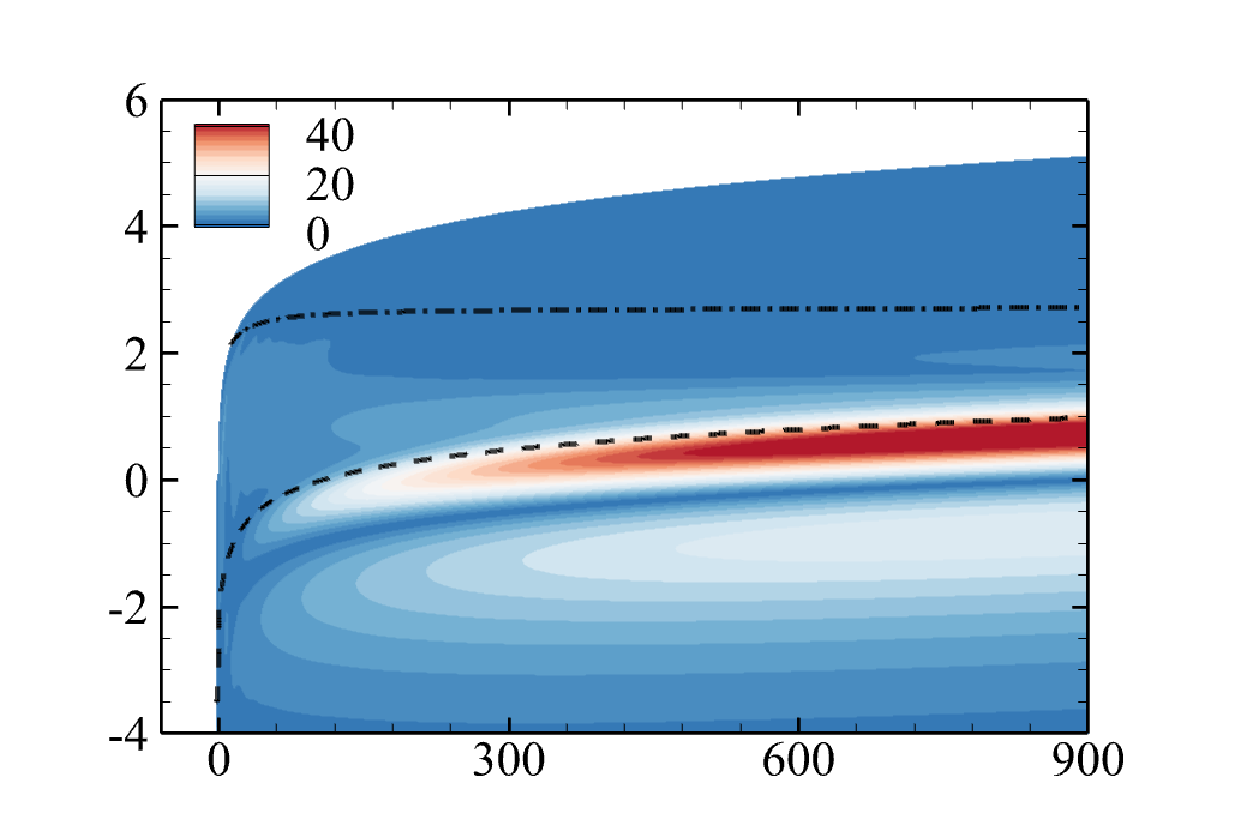}
    \put(-180,50){\rotatebox{90}{$\ln(k_3 y_n)$}}
    \put(-90,0){$k_3 x_s$}
    \put(-120,95){$|\hat T|$}
    \put(-185,110){$(c)$}
    \includegraphics[width = 0.48\textwidth] {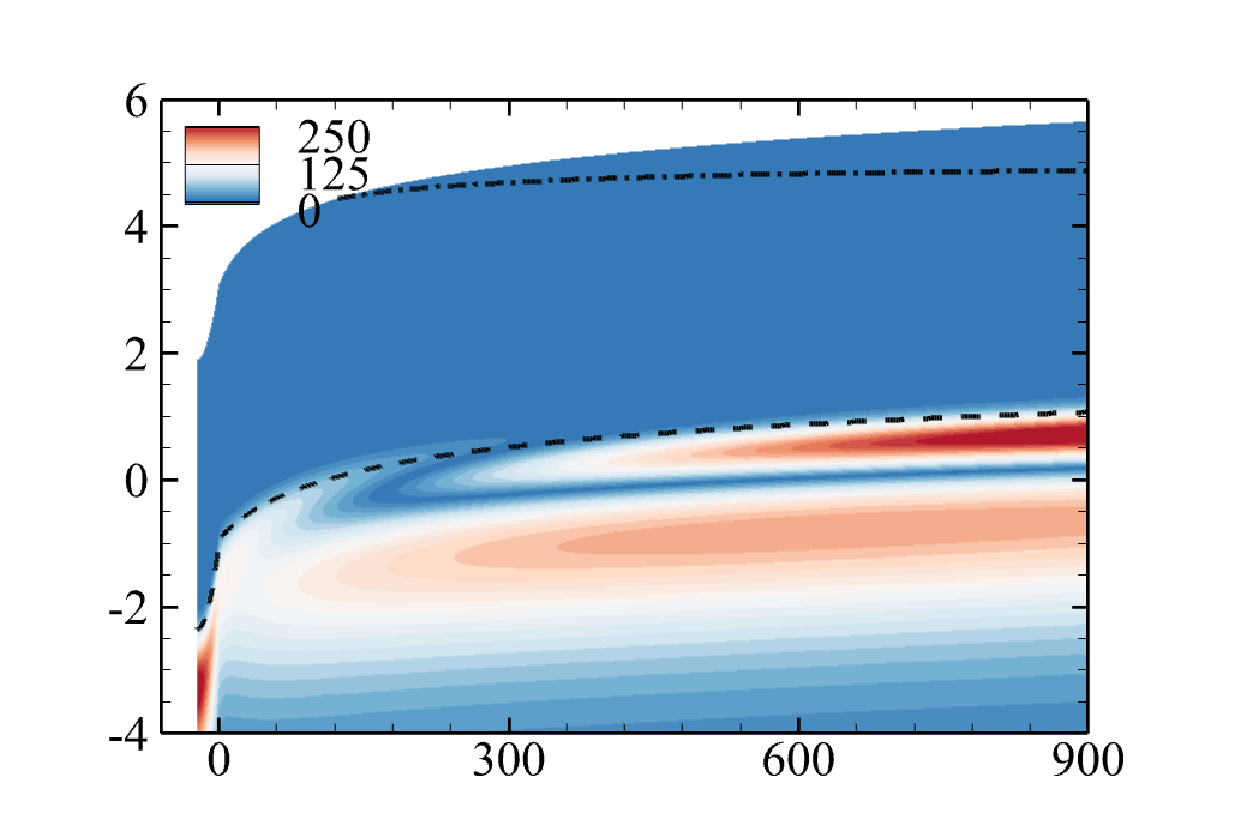}
    \put(-180,50){\rotatebox{90}{$\ln(k_3 y_n)$}}
    \put(-90,0){$k_3 x_s$}
    \put(-120,98){$|\hat T|$}
    \put(-185,110){$(d)$}
    \caption{Contours of the perturbation field for cases with freestream entropy-perturbation forcing, where $\omega=0$, $k_3^*=1.5$ mm$^{-1}$ and $\vartheta=15^{\circ}$. ($a$): $\hat p$ in the $k_3 x$-$k_3 y$ plane for case Ae; ($b$) and ($c$): $\hat u_s$ and $\hat T$ for case Ae in the $k_3 x_s$-$\ln(k_3 y_n)$ plane, respectively; ($d$): $\hat T$ in the $k_3 x_s$-$\ln(k_3 y_n)$ plane for case Ce. The dash-dotted and dashed lines mark the entropy layer and boundary layer, respectively.}
\label{fig:Contour_entropy_r=1_w=0}
    \end{center}
\end{figure}
Figure \ref{fig:Contour_entropy_r=1_w=0} displays the perturbation  field for cases subject to freestream entropy perturbations under the conditions of $\omega=0$, $k_3^*=1.5$ mm$^{-1}$ and $\vartheta=15^\circ$. The perturbation fields of $\hat p$, $\hat u_s$ and $\hat T$ for case Ae, shown in panels ($a,b,c$), closely resemble those for case Af, with the amplitudes differing by a certain factor, as compared to figures \ref{fig:Contour_fast-slow_r=1_w=0_fig1}($a,c,e$). Additionally, comparing panel (d) with figure \ref{fig:Contour_fast_r=5-10_w=0}-(b), we find a notable agreement in the  $\hat T$ field between cases Ce and Cf.

\begin{figure}
    \begin{center}
    \hfill
    \includegraphics[width = 0.48\textwidth] {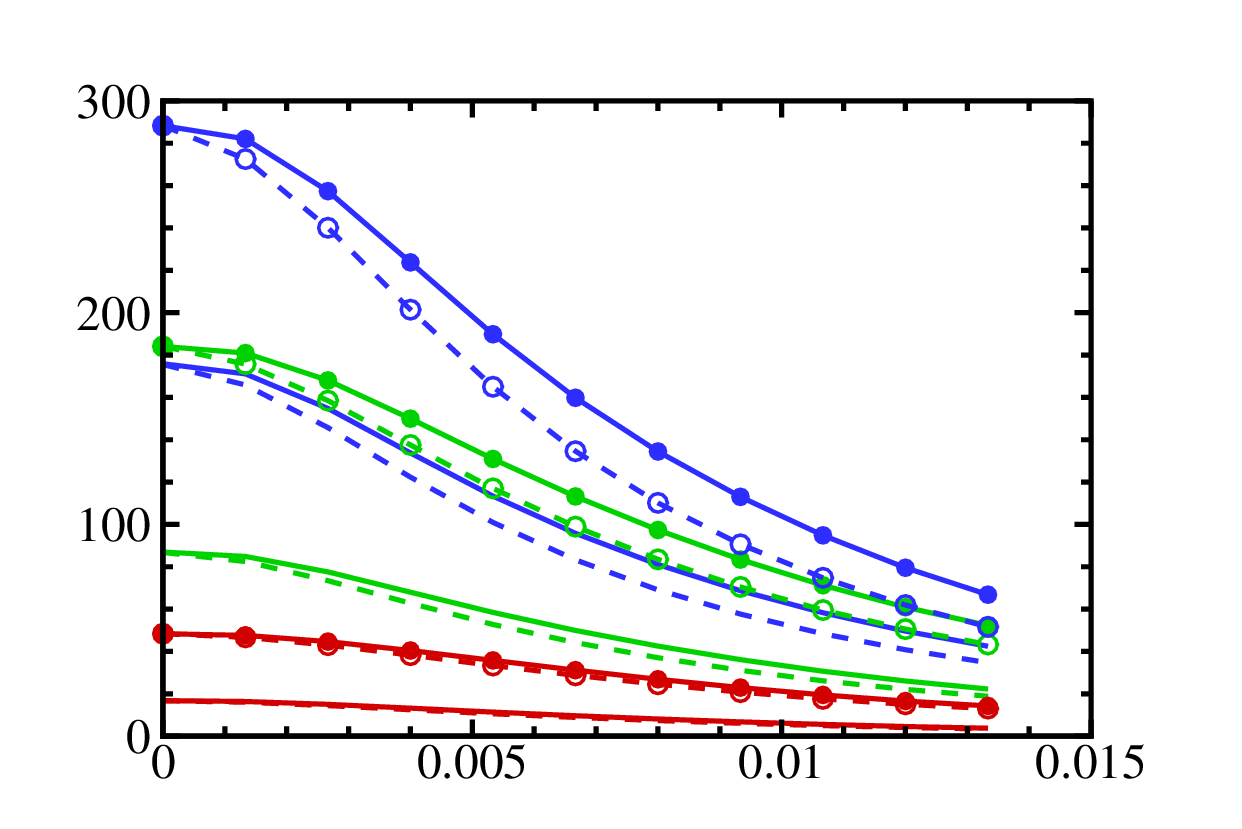}
    \put(-190,110){$(a)$}
    \put(-190,45){\rotatebox{90}{$A_u$, $A_T$}}
    \put(-95,0){$\omega/k_3$}
    \put(-380,70){$A_u$}
    \put(-360,75){$\vartheta=15^\circ$}
    \put(-360,65){$\vartheta=-15^\circ$} 
    \put(-380,40){$A_T$}
    \put(-360,45){$\vartheta=15^\circ$}
    \put(-360,35){$\vartheta=-15^\circ$}
    \put(-365,36){\begin{tikzpicture}
    \draw[decorate, decoration={brace, amplitude=3pt, mirror}, thick](0,0) -- (0,-0.5);
    \draw[decorate, decoration={brace, amplitude=3pt, mirror}, thick](0,-1) -- (0,-1.5);
    \end{tikzpicture}}
    \put(-325,90){Case Ae \,\,\, Case Be \,\, Case Ce}
    \put(-315,35){\begin{tikzpicture}
    \draw[red,thick] (0,0) -- (0.5,0);
    \draw[green,thick] (1.5,0) -- (2.0,0);
    \draw[blue,thick] (3.0,0) -- (3.5,0);
    \draw[red,thick,dashed] (0,-0.4) -- (0.5,-0.4);
    \draw[green,thick,dashed] (1.5,-0.4) -- (2.0,-0.4);
    \draw[blue,thick,dashed] (3.0,-0.4) -- (3.5,-0.4);
    \draw[red,thick] (0,-1.1) -- (0.5,-1.1);
    \node[draw,red,circle,fill,inner sep=1.2pt] at (0.25,-1.1) {};
    \draw[green,thick] (1.5,-1.1) -- (2.0,-1.1);
    \node[draw,green,circle,fill,inner sep=1.2pt] at (1.75,-1.1) {};
    \draw[blue,thick] (3.0,-1.1) -- (3.5,-1.1);
    \node[draw,blue,circle,fill,inner sep=1.2pt] at (3.25,-1.1) {};
    \draw[red,thick,dashed] (0,-1.5) -- (0.5,-1.5);
    \node[draw,red,circle,inner sep=1.2pt] at (0.25,-1.5) {};
    \draw[green,thick,dashed] (1.5,-1.5) -- (2.0,-1.5);
    \node[draw,green,circle,inner sep=1.2pt] at (1.75,-1.5) {};
    \draw[blue,thick,dashed] (3.0,-1.5) -- (3.5,-1.5);
    \node[draw,blue,circle,inner sep=1.2pt] at (3.25,-1.5) {};
    \end{tikzpicture}}\\
    \includegraphics[width = 0.48\textwidth] {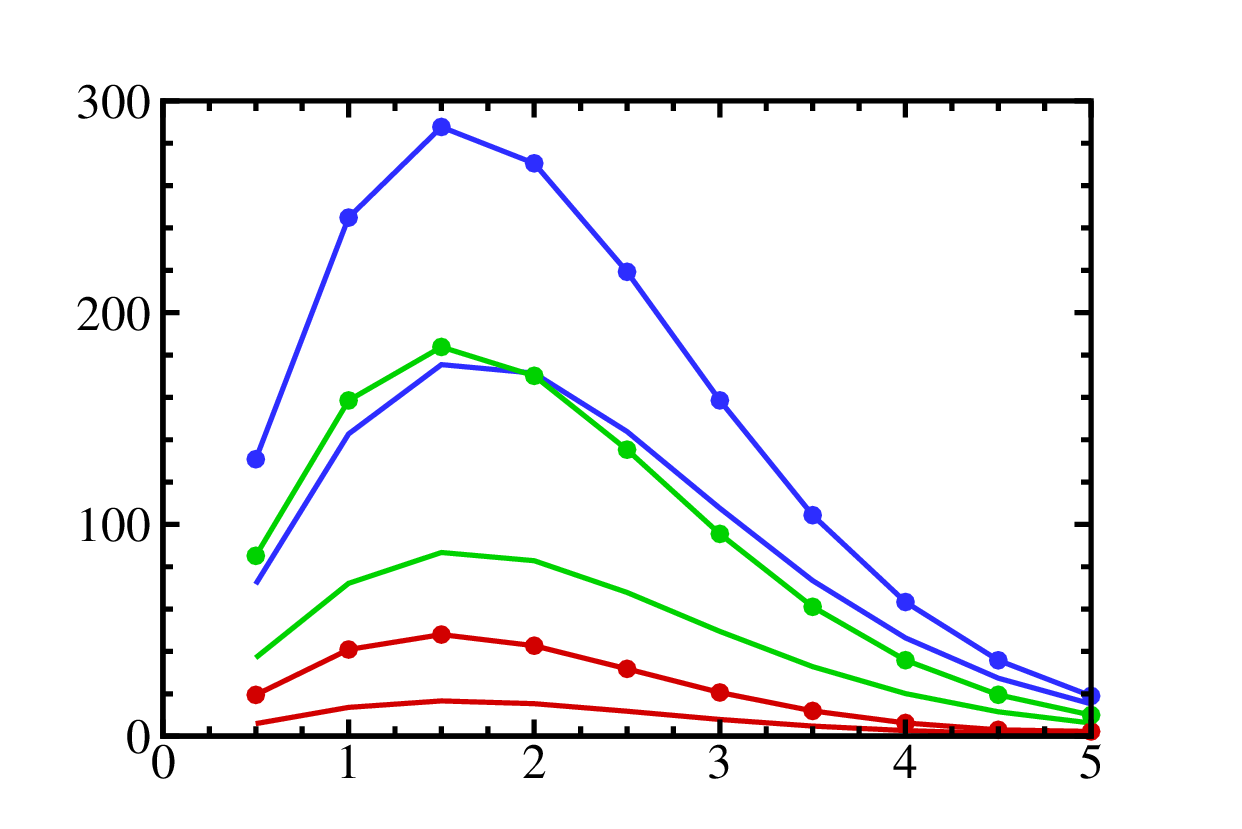}
    \put(-190,110){$(b)$}
    \put(-190,45){\rotatebox{90}{$A_u$, $A_T$}}
    \put(-95,0){$k_3^*$}
    \put(-35,0){(mm$^{-1}$)}
    \includegraphics[width = 0.48\textwidth] {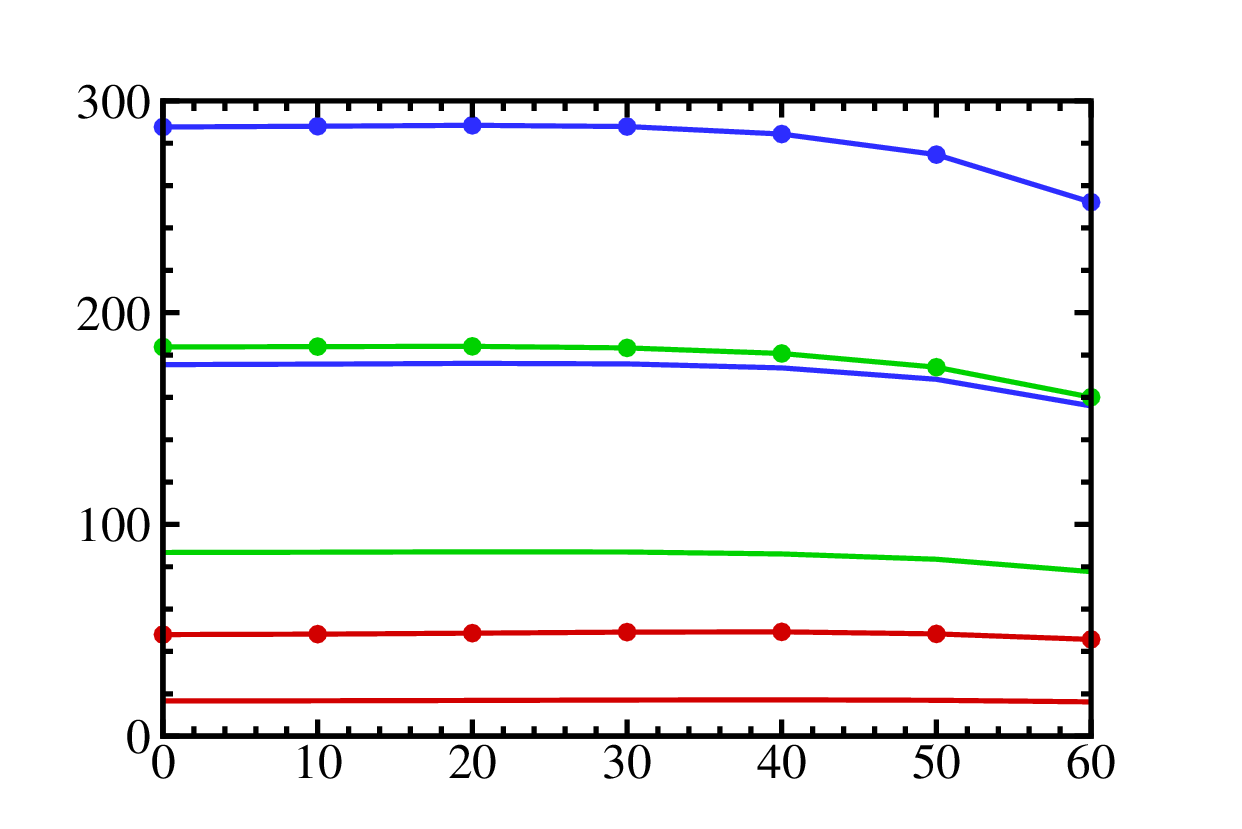}
    \put(-190,110){$(c)$}
    \put(-190,45){\rotatebox{90}{$A_u$, $A_T$}}
    \put(-95,0){$\vartheta$}
    \put(-28,0){($^{\circ}$)}
    \caption{$A_u$ (lines) and $A_T$ (symbolized lines) at $x_s^*=600$ mm for cases Ae, Be and Ce. ($a$): dependence on $\omega/k_3$ for $k_3^*=1.5$ mm$^{-1}$ and $\vartheta=\pm 15^{\circ}$; ($b$): dependence on $k_3^*$ for $\omega=0$ and $\vartheta=0$; ($c$): dependence on $\vartheta$  for $\omega=0$ and $k_3^*=1.5$ mm$^{-1}$.}
    \label{fig:LNS_A-k1_entropy}
    \end{center}
\end{figure}
Figure \ref{fig:LNS_A-k1_entropy} compares the variations  in  the amplitudes $A_u$ and $A_T$ at $x_s^*=600$ mm across cases Ae, Be, and Ce concerning different controlling parameters. 
Panel (a) demonstrates a consistent decline in receptivity efficiency as frequency increases; panel (b) shows an optimal dimensional spanwise wavenumber at  $k_3^*\approx 1.5$ mm$^{-1}$; panel (c) depicts a broad plateau in the dependency of $A_u$ and $A_T$ on $\vartheta$ in the region $\vartheta<50^\circ$.
Notably,  the enhanced receptivity efficiency with an increase in the nose bluntness aligns with those observed for cases subject to freestream fast or slow acoustic  perturbations.

\subsection{Comparison of the receptivity efficiency among various external perturbations}
\label{sec:summary}
In this subsection, we aim to compare the receptivity efficiency concerning various freestream forcings. To ensure a fair comparison, the energy of freestream perturbations across different case studies needs to be rescaled. Drawing inspiration from \citet{mack1969boundary}, we introduce a positive definite energy norm:
{
\begin{equation}
\mathcal{E}(\hat{\pmb\varphi};x_s,y_n)=||\hat {\pmb\varphi}||_\mathcal{E}=\hat {\pmb\varphi}^{\mathrm{H}} \pmb M \hat {\pmb\varphi},
\end{equation}
where the superscript $\mathrm{H}$ presents the conjugate transpose, and
\begin{equation}\pmb M=\mbox{diag}\Big(\frac{\bar T}{\gamma M^2 \bar \rho}, \bar\rho, \bar\rho, \bar\rho, \frac{\bar\rho}{\gamma(\gamma-1) M^2 \bar T}\Big).
\end{equation}
}
In the freestream, the energy norm is denoted by $\mathcal E_\infty\equiv \mathcal E(\hat{\pmb\varphi}_{\infty})$. Thus, for freestream  acoustic, entropy and vortice disturbances illustrated in (\ref{sec:freestream_perturbation}), the corresponding energy norms are  given  by
$$
\mathcal{E}_{\infty}^a=2M^2,\quad \mathcal{E}_{\infty}^e=\frac{1}{(\gamma-1)M^2},\quad 
\mathcal{E}_{\infty}^v=1. 
$$

\begin{figure}
    \includegraphics[width = 0.80\textwidth] {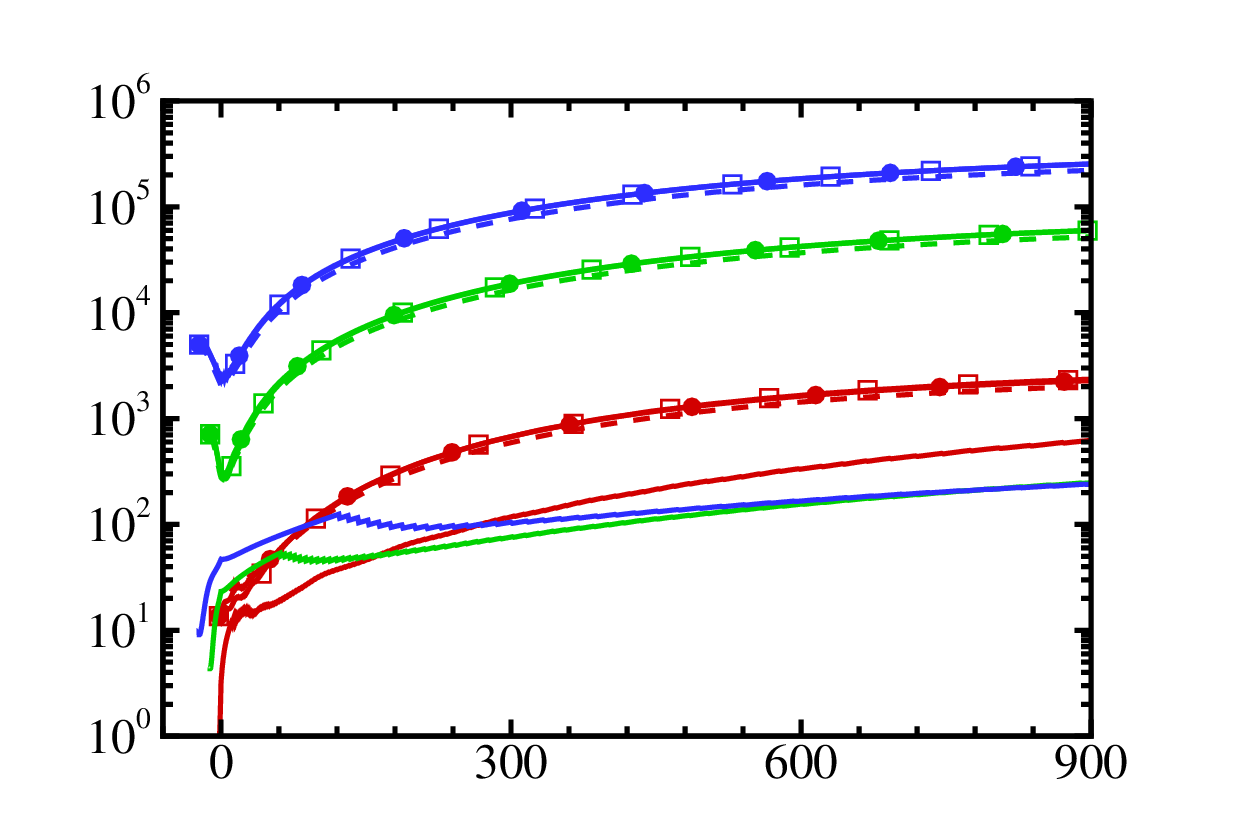}
    \put(-310,90){\rotatebox{90}{\large$\bar{\mathcal E}/{\mathcal E}_\infty$}}
    \put(-160,5){\large$k_3 x_s$}
    \put(0,170){Case Av}
    \put(0,160){Case Bv}
    \put(0,150){Case Cv}
    \put(0,135){Case Af}
    \put(0,125){Case Bf}
    \put(0,115){Case Cf}
    \put(0,100){Case As}
    \put(0,90){Case Bs}
    \put(0,80){Case Cs}
    \put(0,65){Case Ae}
    \put(0,55){Case Be}
    \put(0,45){Case Ce}
    \put(-20,46){\begin{tikzpicture}
    \draw[red,thick] (0,-0.2) -- (0.5,-0.2);
    \draw[green,thick] (0,-0.6) -- (0.5,-0.6);
    \draw[blue,thick] (0,-1.0) -- (0.5,-1.0);
    \draw[red,thick] (0,-1.5) -- (0.5,-1.5);
    \node[draw,red,rectangle,inner sep=1.8pt] at (0.25,-1.5) {};
    \draw[green,thick] (0,-1.9) -- (0.5,-1.9);
    \node[draw,green,rectangle,inner sep=1.8pt] at (0.25,-1.9) {};
    \draw[blue,thick] (0,-2.2) -- (0.5,-2.2);
    \node[draw,blue,rectangle,inner sep=1.8pt] at (0.25,-2.2) {};
    \draw[red,thick] (0,-2.7) -- (0.5,-2.7);
    \node[draw,red,circle,fill,inner sep=1.2pt] at (0.25,-2.7) {};
    \draw[green,thick] (0,-3.1) -- (0.5,-3.1);
    \node[draw,green,circle,fill,inner sep=1.2pt] at (0.25,-3.1) {};
    \draw[blue,thick] (0,-3.5) -- (0.5,-3.5);
    \node[draw,blue,circle,fill,inner sep=1.2pt] at (0.25,-3.5) {};
    \draw[red,dashed,thick] (0,-3.95) -- (0.5,-3.95);
    \draw[green,dashed,thick] (0,-4.35) -- (0.5,-4.35);
    \draw[blue,dashed,thick] (0,-4.7) -- (0.5,-4.7);
    \end{tikzpicture}}
    \caption{Streamwise evolution of the normalised total energy  $\bar {\cal E}/ {\cal E}_\infty$  across all the cases,   
    where $\omega=0$, $k_3^*=1.5$ mm$^{-1}$ and $\vartheta= 15^{\circ}$. }
    \label{fig:LNS_A-AT-xs_check}
\end{figure}
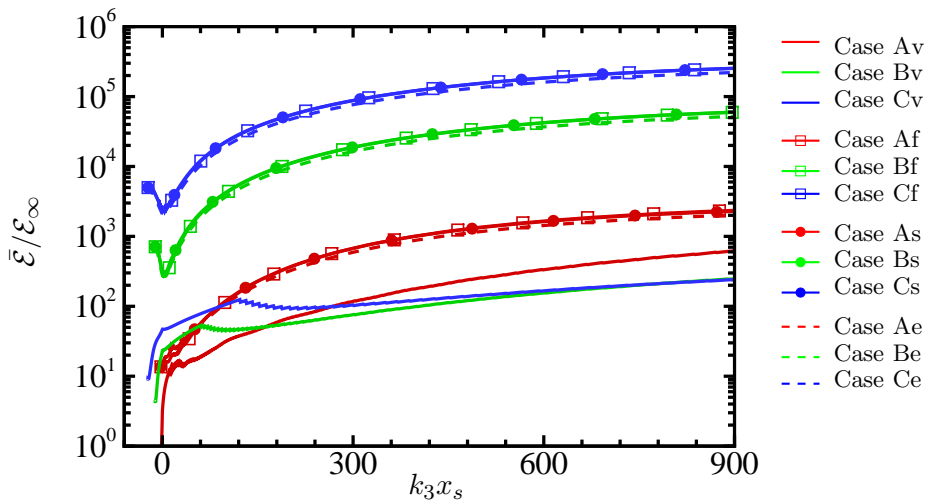
At each streamwise location $x_s$, we define the total energy encompassing  the wall-normal perturbation profile as \begin{equation}
\bar{\mathcal E}(\hat{\pmb\varphi};x_s)=\int_0^{y_s} {\mathcal E}(\hat{\pmb\varphi};x_s,y_n) \mathrm{d}y_n,  
\end{equation}
where the upper band of the integral is selected as the edge of the entropy layer.
Figure \ref{fig:LNS_A-AT-xs_check} compares the streamwise evolution of the normalised total energy $\bar{\mathcal E}(x_s){/\mathcal E}_{\infty}$ across all the cases involving different nose radii and freestream perturbations. The red, green and blue solid lines depict the results for cases Av, Bv and Cv, respectively. It is evident that in the downstream region, the total energy of the  non-modal perturbation excited by freestream vortical perturbations decreases as the nose radius $r^*$ expands, with the decrease for relatively larger $r^*$ values  being rather mild.
Conversely, in cases forced by freestream acoustic and entropy perturbations, depicted by the symboled and dashed lines, the excited perturbations achieve greater downstream energy as the nose radius increases. Upon examining the magnitude of the excited perturbations by unity energy forcing, it is evident that the receptivity of boundary-layer non-modal perturbations in downstream positions to freestream vortical perturbations is rather ineffective, whereas the receptivity to freestream fast or slow  acoustic and entropy perturbations is comparable, with the acoustic receptivity demonstrating slightly superior effectiveness. 
Particularly, the nose-region response to freestream acoustic and entropy forcing is reinforced with  increasing nose radius, which could be the dominant factor contributing to the heightened receptivity efficiency observed in the downstream region.

To gain a deeper insight of the receptivity process, we plot the rescaled perturbation field in the near-nose region for the same nose radius (case A) subject to various freestream forcing in figure \ref{fig:Contour_entropy_r=1_w=0_nose}. The nose-region perturbations induced by freestream fast acoustic, slow acoustic and entropy perturbations, as shown in panels (b,c,d), exhibit comparable features. The pressure perturbation reaches its peak value in the region immediately behind the shock, while a secondary peak forms at the location where the surface curvature shows discontinuity. The latter indicates a radiating Mach wave due to the scattering effect, with the $\hat u$, $\hat v$ and $\hat T$ perturbations also reaching their maxima. Comparing with panel (a), we find that the perturbations induced by freestream vortical perturbations are quite inefficient. Since the perturbation profile here is the initial form of the downstream non-modal streaks, it is understandable that the excited non-modal perturbations in the downstream region are much weaker.
\begin{figure}
    \begin{center}
    \includegraphics[width = 0.48\textwidth] {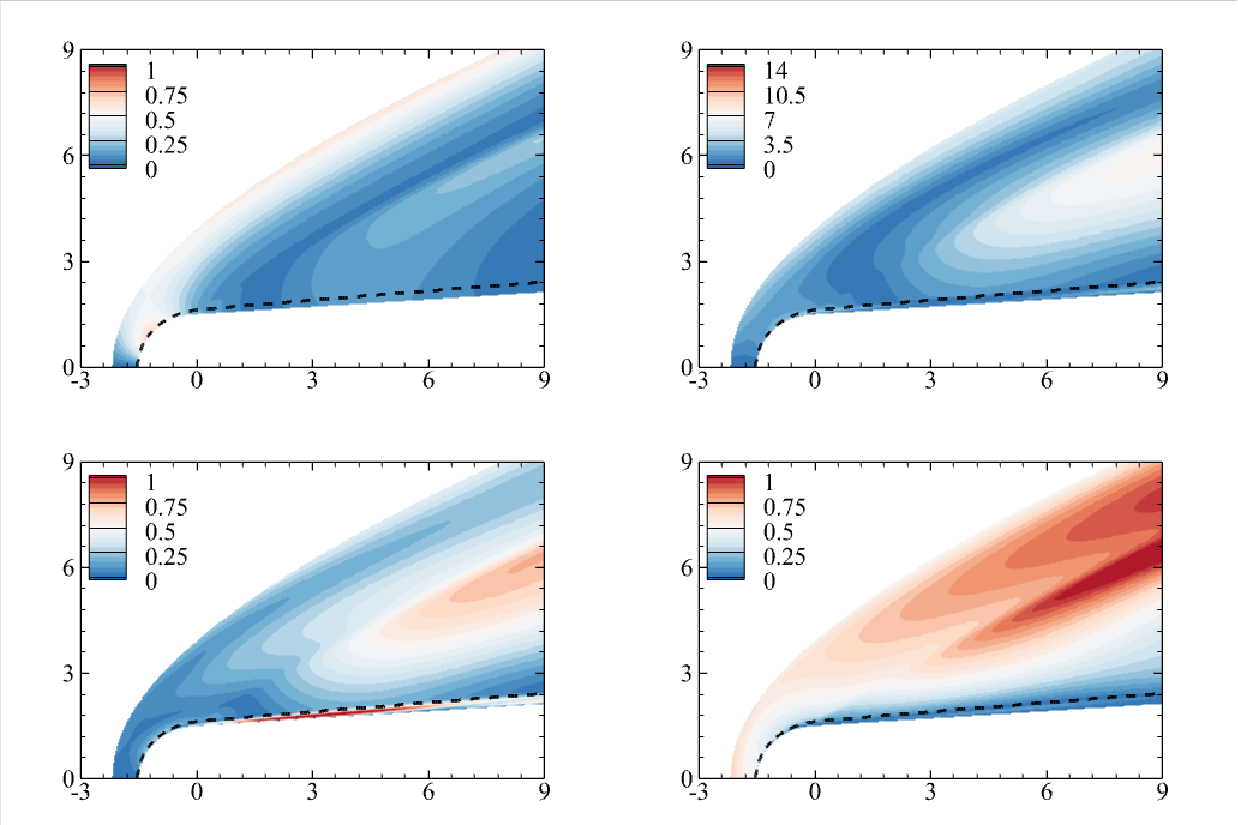}
    \put(-185,120){$(a)$}
    \put(-185,85){\rotatebox{90}{$k_3y$}}
    \put(-185,25){\rotatebox{90}{$k_3y$}}
    \put(-155,108){\fontsize{6pt}{6pt}\selectfont $|\hat p|/\sqrt{\mathcal{E}_{\infty}}$}
    \put(-64,108){ \fontsize{6pt}{6pt}\selectfont $|\hat T|/\sqrt{\mathcal{E}_{\infty}}$}
    \put(-155,47){ \fontsize{6pt}{6pt}\selectfont $|\hat u|/\sqrt{\mathcal{E}_{\infty}}$}
    \put(-64,47){  \fontsize{6pt}{6pt}\selectfont $|\hat v|/\sqrt{\mathcal{E}_{\infty}}$}
    \includegraphics[width = 0.48\textwidth] {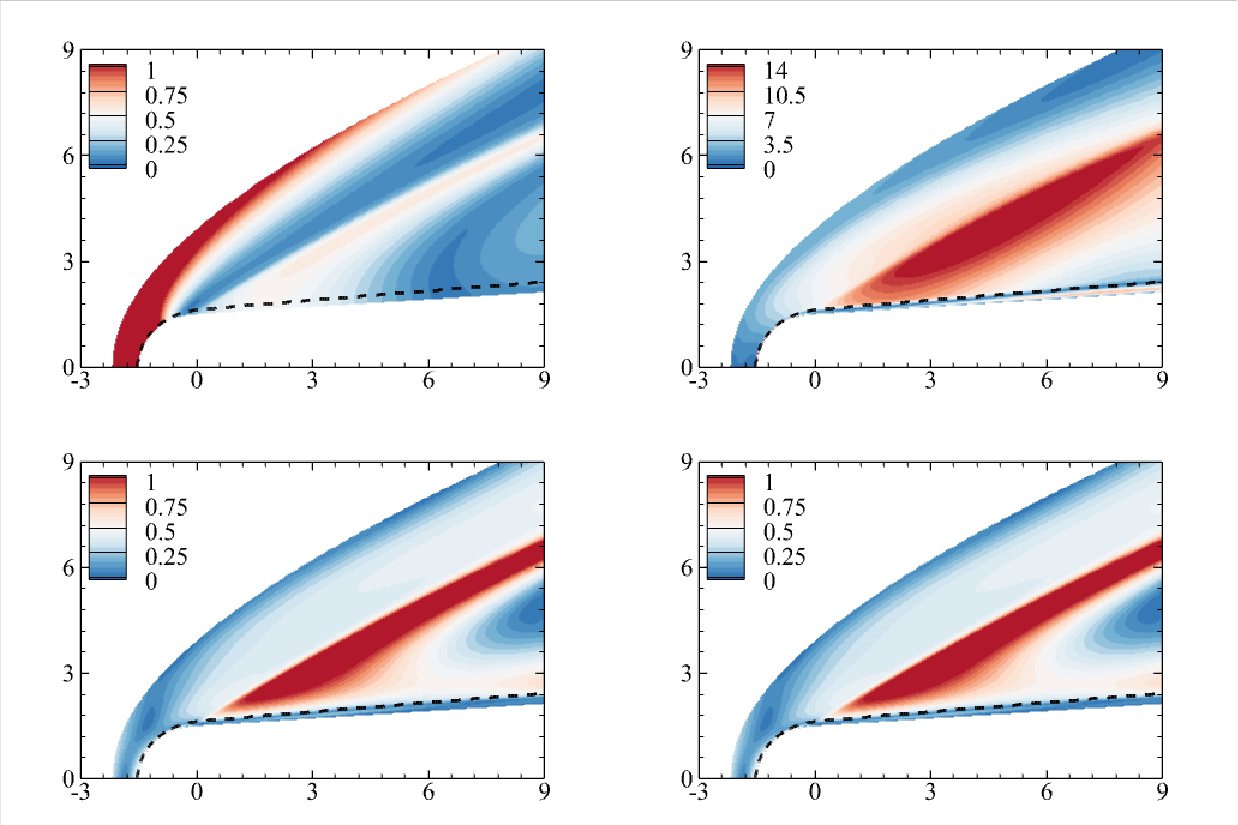}
    \put(-185,120){$(b)$}
    \put(-155,108){\fontsize{6pt}{6pt}\selectfont $|\hat p|/\sqrt{\mathcal{E}_{\infty}}$}
    \put(-64,108){ \fontsize{6pt}{6pt}\selectfont $|\hat T|/\sqrt{\mathcal{E}_{\infty}}$}
    \put(-155,47){ \fontsize{6pt}{6pt}\selectfont $|\hat u|/\sqrt{\mathcal{E}_{\infty}}$}
    \put(-64,47){  \fontsize{6pt}{6pt}\selectfont $|\hat v|/\sqrt{\mathcal{E}_{\infty}}$}
    \\
    \includegraphics[width = 0.48\textwidth] {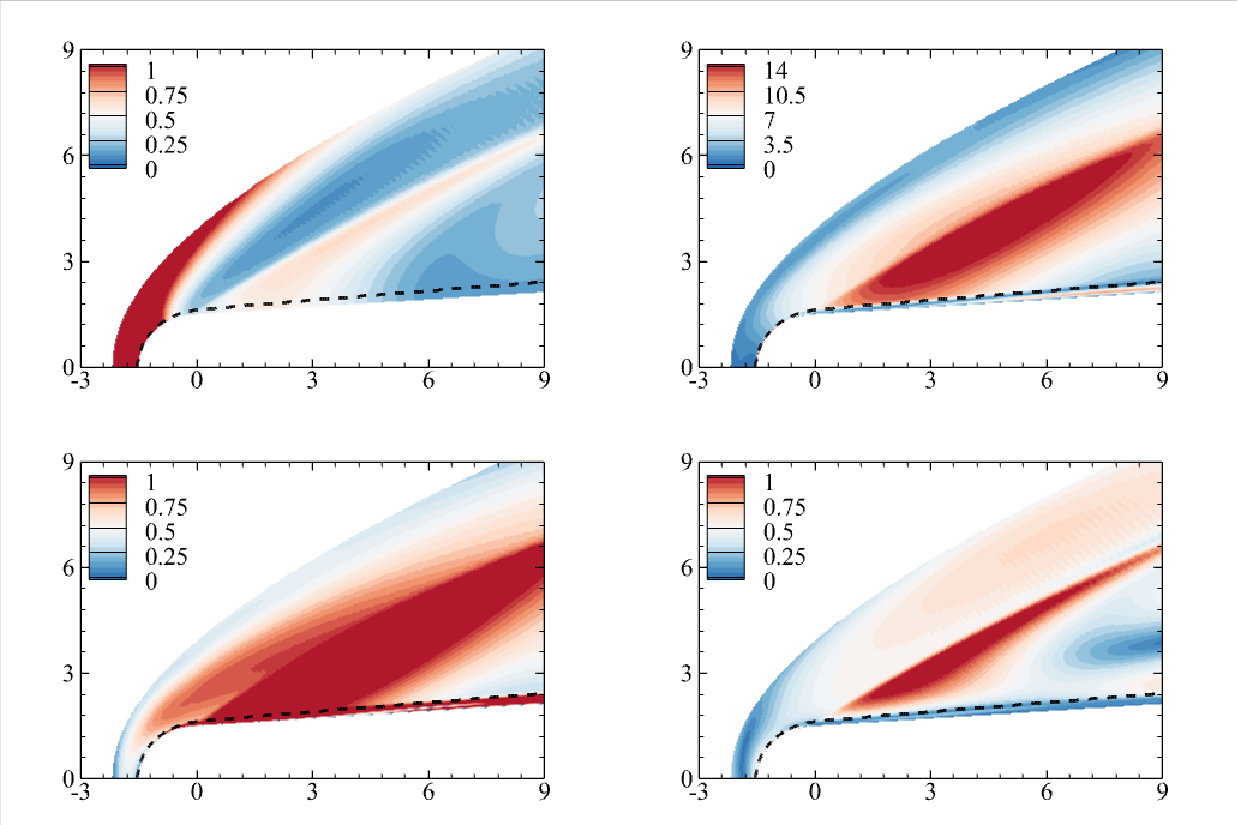}
    \put(-185,120){$(c)$}
    \put(-185,85){\rotatebox{90}{$k_3y$}}
    \put(-185,25){\rotatebox{90}{$k_3y$}}
    \put(-140,-5){$k_3x$}
    \put(-50,-5){$k_3x$}
    \put(-155,108){\fontsize{6pt}{6pt}\selectfont $|\hat p|/\sqrt{\mathcal{E}_{\infty}}$}
    \put(-64,108){ \fontsize{6pt}{6pt}\selectfont $|\hat T|/\sqrt{\mathcal{E}_{\infty}}$}
    \put(-155,47){ \fontsize{6pt}{6pt}\selectfont $|\hat u|/\sqrt{\mathcal{E}_{\infty}}$}
    \put(-64,47){  \fontsize{6pt}{6pt}\selectfont $|\hat v|/\sqrt{\mathcal{E}_{\infty}}$}
    \includegraphics[width = 0.48\textwidth] {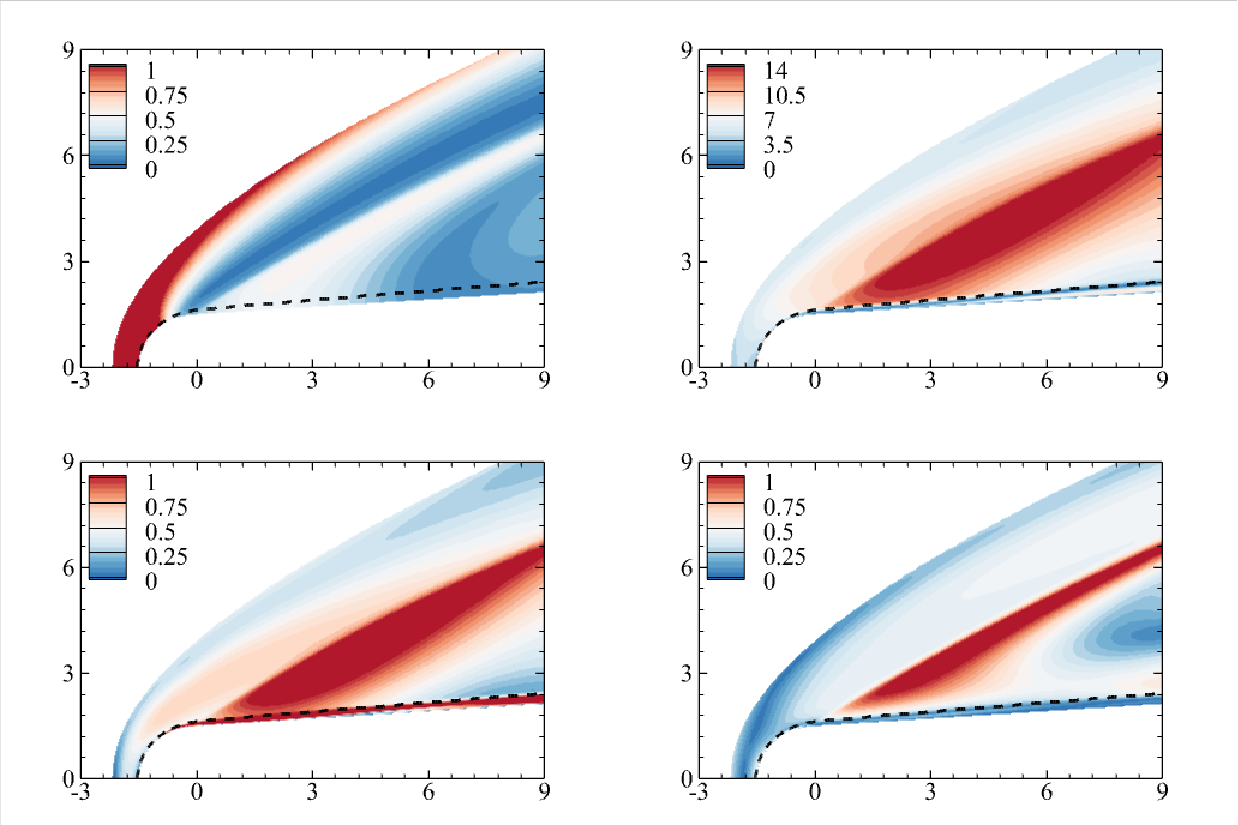}
    \put(-185,120){$(d)$}
    \put(-140,-5){$k_3x$}
    \put(-50,-5){$k_3x$}
    \put(-155,108){\fontsize{6pt}{6pt}\selectfont $|\hat p|/\sqrt{\mathcal{E}_{\infty}}$}
    \put(-64,108){ \fontsize{6pt}{6pt}\selectfont $|\hat T|/\sqrt{\mathcal{E}_{\infty}}$}
    \put(-155,47){ \fontsize{6pt}{6pt}\selectfont $|\hat u|/\sqrt{\mathcal{E}_{\infty}}$}
    \put(-64,47){  \fontsize{6pt}{6pt}\selectfont $|\hat v|/\sqrt{\mathcal{E}_{\infty}}$}
    \caption{Contours of the rescaled $|\hat p|$, $|\hat T|$, $|\hat u|$ and $|\hat v|$ with $\omega=0$, $k_3^*=1.5$ mm$^{-1}$ and $\vartheta=15^{\circ}$ for cases Av ($a$), Af ($b$), As ($c$) and Ae ($d$). The dashed lines mark the boundary-layer edge.}
    \label{fig:Contour_entropy_r=1_w=0_nose}
    \end{center}
\end{figure}

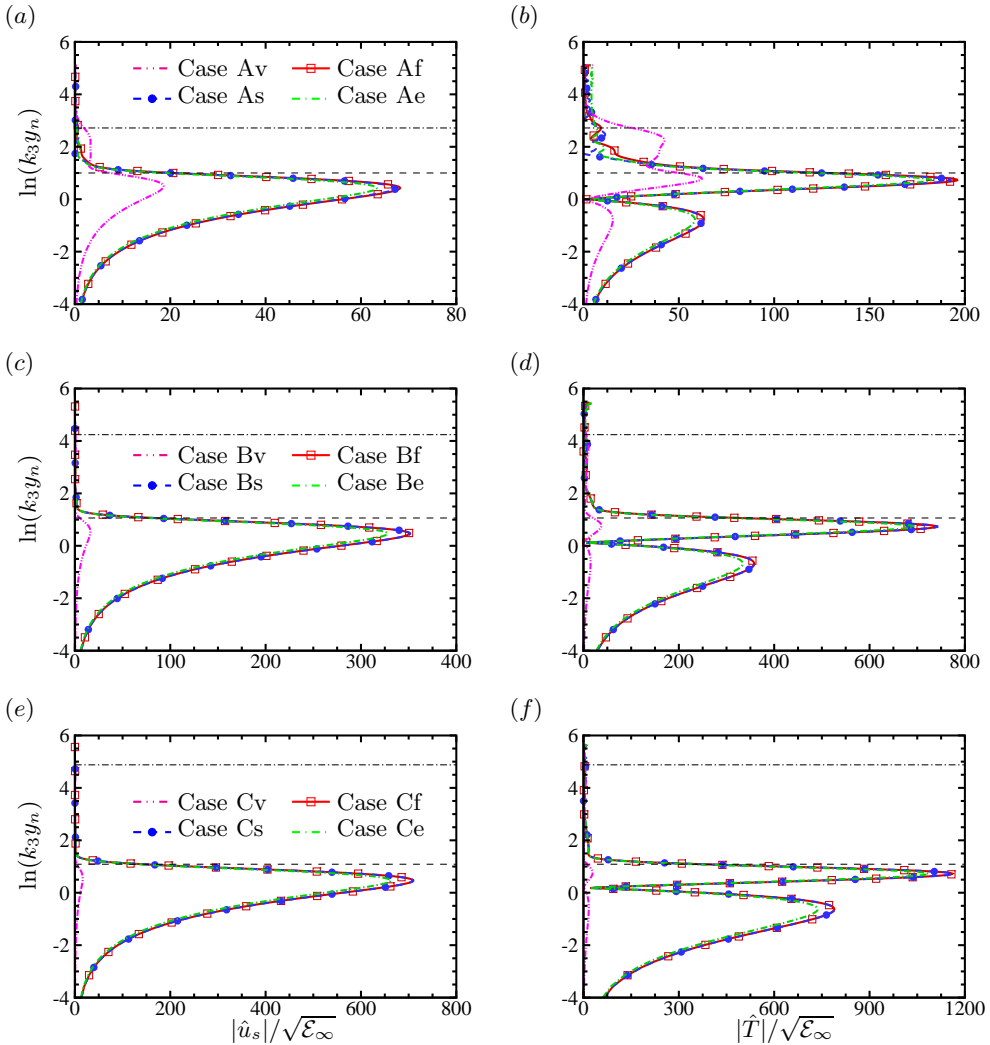
\begin{figure}
    \begin{center}
    \includegraphics[width = \textwidth] {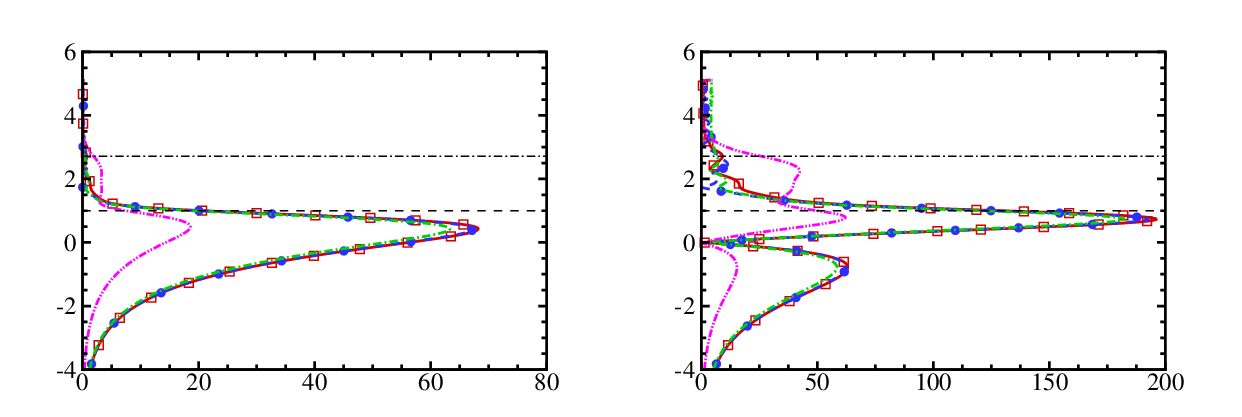}
    \put(-385,120){$(a)$}
    \put(-380,55){\rotatebox{90}{$\ln(k_3 y_n)$}}
    \put(-195,120){$(b)$}
    \put(-320,100){Case Av}
    \put(-260,100){Case Af}
    \put(-320,90){Case As}
    \put(-260,90){Case Ae}
    \put(-337,90){\begin{tikzpicture}
    \draw[magenta,dash dot dot,thick] (0,0) -- (0.5,0);
    \draw[red,thick] (2.1,0) -- (2.6,0);
    \node[draw,red,rectangle,inner sep=1.5pt] at (2.35,0) {};
    \draw[blue,dashed,thick] (0,-0.4) -- (0.5,-0.4);
    \node[draw,blue,circle,fill,inner sep=1pt] at (0.25,-0.4) {};
    \draw[green,dash dot,thick] (2.1,-0.4) -- (2.6,-0.4);
    \end{tikzpicture}}\\
    \includegraphics[width = \textwidth] {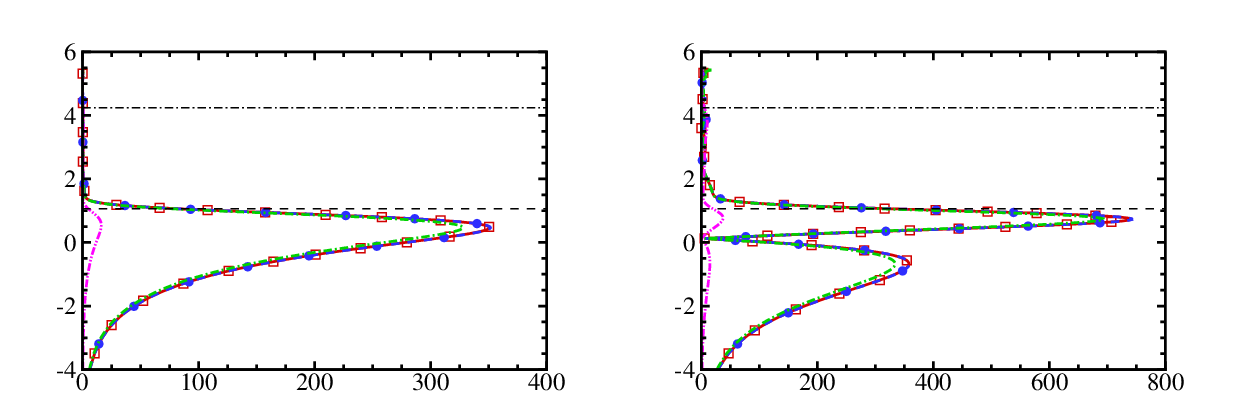}
    \put(-385,120){$(c)$}
    \put(-380,55){\rotatebox{90}{$\ln(k_3 y_n)$}}
    \put(-195,120){$(d)$}
    \put(-320,85){Case Bv}
    \put(-260,85){Case Bf}
    \put(-320,75){Case Bs}
    \put(-260,75){Case Be}
    \put(-337,75){\begin{tikzpicture}
    \draw[magenta,dash dot dot,thick] (0,0) -- (0.5,0);
    \draw[red,thick] (2.1,0) -- (2.6,0);
    \node[draw,red,rectangle,inner sep=1.5pt] at (2.35,0) {};
    \draw[blue,dashed,thick] (0,-0.4) -- (0.5,-0.4);
    \node[draw,blue,circle,fill,inner sep=1pt] at (0.25,-0.4) {};
    \draw[green,dash dot,thick] (2.1,-0.4) -- (2.6,-0.4);
    \end{tikzpicture}}\\
    \includegraphics[width = \textwidth] {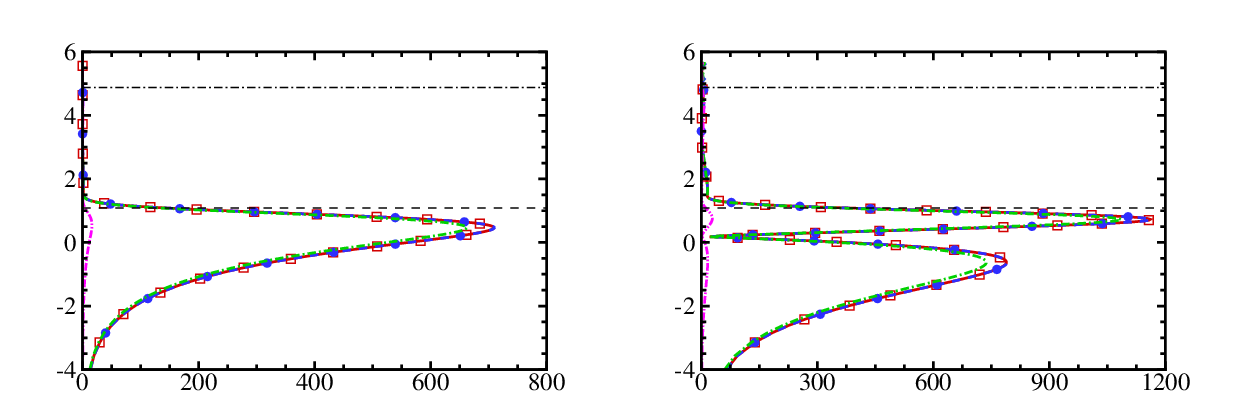}
    \put(-385,120){$(e)$}
    \put(-380,55){\rotatebox{90}{$\ln(k_3 y_n)$}}
    \put(-300,0){$|\hat u_s|/\sqrt{\mathcal{E}_{\infty}}$}
    \put(-195,120){$(f)$}
    \put(-110,0){$|\hat T|/\sqrt{\mathcal{E}_{\infty}}$}
    \put(-320,85){Case Cv}
    \put(-260,85){Case Cf}
    \put(-320,75){Case Cs}
    \put(-260,75){Case Ce}
    \put(-337,75){\begin{tikzpicture}
    \draw[magenta,dash dot dot,thick] (0,0) -- (0.5,0);
    \draw[red,thick] (2.1,0) -- (2.6,0);
    \node[draw,red,rectangle,inner sep=1.5pt] at (2.35,0) {};
    \draw[blue,dashed,thick] (0,-0.4) -- (0.5,-0.4);
    \node[draw,blue,circle,fill,inner sep=1pt] at (0.25,-0.4) {};
    \draw[green,dash dot,thick] (2.1,-0.4) -- (2.6,-0.4);
    \end{tikzpicture}}
    \caption{Wall-normal profiles of the rescaled $\hat u_s$ ($a,c,e$) and $\hat T$ ($b,d,f$) at $x_s^*=600$ mm, with $\omega=0$, $k_3^*=1.5$ mm $^{-1}$ and $\vartheta= 15^{\circ}$. ($a,b$): case A; ($c,d$): case B; ($e,f$): case C.}
    \label{fig:profile-sn-rescale-uT_xs=600_r=1-5-10}
    \end{center}
\end{figure}
Figure \ref{fig:profile-sn-rescale-uT_xs=600_r=1-5-10} presents a comparison of the profiles of $\hat u_s/\sqrt{{\cal E}_\infty}$  and $\hat T/\sqrt{{\cal E}_\infty}$ at $x_s^*=600$ mm across all the cases. Once again,  the receptivity efficiency increases with the nose radius for cases forced by acoustic and entropy perturbations, while it decreases for cases forced by vortical perturbations. Across different nose radii, the perturbation profiles of $\hat u_s$ exhibit notable similarity albeit with different magnitudes, whereas those of $\hat T$ do not show similarity. The secondary peak of $\hat T$ in the near-wall region becomes more prominent  as the nose bluntness increases.

\subsection{Examination of the credibility of the optimal growth theory}
\label{sec:compare_optimal}
In the previous studies \citep{paredes2016optimal,paredes2019nonmodal,paredes2020mechanism}, the excited non-modal perturbations were modeled by the optimal growth theory (OGT), which describes the maximum amplification  within a selected  domain. For linear perturbations of the form (\ref{eq:travelling}a) with the coordinate system transformed from $(\xi,\eta)$ to $(x_s,y_n)$, we define the energy gain from the initial position $x_{s0}$ to the terminal position $x_{sI}$ as
\begin{equation}
J_E(\hat{\pmb\varphi};x_{s0},x_{sI})=\frac{\bar {\mathcal E}(\hat{\pmb\varphi};x_{sI})}{\bar {\mathcal E}(\hat{\pmb\varphi};x_{s0})}.
\end{equation}
To estimate the maximum energy amplification,  
\begin{equation}
    G_E(x_{s0},x_{sI})=\max_{\hat{\pmb\varphi}} J_E(\hat{\pmb\varphi};x_{s0},x_{sI}),
\end{equation}
an optimisation problem is formulated, which can be solved through  an iterative approach. Each iteration step involves two sweeps: one downstream march from the initial position $x_{s0}$  employing the HLNS or the linearised parabolized stability equation (LPSE) approach, and an upstream march from the terminal position $x_{sI}$  employing the adjoint HLNS or LPSE approach. Notably, the  shock region is excluded from the computational domain to ensure the applicability of these approaches. A detailed introduction of the OGT calculation can be found in \citet{paredes2016optimal}.
It is worth noting that the OGT does not take into account the freestream forcing, presenting merely the supremum of the potential energy amplification. Through  the SF-HLNS approach, we are able to examine  the credibility of OGT on the predictions of the energy amplification of the non-modal perturbations in this paper.

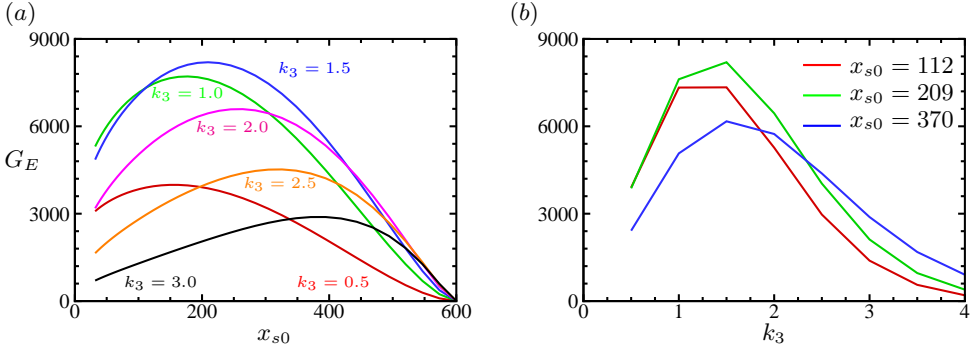
\begin{figure}
    \begin{center}
    \includegraphics[width = \textwidth] {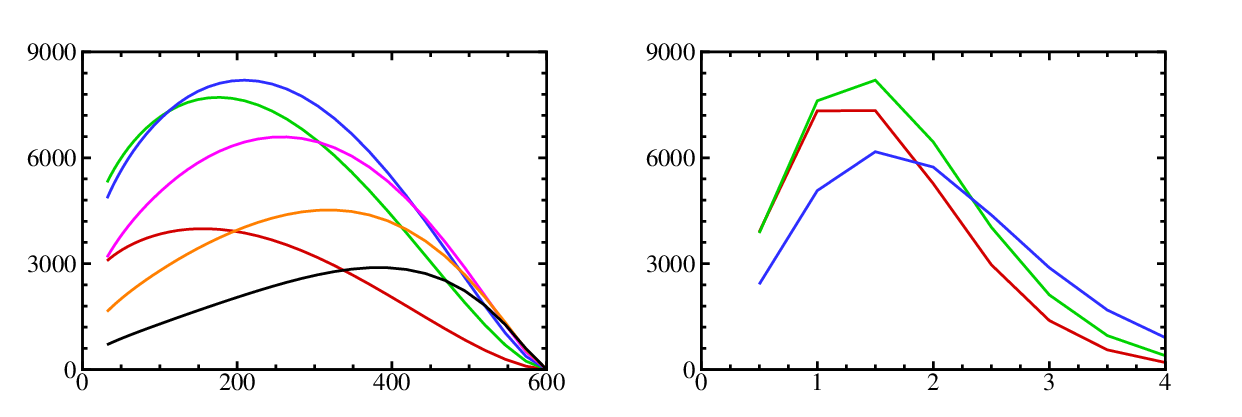}
    \put(-385,120){$(a)$}
    \put(-385,65){$G_E$}
    \put(-290,0){$x_{s0}$}
    \put(-275,20){\fontsize{6pt}{6pt}\color{red}\selectfont $k_3=0.5$}
    \put(-330,91){\fontsize{6pt}{6pt}\color{green}\selectfont $k_3=1.0$}
    \put(-282,100){\fontsize{6pt}{6pt}\color{blue}\selectfont $k_3=1.5$}
    \put(-313,78){\fontsize{6pt}{6pt}\color{magenta}\selectfont $k_3=2.0$}
    \put(-295,57){\fontsize{6pt}{6pt}\color{orange}\selectfont $k_3=2.5$}
    \put(-340,20){\fontsize{6pt}{6pt}\color{black}\selectfont $k_3=3.0$}
    \put(-195,120){$(b)$}
    \put(-100,0){$k_3$}
    \put(-67,100){$x_{s0}=112$}
    \put(-67,90){$x_{s0}=209$}
    \put(-67,80){$x_{s0}=370$}
    \put(-85,80){\begin{tikzpicture}
    \draw[red,thick] (0,0) -- (0.5,0);
    \draw[green,thick] (0,-0.4) -- (0.5,-0.4);
    \draw[blue,thick] (0,-0.8) -- (0.5,-0.8);
    \end{tikzpicture}}
    \caption{Dependence of the optimal energy gain $G_E$ on the initial position $x_{s0}$ ($a$) and the spanwise wavenumber $k_3$ ($b$) for fixed terminal position $x_{sI}=600$. }
    \label{fig:optimaldis_line}
    \end{center}
\end{figure}
By fixing the terminal position $x_{sI}=600$ and the perturbation frequency to be zero, we explore the influence of the initial position $x_{s0}$ and the spanwise wavenumber $k_3$ on the energy gain $G_E$, as illustrated in  figure \ref{fig:optimaldis_line} for case A. In panel ($a$), the curve for each $k_3$ value initially rises with  $x_{s0}$, but a subsequent reversal in trend becomes apparent after reaching a peak position. As $k_3^*$ increases, the peak position shifts downstream. Panel ($b$) illustrates the  dependence of $G_E$ on $k_3$ for fixed $x_{s0}$ values, revealing the optimal spanwise wavenumber at $k_3\approx 1.5$. The maximum energy gain is achieved at  $G_E=8199$, appearing at $x_{s0}=209$ and $k_3=1.5$.

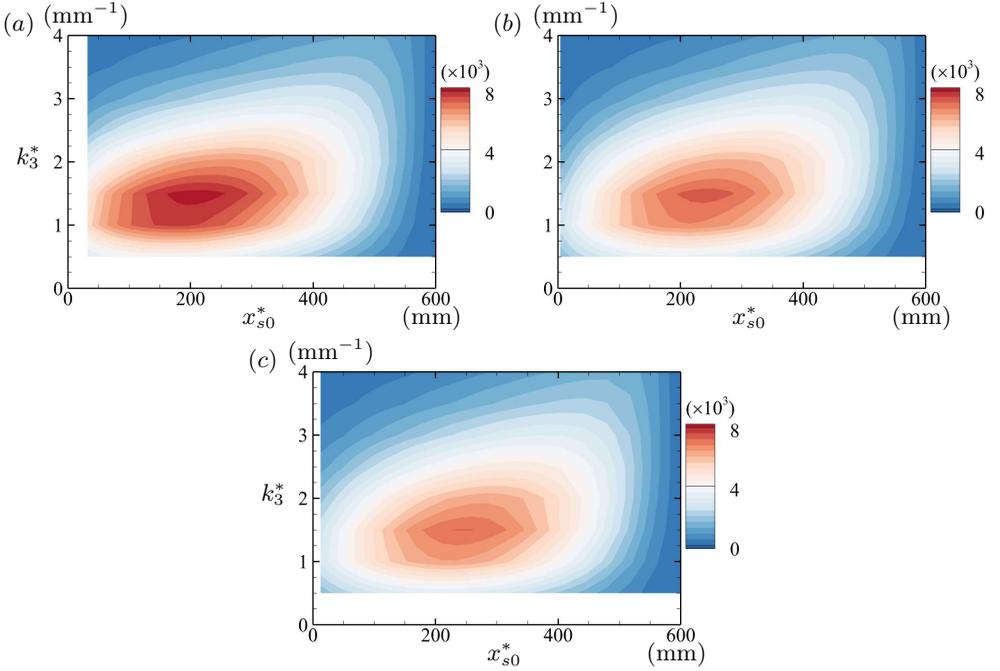
\begin{figure}
    \begin{center}
    \includegraphics[width = 0.48\textwidth] {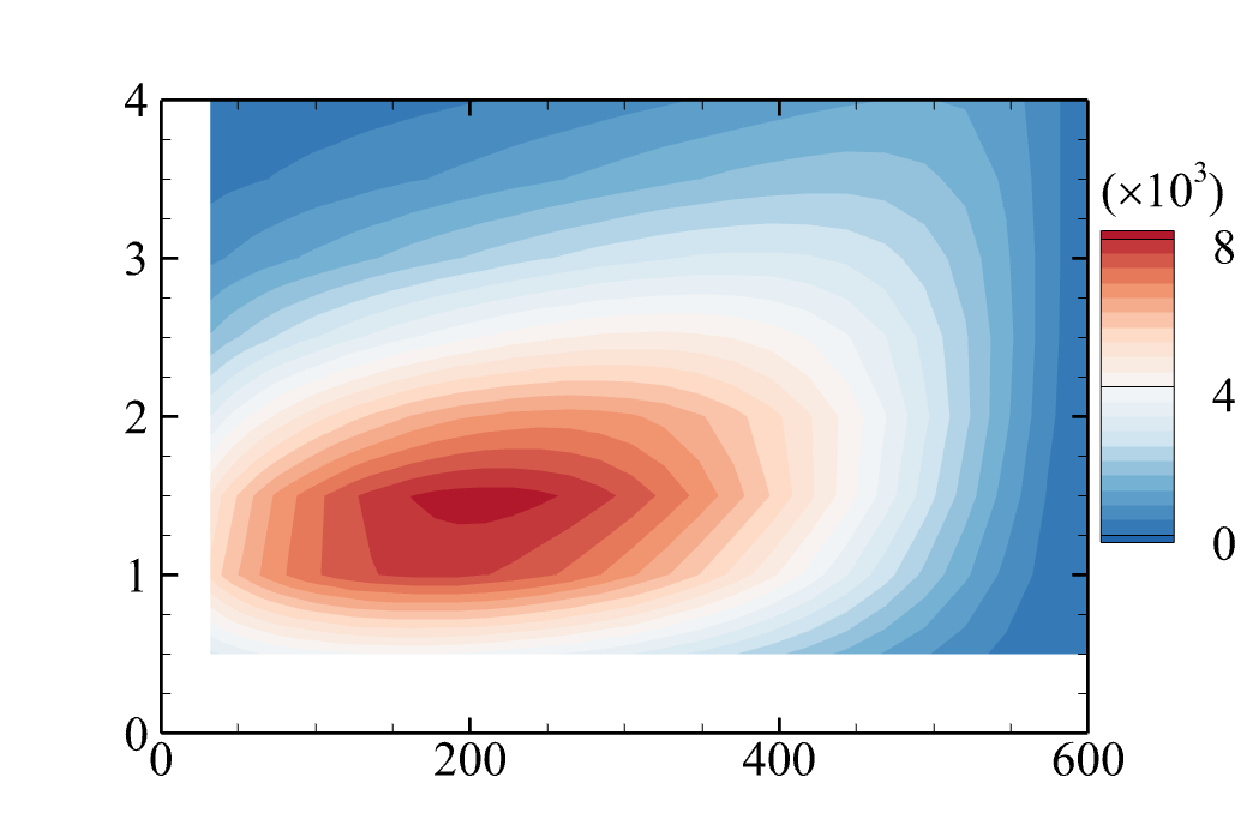}
    \put(-185,110){$(a)$}
    \put(-180,60){$k_3^*$}
    \put(-95,0){$x_{s0}^*$}
    \put(-35,0){(mm)}
    \put(-170,113){(mm$^{-1}$)}
    \includegraphics[width = 0.48\textwidth] {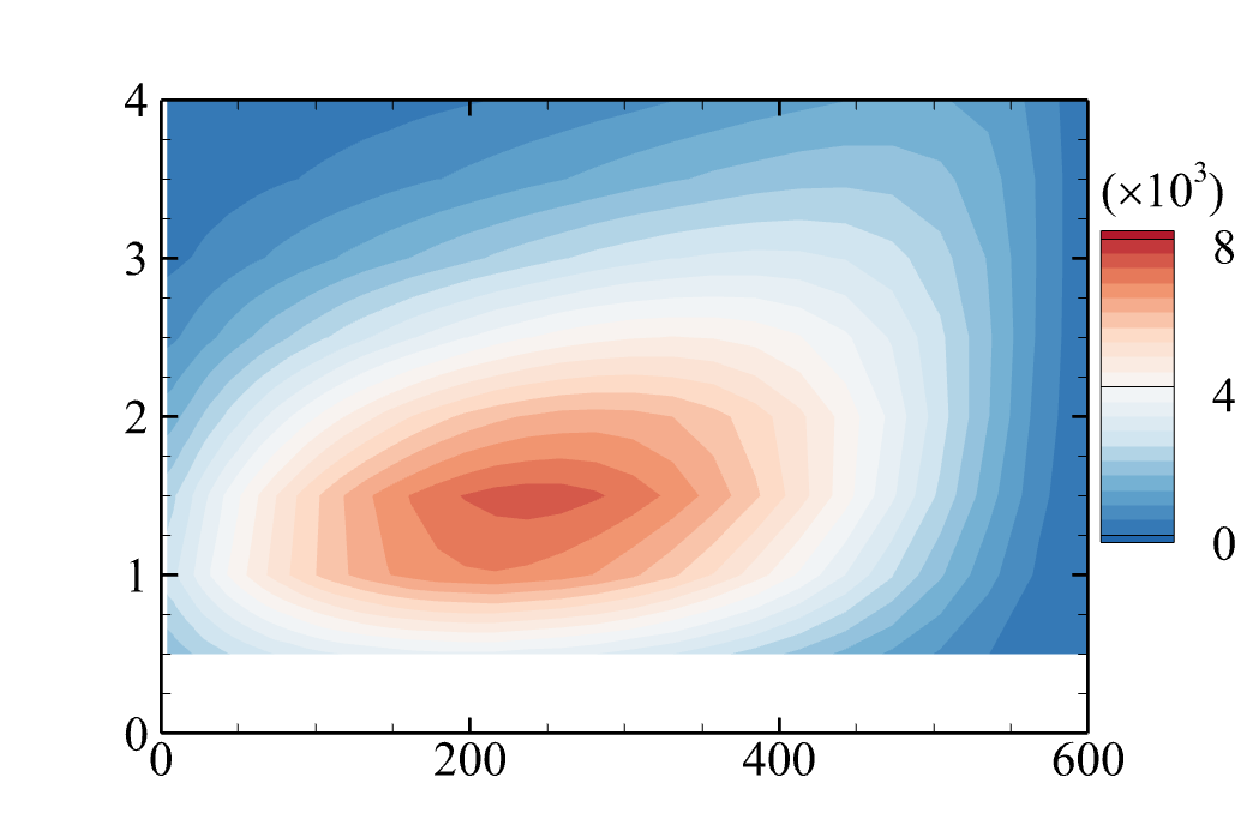}
    \put(-185,110){$(b)$}
    \put(-95,0){$x_{s0}^*$}
    \put(-35,0){(mm)}
    \put(-170,113){(mm$^{-1}$)}\\
    \includegraphics[width = 0.48\textwidth] {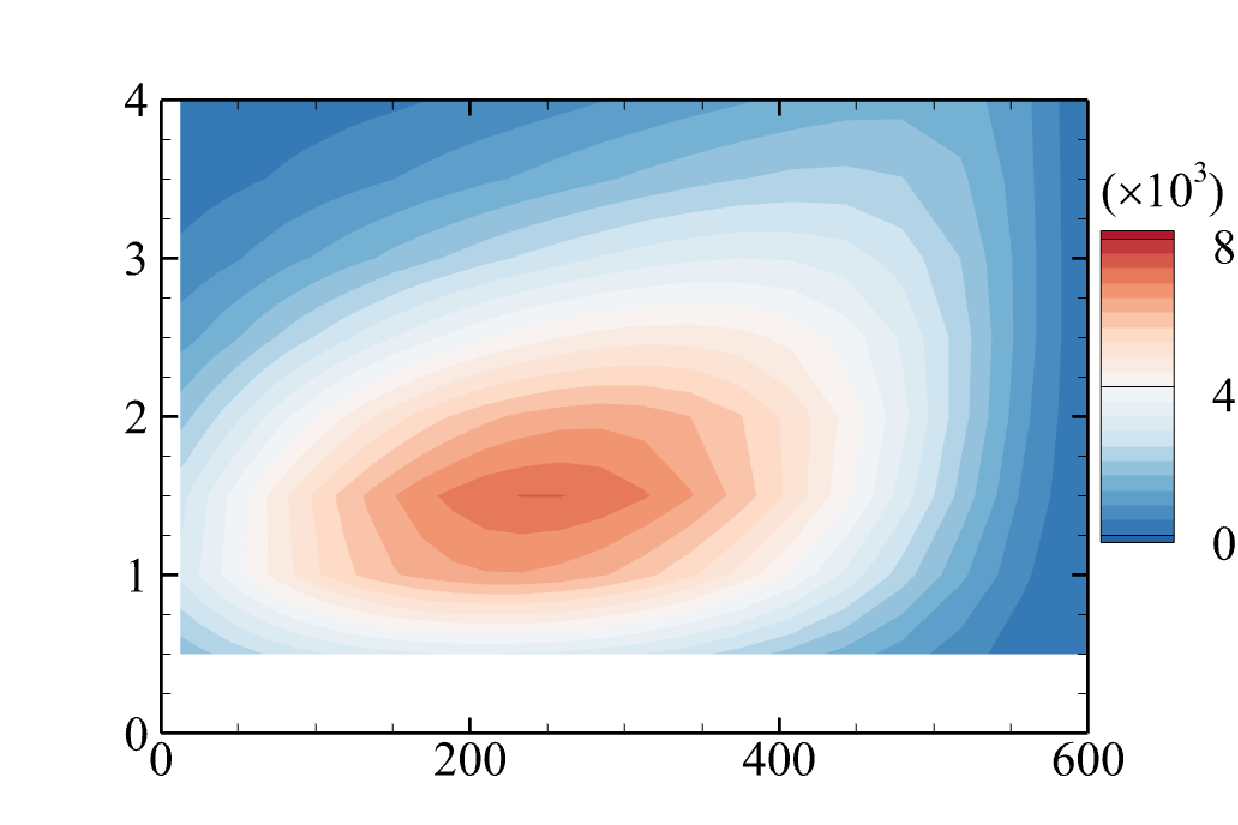}
    \put(-185,110){$(c)$}
    \put(-180,60){$k_3^*$}
    \put(-95,0){$x_{s0}^*$}
    \put(-35,0){(mm)}
    \put(-170,113){(mm$^{-1}$)}
    \caption{Contours of the energy gain $G_E$ in the $k_3^*$-$x_{s0}^*$ plane. ($a$): case A; ($b$): case B; ($c$): case C.}
    \label{fig:optimaldis_cont}
    \end{center}
\end{figure}
\begin{table}
	\begin{center}
		\def~{\hphantom{0}}
		\begin{tabular}{ccccc}
			\vspace{.2cm}
			Case & 
   $\begin{array}{c}
			\mbox{  Spanwise} \\\mbox{wavenumber} \\
			   k_3^* (\mbox{mm}^{-1})
			\end{array} $& 
   $\begin{array}{c}
			\mbox{   Optimal interval} \\
			 \mbox{[}x_{s0}^*
,x_{sI}^*] (\mbox{mm})\end{array} $  & $\begin{array}{c}
			   \mbox{ Optimal growth }\\
			   G_{E,max}
			\end{array}$ &$\begin{array}{c}
			  \mbox{SF-HLNS}   \\ \mbox{amplification}\\
     \mbox{in the same interval}
			     
			\end{array}$  \\ \vspace{.2cm}
			A  &1.5         &  [209,600]&8199& 3.1\\
   \vspace{.2cm}
                B &1.5    &  [237,600]   &7378& 2.6 \\
   \vspace{.2cm}
                C        &1.5 &  [258,600] &7219&   2.2 
		\end{tabular}
		\caption{Summary of the optimal growth calculations for the three cases.}
		\label{tab:optimal}
	\end{center}
\end{table}
Figure \ref{fig:optimaldis_cont} plots the contours of  $G_E$ in the $x_{s0}^*$-$k_3^*$ plane across the three cases. For convenience of comparison, the dimensional length scales are employed, with the terminal position chosen as $x_{sI}^*$=600 mm. The highest energy gain and its corresponding initial position for each case are summarised in  Table \ref{tab:optimal}.
The optimal spanwise wavenumbers across all the  nose radii are $k_3^*\approx$1.5 mm$^{-1}$, consistent  with the SF-HLNS calculations under various freestream forcing conditions. This trend may be linked to the boundary-layer thickness at the terminal position, determining the optimal length scale of non-modal perturbations. 
The optimal initial position $x_0^*$ shifts downstream with increasing $r^*$. 
However, the OGT energy amplification within the optimal interval reduces with increase of the  nose bluntness, which is in contrast to the conclusion of heightened receptivity to acoustic and entropy forcing for larger bluntness, as illustrated in Figure \ref{fig:LNS_A-AT-xs_check}.

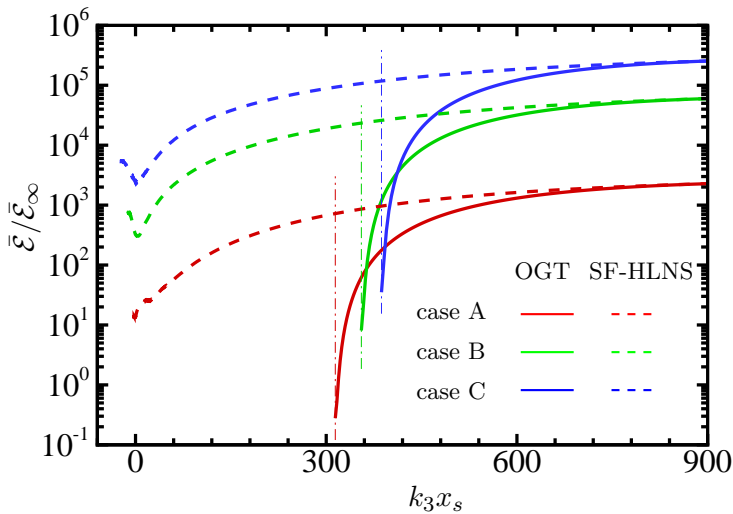
\begin{figure}
    \begin{center}
    \includegraphics[width = 0.8\textwidth] {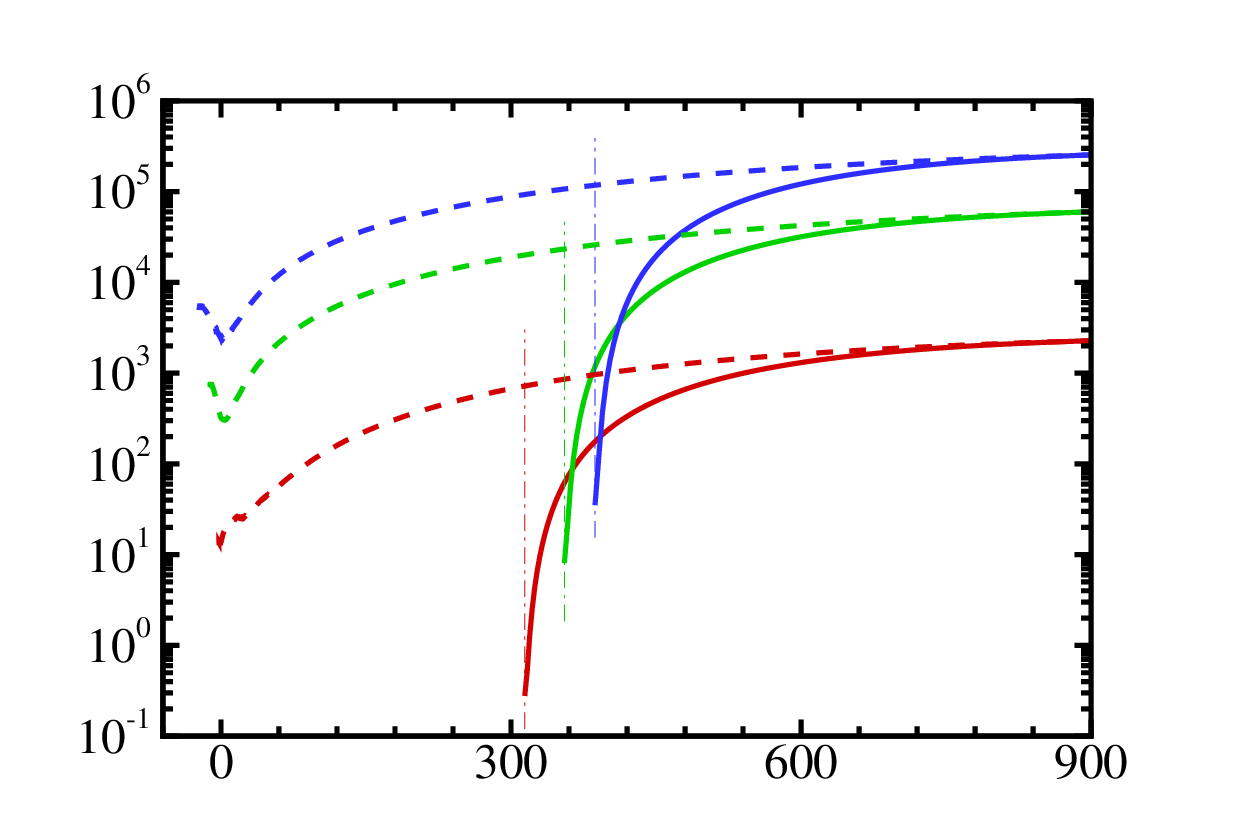}
    \put(-300,95){\rotatebox{90}{\large$\bar{\mathcal E}/\bar{\mathcal E}_\infty$}}
    \put(-110,85){OGT}
    \put(-85,85){ SF-HLNS} 
    \put(-150,70){ case A}
    \put(-150,55){ case B}
    \put(-150,40){ case C}
    \put(-108,43){\begin{tikzpicture}
    \draw[red,thick] (0,0) -- (0.7,0);
    \draw[green,thick] (0,-0.5) -- (0.7,-0.5);
    \draw[blue,thick] (0,-1) -- (0.7,-1);
    \draw[red,dashed,thick] (1.2,0) -- (1.8,0);
    \draw[green,dashed,thick] (1.2,-0.5) -- (1.8,-0.5);
    \draw[blue,dashed,thick] (1.2,-1) -- (1.8,-1);
    \end{tikzpicture}}
    \put(-150,0){{\large$k_3 x_s$}}
    \caption{Comparison of the total  energy evolution $\bar {\cal E}/\bar {\cal E}_\infty(k_3 x_s)$ between the SF-HLNS calculations subject to freestream slow acoustic forcing (cases As, Bs and Cs extracted from figure \ref{fig:LNS_A-AT-xs_check}) and the OGT (cases A, B and C) calculations, where $\omega=0$, $k_3^*=1.5$ mm$^{-1}$ and $\vartheta=15^{\circ}$. The optimal initial positions, marked by the vertical dash-dotted lines, are chosen for the OGT calculations.}
    \label{fig:line_GE_OGT_SFHLNS_slow}
    \end{center}
\end{figure}

Now, we perform a comparison of the streamwise evolution of the total energy $\bar {\cal E}/\bar {\cal E}_\infty(k_3 x_s)$ obtained by the OGT  and the SF-HLNS calculations  for the three cases in Figure \ref{fig:line_GE_OGT_SFHLNS_slow}. This comparison aims to offer a quantitative examination of the credibility of the OGT method.
The initial and terminal positions of the OGT calculations are referenced  from Table \ref{tab:optimal}.
For representative purposes, the cases forced by freestream slow acoustic perturbations (cases As, Bs and Cs), are chosen for the SF-HLNS calculations.  The OGT curves are adjusted to align the amplitudes at $k_3x_s=900$ with those of the SF-HLNS calculations. It is clear that in the downstream region, the growth trend of the perturbation energy  predicted by OGT agrees well with the SF-HLNS calculation for each case. 
However, near the initial position, the perturbation energy predicted by OGT exhibits a much faster growth rate compared to the SF-HLNS calculation. Consequently, the energy amplification of the non-modal perturbation  is greatly over-predicted by OGT, with  the comparison of the energy amplification within the same optimal interval between OGT and SF-HLNS calculations being presented in the third and fourth columns of Table  \ref{tab:optimal}.  The discrepancy of the energy amplification between the two approaches is substantial, reaching up to three orders of magnitude. Insights from the SF-HLNS calculations reveal that the amplification of non-modal perturbations arises from both the rapid energy rise in the nose region and successive growth in the downstream region, rather than solely from a sharp local energy surge in the  downstream region depicted by OGT.

\begin{figure}
    \begin{center}
    \includegraphics[width = 0.48\textwidth] {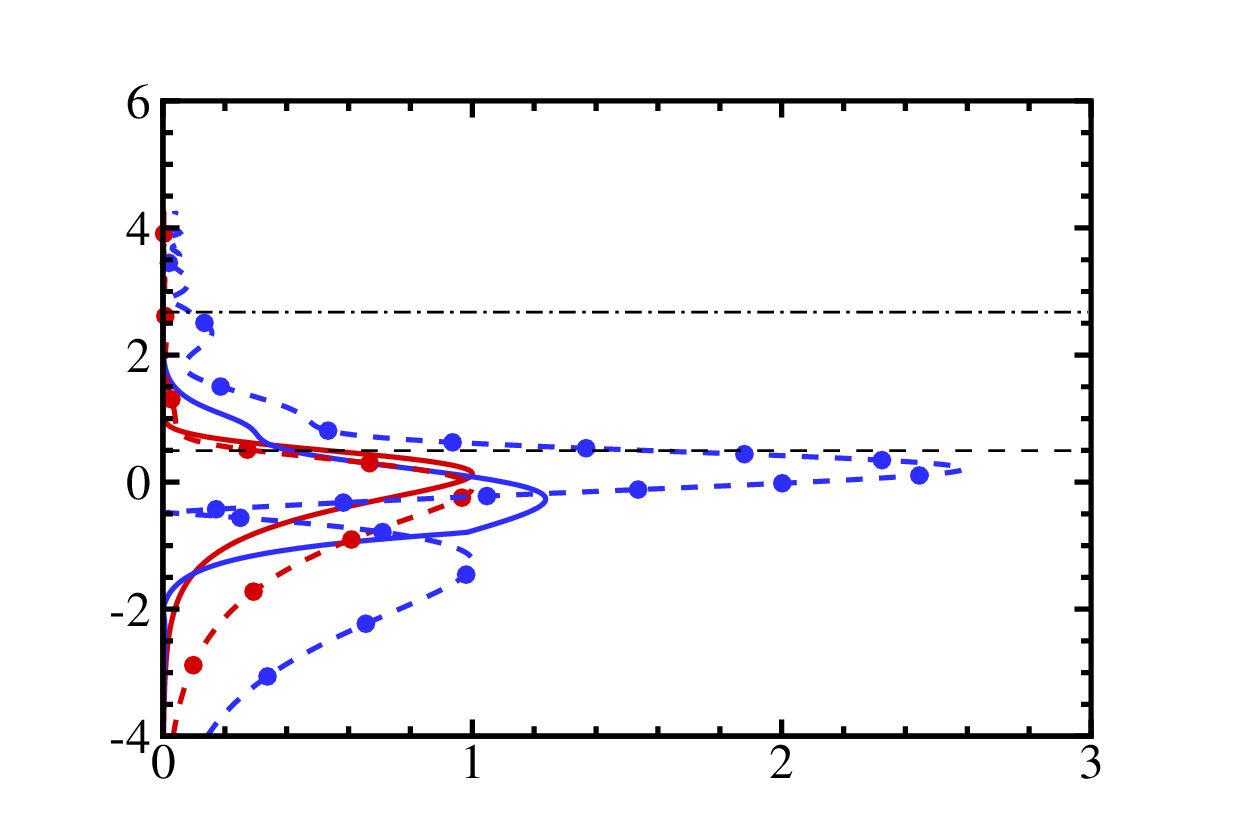}
    \includegraphics[width = 0.48\textwidth] {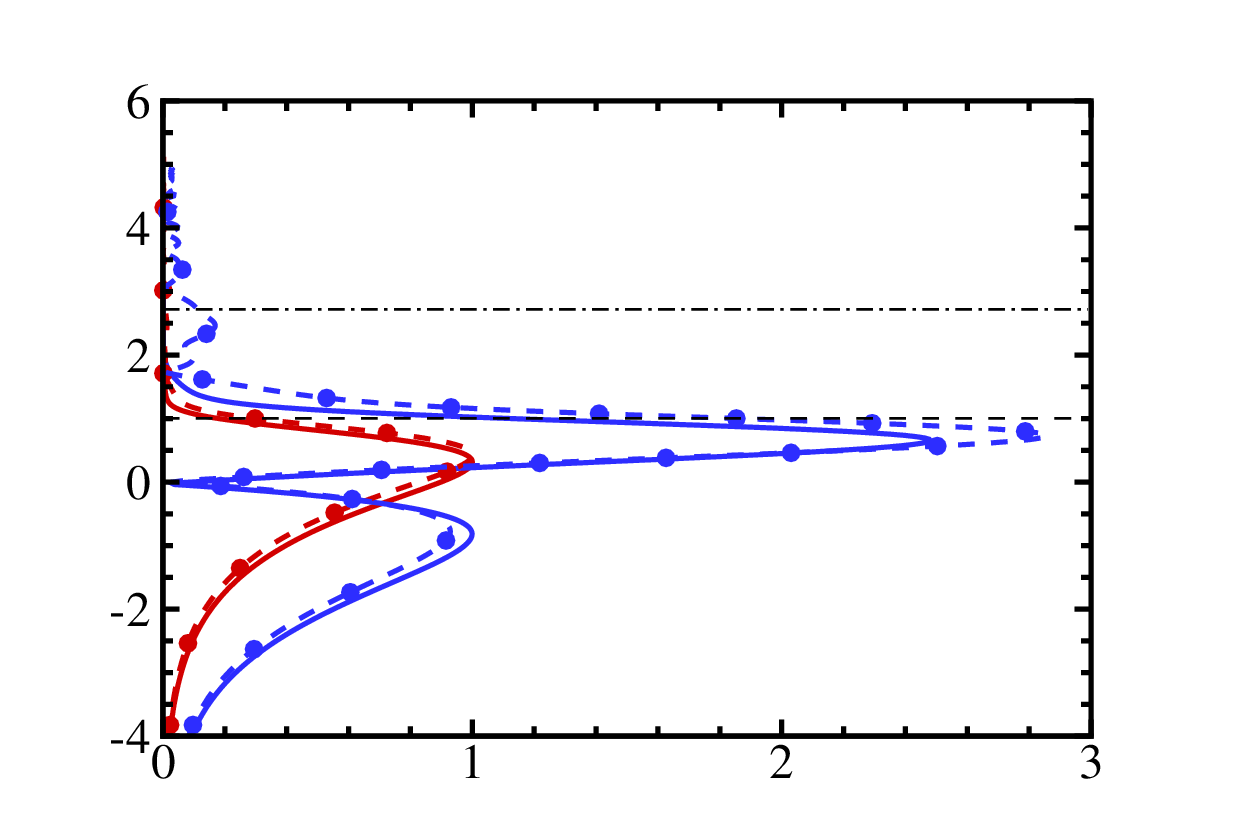}
    \put(-375,110){$(a)$}
    \put(-375,55){\rotatebox{90}{$\ln(k_3 y_n)$}}
    \put(-185,110){$(b)$}
    \put(-340,90){\fontsize{6pt}{6pt}\selectfont OGT}
    \put(-340,82){\fontsize{6pt}{6pt}\selectfont SF-HLNS}  
    \put(-302,98){\fontsize{8pt}{8pt}\selectfont $|\hat u_s|$}
    \put(-280,98){\fontsize{8pt}{8pt}\selectfont $|\hat T|$}
    \put(-258,98){\fontsize{8pt}{8pt}\selectfont $|\hat v_n|$}
    \put(-236,98){\fontsize{8pt}{8pt}\selectfont $|\hat w|$}
    \put(-303,82){\begin{tikzpicture}
    \draw[red,thick] (0,0) -- (0.5,0);
    \draw[blue,thick] (0.75,0) -- (1.25,0);
    \draw[green,thick] (1.5,0) -- (2.0,0);
    \draw[magenta,thick] (2.25,0) -- (2.75,0);
    \draw[red,thick,dashed] (0,-0.3) -- (0.5,-0.3);
    \node[draw,red,circle,fill,inner sep=1pt] at (0.25,-0.3) {};
    \draw[blue,thick,dashed] (0.75,-0.3) -- (1.25,-0.3);
    \node[draw,blue,circle,fill,inner sep=1pt] at (1,-0.3) {};
    \draw[green,thick,dashed] (1.5,-0.3) -- (2.0,-0.3);
    \node[draw,green,circle,fill,inner sep=1pt] at (1.75,-0.3) {};
    \draw[magenta,thick,dashed] (2.25,-0.3) -- (2.75,-0.3);
    \node[draw,magenta,circle,fill,inner sep=1pt] at (2.5,-0.3) {};
    \end{tikzpicture}}\\
    \includegraphics[width = 0.48\textwidth] {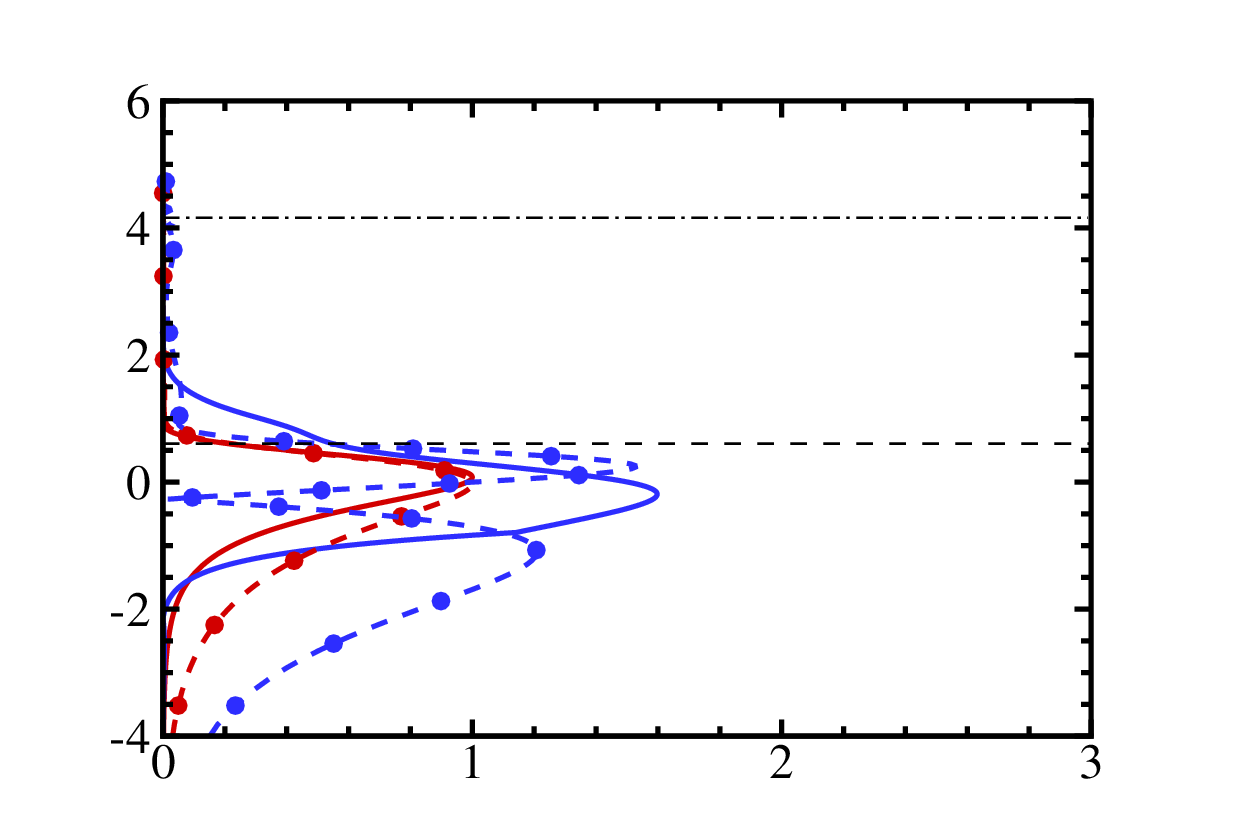}
    \includegraphics[width = 0.48\textwidth] {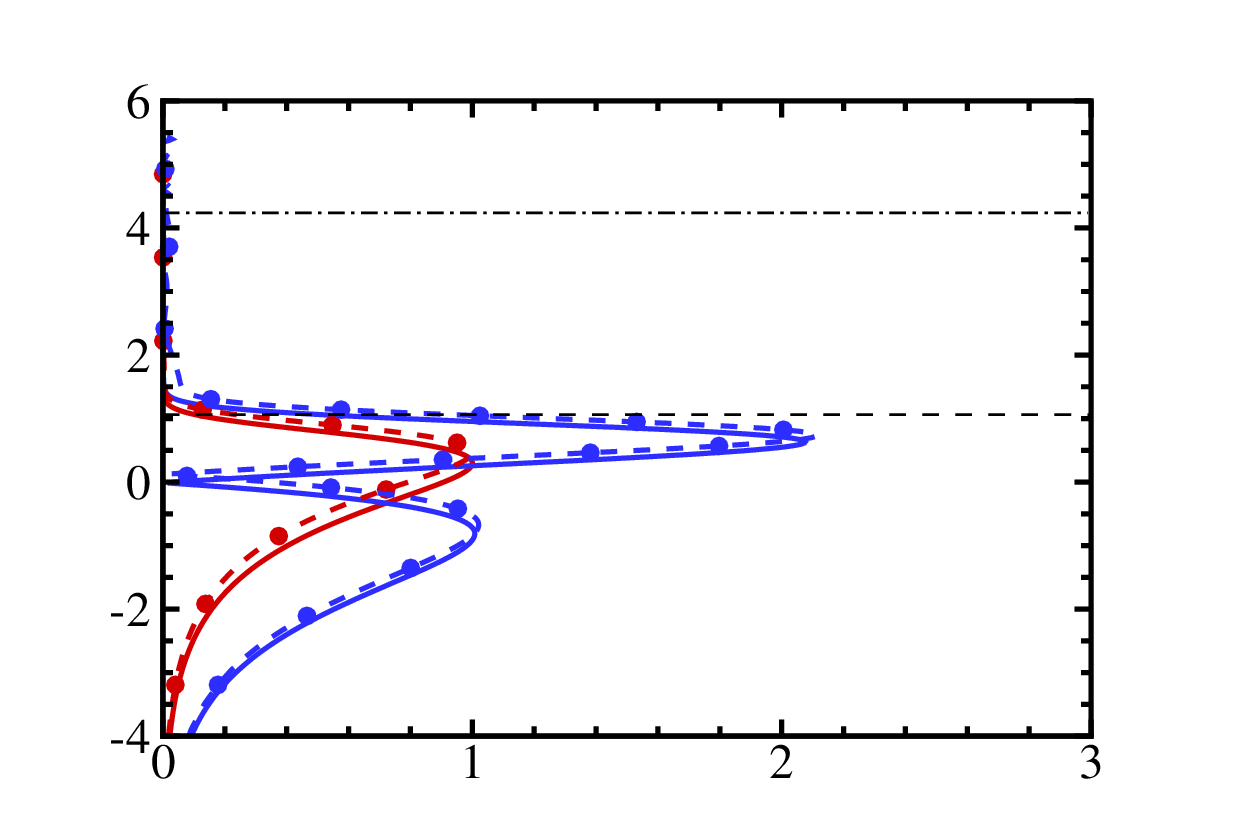}
    \put(-375,110){$(c)$}
    \put(-375,55){\rotatebox{90}{$\ln(k_3 y_n)$}}
    \put(-185,110){$(d)$}\\
    \includegraphics[width = 0.48\textwidth] {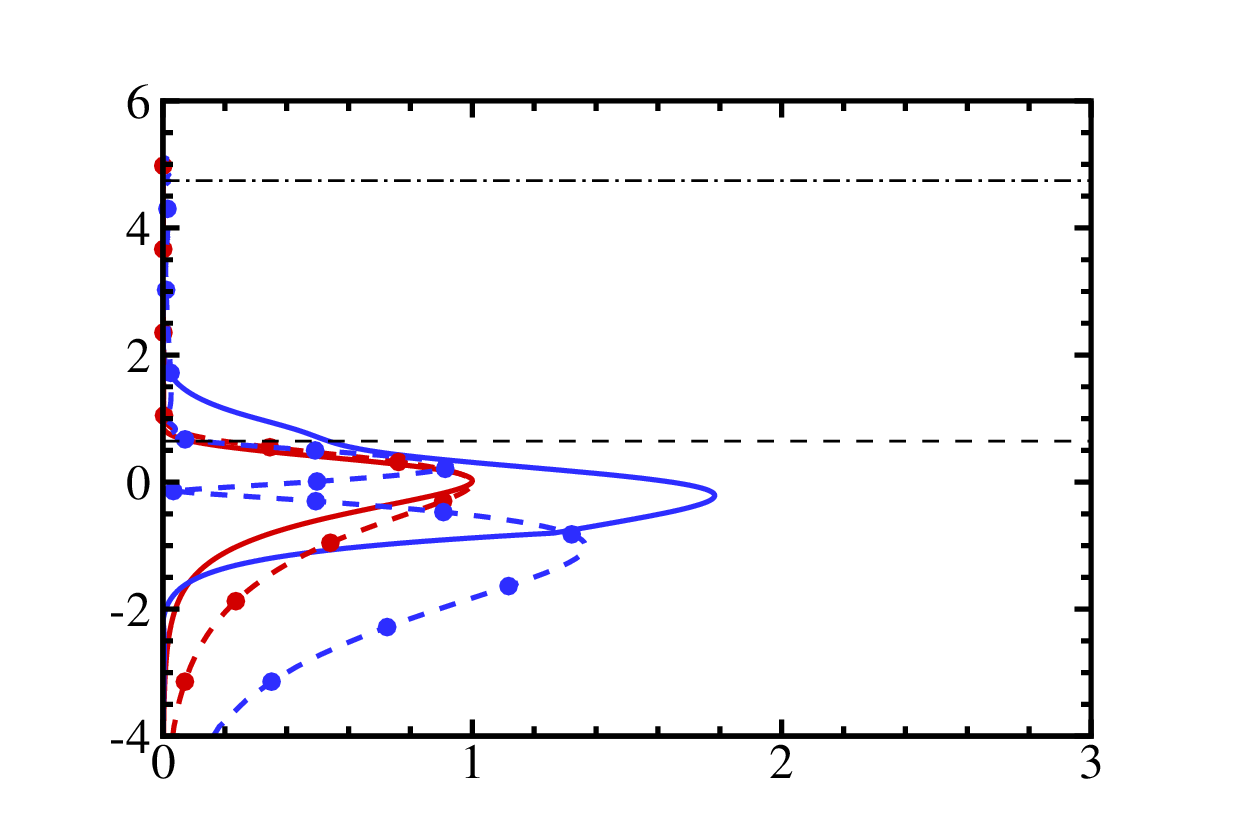}
    \includegraphics[width = 0.48\textwidth] {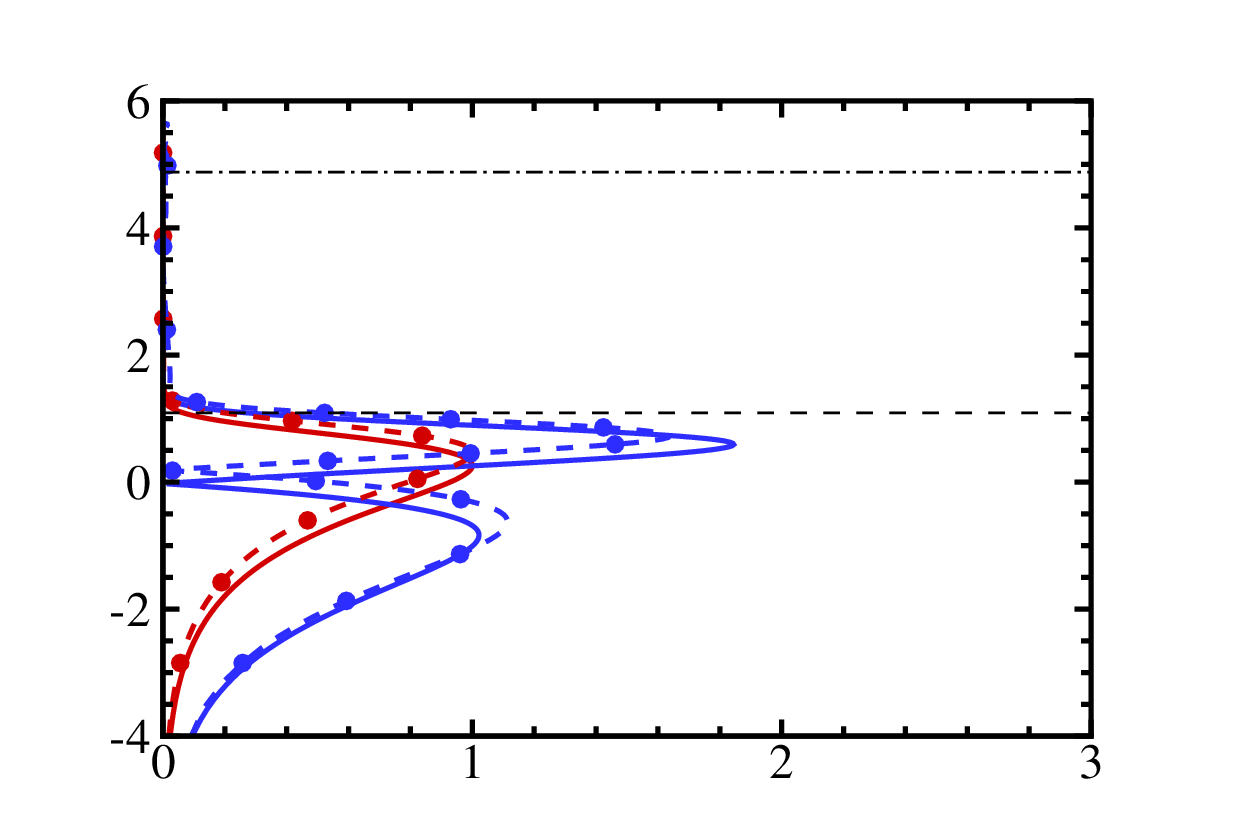}
    \put(-375,110){$(e)$}
    \put(-375,55){\rotatebox{90}{$\ln(k_3 y_n)$}}
    \put(-185,110){$(f)$}
    \put(-295,0){$|\hat u_s|$, $|\hat T|$}
    \put(-108,0){$|\hat u_s|$, $|\hat T|$}\\
    \includegraphics[width = 0.48\textwidth] {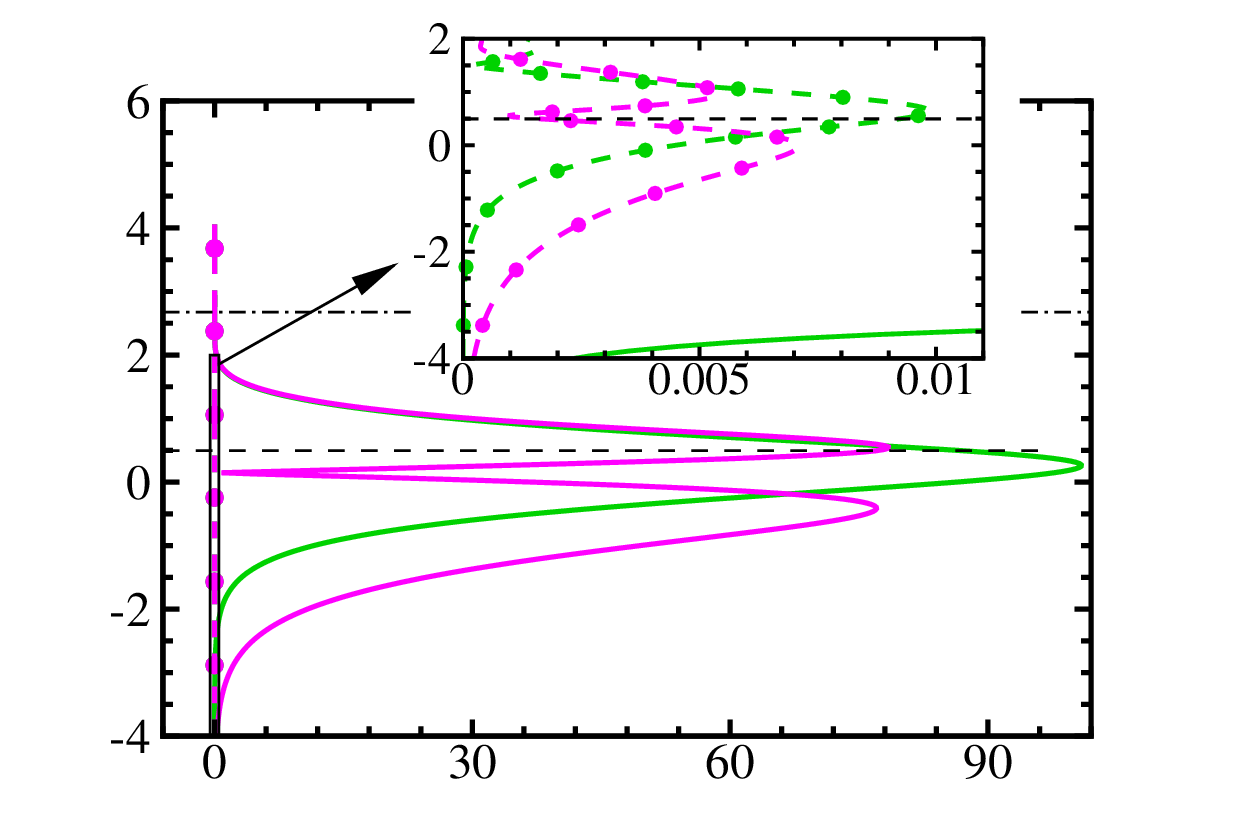}
    \includegraphics[width = 0.48\textwidth] {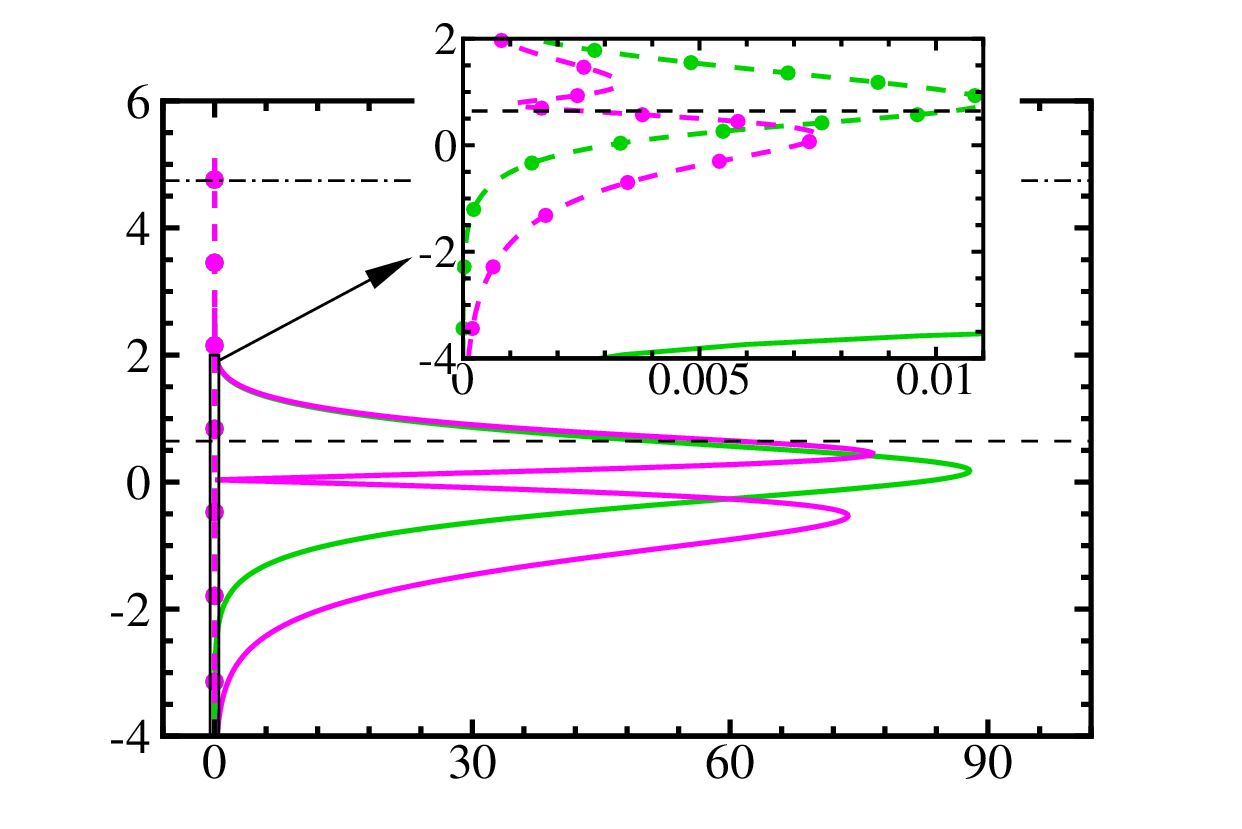}
    \put(-375,110){$(g)$}
    \put(-375,55){\rotatebox{90}{$\ln(k_3 y_n)$}}
    \put(-185,110){$(h)$}
    \put(-298,0){$|\hat v_n|$, $|\hat w|$}
    \put(-110,0){$|\hat v_n|$, $|\hat w|$}
    \caption{Comparison of the perturbation profiles between the OGT and SF-HLNS calculations, with $k_3^*=1.5$ mm$^{-1}$ for cases A ($a,b$), B ($c,d$) and C ($e,f$). ($a,c,e$): profiles of $\hat u_s$ and $\hat T$ at the optimal initial position $x_{s0}^*$ selected from Table \ref{tab:optimal};  ($b,d,f$):  profiles  of $\hat u_s$ and $\hat T$ at the terminal position $x_{sI}^*=600$ mm; ($g,h$):  profiles of $\hat v_n$ and $\hat w$ at $x_{s0}^*$ for cases A and C, respectively. The profiles are normalised by the peak value of $|\hat u_s|$. }
    \label{fig:optimaldis_profile}
    \end{center}
\end{figure}
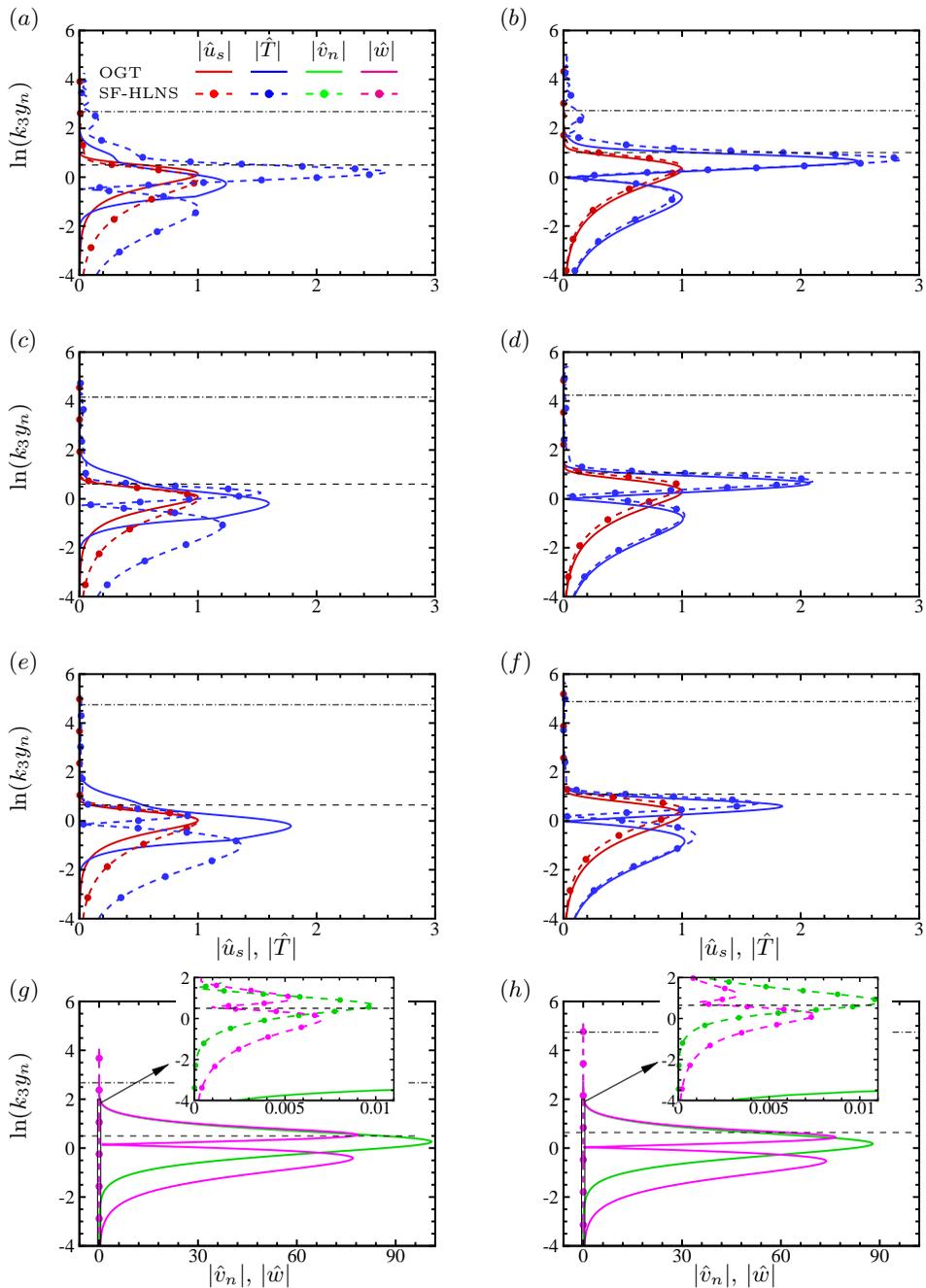
In figure \ref{fig:optimaldis_profile}, we further  conduct a detailed comparison of the optimal perturbation profiles with the SF-HLNS calculations at both the initial and terminal positions. For each nose radius, the profiles of $\hat u_s$ and $\hat T$ obtained by OGT and SF-HLNS at the terminal position $x_{sI}^*=600$ mm, as shown in panels ($b,d,f$), exhibit notable similarity. However, the optimal profiles of  $\hat u_s$ and $\hat T$ at the optimal initial position $x_{s0}^*$, depicted in panels ($a,c,e$), deviate remarkably from the SF-HLNS calculations. In the $y_n$-$z$ plane, the wall-normal and spanwise velocities reveal a distinct roll structure that could facilitate the generation of an elongated streaky pattern due to the lift-up mechanism. The wall-normal velocity  $\hat v_n$ is obtained by projecting the velocity vector in the wall-normal direction in our calculations.
Comparing the perturbation profiles of the wall-normal velocity $\hat v_n$ and spanwise velocity $\hat w$ at $x_{s0}^*$ in panels ($g,h$),  it is evident that the OGT roll structure surpasses the SF-HLNS roll structure by four orders of magnitude. This roll structure proves more effective in seeding energy amplification due to its minimal initial energy, a fundamental principle of the OGT. However, in practical scenarios, the  excited perturbation in the boundary layer is forced by various types of freestream perturbations, and the localized optimal roll structure is unlikely manifest, as confirmed by the SF-HLNS calculations. 
This discrepancy stands as the primary reason for the substantial over-prediction of the energy amplification by  OGT observed in figure \ref{fig:line_GE_OGT_SFHLNS_slow}. Consequently, the  conclusion drawn from the OGT predictions in figure \ref{fig:optimaldis_cont}, suggesting that the perturbation amplification is greater for a smaller bluntness, is questionable. This highlights a notable limitation of the OGT in capturing the growth of non-modal perturbations.

\section{Concluding remarks and discussion}
\label{sec:conclusion}
The motivation of this paper  arises  from a current and prevalent topic concerning with the impact of nose bluntness on hypersonic boundary-layer transition. Experimental findings have shown that an increase in bluntness can delay transition when the bluntness is small, due to its stabilising effect on the Mack-mode instability; however, a reversed scenario is observed when the bluntness exceeds a certain threshold, causing the bypassing of Mack-mode instability amplification in the laminar phase. Attributing the non-modal perturbation as the dominant factor in triggering transition for large bluntness, the optimal growth theory becomes popular. While  this theory qualitatively agrees with DNS on predicting the perturbation profiles in the downstream region, it falls short in establishing a clear connection between optimal perturbations and freestream forcing. To address this gap, we develop a highly efficient  numerical approach, the SF-HLNS approach, designed to quantitatively describe the excitation of boundary-layer non-modal perturbations by freestream perturbations.

The SF-HLNS approach extends the HLNS approach developed in \cite{Zhao2019} by incorporating  the shock-fitting method at the upper boundary of the computational domain, such that the interaction of the freestream perturbations and the detached bow shock  is taken into account. Such an extension  is a non-trivial task as the wavy movement of the bow shock introduces an additional unknown quantity in the linearised system. Consequently, alongside the R-H relation utilized as the upper boundary condition, a compatibility relation at the bow shock  is introduced to close this system.

By choosing hypersonic boundary layers over blunt wedges as the physical model, we investigate the excitation of boundary-layer perturbations by various freestream perturbations. Through linear stability analysis, it is observed that the Mack instability does not manifest in the computational domains under examination. Instead, the entropy-layer instability, with its peak located within the entropy layer, emerges. However, this instability is unlikely to trigger transition due to its rather low growth rate. 
In contrast, non-modal perturbations with higher amplification rates can be stimulated.  
We would like to highlight the following observations from the SF-HLNS calculations:

(1) In all cases, the excited temperature perturbation exhibits a significantly greater magnitude than the velocity perturbation, with their profiles showing notable differences from the entropy-layer instability.

(2) Overall, the receptivity of non-modal perturbations to freestream fast acoustic, slow acoustic and entropy perturbations demonstrates similar efficiency, which is notably more efficient than that to freestream vortical perturbations. 

(3) As the nose bluntness increases, the receptivity efficiency to freestream acoustic and entropy forcing becomes more pronounced, mirroring the transition reversal phenomenon observed in experiments involving configurations with relatively large bluntness. Conversely, the receptivity to freestream vortical perturbations weakens with increasing nose bluntness.

(4) At the downstream position of our concern, receptivity is optimised when the spanwise wavelength of the freestream perturbation is comparable with the local boundary-layer thickness, indicating a distinct characteristic length scale that differs from the nose radius.

(5) The receptivity efficiency decreases with increasing frequency, showing a representative non-modal perturbation feature. Additionally, for cases subject to freestream acoustic and entropy forcing, the receptivity efficiency exhibits less dependence on the vertical wavenumber, unless the declination angle (the ratio of vertical to spanwise wavenumbers) is very large.

(6) Thanks to the SF-HLNS approach, we are able to examine the credibility of the optimal growth theory, a commonly used method for describing non-modal perturbations in the literature. Through a comprehensive comparison, it becomes evident  that while OGT calculations can predict the streaky structure in the downstream region, its predictions of the initial profile   are not reliable, leading to a significant overestimation of the energy amplification within the optimal interval and an incorrect trend of the dependence of the amplification factor on the nose bluntness. To the best knowledge of the authors, this study is the first to examine the reliability of OGT through a comprehensive comparison with numerical calculations of the non-modal perturbation evolution considering the leading-edge receptivity in a hypersonic blunt-body boundary layer.

The SF-HLNS approach developed in this paper  considers the influences of the bow shock, the entropy layer, and their interaction with perturbations, offering advantage  over the traditional LST, PSE and even HLNS approaches. Furthermore, the computational time for each case using the SF-HLNS method in this paper is less than 5 minutes, which is three to four orders of magnitude lower than that required for DNS. Due to its accuracy and efficiency, this approach holds promise as a valuable tool for future investigations of hypersonic blunt-body boundary layer transition, considering the effects of environmental perturbations. While this paper focuses solely on the receptivity of non-modal perturbations, the approach is also applicable for describing the normal-mode receptivity, akin to the DNS studies \citep{balakumar2018transition,cerminara2017acoustic,wan2018response} and the case study in Appendix \ref{Appendix:C1}. Moreover, it can be extended to configurations considering the combined interaction between freestream perturbations and surface vibration, akin to the  study for sharp-leading-edge configurations in \cite{song2024influence}. Although the linearised system is not suitable for describing nonlinear phases, such as nonlinear saturation, secondary instability, and nonlinear breakdown, the SF-HLNS calculations can provide an initial perturbation for the nonlinear PSE approach in subsequent calculations, which will be the focus of our future work.

\appendix
\section{Coefficient matrices of the equation (\ref{eq:LNS_compact})}
\label{Appendix:matrix}

The nonzero elements of matrix $\pmb {\Gamma}$ are
\begin{equation}
    \begin{aligned}
    &\mathit{\Gamma}_{11}=1,\quad \mathit{\Gamma}_{22}=\bar\rho,\quad \mathit{\Gamma}_{33}=\bar\rho,\quad \mathit{\Gamma}_{44}=\bar\rho,\quad \mathit{\Gamma}_{55}=\bar\rho c_v, \quad \mathit{\Gamma}_{51}=\bar e_t,
    \\ &\mathit{\Gamma}_{21}=\bar u,\quad \mathit{\Gamma}_{31}=\bar v,\quad \mathit{\Gamma}_{41}=\bar w,\quad
    \mathit{\Gamma}_{52}=\bar\rho \bar u,\quad \mathit{\Gamma}_{53}=\bar\rho \bar v,\quad \mathit{\Gamma}_{54}=\bar\rho \bar w.
    \end{aligned}
\end{equation}

For convenience of demonstration, we denote the derivatives of the convection and viscous terms as
\begin{equation}
\pmb {\bar{\hat E}}^c_{ \pmb \varphi}= \frac{\partial \pmb {\bar{\hat E}}^c} {\partial \pmb {\bar{\varphi}}}
    ,\quad
    \pmb {\bar{\hat F}}^c_{ \pmb \varphi}= \frac{\partial \pmb {\bar{\hat F}}^c} {\partial \pmb {\bar{\varphi}}}
    ,\quad
    \pmb {\bar{\hat G}}^c_{ \pmb \varphi}= \frac{\partial \pmb {\bar{\hat G}}^c} {\partial \pmb {\bar{\varphi}}}
    ,    
\end{equation}
\begin{equation}
    \begin{aligned}
    \pmb {\bar{\hat E}}^v_{ \pmb \varphi}= \frac{\partial \pmb {\bar{\hat E}}^v} {\partial \pmb {\bar{\varphi}}}
    ,\quad
    \pmb {\bar{\hat E}}^v_{ \pmb \varphi_x}= \frac{\partial \pmb {\bar{\hat E}}^v} {\partial \pmb {\bar{\varphi}}_x}
    ,\quad
    \pmb {\bar{\hat E}}^v_{ \pmb \varphi_y}= \frac{\partial \pmb {\bar{\hat E}}^v} {\partial \pmb {\bar{\varphi}}_y}
    ,\quad
    \pmb {\bar{\hat E}}^v_{ \pmb \varphi_z}= \frac{\partial \pmb {\bar{\hat E}}^v} {\partial \pmb {\bar{\varphi}}_z},
    \\
    \pmb {\bar{\hat F}}^v_{ \pmb \varphi}= \frac{\partial \pmb {\bar{\hat F}}^v} {\partial \pmb {\bar{\varphi}}}
    ,\quad
    \pmb {\bar{\hat F}}^v_{ \pmb \varphi_x}= \frac{\partial \pmb {\bar{\hat F}}^v} {\partial \pmb {\bar{\varphi}}_x}
    ,\quad
    \pmb {\bar{\hat F}}^v_{ \pmb \varphi_y}= \frac{\partial \pmb {\bar{\hat F}}^v} {\partial \pmb {\bar{\varphi}}_y}
    ,\quad
    \pmb {\bar{\hat F}}^v_{ \pmb \varphi_z}= \frac{\partial \pmb {\bar{\hat F}}^v} {\partial \pmb {\bar{\varphi}}_z},
    \\
    \pmb {\bar{\hat G}}^v_{ \pmb \varphi}= \frac{\partial \pmb {\bar{\hat G}}^v} {\partial \pmb {\bar{\varphi}}}
    ,\quad
    \pmb {\bar{\hat G}}^v_{ \pmb \varphi_x}= \frac{\partial \pmb {\bar{\hat G}}^v} {\partial \pmb {\bar{\varphi}}_x}
    ,\quad
    \pmb {\bar{\hat G}}^v_{ \pmb \varphi_y}= \frac{\partial \pmb {\bar{\hat G}}^v} {\partial \pmb {\bar{\varphi}}_y}
    ,\quad
    \pmb {\bar{\hat G}}^v_{ \pmb \varphi_z}= \frac{\partial \pmb {\bar{\hat G}}^v} {\partial \pmb {\bar{\varphi}}_z}.
    \end{aligned}
\end{equation}
Then, upon using the Einstein summation with $i=1,2$ and $3$ respectively denoting $x,y$ and $z$, the matrices $\pmb A$, $\pmb B$, $\pmb C$, $\pmb D$, $\pmb V_{\xi\xi}$, $\pmb V_{\eta\eta}$, $\pmb V_{\zeta\zeta}$, $\pmb V_{\xi\eta}$, $\pmb V_{\xi\zeta}$ and $\pmb V_{\eta\zeta}$ are expressed as
\refstepcounter{equation}
$$
    \pmb A=\pmb {\bar{\hat E}}^c_{\pmb \varphi}-[\pmb {\bar{\hat E}}^v_{ \pmb \varphi}
    +\frac{\partial (\xi_i \pmb {\bar{\hat E}}^v_{ \pmb \varphi_i})}{\partial \xi}
    +\frac{\partial (\xi_i \pmb {\bar{\hat F}}^v_{ \pmb \varphi_i})}{\partial \eta}
    +\frac{\partial (\xi_i \pmb {\bar{\hat G}}^v_{ \pmb \varphi_i})}{\partial \zeta}]
    ,\eqno{(\theequation{\mathit{a}})}
$$
$$
    \pmb B=\pmb {\bar{\hat F}}^c_{\pmb \varphi}-[\pmb {\bar{\hat F}}^v_{ \pmb \varphi}
    +\frac{\partial (\eta_i \pmb {\bar{\hat E}}^v_{ \pmb \varphi_i})}{\partial \xi}
    +\frac{\partial (\eta_i \pmb {\bar{\hat F}}^v_{ \pmb \varphi_i})}{\partial \eta}
    +\frac{\partial (\eta_i \pmb {\bar{\hat G}}^v_{ \pmb \varphi_i})}{\partial \zeta}]
    ,\eqno{(\theequation{\mathit{b}})}
$$
$$
    \pmb C=\pmb {\bar{\hat G}}^c_{\pmb \varphi}-[\pmb {\bar{\hat G}}^v_{ \pmb \varphi}
    +\frac{\partial (\zeta_i \pmb {\bar{\hat E}}^v_{ \pmb \varphi_i})}{\partial \xi}
    +\frac{\partial (\zeta_i \pmb {\bar{\hat F}}^v_{ \pmb \varphi_i})}{\partial \eta}
    +\frac{\partial (\zeta_i \pmb {\bar{\hat G}}^v_{ \pmb \varphi_i})}{\partial \zeta}]
    ,\eqno{(\theequation{\mathit{c}})}
$$
$$
    \pmb D=\frac{\partial \pmb {\bar{\hat E}}^c_{\pmb \varphi}}{\partial \xi}
    +\frac{\partial \pmb {\bar{\hat F}}^c_{\pmb \varphi}}{\partial \eta}
    +\frac{\partial \pmb {\bar{\hat G}}^c_{\pmb \varphi}}{\partial \zeta}
    -(\frac{\partial \pmb {\bar{\hat E}}^v_{ \pmb \varphi}}{\partial \xi}
    + \frac{\partial \pmb {\bar{\hat F}}^v_{ \pmb \varphi}}{\partial \eta}
    + \frac{\partial \pmb {\bar{\hat G}}^v_{ \pmb \varphi}}{\partial \zeta})
    ,\eqno{(\theequation{\mathit{d}})}
$$
$$
    \pmb V_{\xi\xi}=-\xi_i  \pmb {\bar{\hat E}}^v_{ \pmb \varphi_i},\quad
    \pmb V_{\eta\eta}=-\eta_i  \pmb {\bar{\hat F}}^v_{ \pmb \varphi_i},\quad
    \pmb V_{\zeta\zeta}=-\zeta_i  \pmb {\bar{\hat G}}^v_{ \pmb \varphi_i}
    ,\eqno{(\theequation{\mathit{e}})}
$$
$$
    \pmb V_{\xi\eta}=
    -\xi_i  \pmb {\bar{\hat F}}^v_{ \pmb \varphi_i} -
    \eta_i \pmb {\bar{\hat E}}^v_{ \pmb \varphi_i},\quad
    \pmb V_{\xi\zeta}=
    -\xi_i   \pmb {\bar{\hat G}}^v_{ \pmb \varphi_i} -
    \zeta_i \pmb {\bar{\hat E}}^v_{ \pmb \varphi_i},\quad
    \pmb V_{\eta\zeta}=
    -\eta_i  \pmb {\bar{\hat G}}^v_{ \pmb \varphi_i} -
    \zeta_i \pmb {\bar{\hat F}}^v_{ \pmb \varphi_i}
    .\eqno{(\theequation{\mathit{f}})}
$$

At the shock wave, the matrices $\pmb \Gamma^H$, $\pmb A^H$, $\pmb C^H$ and $\pmb D^H$ are expressed as
\begin{equation*}
    \begin{aligned}
        &\pmb \Gamma^H=(\bar y_{\xi}ge_x -\bar x_{\xi}ge_y)\frac{\partial \pmb {\bar U}}{\partial \eta},
        \\&
        \pmb A^H=ge_x\frac{\partial (\pmb {\bar F}-\pmb {\bar F}_v)}{\partial \eta}
                -ge_y\frac{\partial (\pmb {\bar E}-\pmb {\bar E}_v)}{\partial \eta},
        \\&
        \pmb C^H=(\bar y_{\xi}ge_x -\bar x_{\xi}ge_y)\frac{\partial (\pmb {\bar G}-\pmb {\bar G}_v)}{\partial \eta},
        \\&
        \pmb D^H=(ge_x)_{\xi}\frac{\partial (\pmb {\bar F}-\pmb {\bar F}_v)}{\partial \eta}
                -(ge_y)_{\xi}\frac{\partial (\pmb {\bar E}-\pmb {\bar E}_v)}{\partial \eta}
                +g_{\eta}e_y\frac{\partial (\pmb {\bar E}-\pmb {\bar E}_v)}{\partial \xi}
                -g_{\eta}e_x\frac{\partial (\pmb {\bar F}-\pmb {\bar F}_v)}{\partial \xi}.
    \end{aligned}
\end{equation*}

\section{Coefficient matrices in (\ref{eq:n_perturbation_compact}) and (\ref{eq:HLNS_RH_Linear})}
\label{Appendix:RH}
The coefficient matrices in (\ref{eq:n_perturbation_compact}) are expressed as
\begin{equation}\label{eq:ns_per2}
    \begin{split}
    &\pmb A^{n}=\frac{g(\bar y_{\xi}e_x-\bar x_{\xi}e_y)}{\bar\sigma^3}(\bar x_{\xi},\bar y_{\xi},0)^{\mbox T},\quad
    \pmb C^{n} = (0,0,\frac{g(\bar y_{\xi}e_x-\bar x_{\xi}e_y)}{\bar\sigma})^{\mbox T},
    \\
    &\pmb D^{n}=\frac{\bar y_{\xi}(ge_x)_{\xi}-\bar x_{\xi}(ge_y)_{\xi}}{\bar\sigma^3}(\bar x_{\xi},\bar y_{\xi},0)^{\mbox T},\quad
    \Gamma^{v_n}=\frac{g(\bar x_{\xi}e_y-\bar y_{\xi}e_x)}{\bar\sigma},
    \end{split}
\end{equation}
where  $\bar\sigma=\sqrt{\bar x_{\xi}^2+\bar y_{\xi}^2}$. 

In (\ref{eq:HLNS_RH_Linear}), the coefficient matrices $\pmb D_0^{RH}(\bar {\bm \varphi}_0)$ and
$\pmb D_s^{RH}(\bar {\bm \varphi}_s)$ can be uniformly expressed as $\pmb D^{RH}(\bar {\bm \varphi})$, which can be written as
\begin{equation}
    \begin{aligned}
        \begin{bmatrix}
            \bar u_n & \bar \rho\bar n_{sx} & \bar \rho\bar n_{sy} & \bar \rho\bar n_{sz} & 0 \\
            \bar u_n \bar u +p_e\bar T \bar n_{sx} & \bar\rho\bar u \bar n_{sx}+\bar u_n \bar\rho & \bar\rho\bar u \bar n_{sy} & \bar\rho\bar u \bar n_{sz} & p_e\bar\rho \bar n_{sx} \\
            \bar u_n \bar v +p_e\bar T \bar n_{sy} & \bar\rho\bar v \bar n_{sx} & \bar\rho\bar v \bar n_{sy}+\bar u_n \bar\rho & \bar\rho\bar v \bar n_{sz} & p_e\bar\rho \bar n_{sy} \\
            \bar u_n \bar w +p_e\bar T \bar n_{sz} & \bar\rho\bar w \bar n_{sx} & \bar\rho\bar w \bar n_{sy} & \bar\rho\bar w \bar n_{sz}+\bar u_n \bar\rho & p_e\bar\rho \bar n_{sz} \\
            \bar u_n\bar h_t & \bar\rho\bar h_t\bar n_{sx}+\bar u_n\bar\rho\bar u & \bar\rho\bar h_t\bar n_{sy}+\bar u_n\bar\rho\bar v & \bar\rho\bar h_t\bar n_{sz}+\bar u_n\bar\rho\bar w & \bar u_n c_p\bar\rho
        \end{bmatrix}
    \end{aligned},
\end{equation}
with $p_e=1/(\gamma M^2)$.

\section{Code validation}
\label{Appendix:code_validation}
\subsection{Comparison with \cite{zhong1998high}}
\label{Appendix:C1}
To verify our numerical code, we first repeat the DNS case in \cite{zhong1998high} and do comparison. The Mach number and the temperature of the oncoming stream are set to be 15 and 193.0K, respectively, while the temperature at the wall is 1000K. The shape of the blunt body is described by $x_w=4 y_w^2-1$, with the Reynolds number being 6026.6. Figure \ref{fig:pressure_zhong} illustrates the comparison of the streamwise distribution of mean pressure and the wall-normal profiles between our SF-DNS calculations and the results in  \cite{zhong1998high}. Excellent agreement is obtained, confirming the reliability of our base-flow calculation.

\begin{figure}
	\begin{center}
		\includegraphics[width = 0.48\textwidth] {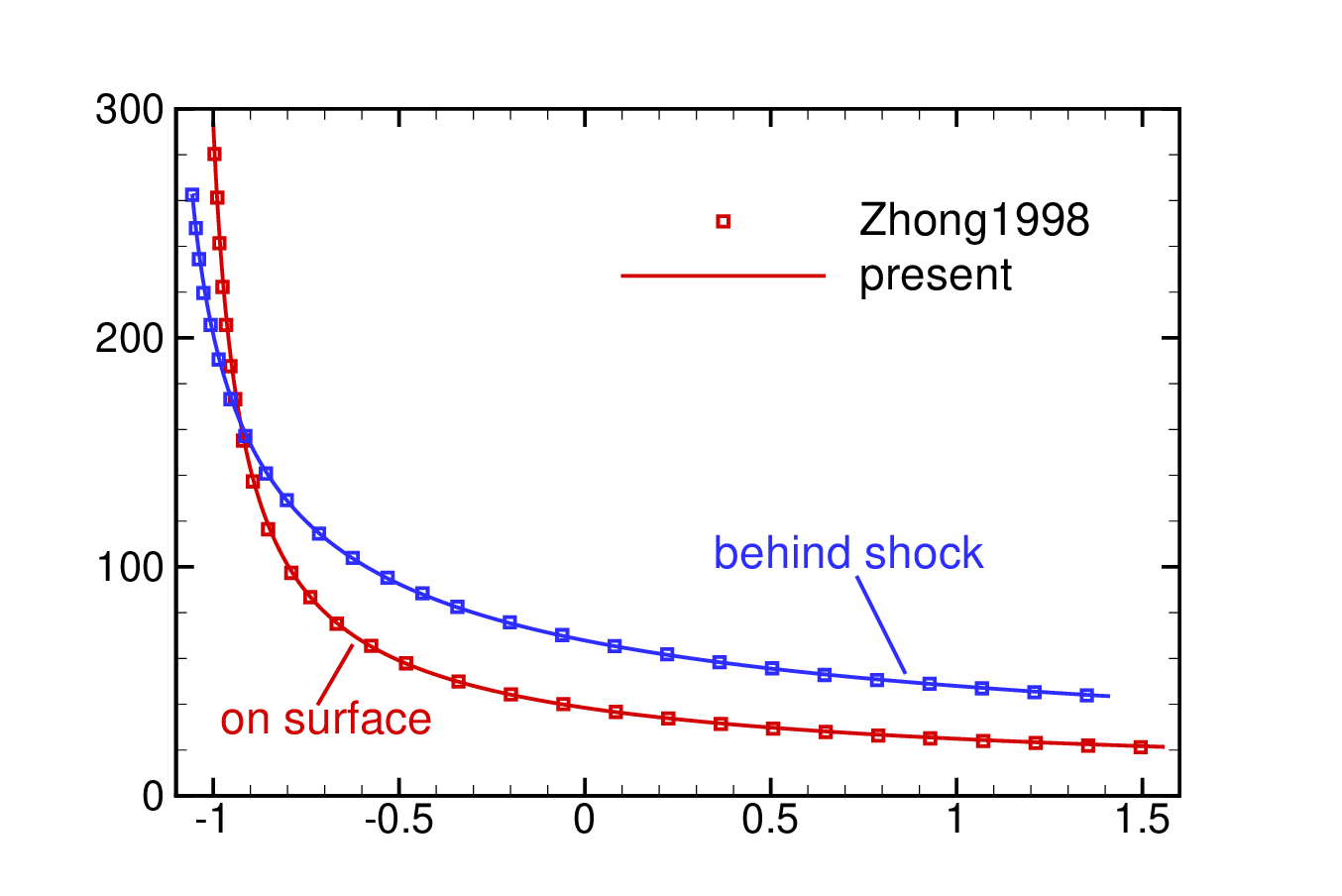}
		\put(-185,100){$(a)$}
		\put(-180,60){$\bar p$}
		\put(-95,0){$x$}
		\includegraphics[width = 0.48\textwidth] {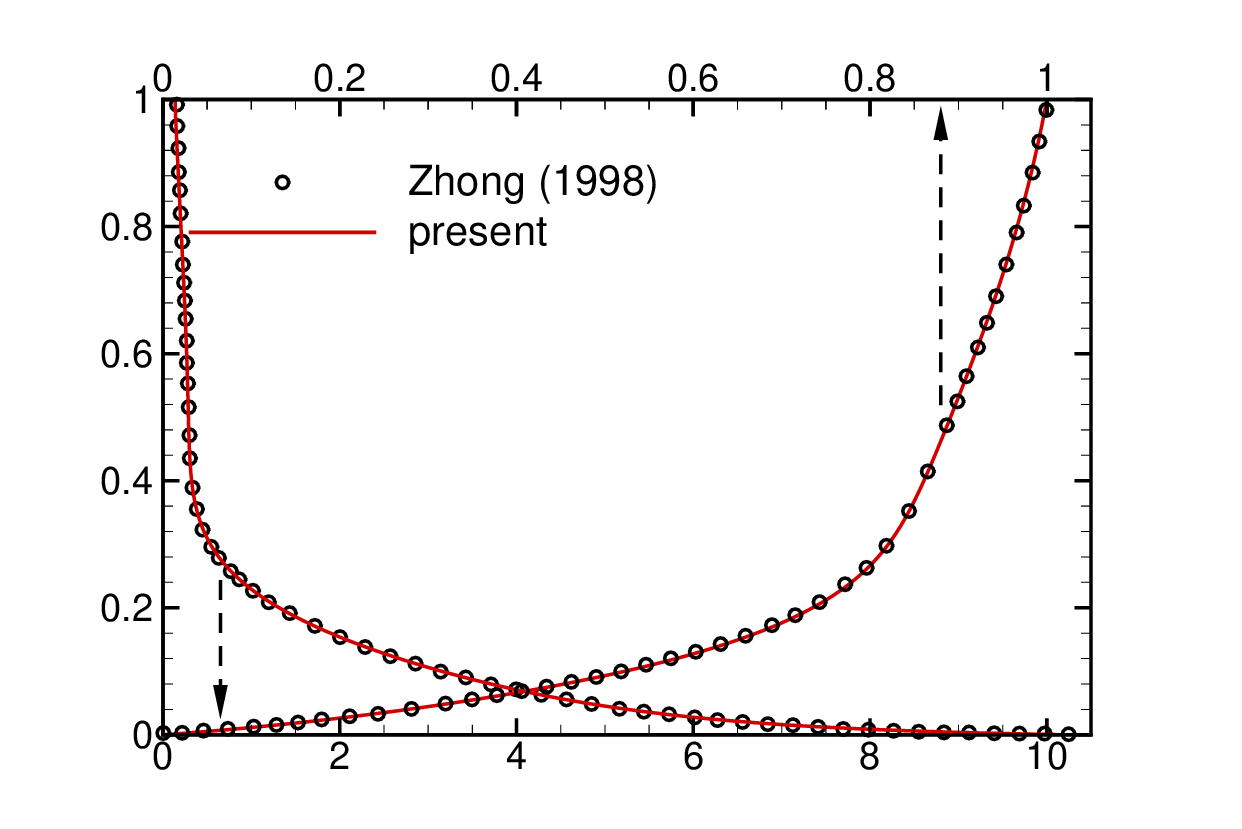}
		\put(-185,100){$(b)$}
		\put(-180,60){$\tilde y_n$}
		\put(-100,117){$\tilde u_s$}
		\put(-105,0){$\partial \tilde u_s/\partial \tilde y_n$}
		\caption{Comparison of the base-flow calculations with \cite{zhong1998high}.  ($a$) Streamwise distribution of the mean pressure at the shock  and wall; ($b$) wall-normal profiles of the tangential velocity $\tilde u_s$ and its first derivative   at  $x_w=-1.4$. In ($b$), $\tilde y_n$ denotes the wall-normal coordinate $y_n$ rescaled by the distance from the wall to the shock, and $\tilde u_s$ denotes the tangential velocity  rescaled by its value immediately behind the shock.}
		\label{fig:pressure_zhong}
	\end{center}
\end{figure}
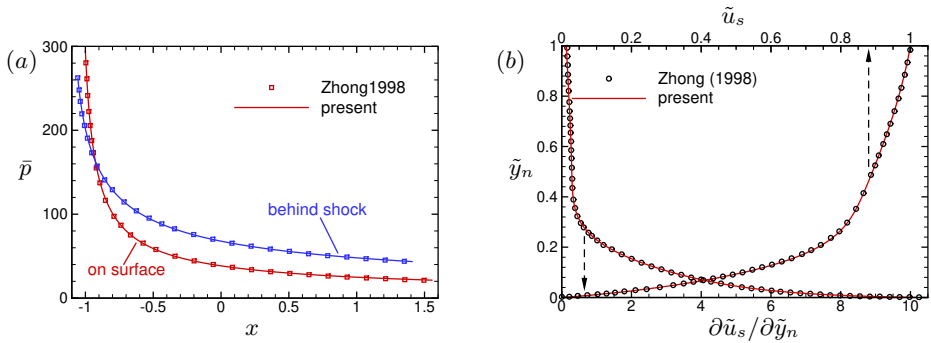

Following \cite{zhong1998high}, we introduce a  fast acoustic wave with zero incident angle  in the free stream, which, according to (\ref{eq:gust}) and (\ref{eq:acoustic_fun}), can be expressed as
\begin{equation}
	\begin{aligned}
		(\rho',u',v',w',T',p')=\frac{\epsilon}{2}(M_{\infty}^2,M_{\infty},0,0,(\gamma-1)M_{\infty}^2,1)\re^{\ri (kx-\omega t)}+c.c..
	\end{aligned}
\end{equation}
In the calculations, we choose  $k=15$,  $\omega=16$ and  $\epsilon=\frac 13 \times 10^{-4}$. To achieve a  comprehensive  comparison, we conduct the SF-HLNS calculations along with two types of DNSs: the SF-DNS introduced in $\S$\ref{sec:SFDNS} and the shock-capturing DNS (SC-DNS) with the same code as in  \cite{peicheng2022study} employed. Figure \ref{fig:entropy_temperature_zhong}($a$) presents a comparison of the streamwise evolution of the entropy perturbation   $(\hat s=\frac{\hat T}{\bar T}-\frac{\gamma-1}{\gamma}\frac{\hat p}{\bar p})$ at the wall among these calculations and the data extracted from \cite{zhong1998high}, demonstrating excellent agreement. In panel ($b$), we further show the comparison of the contours of the temperature perturbation  obtained from both SF-HLNS and our DNS calculations, which also exhibit good agreement.
\subsection{Comparison with our SF-DNS for case Ce}
\begin{figure}
	\begin{center}
		\includegraphics[width = 0.48\textwidth] {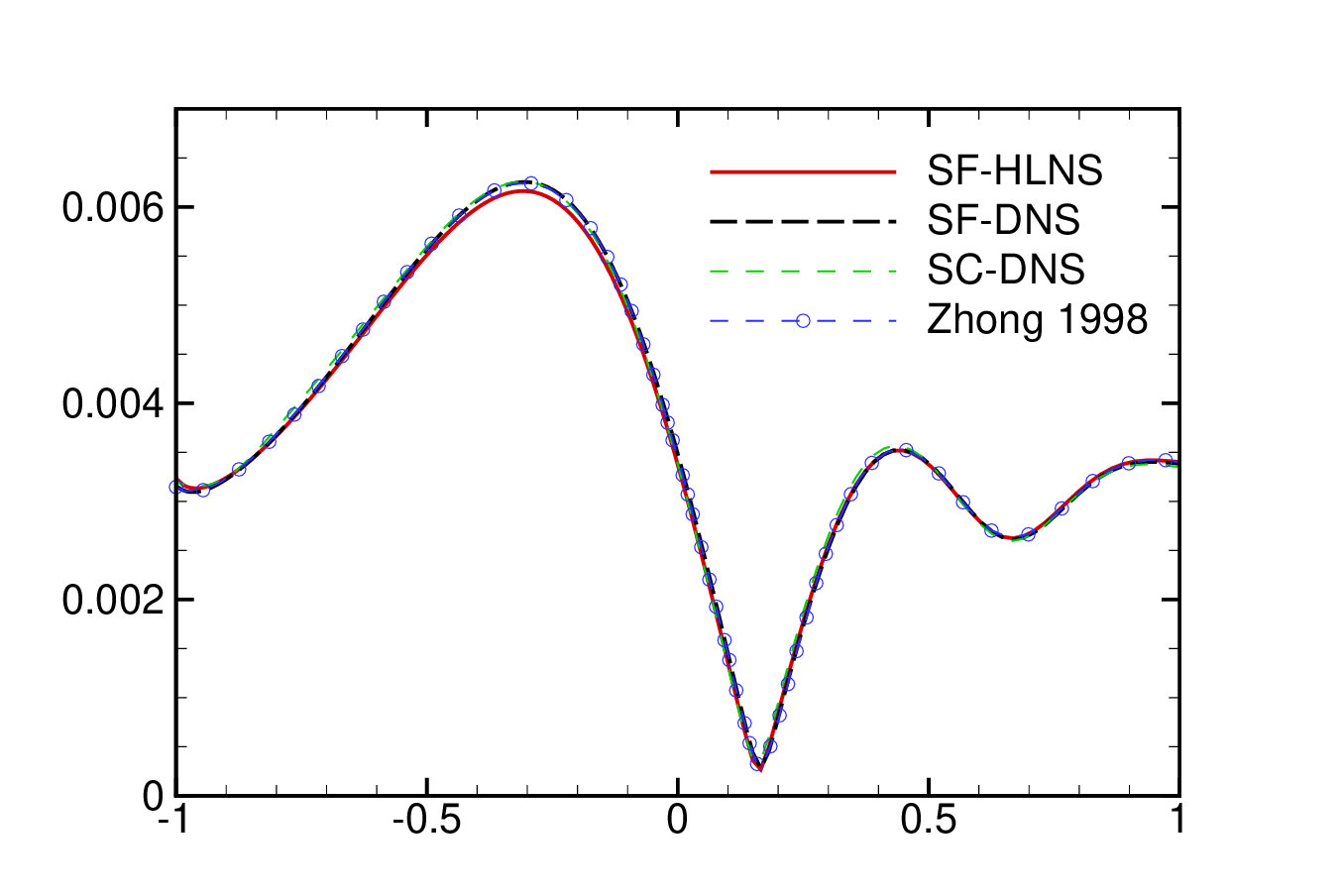}
		\put(-185,100){$(a)$}
		\put(-180,55){$|\hat s|$}
		\put(-95,0){$x$}
		\includegraphics[width = 0.48\textwidth] {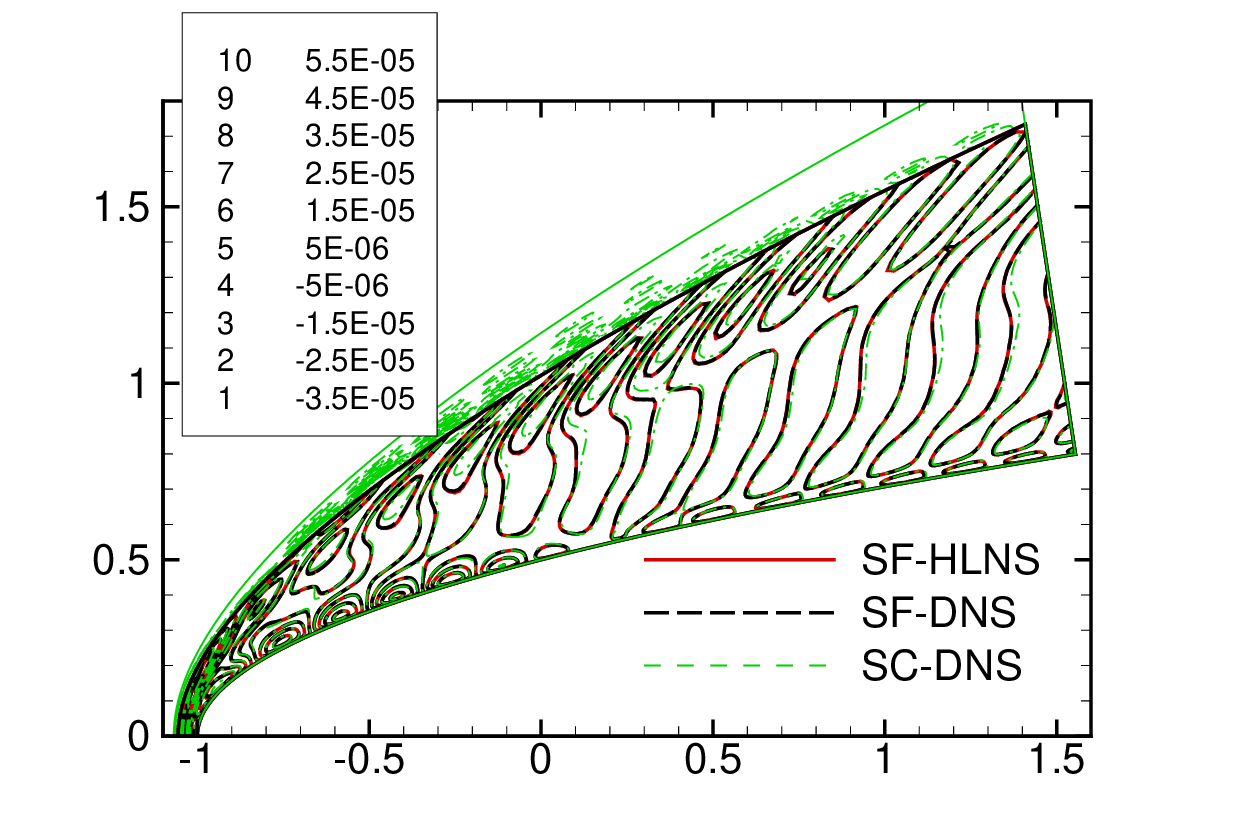}
		\put(-185,100){$(b)$}
		\put(-180,55){$y$}
		\put(-95,0){$x$}
		\caption{Comparison of the perturbation field obtained by different calculations. ($a$) Streamwise evolution of the entropy perturbation at the wall; ($b$) contours of the temperature perturbation.}
		\label{fig:entropy_temperature_zhong}
	\end{center}
\end{figure}
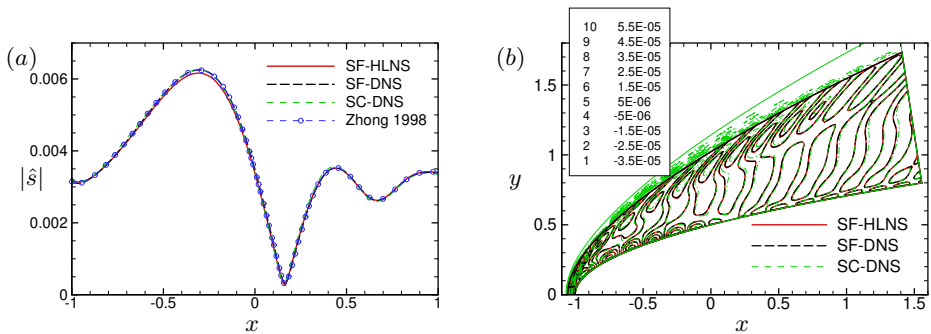
Now we verify our SF-HLNS code on computing the excitation  of non-modal mode by comparing with the SF-DNS result. For demonstration, case Ce with $\omega=0$, $k_3=15$ and $k_2=0$ is selected.   The  amplitude of the freestream entropy perturbation for SF-DNS is set to be $2\times 10^{-8}$. 
The contours  of $|\hat u|$ in the nose region and the wall-normal profiles of $|\hat u_s|$ and $|\hat T|$ at $x_s=3.1$ are compared in figure \ref{fig:validation_entrop_steak}, showing excellent agreement.

\begin{figure}
	\begin{center}
		\includegraphics[width = 0.48\textwidth] {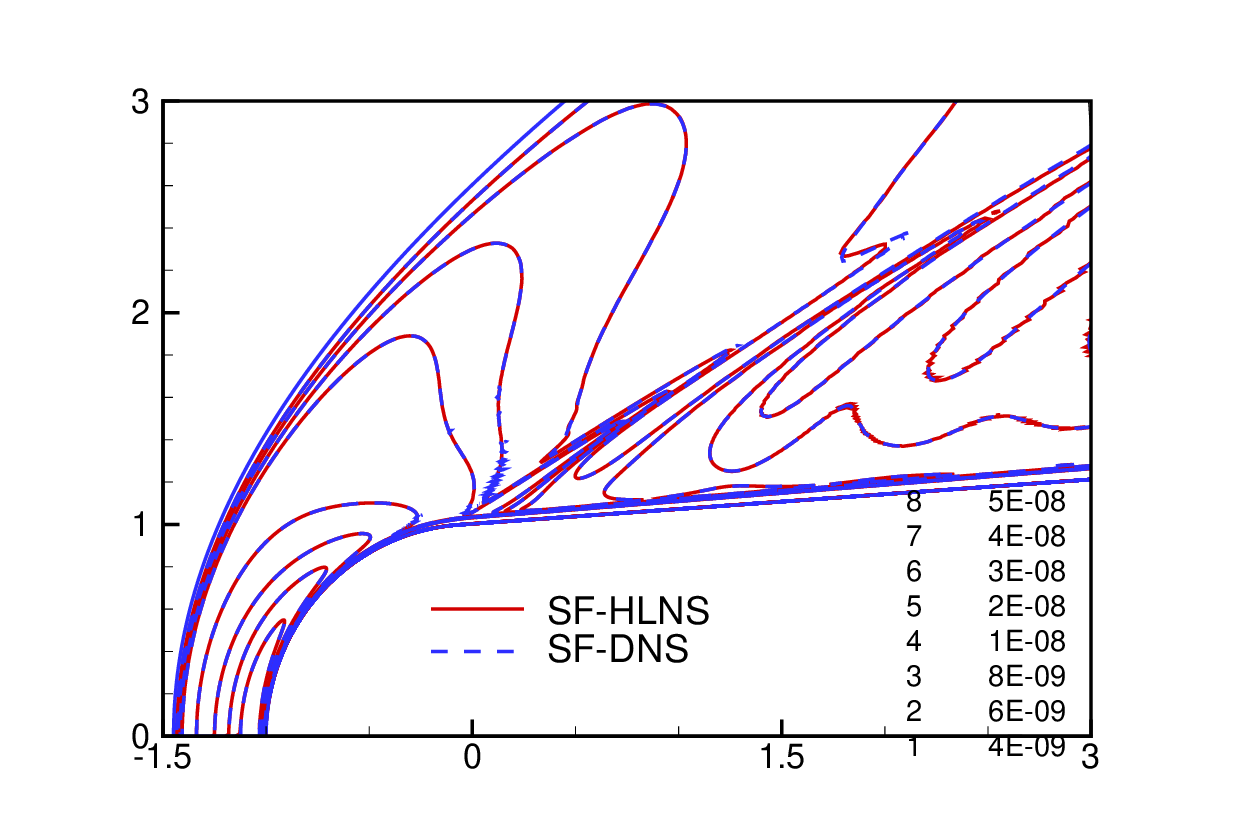}
		\put(-185,100){$(a)$}
		\put(-180,60){$y$}
		\put(-95,0){$x$}
		\includegraphics[width = 0.48\textwidth] {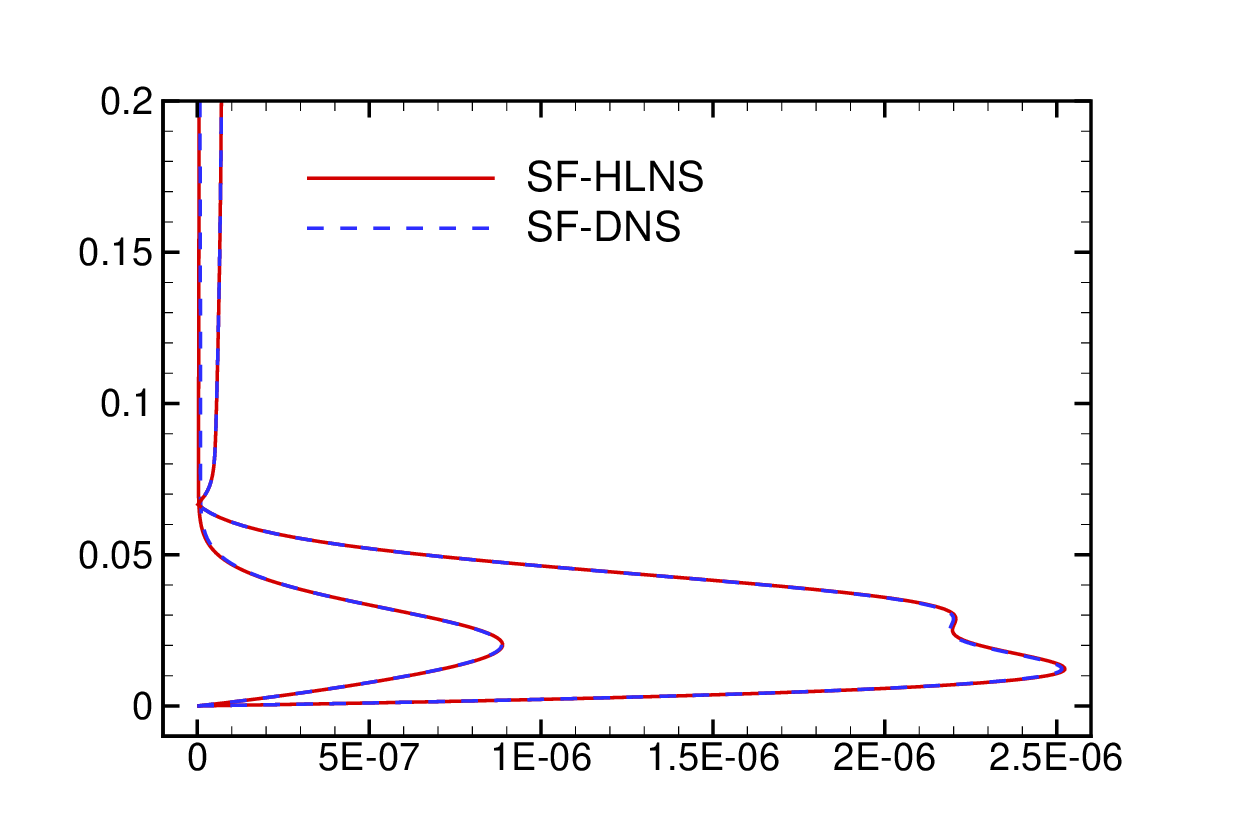}
		\put(-185,100){$(b)$}
		\put(-180,60){$y_n$}
		\put(-103,-2){$|\hat u_s|,|\hat T|$}
		\put(-109,25){$|\hat u_s|$}
		\put(-70,40){$|\hat T|$}
		\caption{Comparison between the SF-HLNS and SF-DNS results. ($a$): Contours of $|\hat u|$ in $(x-y)$ plane; ($b$) wall-normal profiles of $|\hat u_s|$ and $|\hat T|$ at $x_s=3.1$.}
		\label{fig:validation_entrop_steak}
	\end{center}
\end{figure}
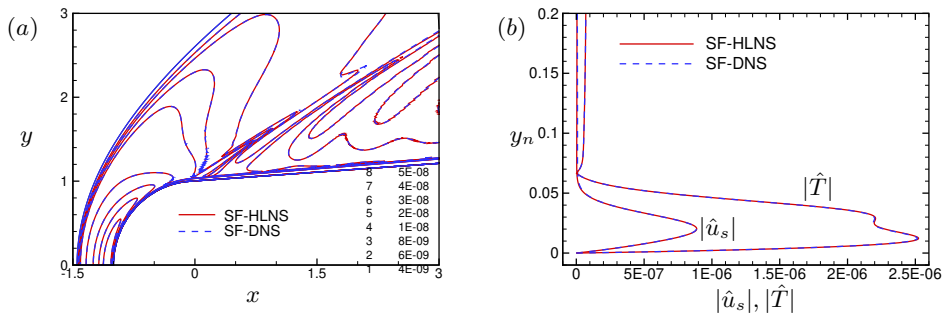

 \vspace{.4cm}
 \noindent\textbf{Acknowledgements}
 The work is supported by National Science Foundation of China (grant nos. 12372222, 92371104,  U20B2003, 12002235), the Strategic Priority Research Program, CAS (no. XDB0620102) and CAS project for Young Scientists in Basic Research (YSBR-087). The authors would also acknowledge PhD candidate Qinyang Song for kindly assisting in the generation of a few figures. 

  \vspace{.4cm}
  \noindent\textbf{Declaration of interests}

  The authors report no conflict of interest.

\appendix

\bibliographystyle{jfm}
\bibliography{shock_fitting_R2}

\end{document}